\documentclass[preprint,amsmath,amssymb,aps]{revtex4-1}

\usepackage{graphicx}
\usepackage{dcolumn}
\usepackage{bm}
\usepackage{color}
\definecolor{dgreen}{rgb}{0,0.4,0}

\def\Im{\mathop{\mathrm{Im}}}

\def\sub#1{_{\mathrm{#1}}}
\def\up#1{^{\mathrm{#1}}}
\def\Vec#1{\boldsymbol #1}
\def\3He{\mbox{$^3$He}}
\def\4He{\mbox{$^4$He}}
\def\87Rb{\mbox{$^{87}$Rb}}

\def\dps{\displaystyle}

\begin{document}



\title{
Quench dynamics of the three-dimensional U(1) complex field theory:
geometric and scaling characterisation of the vortex tangle}

\author{Michikazu Kobayashi$^1$ and Leticia F. Cugliandolo$^{2,3}$}

\affiliation{$^1$Department of Physics, Kyoto University, Oiwake-cho, Kitashirakawa, Sakyo-ku, Kyoto 606-8502, Japan}
\affiliation{$^2$Sorbonne Universit\'es, Universit\'e Pierre et Marie Curie - Paris 6,\\
Laboratoire de Physique Th\'eorique et Hautes Energies UMR 7589,  4 Place Jussieu, 75252 Paris Cedex 05, France}

\date{\today}

\begin{abstract}
We present a detailed study of the equilibrium properties and stochastic dynamic evolution 
of the  U(1)-invariant relativistic complex field theory in three dimensions.
This model has been used to describe, in various limits, 
properties of relativistic bosons at finite chemical potential, 
type II superconductors, magnetic materials and aspects of cosmology.
We characterise the thermodynamic second-order phase transition
in different ways. We study the equilibrium 
vortex configurations and their statistical and geometrical properties in equilibrium at all temperatures. 
We show that at very high temperature the statistics of the filaments is the one of fully-packed loop models.
We identify the temperature, within the ordered phase, at which the number density of vortex lengths
falls-off algebraically and we associate it to a geometric percolation transition that we characterise in various ways.  
We measure the fractal properties of the vortex tangle at this threshold.
Next, we perform  infinite rate quenches from equilibrium in the disordered phase, across the 
thermodynamic critical point, and deep into the ordered phase. We show that three time regimes can be distinguished: a 
first approach towards a state that, within numerical accuracy, shares many features with the one
at the percolation threshold, a later coarsening process that does not alter, at sufficiently low temperature, 
the fractal properties of the long vortex loops, and a final approach to equilibrium. 
These features are independent of the reconnection rule
used to build the vortex lines. In each of these regimes we identify the various length-scales of the vortices in the system.
We also study the scaling properties of the ordering process and the progressive annihilation of topological defects
and we prove that the time-dependence of the time-evolving vortex tangle can be described within the 
dynamic scaling framework.
\end{abstract}


\maketitle

\section{Introduction} \label{sec-intro}

Three dimensional field theories with continuous symmetry breaking are relevant to describe a host
of physical systems. These theories are used to model 
superfluid systems~\cite{Ahlers,Griffin,Minnhagen,Nemirovskii13},
superconductors of type II~\cite{Minnhagen,Vinokur}, 
nematic liquid crystals~\cite{deGennesProst}, 
magnetic samples~\cite{Magnets}, as well as
phase transitions in the early universe~\cite{HindmarshKibble}. 

Phase transitions with spontaneous symmetry breaking lead
to the formation of topological defects of different kind: domain walls, strings or vortices, 
monopoles, etc. depending on the type of symmetry that is broken.
The topological defects we will be interested in are line objects, be them vortices, 
disclinations or cosmic strings~\cite{Godfrin,Vilenkin}. These occur in, e.g., a
field theory with global U(1) symmetry in $d=3$ dimensions. A field configuration 
has a vortex centred at a given point in space if the field vanishes at this 
point and  the phase of the field changes by $2\pi n$, with $n$ a non-vanishing integer, 
along a contour around  this point. The field configuration deviates appreciably form the asymptotic value within 
the finite-width core of the vortex. Therefore, thin tubes of the vanishing field, {\it i.e.} the false vacuum, 
are enclosed within the core. This is most
clearly understood in the context of liquid crystals where the orientation of the 
molecules rotates by such an angle when following a closed path around a line 
disclination~\cite{deGennesProst}. Line-type topological effects are also of importance in the other branches of 
physics mentioned in the first paragraph. For example, topological defects were predicted to form in the Universe via the Kibble 
mechanism and strings were proposed to act as the source for density fluctuations at the origin of 
galaxy formation and other potentially observable effects~\cite{HindmarshKibble}. They also appear in quantum turbulence~\cite{Nemirovskii13,Tsubota-Kobayashi}, 
complex-valued
random wave fields~\cite{Taylor08,Taylor14} used to model wave chaos~\cite{Berry77} and random optical fields~\cite{Taylor08,Taylor14}. 

In this paper we study the statics and stochastic dynamics 
of a three-dimensional relativistic field theory with global U(1) symmetry.
This model serves to describe, in different limits, the physical systems mentioned in the 
previous paragraph as well as relativistic bosons at finite chemical 
potential~\cite{Blaizot,Arnold,Gardiner,Aarts}.
We mimic the coupling to an equilibrium bath by adding dissipation and noise terms in the 
equations of motion. 
We use four slightly different dynamic equations for the evolution of the fields that we call 
over-damped, under-damped or relativistic - the Goldstone model, ultra-relativistic, and non-relativistic - the time-dependent 
Gross Pitaevskii model. The resulting Langevin-like 
equations do not conserve the (complex) order parameter. Similar non-linear equations have been studied in the literature~\cite{Aronson,Bohr,Pismen,Tsubota-Kobayashi,Paoletti}. We show in an appendix that they lead to thermal equilibrium.
We solve them with numerical methods. 

The equilibrium  phase diagram and critical phenomenon of the $3d$ U(1) complex field theory are well documented 
in the literature. In particular, the static critical exponents have been estimated with Monte Carlo simulations 
combined with high temperature expansions~\cite{Vicari,Campostrini,Hasenbusch,Ballesteros},
and the $\epsilon$ expansion~\cite{Vicari,GuidaZJ}. Still, we revisit the equilibrium behaviour of the 
system with our numerical 
algorithm with a double purpose. On the one hand, we validate it by showing that it takes the 
system to thermal equilibrium and captures the expected equilibrium properties. 
On the other, an important part of our analysis will be devoted to the study of the 
vortex tangle in, but also out of, equilibrium.  As the topological stable strings must have no free ends and 
be closed in a space with periodic boundary conditions, we will be talking about vortex loops.
We use a cubic lattice discretisation of the field theory. The 
construction of the vortex network on a lattice involves some ambiguity. Indeed, when a branching point 
at which more than one vortex line enter and exit, some criterium has to be used to decide 
upon the way the reconnection is done. We use here two well-documented rules~\cite{Kajantie,Bittner}:

\noindent
--
The stochastic criterium (S) in which  the vortex line elements are reconnected at random.

\noindent
-- 
The maximal criterium (M) in which the vortex line elements are reconnected in such a way that one among the 
resulting vortex loops has the maximal possible length.

At each step of our analysis we compare the results obtained for the two rules. We pay special attention 
to the geometric transition between a phase in which all loops are finite, and another one in which 
some loops are infinitely extended. We also  characterise in detail the shape and statistics of the 
loops on both sides of the threshold and at the geometric transition that, we show, does not 
coincide with the thermodynamic instability of the U(1) complex field theory. We relate the statistical properties 
of the closed loops in their extended phase to the one of fully-packed loop models in which 
each link on a lattice is covered by part of one and only one loop~\cite{Nahum-book}. These configurations 
will be the initial states for the dynamics.

The relaxation dynamics across a second order phase transition proceeds by coarsening and 
annihilation of topological defects~\cite{Bray,Corberi}.
Analytic approximations to characterize the scaling properties of the relaxation dynamics of models with 
continuous symmetry, after an infinitely rapid quench into the ordered phase,
were developed in~\cite{BrayHumayun,LiuMazenko,Rutenberg,ToyokiHonda,Toyoki-analytic},  see~\cite{Bray} for a review.
The dynamic exponent $z\sub{d}=2$ was predicted analytically~\cite{BrayHumayun,LiuMazenko,ToyokiHonda,Toyoki-analytic,Rutenberg}, 
and a value close to this analytic prediction was 
measured numerically~\cite{Mondello,Toyoki,Abriet} and experimentally in bulk nematic 
liquid crystals~\cite{Wang,Yurke} from the analysis of space-time correlation functions
and dynamic structure  factors. As far as we know, there is no detailed study of the 
dynamics from the point of view of the topological defects themselves and we also develop it here.

Some details about the methodology that we use to investigate this problem are in order. 
In the analysis of the phase transition and static vortex statistics we ensure that the 
system reaches thermal equilibrium. 
In the analysis  of the evolution after a deep instantaneous quench below the ordering transition  
temperature we simply let the system evolve from a chosen initial state under subcritical 
conditions. The vortex string network already present in the initial state evolves after the quench
and we characterise its evolution in full detail. We identify various dynamic regimes and we 
explain what determines them in terms of the changing vortex configurations. 
We base this analysis on the work in~\cite{Arenzon07,Sicilia07,Sicilia08,Sicilia09,Barros09,Loureiro10,Olejarz12,Blanchard14,Arenzon14,Tartaglia15,Takeuchi15,Tartaglia16}
where the stochastic ordering dynamics of $2d$ spin models were analysed from a geometric perspective.

The paper is organised as follows. In Sec.~\ref{sec:model} we introduce the model.
In Sec.~\ref{sec:equilibrium} we describe the equilibrium properties and phase transition in the model; 
it can be read independently from the rest of the paper.  
In Sec~\ref{sec:equilibrium-vortex}  we discuss the properties of the vortex network
in equilibrium.
Section~\ref{sec:dynamics} is devoted to the analysis of the fast quench dynamics. 
Finally, in Sec.~\ref{sec:conclusions} we present our conclusions and some lines for 
future research. A short account of some of our results appeared in~\cite{KoCu16}.

\section{The model}
\label{sec:model}

The Lagrangian density  for  relativistic bosons with finite chemical potential reads~\cite{KobayashiNitta}, in terms of a scalar complex field $\psi$
and its time derivative $\dot \psi = \partial_t \psi$,
\begin{align}
\begin{split}
&\mathcal{L}[\dot{\psi}(\Vec{x},t),\dot{\psi}^\ast(\Vec{x},t), \psi(\Vec{x},t),\psi^\ast(\Vec{x},t)] \\
& \qquad 
= \frac{1}{c^2} |\dot{\psi}(\Vec{x},t)|^2 
+ i \mu \{ \psi^\ast(\Vec{x},t) \dot{\psi}(\Vec{x},t) - \dot{\psi}^\ast(\Vec{x},t) \psi(\Vec{x},t) \} - |\nabla \psi(\Vec{x},t)|^2 \\
&\qquad \quad + g \rho |\psi(\Vec{x},t)|^2 - \frac{g}{2} |\psi(\Vec{x},t)|^4, \label{eq:original-Lagrangian}
\end{split}
\end{align}
where $c$ is the speed of light and $g$ and  $\rho$ are real parameters, $g, \rho \in \mathbb{R}$,
in the potential energy density with Mexican hat form and a degenerate circle of minima at $|\psi|^2= \rho$.
The parameters $\mu \in \mathbb{R}$ and $\rho$ receive different interpretations in different communities. 
The term proportional to $\mu$ breaks the particle-antiparticle symmetry and in the high-energy literature it is associated to a chemical potential, 
while $\rho$ fixes the vacuum expectation value of the U(1) symmetry breaking.
In the condensed matter literature instead the chemical potential is associated to $\rho$ that fixes the particle 
density $|\psi|^2$ in the system and $\mu$ is related to the mass of the particles.

This theory is global U(1)-invariant as the Lagrangian density remains unchanged 
under the global transformation, 
$\psi'(\Vec{x},t) = e^{i\delta} \psi(\Vec{x},t)$, and accordingly for $\psi^*$, with $\delta$ a space and time independent real parameter. Moreover, $\mathcal{L}^* = \mathcal{L}$.

The Lagrangian density \eqref{eq:original-Lagrangian} is also invariant under the Lorentz boost
\begin{align}
\begin{split}
& t^\prime = \gamma_c \bigg(t - \frac{\Vec{v} \cdot \Vec{x}}{c^2}\bigg), \quad\qquad
\Vec{x}^\prime = 
\Vec{x} - \Vec{v} \left[ 
\frac{(1-\gamma_c)}{v^2} (\Vec{v} \cdot \Vec{x}) + \gamma_c t
\right] 
,
\end{split}
\end{align}
with Lorentz factor $\gamma_c = (1-v^2/c^2)^{-1/2}$,
if the field transforms as
\begin{align}
\begin{split}
& \psi^\prime(\Vec{x}', t') = \exp\bigg[ - i \mu c^2 \bigg\{ (1 - \gamma_c) t + \frac{\gamma_c \Vec{v} \cdot \Vec{x}}{c^2} \bigg\} \bigg] 
\psi(\Vec{x},t). 
\label{eq:Lorentz-transformation}
\end{split}
\end{align}
Indeed, under these transformations,
$\mathcal{L}[\partial_t{\psi}(\Vec{x},t),\partial_t{\psi}^\ast(\Vec{x},t), \psi(\Vec{x},t), \psi^\ast(\Vec{x},t)] = $
\linebreak
$\mathcal{L}[\partial_{t^\prime} \psi'(\Vec{x}^\prime,t^\prime), 
\partial_{t^\prime} {\psi'}^\ast(\Vec{x}^\prime,t^\prime),\psi^\prime(\Vec{x}^\prime,t^\prime),\psi^{\prime\ast}(\Vec{x}^\prime,t^\prime)]$. The inverse transformations are
$t= \gamma_c (t' + \Vec{v} \cdot \Vec{x}'/c^2)$ 
and 
$\Vec{x} = \Vec{x}' + \Vec{v} \left[ \gamma_c  t' - (1-\gamma_c) (\Vec{v} \cdot \Vec{r}')/v^2\right]$
and these imply 
\begin{align}
\begin{split}
\partial_{t^\prime} = \gamma_c (\partial_t + \Vec{v} \cdot \nabla), \quad\qquad
\nabla^\prime = 
\nabla - \Vec{v} \left[ \frac{(1-\gamma_c)}{v^2}  (\Vec{v} \cdot \nabla) 
+
\frac{\gamma_c}{c^2}  \partial_t \right]
.
\end{split}
\end{align}

The equation of motion for $\psi$ follows  from the Euler-Lagrangian equation and reads
\begin{align}
- \bigg( \frac{1}{c^2} \frac{\partial^2}{\partial t^2} - \nabla^2 \bigg) \psi + 2 i \mu \dot{\psi} \equiv - \Box \psi + 2 i \mu \dot{\psi} = g (|\psi|^2 - \rho) \psi. \label{eq:original-Euler}
\end{align}
We consider two opposite limits: $\mu \to 0$ and $c \to \infty$.
In the former case, the complex field $\psi$ does not change under the Lorentz transformation \eqref{eq:Lorentz-transformation}, 
and we call it the ``ultra-relativistic" limit. The latter case is ``non-relativistic".
Under these two limits, Eq.~\eqref{eq:original-Euler} becomes
\begin{subequations}
\begin{align}
- \Box \psi &= g (|\psi|^2 - \rho) \psi,  \qquad\qquad\qquad\qquad \mu\to 0 , \label{eq:Goldstone} \\
 2 i \mu \dot{\psi} &= - \nabla^2 \psi + g (|\psi|^2 - \rho) \psi, \qquad \qquad c \to \infty , \label{eq:Gross-Pitaevskii}
\end{align}
\end{subequations}
which are known as the Goldstone and the Gross-Pitaevskii models respectively.
The latter describes  the dynamics of  gaseous Bose-Einstein condensates~\cite{Griffin}.
The former, on the other hand, describes the dynamics of condensates in optical lattices which 
are close to the critical point to the Mott insulator phase with integer fillings~\cite{Altman}.

Under the time evolution in Eq. \eqref{eq:original-Euler}, the energy functional
\begin{align}
E(\dot{\psi}, \dot{\psi}^\ast, \psi,\psi^\ast) = \int d\Vec{x}\: \bigg\{ \frac{|\dot{\psi}|^2}{c^2} + |\nabla \psi|^2 - g \rho |\psi|^2 + \frac{g}{2} |\psi|^4 \bigg\}
 \label{eq:energy-functional}
\end{align}
is conserved. 

The static solutions to \eqref{eq:original-Euler} that minimize the energy are $\psi =\sqrt{\rho} \ e^{i \chi}$ with $\chi$ a
constant. The choice of $\chi$ breaks the U(1) symmetry. Vortex static solutions are also supported by this equation~\cite{Pismen,Tsubota-Kobayashi}.
One such $z$-directed string is given by the axisymmetric field configuration
 $\psi(\Vec x) = f(r) e^{i n \theta}$ with $f(r)$ a smooth function of $r$, the radial direction 
 on the plane perpendicular to the tube. It takes the extreme values 
 $f(0)=0$ and $f(r\to\infty)=\sqrt{\rho}$ and varies over a typical length scale $\simeq (g\rho)^{-1/2}$ 
 that determines the core of the vortex.
 $n$ is the winding number.

Next, we consider the statistical properties of the system described by the U(1) complex field at finite temperatures.
In canonical equilibrium the statistical average of  a real finite functional of the fields $f = f(\psi,\psi^\ast, \dot{\psi}, \dot{\psi}^\ast)$
reads
\begin{align}
\begin{split}
 \langle f \rangle\sub{eq} & \equiv \int D\dot{\psi}\: D\dot{\psi}^\ast\: D\psi\: D\psi^\ast\: f \ \frac{e^{- E/T}}{Z}, 
 \\
Z & = \int D\dot{\psi}\: D\dot{\psi}^\ast\: D\psi\: D\psi^\ast\: e^{- E/T}, \label{eq:statistical-average-original}
\end{split}
\end{align}
where
$T$ is temperature. We set the Boltzmann constant to $k\sub{B} = 1$ in this paper.
When $f$ depends only on $\psi$ and $\psi^\ast$, i.e., 
$f(\psi,\psi^\ast, \dot{\psi}, \dot{\psi}^\ast) \to f(\psi,\psi^\ast)$, 
Eq. \eqref{eq:statistical-average-original} can be simplified to
\begin{align}
\begin{split}
\langle f \rangle\sub{eq} & = \int D\psi\: D\psi^\ast\: f \ \frac{e^{- E_0/T}}{Z_0}, \\
Z_0 & = \int D\psi\: D\psi^\ast\: e^{- E_0/T}, 
\end{split}
\end{align}
with 
\begin{equation}
E_0(\psi,\psi^\ast) = \int d\Vec{x}\: \bigg\{ |\nabla \psi|^2 - g \rho |\psi|^2 + \frac{g}{2} |\psi|^4 \bigg\} . 
\label{eq:statistical-average}
\end{equation}
These are the kind of functionals that we consider in this paper, unless we specify a different dependence.

At $T = 0$ the ground state is $|\psi|^2 = \rho$ and the U(1) symmetry of the energy functional $E_0$ for the phase shift of $\psi$ is 
spontaneously broken.
We therefore expect the temperature at which the spontaneous breaking of this symmetry occurs to be the one 
for  Bose-Einstein condensation.

A simple and efficient method to sample the thermal averages defined above is to use the over-damped Langevin equation
\begin{align}
\begin{split}
& \gamma\sub{L} \dot{\psi} = - \frac{\delta E_0}{\delta \psi^\ast} 
+ \sqrt{\gamma\sub{L} T} (\xi_1 + i \xi_2) = \nabla^2 \psi - g(|\psi|^2 - \rho) \psi + \sqrt{\gamma\sub{L} T} (\xi_1 + i \xi_2), \\
& \langle \xi_a(\Vec{x},t) \rangle =0, \qquad
 \langle \xi_a(\Vec{x},t) \xi_b(\Vec{x}^\prime,t^\prime) \rangle = \delta(t-t^\prime) \delta(\Vec{x} - \Vec{x}^\prime) \delta_{a,b},
\end{split} \label{eq:over-damped-Langevin}
\end{align}
that ensures
\begin{align}
\frac{1}{t} \int_0^t dt^\prime \: f[\psi(\Vec{x},t^\prime), \psi^\ast(\Vec{x},t^\prime)] 
\stackrel{t \to \infty}{\longrightarrow} \langle f \rangle\sub{eq}.
\end{align}

An alternative dynamic approach uses, instead of the energy functional in Eq. \eqref{eq:energy-functional},  the 
Hamiltonian associated to the Lagrangian density \eqref{eq:original-Lagrangian}
\begin{align}
\begin{split}
& H(\psi,\psi^\ast,\phi,\phi^\ast) = \int d\Vec{x}\: \bigg\{ c^2 ( \phi^\ast - i \mu \psi^\ast ) ( \phi + i \mu \psi ) + |\nabla \psi|^2 - g \rho |\psi|^2 + \frac{g}{2} |\psi|^4 \bigg\} \\
& \phantom{H(\psi,\psi^\ast,\phi,\phi^\ast)} = E_0(\psi,\psi^\ast) + c^2 \int d\Vec{x}\: ( \phi^\ast - i \mu \psi^\ast ) ( \phi + i \mu \psi )
\end{split}
\end{align}
with the generalized momentum $\phi$
\begin{align}
\begin{split}
& \phi = \frac{\partial \mathcal{L}}{\partial \dot{\psi}^\ast} = \frac{\dot{\psi}}{c^2} - i \mu \psi, \label{eq:Hamiltonian}
\end{split}
\end{align}
and its complex conjugate $\phi^*$.
The under-damped Langevin equation
\begin{align}
\begin{split}
\dot{\phi} = - \frac{\delta H}{\delta \psi^\ast} - \gamma\sub{L} c^2 (\phi + i \mu \psi) + \sqrt{\gamma\sub{L} T} (\xi_1 + i \xi_2) 
\label{eq:under-damped-Langevin-general}
\end{split}
\end{align}
becomes
\begin{align}
\begin{split}
- \Box \psi + (2 i \mu - \gamma\sub{L}) \dot{\psi}
= g(|\psi|^2 - \rho) \psi - \sqrt{\gamma\sub{L} T} (\xi_1 + i \xi_2). \label{eq:under-damped-Langevin}
\end{split}
\end{align}
See Refs.~\cite{Kasamatsu03} for an alternative derivation of the dissipative term proportional to $-\gamma\sub{L} \dot \psi$.
In the ultra-relativistic and non-relativistic limits, the Langevin equation \eqref{eq:under-damped-Langevin} approaches
\begin{subequations}
\begin{align}
- \Box \psi &= g(|\psi|^2 - \rho) \psi + \gamma\sub{L} \dot{\psi} - \sqrt{\gamma\sub{L} T} (\xi_1 + i \xi_2), \label{eq:under-damped-ultra-relativistic} \\
(2 i \mu - \gamma\sub{L}) \dot{\psi} & = - \nabla^2 \psi + g(|\psi|^2 - \rho) \psi - \sqrt{\gamma\sub{L} T} (\xi_1 + i \xi_2). \label{eq:under-damped-non-relativistic}
\end{align} \label{eq:under-damped-limits}
\end{subequations}
Equation \eqref{eq:under-damped-ultra-relativistic} is the conventional under-damped Langevin equation corresponding 
to the over-damped Langevin equation \eqref{eq:over-damped-Langevin}, whereas Eq. \eqref{eq:under-damped-non-relativistic} is known as 
the stochastic Gross-Pitaevskii equation describing Bose-Einstein condensates at finite temperatures~\cite{Gardiner}. The 
Hamiltonian \eqref{eq:Hamiltonian} and the Langevin equation \eqref{eq:under-damped-Langevin} give the 
same ensemble averages for equilibrium states as those obtained using 
Eqs.~\eqref{eq:energy-functional} and \eqref{eq:statistical-average}, 
\begin{align}
\begin{split}
 \langle f \rangle\sub{eq} & = \int D\psi\: D\psi^\ast\: D\phi\: D\phi^\ast\: f \ \frac{e^{- H/T}}{Z_H} = \int D\psi\: D\psi^\ast\: f \ \frac{e^{- E_0/T}}{Z_0} , \\
Z_H & = \int D\psi\: D\psi^\ast\: D\phi\: D\phi^\ast\: e^{- H/T}, 
\label{eq:statistical-average-under-damped}
\end{split}
\end{align}
see App.~\ref{app:FP}.

Although the above discussion applies in all dimensions, we concentrate in three-dimensional 
systems in this paper. We close this section by making explicit the numerical procedure that we used to solve 
eqs. \eqref{eq:over-damped-Langevin}, \eqref{eq:under-damped-Langevin}, \eqref{eq:under-damped-ultra-relativistic}, 
and \eqref{eq:under-damped-non-relativistic} numerically.
We first collect time and space coordinates into a four component vector that we call 
$\mbox{\bf x} = (x_0, x_1, x_2, x_3)$ with $x_0=t$ and $\Vec{x} = (x_1,x_2,x_3)$ the space-like components.
We discretize  $\mbox{\bf x}$ 
using a different mesh in the space and time directions:
$x_1 = j \Delta x$ with $j=0, \dots ,  L-1$,
$x_2= k \Delta x$ with $k=0, \dots ,  L-1$, 
$x_3 = l \Delta x$ with $l=0, \dots ,  L-1$, 
and $x_0 = i \Delta t$ with $i=0, \dots $
We define the complex field $\psi$ on the discretised space-time as 
$\psi_{\mbox{\bf x}} \equiv \psi(t,\Vec{x})$ with $\mbox{\bf x}  \equiv (i\Delta t ,j \Delta x , k\Delta x ,l \Delta x)$.
We call $ \hat e_0, \hat e_1, \hat e_2, \hat e_3$ the orthonormal basis of unit vectors on the vector $\mbox{\bf x}$.
The spatial gradient then becomes
\begin{align}
|\nabla \psi|^2 &\to \sum_{a=1}^3 \frac{|\psi_{\mbox{\bf x}} - \psi_{\mbox{\bf x}-\Delta x\hat e_a}|^2}{(\Delta x)^2}, \\
\nabla^2 \psi &\to \sum_{a=1}^3 \frac{\psi_{\mbox{\bf x}+\Delta x \hat e_a} + \psi_{\mbox{\bf x}-\Delta x \hat e_a} - 
2 \psi_{\mbox{\bf x}}}{(\Delta x)^2},
\end{align}
and the spatial integral $\int d\Vec{x}\: \to (\Delta x)^3 \sum_{j,k,l}$,
with $\Delta x$ the lattice spacing.
We use periodic cubes with linear sizes $L = 40$, $60$, $80$, and $100$.
For the time evolution, we use the lowest-ordered stochastic 
Runge-Kutta method for Eq. \eqref{eq:over-damped-Langevin},
\begin{align}
\begin{split}
& \psi_{(1)\mbox{\bf x}} = \psi_{\mbox{\bf x}} - \frac{\partial E_0[\psi_{\mbox{\bf x}},\psi^\ast_{\mbox{\bf x}}]}{\partial \psi^\ast_{\mbox{\bf x}}} \frac{\Delta t}{\gamma\sub{L}} + \sqrt{\frac{T \Delta t}{\gamma\sub{L}}} (\Delta W_{\mbox{\bf x},1} + i \Delta W_{\mbox{\bf x},2}), \\
& \psi_{(2)\mbox{\bf x}} = \psi_{\mbox{\bf x}} - \frac{\partial E_0[\psi_{\mbox{\bf x}},\psi^\ast_{\mbox{\bf x}}]}{\partial \psi^\ast_{\mbox{\bf x}}} \frac{\Delta t}{2 \gamma\sub{L}} + \sqrt{\frac{T \Delta t}{\gamma\sub{L}}} (\Delta W_{\mbox{\bf x},1} + i \Delta W_{\mbox{\bf x},2}), \\
& \psi_{\mbox{\bf x}+\Delta t\hat{e}_0} = \psi_{(2)\mbox{\bf x}} - \frac{\partial E_0[\psi_{(1)\mbox{\bf x}},\psi_{(1)\mbox{\bf x}}^\ast]}{\partial \psi_{(1)\mbox{\bf x}}^\ast} \frac{\Delta t}{2 \gamma\sub{L}},
\end{split} \label{eq:over-damped-Langevin-discrete}
\end{align}
and for Eq. \eqref{eq:under-damped-non-relativistic},
\begin{align}
\begin{split}
& \psi_{(1)\mbox{\bf x}} = \psi_{\mbox{\bf x}} + \frac{\partial E_0[\psi_{\mbox{\bf x}},\psi^\ast_{\mbox{\bf x}}]}{\partial \psi^\ast_{\mbox{\bf x}}} \frac{\Delta t}{2 i \mu - \gamma\sub{L}} + \frac{\dps \sqrt{\gamma\sub{L} T \Delta t}}{2 i \mu - \gamma\sub{L}} (\Delta W_{\mbox{\bf x},1} + i \Delta W_{\mbox{\bf x},2}), \\
& \psi_{(2)\mbox{\bf x}} = \psi_{\mbox{\bf x}} + \frac{\partial E_0[\psi_{\mbox{\bf x}},\psi^\ast_{\mbox{\bf x}}]}{\partial \psi^\ast_{\mbox{\bf x}}} \frac{\Delta t}{2 (2 i \mu - \gamma\sub{L})} + \frac{\dps \sqrt{\gamma\sub{L} T \Delta t}}{2 i \mu - \gamma\sub{L}} (\Delta W_{\mbox{\bf x},1} + i \Delta W_{\mbox{\bf x},2}), \\
& \psi_{\mbox{\bf x}+\Delta t\hat{e}_0} = \psi_{(2)\mbox{\bf x}} + \frac{\partial E_0[\psi_{(1)\mbox{\bf x}},\psi_{(1)\mbox{\bf x}}^\ast]}{\partial \psi_{(1)\mbox{\bf x}}^\ast} \frac{\Delta t}{2 (2 i \mu - \gamma\sub{L})},
\end{split} \label{eq:under-damped-non-relativistic-discrete}
\end{align}
and the lowest-ordered symplectic method for Eq. \eqref{eq:under-damped-ultra-relativistic}
\begin{align}
\begin{split}
& \psi_{\mbox{\bf x}+\Delta t\hat{e}_0} = \psi_{\mbox{\bf x}} + c^2 \phi_{\mbox{\bf x}} \Delta t, \\
& \phi_{\mbox{\bf x}+\Delta t\hat{e}_0} = \phi_{\mbox{\bf x}} - \bigg( \frac{\partial E_0[\psi_{\mbox{\bf x}},\psi^\ast_{\mbox{\bf x}}]}{\partial \psi^\ast_{\mbox{\bf x}}} + c^2 \gamma\sub{L} \phi_{\mbox{\bf x}} \bigg) \Delta t + \sqrt{\gamma\sub{L} T \Delta t} (\Delta W_{\mbox{\bf x},1} + i \Delta W_{\mbox{\bf x},2}),
\end{split} \label{eq:under-damped-ultra-relativistic-discrete}
\end{align}
and for Eq. \eqref{eq:under-damped-Langevin},
\begin{align}
\begin{split}
& \psi_{\mbox{\bf x}+\Delta t\hat{e}_0} = \psi_{\mbox{\bf x}} + c^2 \phi_{\mbox{\bf x}} \Delta t, \\
& \phi_{\mbox{\bf x}+\Delta t\hat{e}_0} = \phi_{\mbox{\bf x}} - \bigg\{ \frac{\partial E_0[\psi_{\mbox{\bf x}},\psi^\ast_{\mbox{\bf x}}]}{\partial \psi^\ast_{\mbox{\bf x}}} + c^2 (\gamma\sub{L} - 2 i \mu) \phi_{\mbox{\bf x}} \bigg\} \Delta t \\
& \phantom{\phi_{\mbox{\bf x}+\Delta t\hat{e}_0} =}
+ \sqrt{\gamma\sub{L} T \Delta t} (\Delta W_{\mbox{\bf x},1} + i \Delta W_{\mbox{\bf x},2}),
\end{split} \label{eq:under-damped-Langevin-discrete}
\end{align}
with $\langle \Delta W_{\mbox{\bf x},a} \rangle = 0$ and $\langle \Delta W_{\mbox{\bf x},a} \Delta W_{\mbox{\bf x}^\prime,b} \rangle = \delta_{\mbox{\bf x},\mbox{\bf x}^\prime} \delta_{a,b}$.
As regards the numerical parameters, we use $c = 1$, $\rho = 1$, $g = 1$, 
$\mu = 0.5$, $\gamma\sub{L} = 1$, $\Delta x = 1$ in all cases, and $\Delta t = 0.005$ for Eq. \eqref{eq:over-damped-Langevin-discrete} and 
$\Delta t = 0.01$ for Eqs. \eqref{eq:under-damped-non-relativistic-discrete}, \eqref{eq:under-damped-ultra-relativistic-discrete}, and \eqref{eq:under-damped-Langevin-discrete}.
The number of spatial grid points is $N = L^3$.
With these parameters, space and time are measured in units of the equilibrium correlation length at $T=0$ in the mean-field approximation,
$x_0 \equiv 1 / \sqrt{g \rho}$,  and the corresponding correlation time $t_0 \equiv \gamma\sub{L} / x_0^2$.
The order parameter is further measured in units of $\sqrt{\rho}$ with $\rho$ the zero-temperature  density.
Then the velocity $c$, parameter $\mu$, temperature $T$, and the Langevin noise $\xi_a$ are measured in 
units of $x_0 / t_0$, $t_0 / x_0^2$, $x_0^{-2}$, and $\sqrt{\rho / t_0}$, respectively.

\section{Equilibrium properties}
\label{sec:equilibrium}

In this section, we focus on the equilibrium properties attained by the dynamical model. As we will 
show, this model undergoes a second order phase transition at a critical point below which the 
U(1) continuous symmetry is spontaneously broken.  We confirm numerically that the 
universality class it the same as the one of the usual U(1) complex field 
theory~\cite{Sieberer13,Sieberer14,Tauber14}. This critical phenomenon has been studied with 
equilibrium methods in the past and the best estimates for the critical exponents are~\cite{Campostrini}
\begin{eqnarray}
\begin{array}{llll}
&
\alpha= -0.0151(3), 
\qquad\qquad
&
\nu=0.6717(1), 
\qquad\qquad
&
 \eta=0.0381(2), 
\\
&
\gamma=1.3178(2), 
\qquad\qquad
&
\beta=0.3486(1), 
\qquad\qquad
&
\delta=4.780(1).
\end{array}
\label{eq:crit-exp-Campostrini}
\end{eqnarray}
These have been obtained with finite-size scaling of Monte Carlo data for system sizes up to $L=128$ 
and high-temperature expansions.
The  $\epsilon$-expansion RG method yields~\cite{GuidaZJ}
\begin{eqnarray}
\begin{array}{llll}
&
\alpha= -0.011(4), 
\qquad\qquad
&
\nu=0.6703(15), 
\qquad\qquad
&
 \eta=0.0354(25), 
\\
&
\gamma=1.3169(20), 
\qquad\qquad
&
\beta=0.3470(16), 
\qquad\qquad
&
\delta=4.795(14).
\end{array}
\end{eqnarray}
These values are compatible with the ones given above within numerical error
and also with other numerical evaluations~\cite{Ballesteros}.
Since, the spontaneous breaking of the U(1) symmetry 
is isomorphic to the one in the $3d$ $XY$-spin model, these critical exponents 
are expected to be the same as the ones of this spin model.
Experimental measurements in the superfluid transition of $^4$He yield $\alpha=-0.0127(3)$~\cite{Lipa}. Other experimental 
results for this and other critical exponents can be found in~\cite{Vicari}.
The dynamic critical exponent $z_{\rm eq}$ was estimated to be $z_{\rm eq} \simeq 2.1$ 
for periodic boundary conditions with numerical methods in~\cite{Jensen,Roma}, see 
also~\cite{Tauber14} for the RG prediction, and we will further discuss it below.

Our aim in this section is twofold.
On the one hand, we test whether the  three dynamic formulations, 
given by the Langevin equations \eqref{eq:over-damped-Langevin}, and \eqref{eq:under-damped-Langevin} in its 
two limits \eqref{eq:under-damped-limits}, capture the equilibrium and critical phenomena correctly. 
In order to avoid long relaxation times and reach equilibrium more 
quickly, we start all simulations from the completely ordered initial state $\psi(t = 0) = \sqrt{\rho}$.
We estimate the relaxation time to be $\tau \lesssim 900$ for the parameters explored, and we verify 
that the system reaches its asymptotic regime for $t > 1000$, see Fig.~\ref{fig:tau}. 
On the other hand, we characterize the equilibrium configurations at off-critical and critical 
temperatures. We confirm that the percolation of the longest vortex loops in 
equilibrium occurs at a temperature 
that is within the ordered phase and different from the critical one~\cite{Kajantie,Bittner,KoCu16}. We 
pay special attention to  the geometric and statistical properties of the vortex 
lines at the critical percolation point, 
as these are going to be of relevance to our analysis of the relaxation dynamics.
We also distinguish the statistical and geometrical properties of the string networks obtained with the
two reconnection conventions.

In this section, we perform a double average of data: we take an ensemble  
average by taking the mean over $100$ noise realisations for the same initial state $\psi(t = 0) = \sqrt{\rho}$, 
and we average over $3000$ different times at $t = (1000 + i) \Delta t$ with $1 \leq i \leq 3000$  for each dynamical run.
The total number of data contributing to each data point shown is, therefore, $100 \times 3000$.
We write this ensemble average as $\langle \cdots \rangle\sub{stat}$ and we use this as the 
equilibrium ensemble average $\langle \cdots \rangle\sub{eq}$. The lattice linear sizes used are comparable to the 
ones utilized in Monte Carlo studies of the phase transition and critical phenomena~\cite{Campostrini}.

\vspace{0.25cm}

\begin{figure}[tbh]
\centering
\begin{minipage}{0.49\linewidth}
\centering
(a)\\
\includegraphics[width=0.95\linewidth]{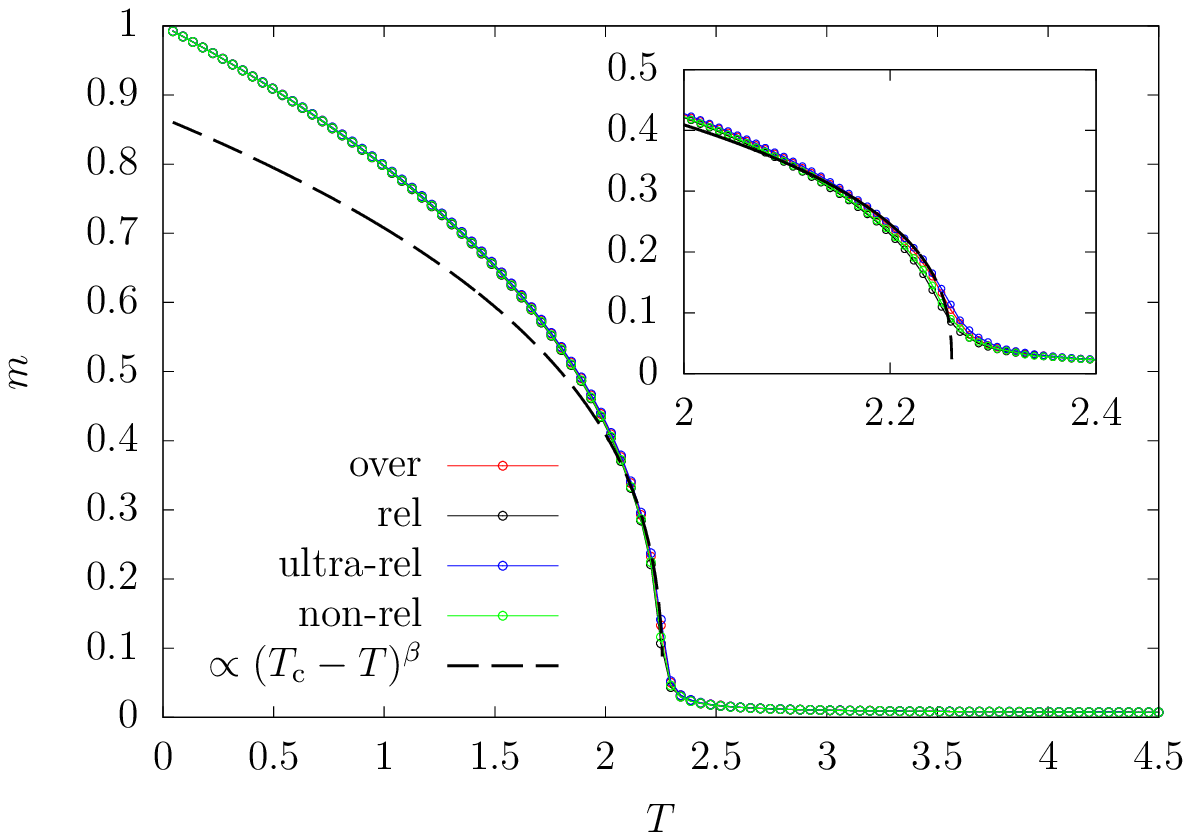}
\end{minipage}
\begin{minipage}{0.49\linewidth}
\centering
(b)\\
\includegraphics[width=0.95\linewidth]{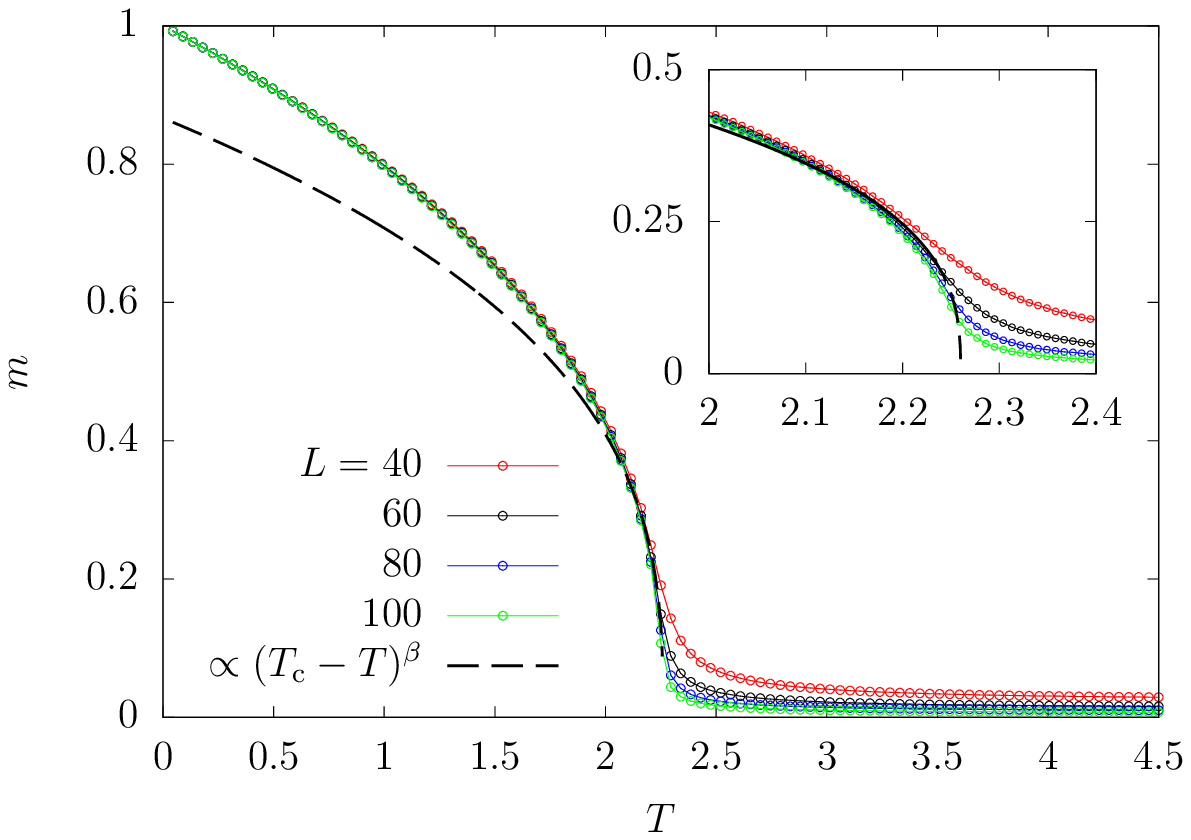}
\end{minipage}
\caption{\label{fig:m-all} (Color online.)
Temperature dependence of the order parameter $m$. (a) Comparison between the estimate for $m(T)$ obtained with different stochastic dynamics: over-damped Langevin equation \eqref{eq:over-damped-Langevin} (over), under-damped Langevin equation \eqref{eq:under-damped-Langevin} (rel), ultra-relativistic limit of the under-damped Langevin equation \eqref{eq:under-damped-ultra-relativistic} (untra-rel), and non-relativistic limit of the under-damped Langevin equation \eqref{eq:under-damped-non-relativistic} (non-rel).
Systems with $L = 100$ linear size are used in all cases. (b) Finite size dependence of $m(T)$ as 
obtained from the under-damped Langevin equation \eqref{eq:under-damped-Langevin}.
In both panels, we plot with a dotted line the critical decay of the order parameter, $m(T) \propto (T\sub{c} - T)^\beta$, at temperatures $T < T\sub{c}$ with $T\sub{c} = 2.26$ 
and $\beta = 0.347$.
For details on these values see Figs.~\ref{fig:U1-U2-rel} and \ref{fig:m-chi-c1-xi-scale}~(a),
and the corresponding discussion. In the inserts: zoom over the critical region.
}
\end{figure}

\subsection{The order parameter}

There are three well-known definitions of the order parameter $m$ for the spontaneous breaking of the U(1) symmetry.
In the usual one is statistical physics, $m$ is calculated 
as $m \equiv \lim_{h\to 0} \langle \psi_h \rangle\sub{eq}$ by applying a perturbation $h$
that couples linearly to the field. In the case of the model with Hamiltonian $H$ this is achieved as
\begin{align}
H = E_0(\psi_h, \psi^\ast_h) + c^2 \int d\Vec{x}\: \{ ( \phi^\ast_h - i \mu \psi^\ast_h ) ( \phi_h + i \mu \psi_h ) - h (\psi_h + \psi^\ast_h) \},
\end{align}
where the sub-index $_h$ recalls that the complex fields $\psi_h$ and $\phi_h$ are computed 
under the external field $h$.
The resulting Langevin equation becomes
\begin{align}
\begin{split}
- \Box \psi_h + (2 i \mu - \gamma\sub{L}) \dot{\psi}_h
= g(|\psi_h|^2 - \rho) \psi_h - h - \sqrt{\gamma\sub{L} T} (\xi_1 + i \xi_2).
\end{split}
\end{align}

In the context of Bose-Einstein condensation and superfluidity, however, the external field $h$ does not find an easy implementation and, instead, the following definition is considered: 
\begin{align}
m \propto \sqrt{|C(r \to \infty)|}, \quad
C(r) = \Bigg\langle \frac{1}{4 \pi r^2} \int d\Omega_{\Vec{r} = r}\: \int d\Vec{x}\: \psi^\ast(\Vec{x}) \psi(\Vec{x} + \Vec{r}) \Bigg\rangle\sub{eq},
\end{align}
where the integral $1/(4 \pi r^2) \int d\Omega_{\Vec{r} = r}$  is the average over the solid angle at $\Vec{r} = r$.
Numerically, $m$ is computed as $m \propto \sqrt{|C(r = L/2)|}$,
 and in order to improve the numerical accuracy of the measurement of the integral over the spherical 
 surface, $1/(4 \pi r^2) \int d\Omega_{\Vec{r} = r}$,  we average over all lattice sites falling within the shell 
 $[|\Vec{r}|, |\Vec{r}|+1)$. 
This definition corresponds to the off-diagonal long-range order $\lim_{\Vec{r} \to \infty} \langle \psi^\dagger(\Vec{x}) \psi(\Vec{x}+\Vec{r}) \rangle$ for the presence of the Bose-Einstein condensate in a quantum-boson system.

In many numerical works, the definition
\begin{align}
m \equiv \big\langle \big| \overline{\psi} \big| \big\rangle\sub{eq}, \quad\qquad
\overline{\psi} = \frac{1}{L^3} \sum_{j,k,l} \psi_{\mbox{\bf x}}.
\label{eq:m-def}
\end{align}
is instead used due to its numerical simplicity.

All definitions of  $m$  should give the same temperature dependence 
for sufficiently large system size $L$. We adopt the definition~(\ref{eq:m-def}) in this paper.

Figure \ref{fig:m-all}  shows the temperature dependence of the order parameter, $m$, 
for the spontaneous breaking of the U(1) symmetry
for different Langevin equations [Fig.~\ref{fig:m-all} (a)] and for different system sizes $L$ [Fig.~\ref{fig:m-all} (b)].
Different Langevin equations give the same ensemble averages, within numerical accuracy.
From  this figure,  we roughly calculate  that the spontaneous symmetry breaking occurs at 
$
T\sub{c} \simeq 2.3
$
in the $L \to \infty$ limit. The dotted blue lines in both panels represent the 
critical behaviour, $m(T) \simeq |T-T_c|^\beta$ with $T_c=2.26$ [see Eq.~(\ref{eq:critical-temp})] and $\beta = 0.347$.
They are very close to the numerical results. In the inset we zoom over the critical region.

We end with a note on the difference between $m$ and the particle number observable, $L^{-3} \sum_{jkl} |\psi_{\mbox{\bf x}}|$, 
that is different from zero at all $T$ (see the inset in Fig.~1 in~\cite{KoCu16}).

\subsection{The critical temperature}

The critical temperature can be better evaluated from the Binder ratio, $U_1$,
and the ratio of correlation functions, $U_2$, 
\begin{align}
& U_1 \equiv \frac{\dps \big\langle \big| \overline{\psi} \big|^4 \big\rangle\sub{eq}}{\dps \big\langle \big| \overline{\psi} \big|^2 \big\rangle\sub{eq}^2},
\qquad\qquad\qquad
U_2  \equiv \frac{C(r=L/2)}{C(r=L/4)}
.
\end{align}
$U_1$ and $U_2$ are expected to take fixed values independently of the system size $L$ at the critical temperature $T\sub{c}$.

\vspace{0.25cm}

\begin{figure}[tbh]
\centering
\begin{minipage}{0.49\linewidth}
\centering
(a)\\
\includegraphics[width=0.95\linewidth]{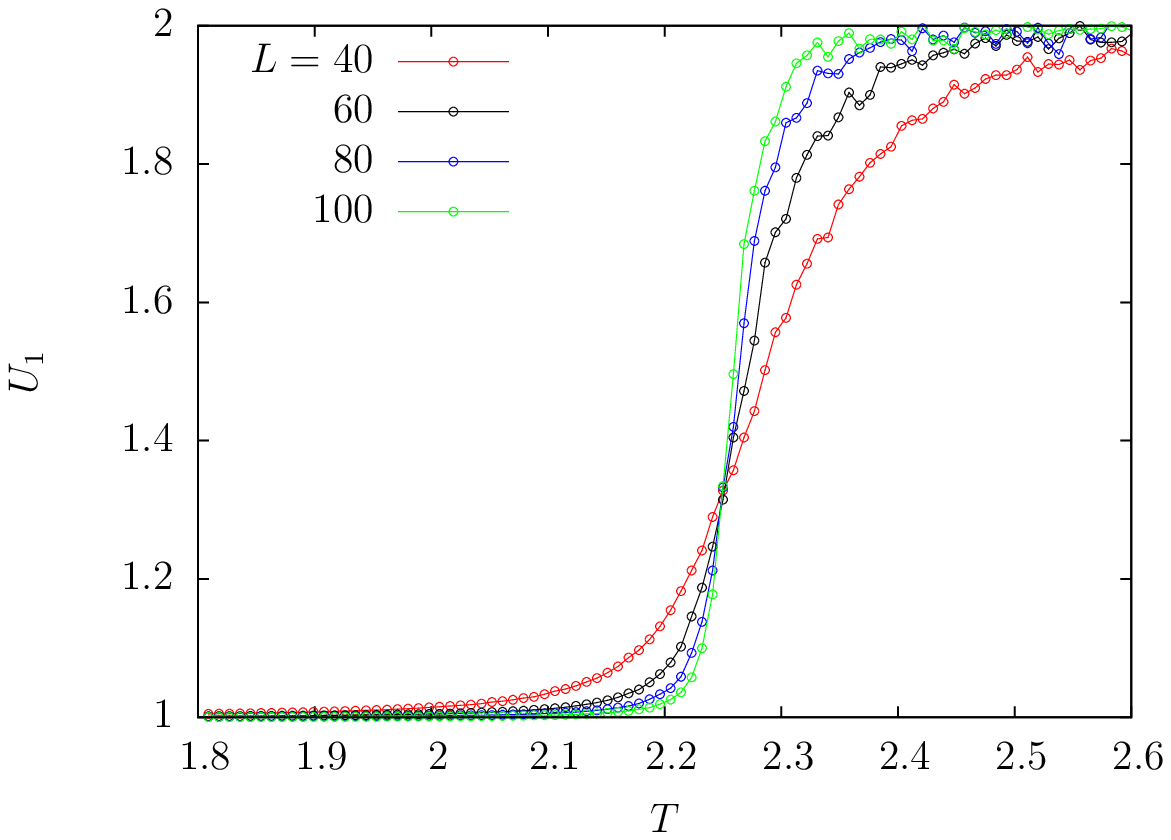}
\end{minipage}
\begin{minipage}{0.49\linewidth}
\centering
(b)\\
\includegraphics[width=0.95\linewidth]{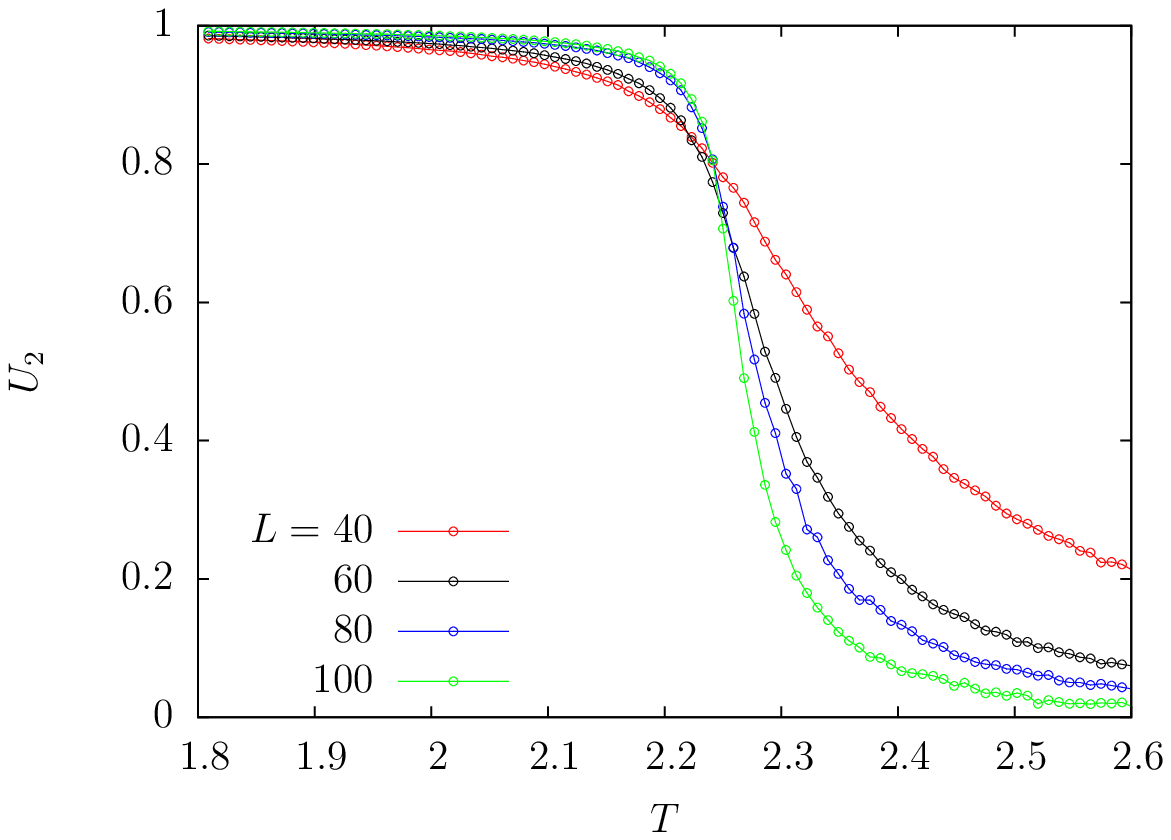}
\end{minipage}
\caption{\label{fig:U1-U2-rel} 
(Color online.)
Temperature dependence of the Binder ratio, $U_1$, and the ratio of the correlation function
at different distances, $U_2$, as obtained from the under-damped Langevin equation \eqref{eq:under-damped-Langevin}, for different 
system sizes $L$ given in the 
keys.
}
\end{figure}

Figure~\ref{fig:U1-U2-rel} shows the temperature dependence of $U_1$ (a) and $U_2$ (b) obtained 
from the under-damped Langevin equation \eqref{eq:under-damped-Langevin}.
The estimated critical temperature is 
\begin{equation}
T\sub{c} = 2.26 \pm 0.02
\label{eq:critical-temp}
\; . 
\end{equation}
As well as for $m$, the other Langevin equations \eqref{eq:over-damped-Langevin}, \eqref{eq:under-damped-ultra-relativistic}, and \eqref{eq:under-damped-non-relativistic} give almost the same values of $U_1$, $U_2$ and $T\sub{c}$ (not shown).

\vspace{0.25cm}

\begin{figure}[tbh]
\centering
\begin{minipage}{0.49\linewidth}
\centering
(a)\\
\includegraphics[width=0.95\linewidth]{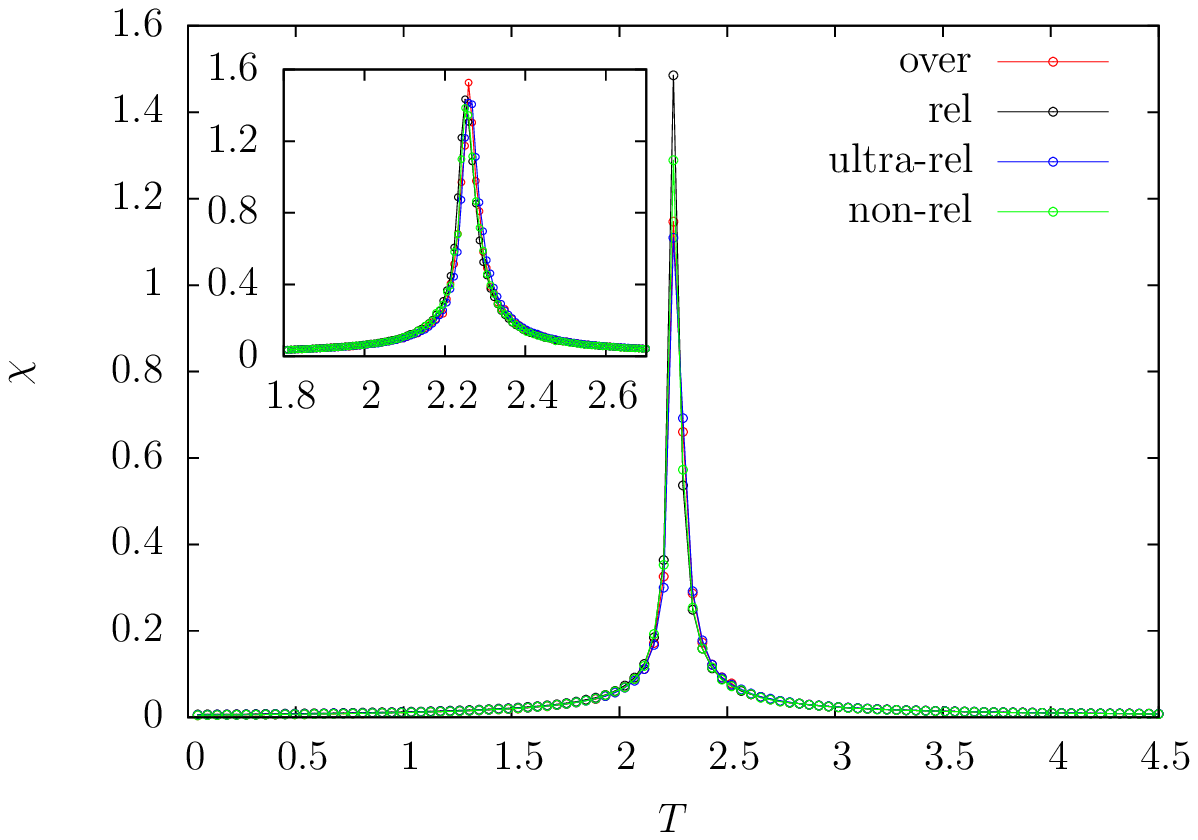}
\end{minipage}
\begin{minipage}{0.49\linewidth}
\centering
(b)\\
\includegraphics[width=0.95\linewidth]{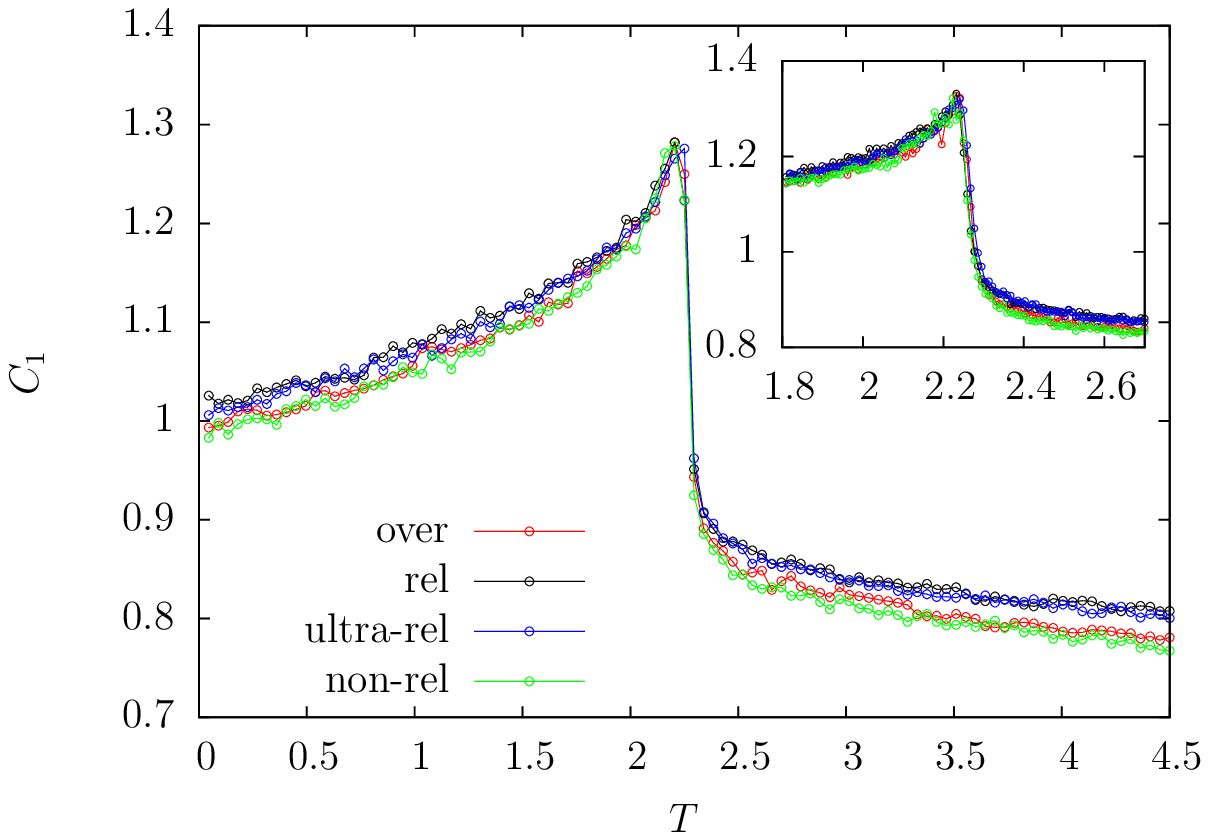}
\end{minipage}
\begin{minipage}{0.49\linewidth}
\centering
(c)\\
\includegraphics[width=0.95\linewidth]{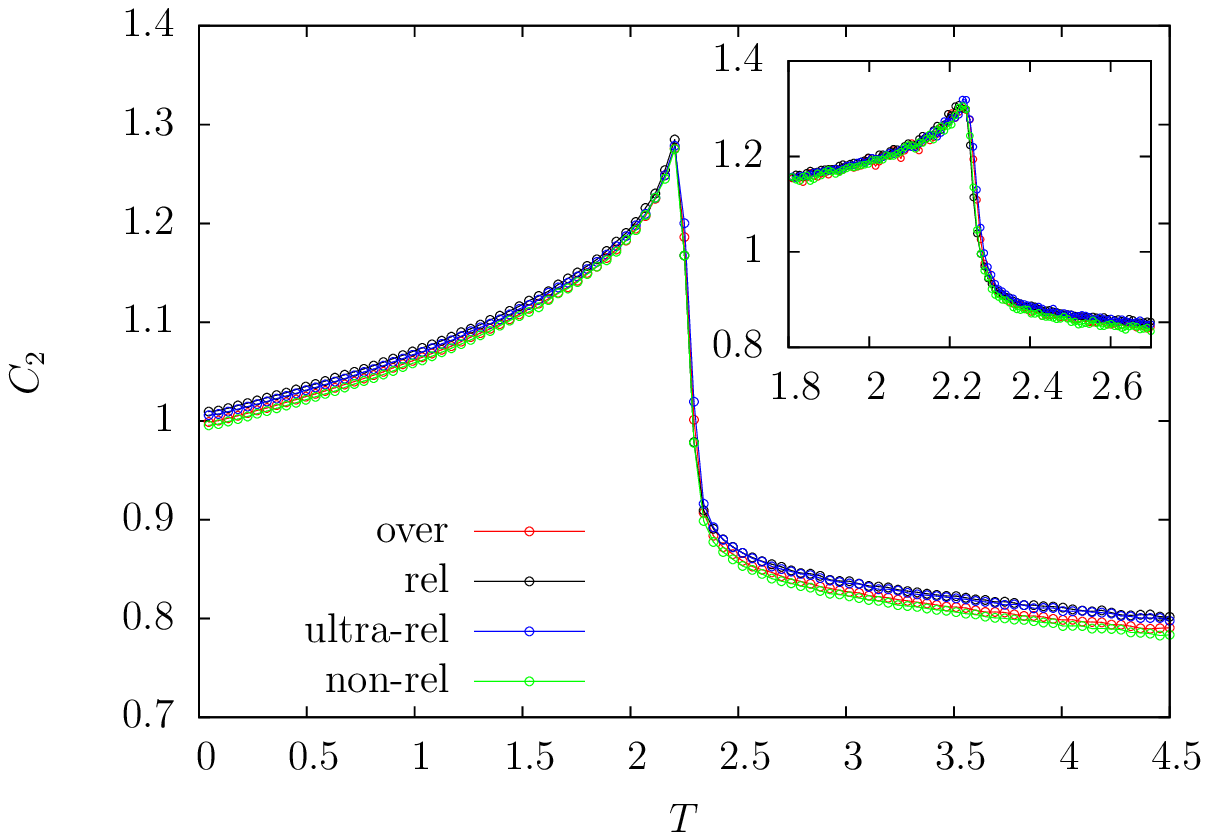}
\end{minipage}
\begin{minipage}{0.49\linewidth}
\centering
(d)\\
\includegraphics[width=0.95\linewidth]{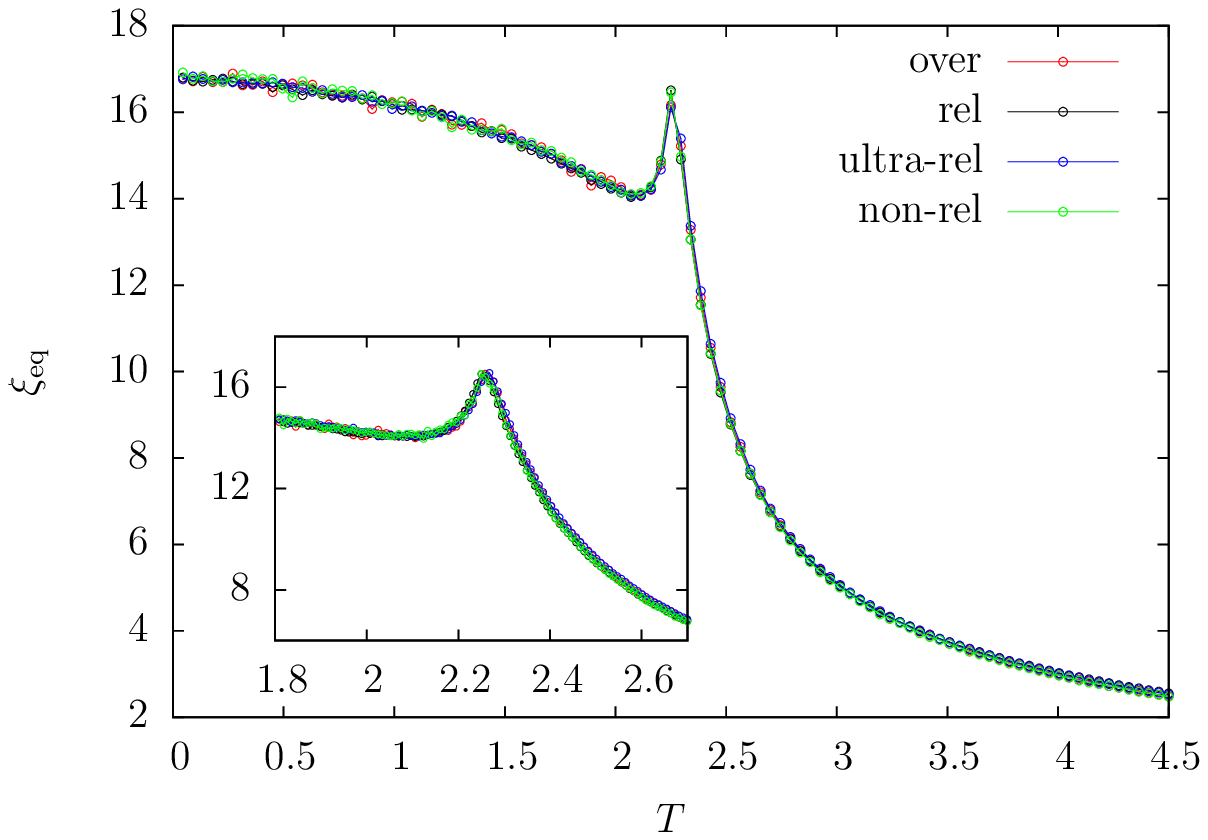}
\end{minipage}
\caption{\label{fig:chi-c1-c2-xi-all} 
(Color online.)
Temperature dependence of the susceptibility, $\chi$, the two definitions of the 
specific heat, $C_1$ and $C_2$, and the correlation length, $\xi\sub{eq}$, for a system with linear size, $L = 100$,
and the four stochastic dynamic equations. Inserts: zoom over the critical region. }
\end{figure}

We also calculated the susceptibility
\begin{align}
\chi \equiv \frac{\dps \big\langle \big| \overline{\psi} \big|^2 \big\rangle\sub{eq} - \big\langle \big| \overline{\psi} \big| \big\rangle\sub{eq}^2}{T}
\end{align}
and the specific heat
\begin{align}
C \equiv \frac{d \langle E_0 \rangle\sub{eq}}{d T}
,
\end{align}
that assuming the equilibrium ensemble average shown in Eq. \eqref{eq:statistical-average}, can also be written as
\begin{align}
C = \frac{\langle E_0^2 \rangle\sub{eq} - \langle E_0 \rangle\sub{eq}^2}{T^2}.
\end{align}
We implemented these definitions numerically as
\begin{align}
C_1 \equiv \frac{\langle E_0^2 \rangle\sub{stat} - \langle E_0 \rangle\sub{stat}^2}{T^2}, \quad
C_2 \equiv \frac{\langle E_0(T + \Delta T) \rangle\sub{stat} - \langle E_0(T - \Delta T) \rangle\sub{stat}}{2 \Delta T}.
\end{align}
The equilibrium correlation length $\xi\sub{eq}$ was calculated by assuming that the connected correlation length decays exponentially as
$C( r ) - m^2 \to r^{- r / \xi\sub{eq}}$, from the corresponding small-$k$ behaviour or the structure factor
\begin{equation}
S(k) \equiv \Bigg\langle \frac{1}{4 \pi k^2 L^3} \int d\Omega_{\Vec{k} = k}\: |\tilde{\psi}(\Vec{k})|^2 \Bigg\rangle\sub{eq},  
\end{equation}
with $\tilde{\psi}(\Vec{k}) = \int d\Vec{x}\: \psi(\Vec{x}) e^{- i \Vec{k} \cdot \Vec{x}}$  the Fourier transformation of the field,
estimated numerically from
\begin{align}
\frac{S(k = 2 \pi / \xi\sub{eq})}{S(k \to 0)} = 10^{-1}, \quad
S(k \to 0) \equiv 2 S(\Delta k) - S(2 \Delta k), \label{eq:correlation-length}
\end{align}
where $S(k \to 0)$ is the linear interpolation from $S(\Delta k)$ and $S(2 \Delta k)$ to the value at $k = 0$ with $\Delta k = 2 \pi / L$.


The panels (a)-(d) in Fig.~\ref{fig:chi-c1-c2-xi-all} show the temperature dependences of $\chi$, $C_1$, $C_2$, and $\xi\sub{eq}$ for $L = 100$.
Their values are also almost independent of the type of 
Langevin equation used even close to criticality at $T\simeq T_c$.
The two specific heats $C_1$ and $C_2$ are almost the same except for the slightly more jagged shape of $C_1$. 
We note that both $C_1$ and $C_2$ converge to the finite values $C_1 \simeq 1$ and $C_2 \simeq 1$ in the 
zero temperature limit, because of the continuous U(1) symmetry breaking and the resulting 
Nambu-Goldstone modes.
This unphysical result can be cured by taking into account quantum effects.

\subsection{Critical scaling}
 
Finite-size scaling~\cite{Barber} states that 
$m / L^{- \beta / \nu}$, $\chi / L^{\gamma / \nu}$, $C_1 / L^{\alpha / \nu}$, $C_2 / L^{\alpha / \nu}$, and $\xi\sub{eq} / L$ 
should be universal functions of $L^{1/\nu} [(T - T\sub{c}) / T\sub{c}]$ independently  of $L$ near the 
critical temperature~$T\sub{c}$.
\begin{figure}[tbh]
\centering
\begin{minipage}{0.49\linewidth}
\centering
(a)\\
\includegraphics[width=0.95\linewidth]{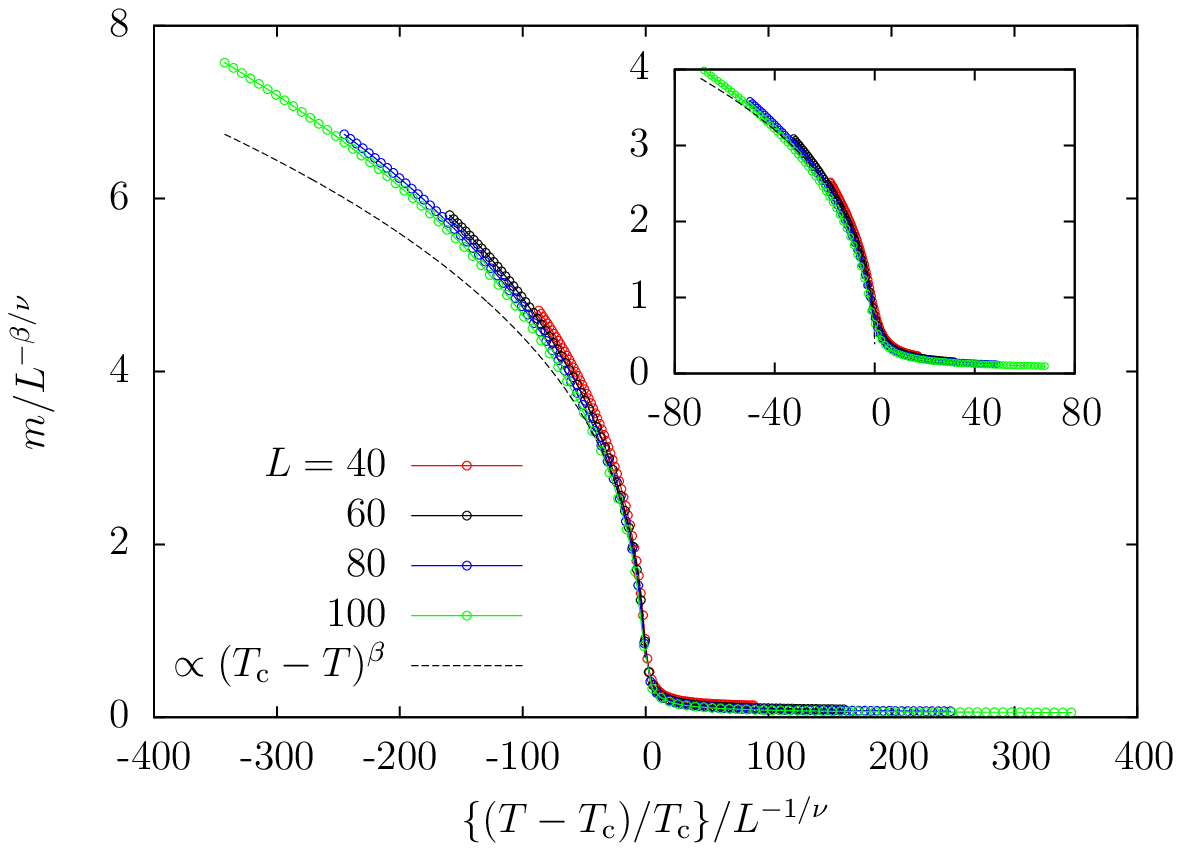}
\end{minipage}
\begin{minipage}{0.49\linewidth}
\centering
(b)\\
\includegraphics[width=0.95\linewidth]{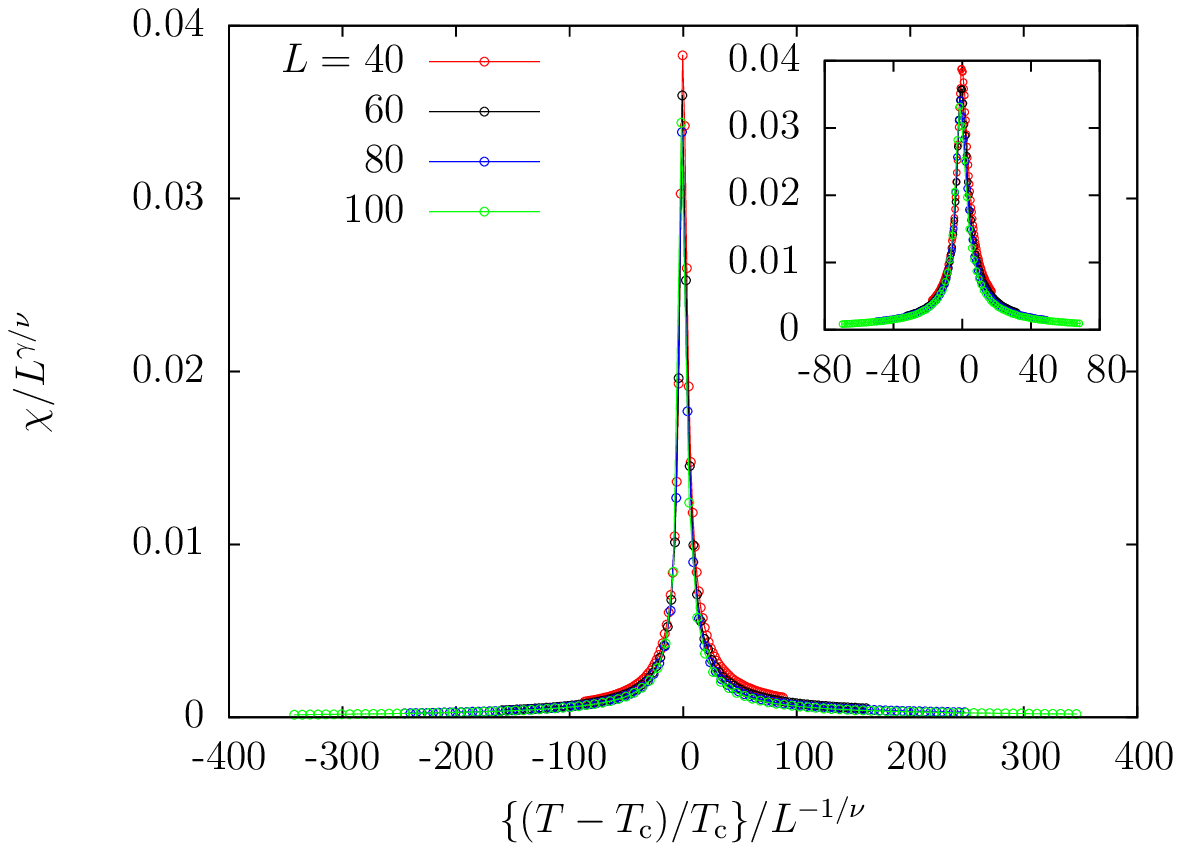}
\end{minipage}
\begin{minipage}{0.49\linewidth}
\centering
(c)\\
\includegraphics[width=0.95\linewidth]{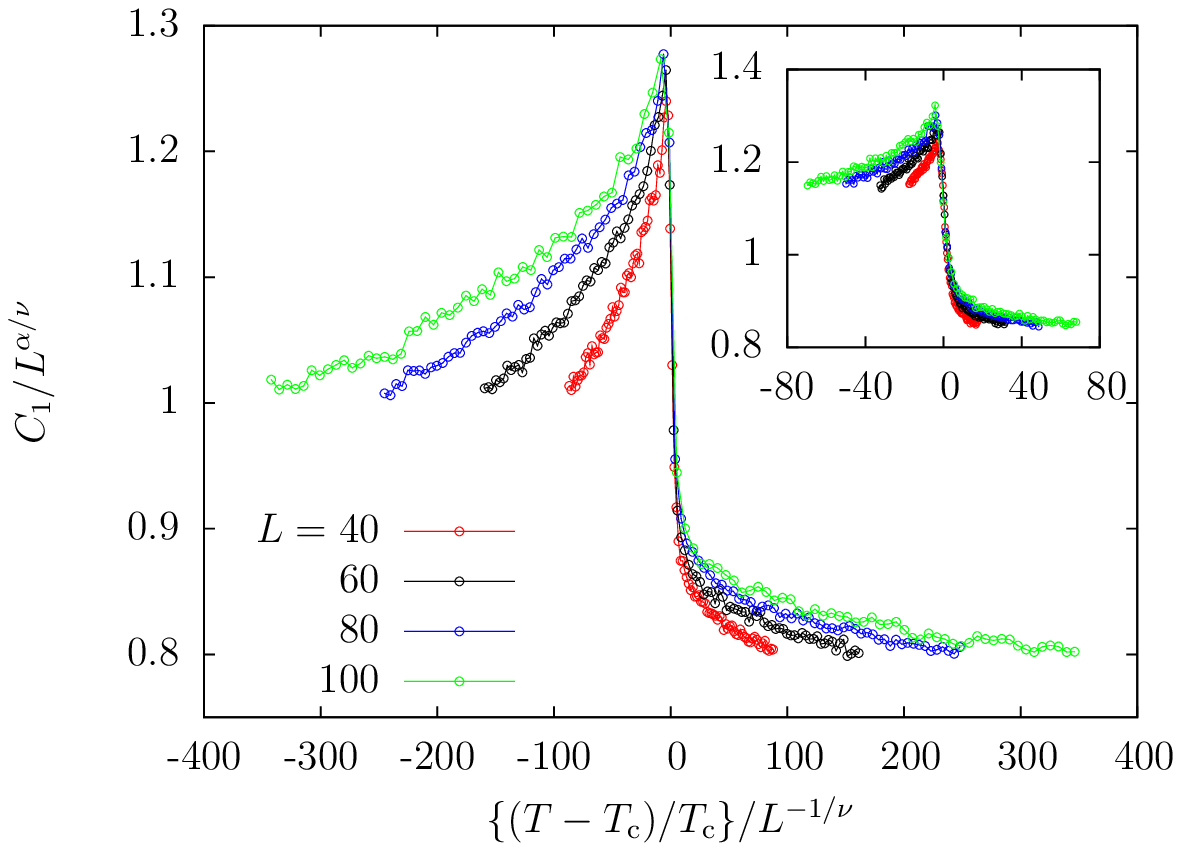}
\end{minipage}
\begin{minipage}{0.49\linewidth}
\centering
(d)\\
\includegraphics[width=0.95\linewidth]{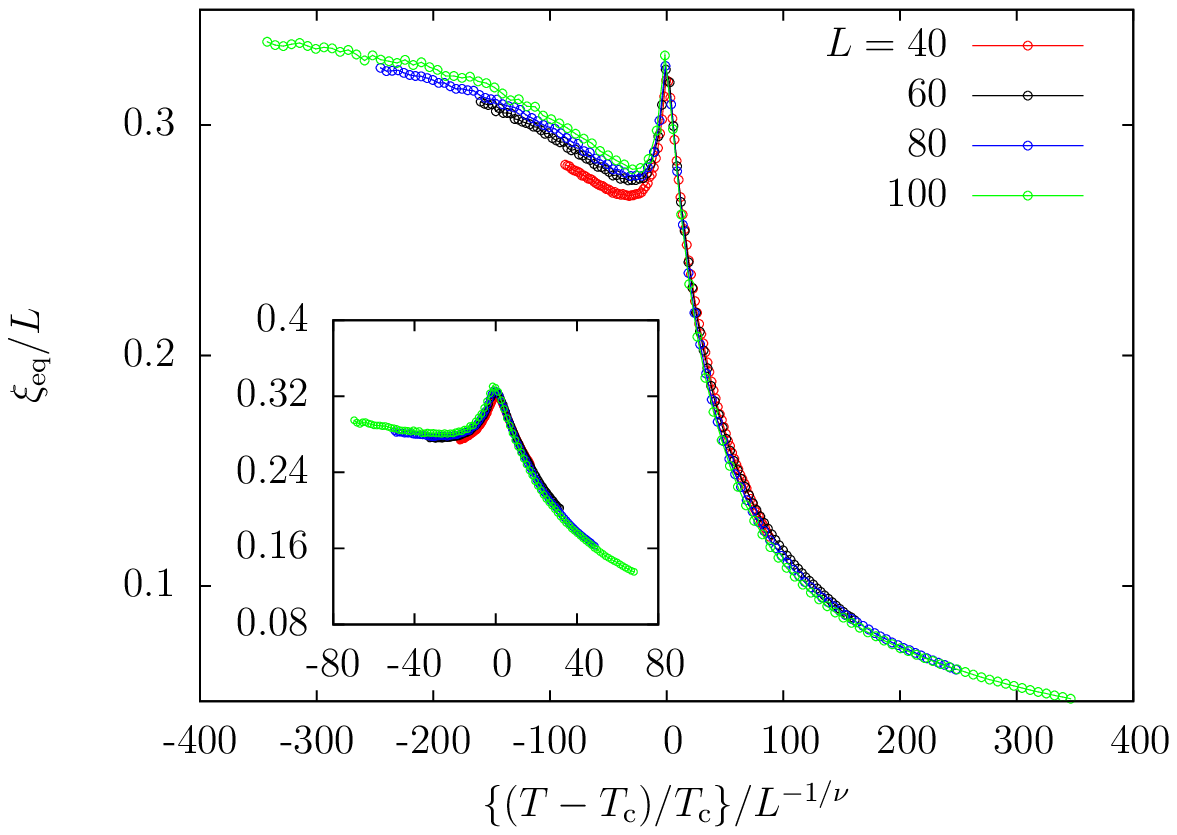}
\end{minipage}
\caption{\label{fig:m-chi-c1-xi-scale} 
(Color online.)
Finite-size scaling of the order parameter, $m$, the susceptibility, $\chi$, the specific heat, $C_1$, and the correlation 
length, $\xi\sub{eq}$, all obtained from the under-damped Langevin equation \eqref{eq:under-damped-Langevin}.
The system sizes and colour code are given in the key. Inserts: zoom over the critical region.
}
\end{figure}

Figures \ref{fig:m-chi-c1-xi-scale} (a)-(d) show the expected universal functions for $m$, $\chi$, $C_1$, and $\xi\sub{eq}$, 
where we used the critical exponents obtained from the $\epsilon$-expansion~\cite{GuidaZJ}: 
$\alpha = - 0.011$, $\beta = 0.3470$, $\gamma = 1.3169$, and $\nu = 0.6703$.
The scaling of $m$, $\chi$, and $\xi\sub{eq}$  are very satisfactory. 
The scaling of $C_1$ is not as good because $|\alpha|$ is so small that the logarithmic correction to 
the power-law behaviour cannot be neglected, i.e., $C_1$ behaves as $C_1 \propto |T - T\sub{c}|^{-\alpha}$ 
only at temperatures very close to $T\sub{c}$ and it behaves as $C_1 \propto \log|T - T\sub{c}| + \mathrm{const}$ otherwise.
The logarithmic behaviour of the specific heat near the critical temperature has been confirmed in liquid \4He~\cite{Ahlers}. 

\begin{figure}[tbh]
\centering
\includegraphics[width=0.49\linewidth]{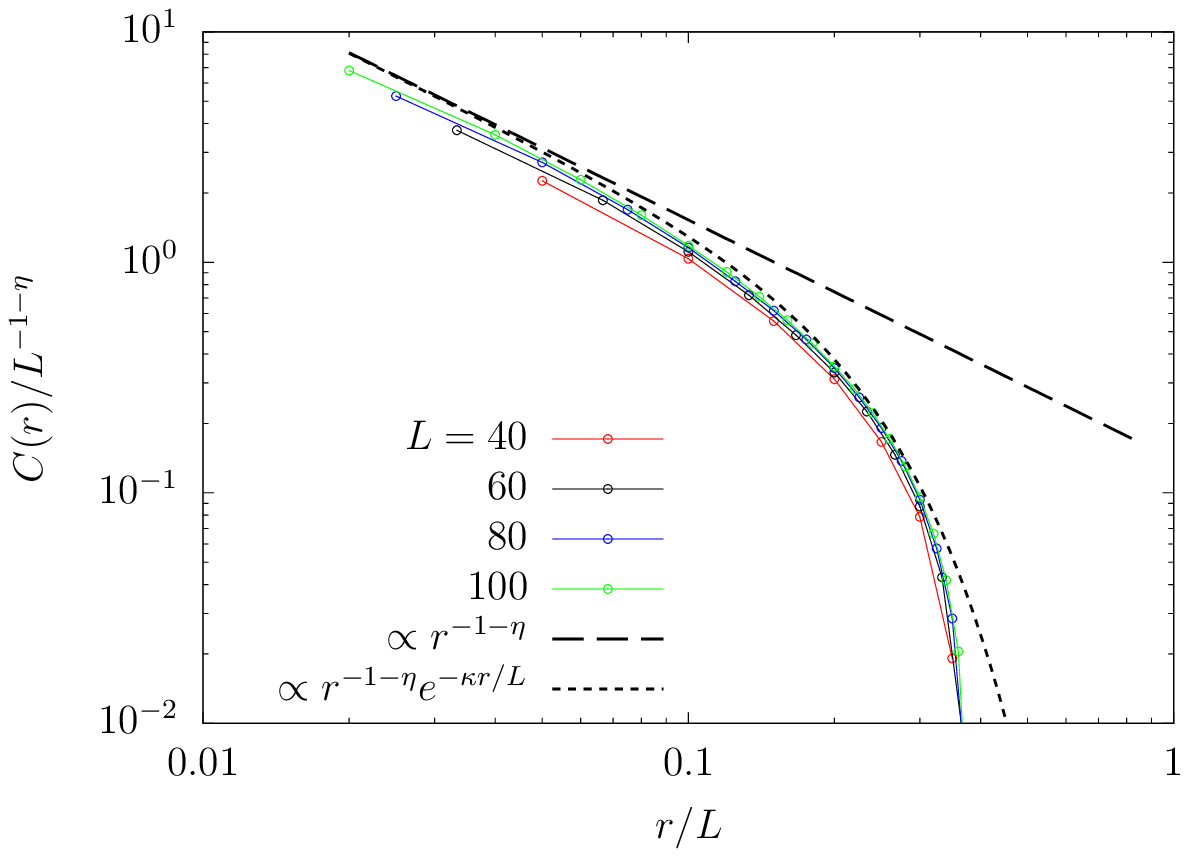}
\caption{\label{fig:gr-scale} 
(Color online.)
Finite-size scaling of the correlation function $C(r)$ near the 
critical temperature $T\sub{c} = 2.26$ obtained using the under-dampled Langevin equation 
\eqref{eq:under-damped-Langevin}. The dotted line is the analytic prediction $(r/L)^{-1-\eta}$
with $\eta=0.038$ and the solid line includes the exponential cut-off.}
\end{figure}

The correlation function $C(r) / L^{-1 - \eta}$ is also expected to be a universal function on $r / L$ at the critical temperature $T\sub{c}$.
We can see this universality in Fig.~\ref{fig:gr-scale} with $\eta = 0.038$, together with the algebraic and complete
analytic forms including the exponential cut-off due to the finite size of the sample.

\vspace{0.25cm}

We next consider the helicity modulus $\Upsilon$ defined as~\cite{Fisher73}
\begin{align}
\Upsilon \equiv \lim_{\Delta \to 0} \frac{F(\Delta) - F(0)}{\Delta^2},
\end{align}
where $F(\Delta)$ is the free energy $- T \log Z_0(\Delta)$ and  $Z_0(\Delta)$  the partition function
under the twisted boundary condition along $x$-direction:
\begin{align}
\psi(t,x+L,y,z) = \psi(t,x,y,z) e^{i \Delta}.
\end{align}
We calculated  $Z_0 = \langle e^{- E_0 / T} \rangle\sub{stat}$ and we  used $\Delta = 0.01 / L$.
We confirmed that $\Upsilon$ takes almost same value for $\Delta = 0.02 / L$ and $\Delta = 0.005 / L$.

\vspace{0.25cm}

\begin{figure}[tbh]
\centering
\begin{minipage}{0.49\linewidth}
\centering
(a)\\
\includegraphics[width=0.95\linewidth]{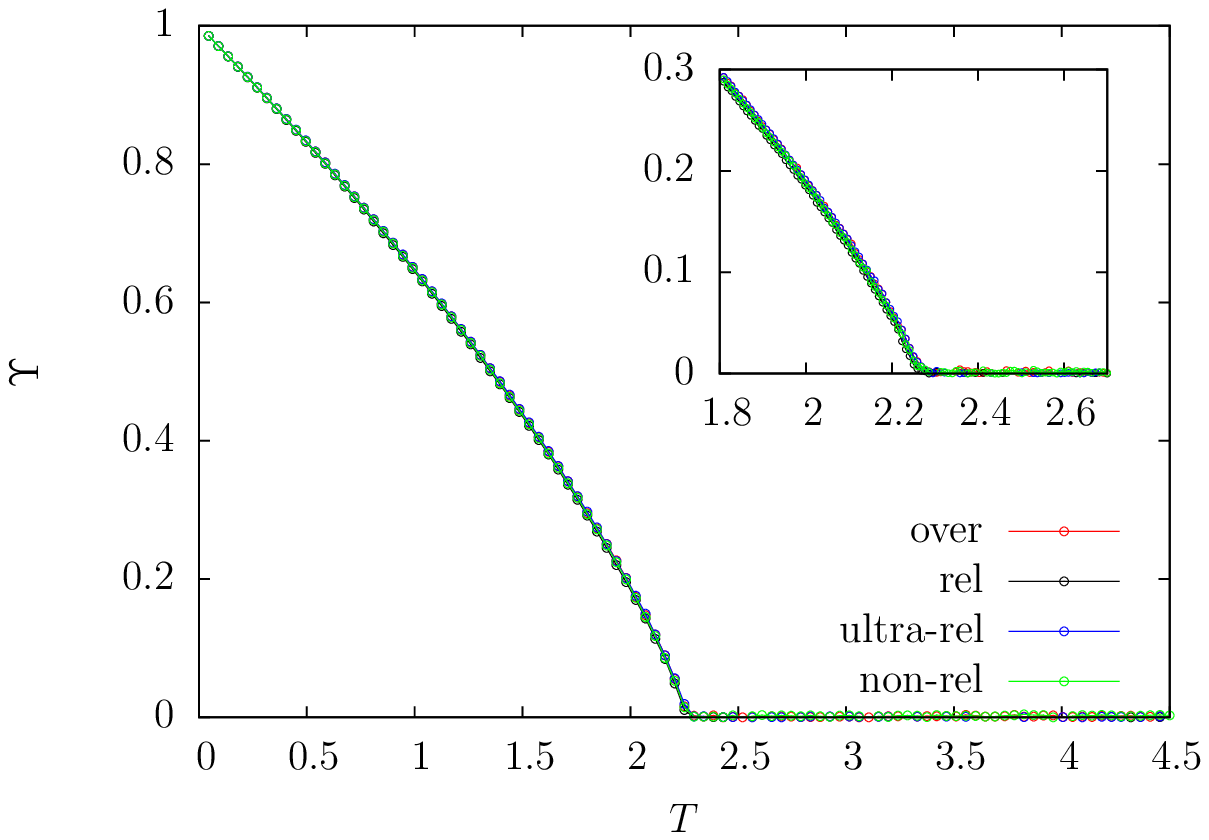}
\end{minipage}
\begin{minipage}{0.49\linewidth}
\centering
(b)\\
\includegraphics[width=0.95\linewidth]{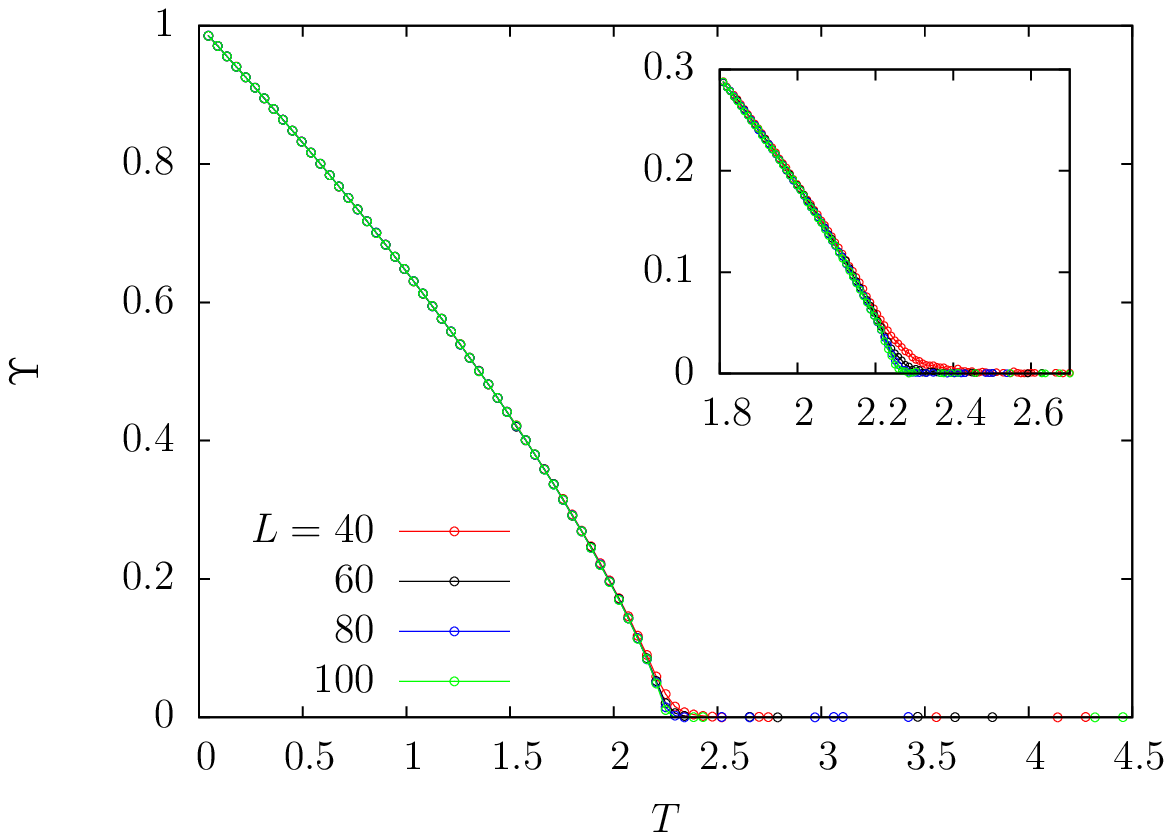}
\end{minipage}
\caption{\label{fig:rhos} 
(Color online.)
Temperature dependence of the helicity modulus $\Upsilon$. 
(a) Data for different Langevin equations with $L = 100$.
(b) Data for different system sizes $L$ as obtained from the under-damped Langevin equation \eqref{eq:under-damped-Langevin}.
Inserts: zoom over the critical region.
}
\end{figure}

Figure~\ref{fig:rhos}~(a) and~(b) shows the dependence of  $\Upsilon$ on $T$ for different 
Langevin equations [Fig.~\ref{fig:rhos} (a)] and 
different system sizes $L$  [Fig.~\ref{fig:rhos} (b)].
The helicity modulus is not affected by the Langevin dynamics. 
Compared to the order parameter $m$ shown in Figs.~\ref{fig:m-all}~(a) and~(b),  
$\Upsilon$ has less dependence on the system size $L$ and 
completely vanishes around the critical temperature $T \simeq T\sub{c}$.

\begin{figure}[tbh]
\centering
\includegraphics[width=0.49\linewidth]{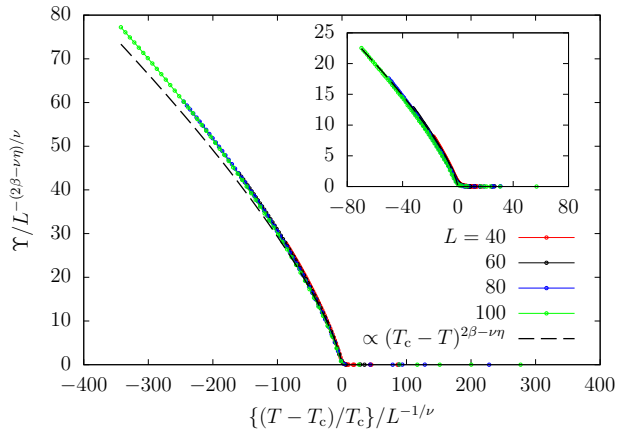}
\caption{\label{fig:rhos-scale} 
(Color online.)
Finite-size scaling of the helicity modulus $\Upsilon$ obtained from the underdampled 
Langevin equation \eqref{eq:under-damped-Langevin}. The sizes used are given in the key and the dotted (blue) line represents the 
critical behaviour close to the transition, see the text for a discussion. 
}
\end{figure}

The Noether current for the phase shift $\psi \to \psi e^{i \delta}$ given from the Lagrangian \eqref{eq:original-Lagrangian} is
\begin{align}
j^\mu
&= \frac{i}{2} \bigg\{ \frac{\partial \mathcal{L}}{\partial (\partial_\mu \psi^\ast)} \psi^\ast - \frac{\partial \mathcal{L}}{\partial (\partial_\mu \psi)} \psi \bigg\}
= f^2 \bigg( \mu -  \frac{\dot{\theta}}{c^2}, 
\nabla \theta \bigg), \label{eq:Noether-theorem} 
\end{align}
where $f$ and $\theta$ are defined from $\psi = f e^{i \theta}$.
Equation \eqref{eq:Noether-theorem} indicates that the twisted phase 
$\nabla \theta$ induces the current density $f^2 \nabla \theta$ for the charge density $f^2 (\mu - \dot{\theta}/c^2)$.
For non-zero $\mu$, $\mu f^2$ and $f^2 \nabla \theta$ are regarded as the density and the 
supercurrent density of bosons respectively. A non-vanishing helicity modulus $\Upsilon$ 
induced by the twisted phase implies a finite free-energy cost for a finite supercurrent and 
the system enters the superfluid phase.
Our results in Fig.~\ref{fig:rhos} show that superfluidity appears at the same critical temperature 
$T\sub{c} \simeq 2.26$ as the one for the spontaneous symmetry breaking.
In the ultra-relativistic limit, $\mu =0$, the charge density induced by the 
twisted phase is the conventional Noether charge 
$- f^2 \dot{\theta} / c^2$, and there is no relationship 
between the helicity modulus $\Upsilon$ and superfluidity.
The helicity modulus $\Upsilon$ is also expected to show critical 
behaviour characterised by the Josephson scaling relation $\Upsilon \propto (T\sub{c} - T)^{2 \beta - \nu \eta}$ at $T < T\sub{c}$. 
For finite sizes one 
therefore expects universal scaling of  
$\Upsilon / L^{-(2 \beta - \nu \eta)/\nu}$ as a function of $(T - T\sub{c}) / L^{-1/\nu}$ independently of $L$.
We can see this universality in Fig.~\ref{fig:rhos-scale}.

\vspace{0.25cm}

\subsection{The equilibrium relaxation time}
\label{sec:equilibrium-relax-time}

The equilibrium relaxation time $\tau$ is defined as
\begin{align}
\frac{\dps \big\langle |\overline{\psi}(t)|^2 \big\rangle\sub{relax} - \big\langle |\overline{\psi}|^2 \big\rangle\sub{stat} }{\rho} 
\stackrel{t \to \infty}{\longrightarrow} e^{- t / \tau}, \qquad
\psi(t=0) = \sqrt{\rho}, \label{eq:definition-tau}
\end{align}
where $\langle f(t) \rangle\sub{relax}$ is the noise average of $f(t)$ at time $t$ of evolution from 
the fully ordered initial state $\psi(t = 0) = \sqrt{\rho}$.

\begin{figure}[tbh]
\centering
\begin{minipage}{0.49\linewidth}
\centering
(a)\\
\includegraphics[width=0.95\linewidth]{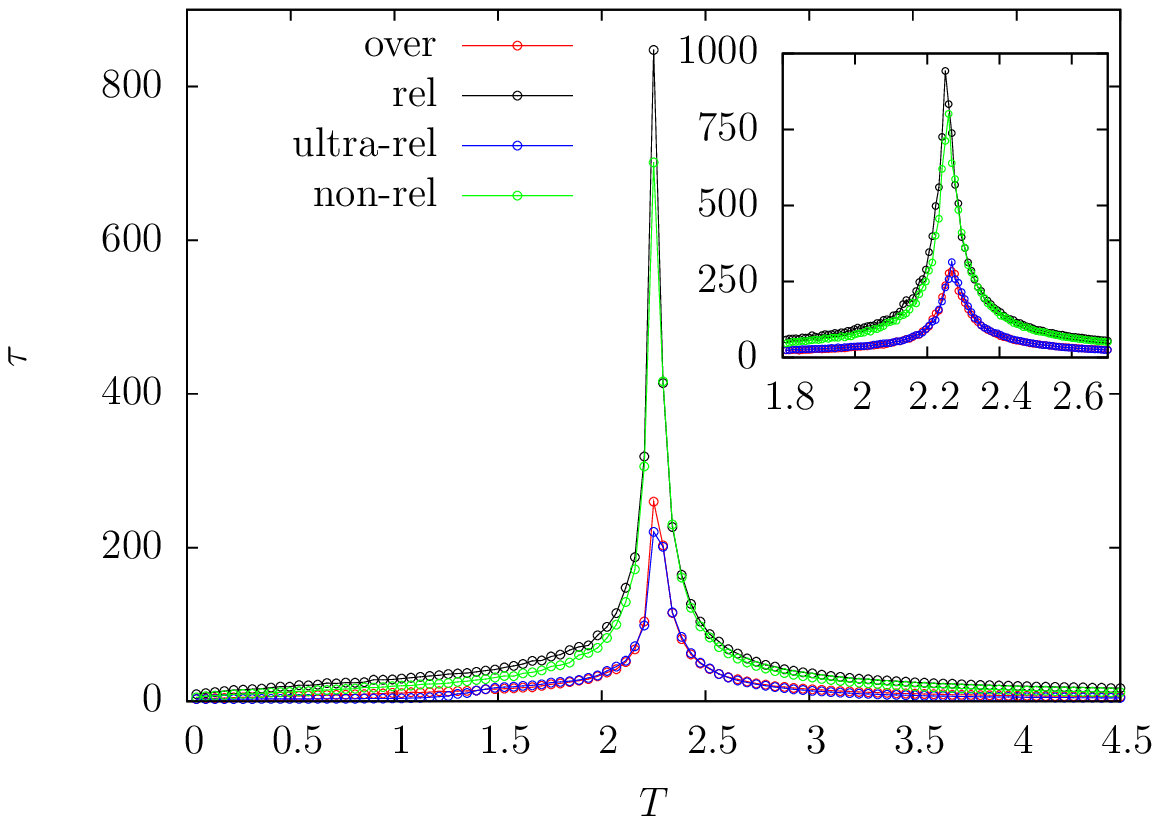}
\end{minipage}
\begin{minipage}{0.49\linewidth}
\centering
(b)\\
\includegraphics[width=0.95\linewidth]{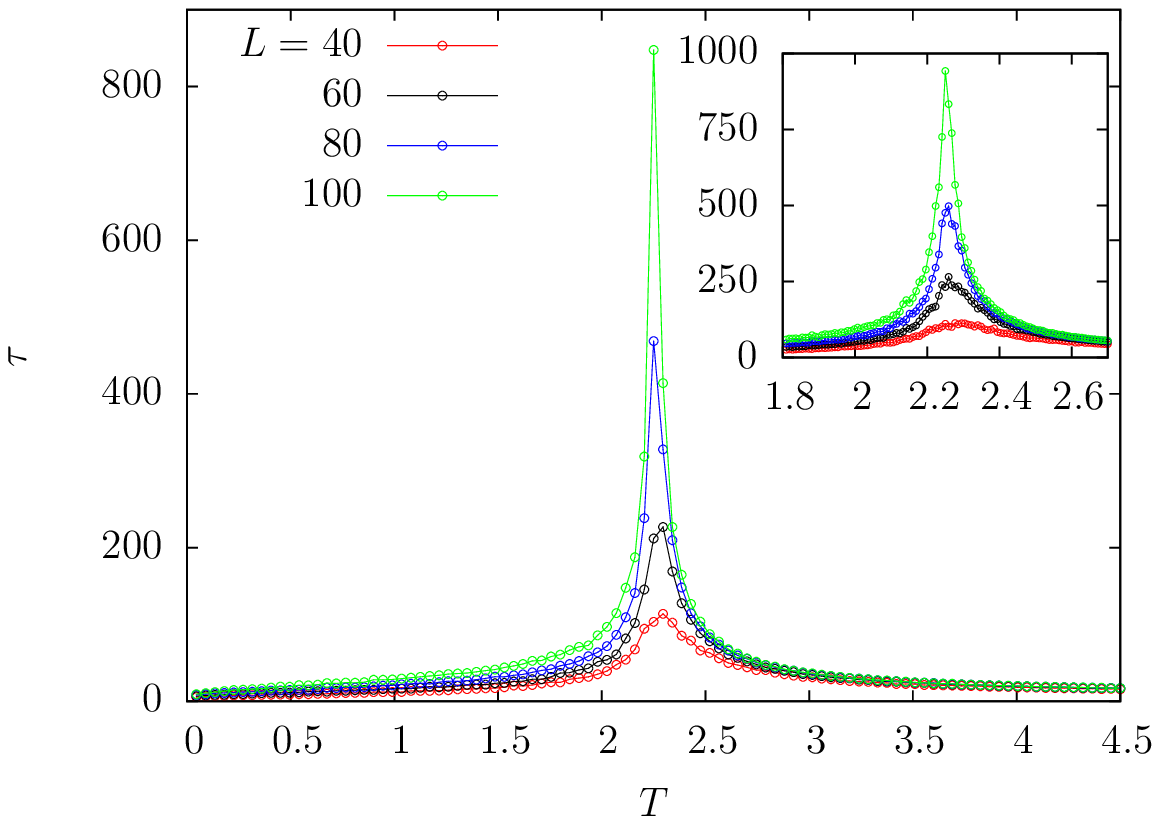}
\end{minipage}
\caption{\label{fig:tau} 
(Color online.)
Temperature dependence of the relaxation time $\tau$ for different Langevin equations with $L = 100$ (a) and 
for different system sizes $L$ obtained from the under-damped Langevin equation \eqref{eq:under-damped-Langevin} (b).
}
\end{figure}

Being a dynamic parameter, the numerical $\tau$ can depend on the type of Langevin equation used 
[Eqs.~\eqref{eq:over-damped-Langevin}, \eqref{eq:under-damped-Langevin}, \eqref{eq:under-damped-ultra-relativistic}, and 
\eqref{eq:under-damped-non-relativistic}]. 
We measure  $\tau$ numerically by using the criterium
\begin{align}
\frac{\dps \big\langle |\overline{\psi}(t = \tau)|^2 \big\rangle\sub{relax} - \big\langle |\overline{\psi}|^2 \big\rangle\sub{stat}}{\rho} = 10^{-3}.
\end{align}
Figure~\ref{fig:tau} shows the dependence of the relaxation time $\tau$ for different Langevin 
equations and $L = 100$ (a), and for different system sizes and one dynamic rule (b).
The relaxation time $\tau$ for the over-damped Langevin equation \eqref{eq:over-damped-Langevin} and the ultra-relativistic limit of the under-damped 
Langevin equation \eqref{eq:under-damped-ultra-relativistic} with $\mu = 0$, and those for the under-damped Langevin equation 
\eqref{eq:under-damped-Langevin} and its non-relativistic limit \eqref{eq:under-damped-non-relativistic} with $\mu = 1$ take similar values, and the latter ones 
are larger than the former ones.

We can evaluate $\tau$ within an approximation in which the noise term is
``renormalised" into the linear term originating in the potential energy by the 
replacement  $\rho \to m^2(T)$.
Equation~\eqref{eq:under-damped-Langevin} then becomes
\begin{align}
\begin{split}
- \Box \psi + (2 i \mu - \gamma\sub{L}) \dot{\psi} = g(|\psi|^2 - m^2) \psi
,
\label{eq:renormalized-Langevin}
\end{split}
\end{align}
and admits the stationary solution $|\psi|^2 = m^2$. Proposing a linear perturbation $\delta \psi$ 
on top of the background $m$,  $\psi = m + \delta \psi$, the equation governing $\psi$ becomes
\begin{align}
- \Box \delta \psi + (2 i \mu - \gamma\sub{L}) \delta \dot{\psi} = g m^2 (\delta \psi + \delta \psi^\ast) + O(\delta \psi^2).
\end{align}
We now assume $\delta \psi \ll m$ at the late stage $t \to \infty$ of the relaxation 
and we neglect the term $O(\delta \psi^2)$ in the right-hand-side of this equation.
Further rewriting the unknown as $\delta \psi = u e^{i (\Vec{k} \cdot \Vec{x} - \omega t)} + v^\ast e^{- i (\Vec{k} \cdot \Vec{x} - \omega^\ast t)}$, we obtain the Bogoliubov-de Gennes equation:
\begin{align}
\begin{pmatrix}
\omega^2 / c^2 - k^2 + (2 \mu + i \gamma\sub{L}) \omega - g m^2 & - g m^2 \\
- g m^2 & \omega^2 / c^2 - k^2 - (2 \mu - i \gamma\sub{L}) \omega - g m^2
\end{pmatrix}
\begin{pmatrix}
u \\ v
\end{pmatrix} = 0. \label{eq:Bogoliubov-de-Gennes}
\end{align}
We now consider the $k \to 0$ mode.
Equation \eqref{eq:Bogoliubov-de-Gennes} has four solutions for the frequency, 
$\omega = \{ \omega_\pm\up{N}, \omega_\pm\up{H} \}$, and they are
\begin{align}
\begin{split}
&
\omega_+\up{N} = 0, \quad\qquad\qquad
\omega_-\up{N} = - \frac{2 i g \gamma_L m^2}{\gamma_L^2 + 4 \mu^2} + O(m^3), \\
& \omega_\pm\up{H} = \pm 2 \mu \bigg(c^2 + \frac{g m^2}{\gamma_L^2 + 4 \mu^4} \bigg) - i \gamma_L \bigg( c^2 - \frac{g m^2}{\gamma_L^2 + 4 \mu^2} \bigg) + O(m^3),
\end{split}
\end{align}
where we implicitly assumed $m \ll 1$ near the critical temperature.
In the dissipation-less limit $\gamma_L \to 0$, $\omega_{\pm}\up{N}$ become gapless Nambu-Goldstone modes
while $\omega_{\pm}\up{H}$ remain gapful Higgs modes.
For finite $\gamma_L$, $\omega_-\up{N}$ is the slowest relaxation mode and the relaxation time $\tau$ is evaluated as
\begin{align}
\tau \propto \frac{\gamma_L^2 + 4 \mu^2}{2 g \gamma_L m^2} 
\simeq 
\frac{\gamma_L^2 + 4 \mu^2}{2 g \gamma_L} \ |T-T_c|^{-2\beta}.
\label{eq:tau-meanfield}
\end{align}

\vspace{0.25cm}

However, close to $T_c$ the approximation used to derive (\ref{eq:tau-meanfield}) breaks down and 
the relaxation time $\tau$ is expected to show 
critical behaviour, 
\begin{equation}
\tau \propto |T - T\sub{c}|^{- \nu z\sub{eq}} ,
\end{equation} 
with a new 
dynamical critical exponent $z\sub{eq}$~\cite{HH}. 
The numerical simulations in~\cite{Mondello}
suggest $z_{\rm eq} \simeq 2.2$ while the ones in~\cite{Jensen} yield $z_{\rm eq} \simeq 2.1$ for periodic boundary conditions, see also~\cite{Roma}.
The equilibrium critical dynamical exponent of the classical $O(N)$ model  in $d$ dimensions 
with relaxational dynamics has been
computed with an $\epsilon=4-d$ expansion and reads~\cite{Bausch}
\begin{equation}
z_{\rm eq} = 2 + \frac{N+2}{(N+8)^2} \left( 3\ln \frac{4}{3}-\frac{1}{2} \right) \epsilon^2 + O(\epsilon^3)
\end{equation}
For $N=2$ in $d=3$ one finds $z_{\rm eq}=2.0145$.  If one uses $
z_{\rm eq} = 2+\eta$, then $z_{\rm eq} \approx 2.038$, using the value of $\eta$ given in Eq.~(\ref{eq:crit-exp-Campostrini}).

We wish to have our own estimate for $z_{\rm eq}$. 
Since it is very hard to determine $z\sub{eq}$ from the direct measurement of $\tau$,
we fix it from the universal scaling 
behaviour of $\tau / L^{z\sub{eq}}$ as a function of $\{ (T - T\sub{c}) / T\sub{c} \} / L^{- 1 / \nu}$ that should be independent 
of $L$. We then define 
\begin{align}
\Delta(z\sub{eq}) = \sum_{L_1, L_2} 
\int_{- T^\prime\sub{max}}^{T^\prime\sub{max}} dT^\prime\: \Bigg| \frac{\tau(L_1,T^\prime(L_1))}{L_1^{z\sub{eq}}} - \frac{\tau(L_2,T^\prime(L_2))}{L_2^{z\sub{eq}}} \Bigg|,
\end{align}
where $T^\prime(L) \equiv \{ (T - T\sub{c}) / T\sub{c} \} / L^{- 1 / \nu}$, and $\tau(L,T^\prime)$ is the numerically 
obtained relaxation time for a system with size $L$ at temperature $T = T\sub{c} (1 + T^\prime L^{- 1 / \nu})$.
Due to the scaling argument for $\tau(L,T^\prime) / L^{z\sub{eq}}$, $\Delta(z\sub{eq})$ should be minimized for the exact $z\sub{eq}$.
We calculate $\Delta(z\sub{eq})$, summing over all pairs of systems sizes $L_1, \ L_2 = 40, \ 60, \ 80, \ 100$, 
with $z\sub{eq} = 1.9$, $2.0$, $2.1$, $2.2$, $2.3$ and 
$T^\prime\sub{max} = 0.05 / 20^{-1 / \nu}$, $0.1 / 20^{-1 / \nu}$, $0.2 / 20^{-1 / \nu}$ $(\nu = 0.6703)$.
We find that $\Delta(z\sub{eq})$ takes minimal values for $z\sub{eq} = 2.1$ independently of $T^\prime\sub{max}$ and the type of 
Langevin equation used.

\vspace{0.25cm}

\begin{figure}[tbh]
\centering
\begin{minipage}{0.49\linewidth}
\centering
(a)\\
\includegraphics[width=0.95\linewidth]{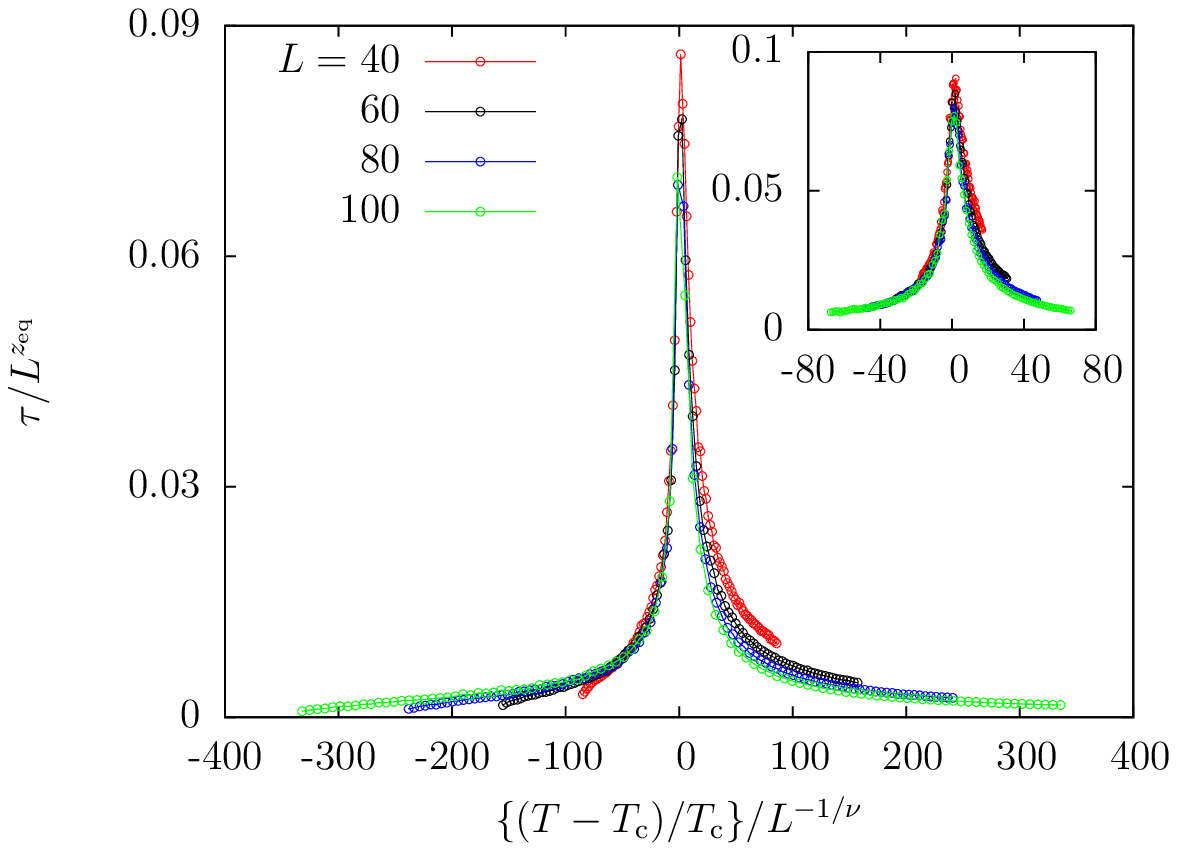}
\end{minipage}
\begin{minipage}{0.49\linewidth}
\centering
(b)\\
\includegraphics[width=0.95\linewidth]{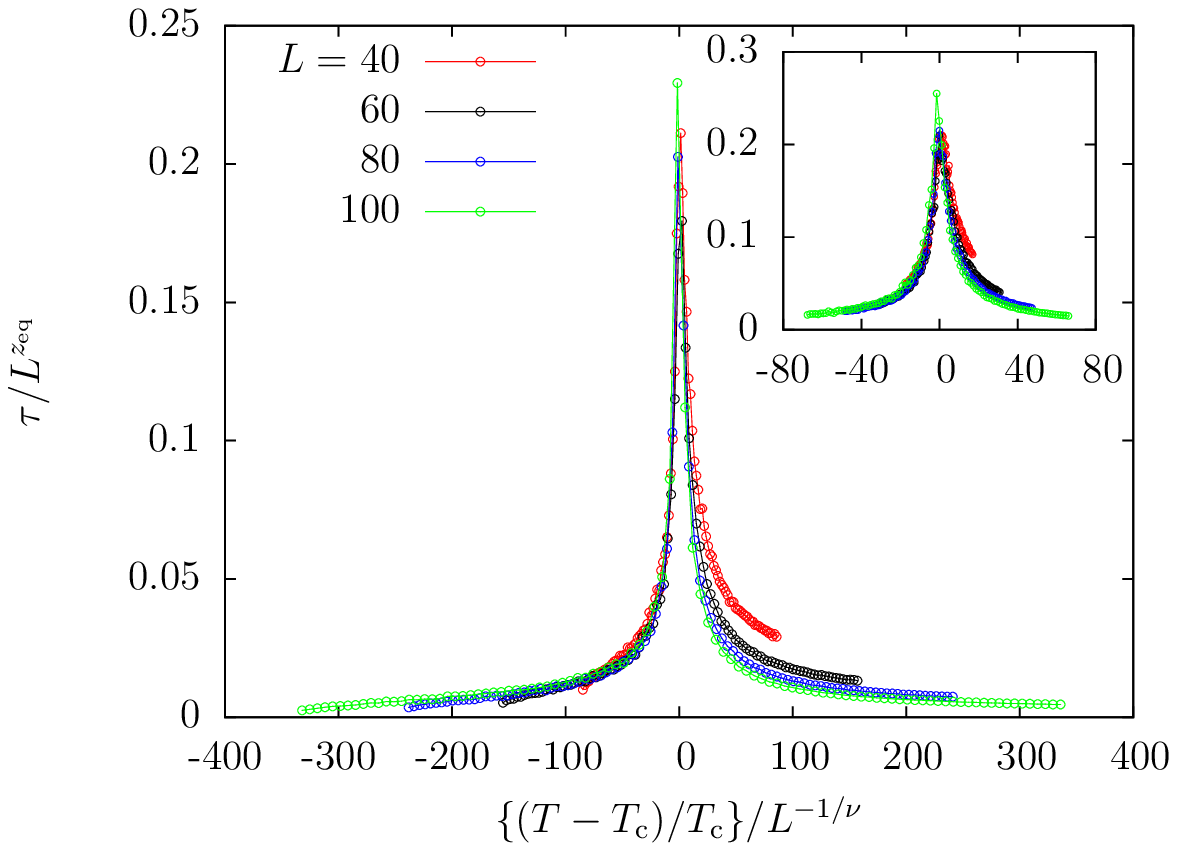}
\end{minipage}
\begin{minipage}{0.49\linewidth}
\centering
(c)\\
\includegraphics[width=0.95\linewidth]{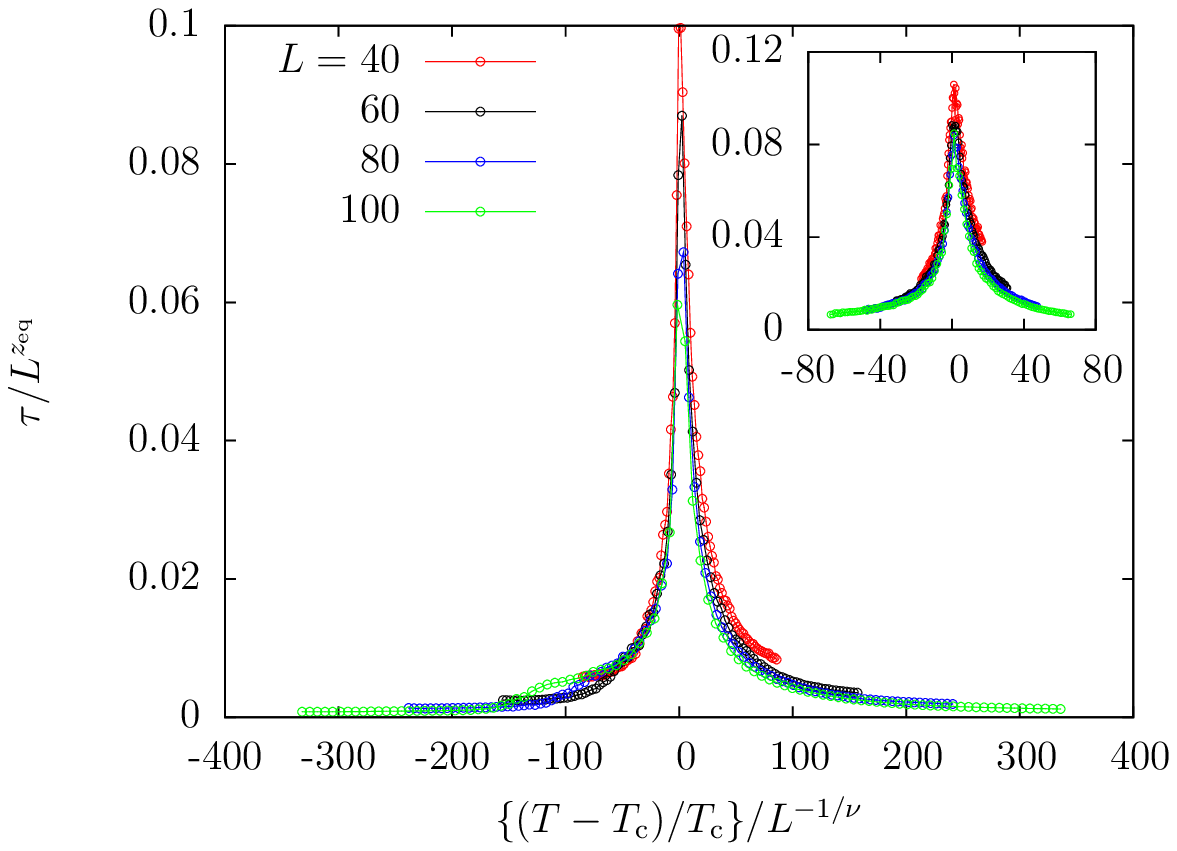}
\end{minipage}
\begin{minipage}{0.49\linewidth}
\centering
(d)\\
\includegraphics[width=0.95\linewidth]{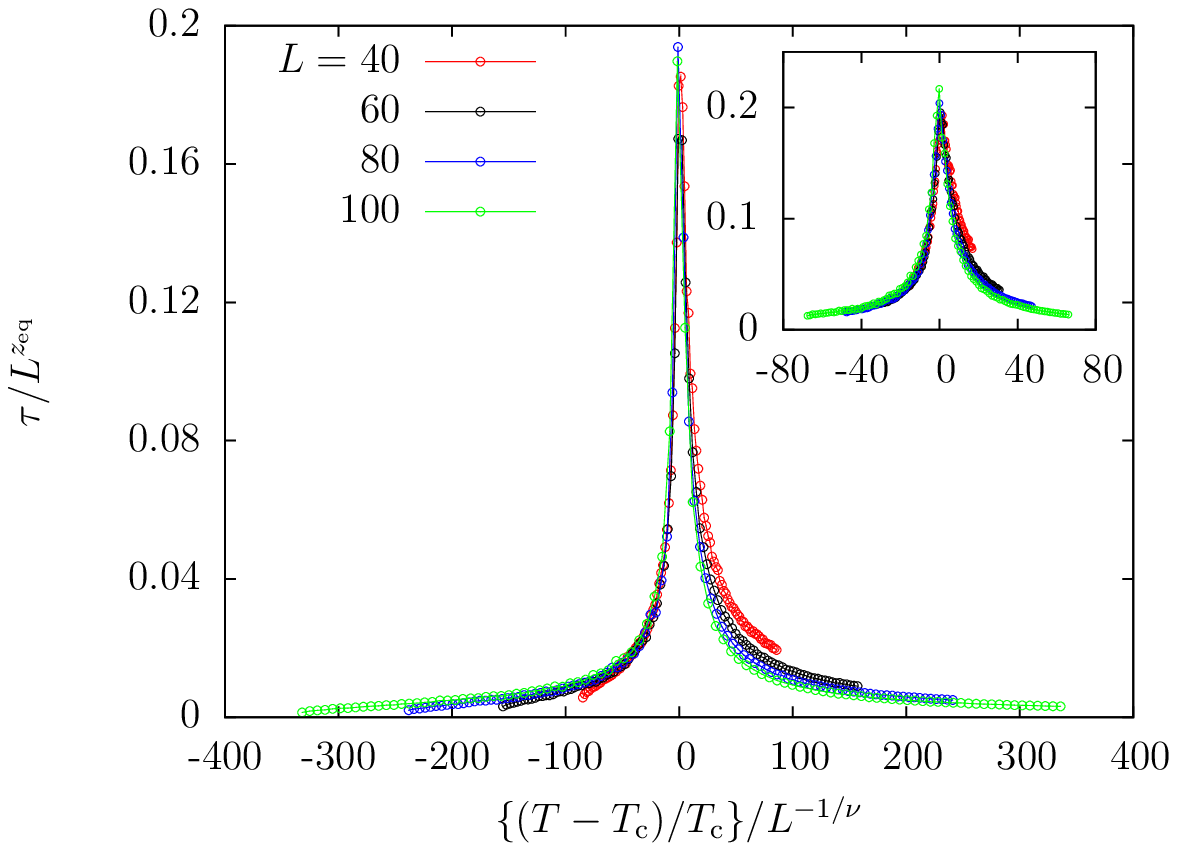}
\end{minipage}
\caption{\label{fig:tau-scale} 
(Color online.)
Finite-size scaling of the relaxation time $\tau$ obtained from (a) the over-damped Langevin 
equation \eqref{eq:over-damped-Langevin}, (b) the under-damped Langevin equation \eqref{eq:under-damped-Langevin}, (c) the ultra-relativistic limit of the 
under-damped Langevin equation \eqref{eq:under-damped-ultra-relativistic}, and (d) the non-relativistic limit of the under-damped Langevin 
equation \eqref{eq:under-damped-non-relativistic}. $z_{\rm eq}=2.1$ and $\nu=0.6703$.
The system sizes are given in the keys. Inserts: zoom over the critical region.}
\end{figure}

Figure~\ref{fig:tau-scale} (a)-(d) shows the finite size scaling of $\tau$ as obtained from 
Eqs.~\eqref{eq:over-damped-Langevin}, \eqref{eq:under-damped-Langevin}, \eqref{eq:under-damped-ultra-relativistic}, 
and \eqref{eq:under-damped-non-relativistic}, respectively.
In these plots we use  $z\sub{eq} \simeq 2.1$ 
and we find acceptable scaling behaviour.  The scaling seems to be 
less dependent on the type of Langevin dynamics than the relaxation time $\tau$ itself, cfr. Fig.~\ref{fig:tau}. 

\section{The vortex observables in equilibrium}
\label{sec:equilibrium-vortex}

In this section we  study the statistical properties of the vortex-loop network in equilibrium.
We start by recalling a number of known results on the statistical properties of line ensembles 
under different conditions. Although we present data for equilibrium configurations generated with the under-damped Langevin 
equation~\eqref{eq:under-damped-Langevin} only, the following results are common to equilibrium data generated with 
all Langevin dynamics.

\subsection{Random geometry of the vortex tangle - background}

The relation between second order thermodynamic phase transitions and
percolation phenomena was established in the late 70s, by using the finite dimensional Ising model of magnetism as a working example. 
In this system the most natural objects to consider are the domains of neighbouring aligned spins. 
Although these percolate and become critical at a threshold,  their critical point  does not necessarily coincide
with the thermodynamic transition~\cite{Muller}, and their scaling properties
do not capture the thermodynamic critical properties of the magnetic system. Instead, 
the thermodynamic instability coincides with a percolation one, and the various critical exponents are linked to those of the 
geometric construction~\cite{Coniglio}, {\it only if} the spin clusters are constructed in a very specific way. The receipt demands
to break the bonds between parallel  spins with a temperature dependent probability, and thus build the so-called Fortuin-Kasteleyn 
clusters~\cite{FK}, with which one can fully characterise the thermodynamic 
phase transition. The extension of this construction to models with a continuous $O(N)$ symmetry 
has been discussed in~\cite{Blanchard00,Fortunato03}.
Apart from providing an alternative way to attack critical phenomena, 
the language of random geometry has been very fruitful in many different contexts, notably in polymer science~\cite{deGennes}, 
and it has helped reaching a better understanding of the behaviour of many physical and mathematical problems
at and away from criticality.

In $3d$ models with continuous symmetry breaking as the one we study here, 
closed line defects or closed vortices are the natural topological objects to consider.
In this and other models with loops the lines undergo a  geometric transition between a ``localised'' phase, 
with only finite length lines, and an ``extended'' phase,  in  which 
a finite fraction of the lines have diverging length in the thermodynamic limit. The actual scaling of their length with the 
system size depends on the boundary conditions. For periodic boundary conditions lines can wrap around
the system many times. As in the Ising model, 
the line-defect geometric transition does not inevitably coincide with the 
thermodynamic one. 
In our study we will confirm that this is not the case for the U(1) relativistic field theory, 
as already shown in~\cite{Kajantie} for the $3d$ XY model and \cite{Bittner} for the O(2) non-relativistic field theory, 
contrary to claims in~\cite{Kohring} in general, in~\cite{AntunesBettencourt,Schakel} in the context of cosmological studies, and 
in~\cite{Nguyen,Camarda} in the field of superfluidity and superconductivity of type II.

The tools to perform a geometric analysis of individual lines and ensembles of lines are well established
and have been very successful in the field of polymer science, see for example~\cite{deGennes}. 
At a critical point, be it thermodynamic or geometric, the clusters or lines that characterise criticality 
satisfy several scaling relations. We recall some of them below.

The linear length along the loop, $l$, and the radius of the smallest sphere that contains the loop, $R$,
are related by~\cite{deGennes,Mandelbrot}
\begin{equation}
l \simeq R^{D}
\label{eq:def-fractal}
\end{equation}
in the limit $l\gg a$ with $a$ a microscopic length-scale. 
$R$ can also be the mean-square end-to-end distance or the radius of gyration of the loop.
$D$ is the fractal Hausdorff dimension of the line ($D=1$ for a smooth line). 
In the thermodynamic limit the number density of vortex loops with length $l$ should behave as~\cite{Stauffer} 
\begin{equation}
P(l) \simeq  l^{-\alpha\sub{L}} e^{- l m\sub{L}},
\label{eq:dist-threshold}
\end{equation}
with $m\sub{L}$ the line tension and $\alpha\sub{L}$ the so-called Fisher exponent. (This form should be corrected
by system-size dependent terms to capture finite size corrections.)
The line tension vanishes at criticality as 
\begin{equation}
m\sub{L} \simeq |T-T\sub{L}|^{\beta\sub{L}}
\end{equation}
with $\beta\sub{L}$ another characteristic exponent.
By requiring that the average number of loops per unit area, with radius of the order of $R$, scales as
$n(R) \simeq R^{-d}$ and equating this law to the result of computing $n(R)\simeq \int_{R^{D}}^\infty ds \ s^{-\alpha\sub{L}}$
one finds~\cite{KondevHenley}
\begin{equation}
\alpha\sub{L} = 1+\frac{d}{D}
\; . 
\label{eq:scaling-relation}
\end{equation} 
Other scaling arguments, that use the algebraic decay of correlation functions at criticality,
allow one to relate $D$ and $\alpha\sub{L}$ to the anomalous dimension of the field in the field theory 
that characterises the statistical properties at criticality. More precisely,
$D=(5-\eta)/2$ and  $\alpha\sub{L}=(11-\eta)/(5-\eta)$~\cite{NahumPRE}, satisfying 
(\ref{eq:scaling-relation}). (Another quantity that is often used in the literature 
is the probability that a line that passes through a chosen link
had length $l$, and it is given by $P_{\rm link}(l) \simeq l^{1-\alpha\sub{L}}$ at criticality.)

Some known values of the fractal dimension and exponent $\alpha\sub{L}$ in three dimensions are:
\begin{itemize}
\item {\it Gaussian random walks}. $D=2$ and $\alpha\sub{L}=5/2$.
This result was found in dense polymer solutions~\cite{deGennes} 
and the initial state of a cosmic string network as modelled in~\cite{Vachaspati,Strobl}.
\item
{\it Self-avoiding random walks}. In $d=3$ the Flory approximation~\cite{Flory} yields $D=5/3$ [$D=3/(d+2)$ in general $d$]
and $\alpha\sub{L}=14/5$. The numerical values for this problem are very close to these $D\simeq 1.7$ and  
$\alpha\sub{L} \simeq 2.76 > 5/2$~\cite{Havlin,Sokal}. 
\item 
{\it Self-seeking random walks}.
These are walks such that $\alpha\sub{L} < 5/2$. 
\item
{\it Coulomb phase in spin-ice}. 
Loops that are shorter than $L^2$ behave as Gaussian random walks.
Loops that are longer than $L^2$ wrap around the system many times, occupy a finite fraction of the system's volume, and 
for them $\alpha\sub{L}=1$~\cite{Jaubert}.
\item
{\it Fully-packed loop models}. These are general models on a lattice 
with various symmetries and loop fugacity (a colour variable), $n$, as a free parameter.
Their field theory representation 
is given by CP$^{n-1}$ models for oriented loops~\cite{Nahum-book,Nahum,NahumPRE,Nahum2,Ortuno09,Barp}. These models also 
present a crossover from Gaussian statistics for $l \ll L^2$ to a more complex function of 
$l$ and $L$ for $l\gg L^2$ 
that depends on the symmetry of the model and $n$. 
For $n=1$ and $l$ not too close to $L^3$, $\alpha\sub{L}=1$.
\end{itemize}

Interestingly enough, we will see some of these statistics emerging in different length and time regimes of the 
U(1) model.

\subsection{Plaquette vorticity}

We consider all $3 L^3$ unit plaquettes  in the cube:
$L^2$ plaquettes along the $yz$-plane with four corners at
 $(\Vec{x}, \Vec{x}+\Delta x \, \hat e_2, \Vec{x}+\Delta x \, \hat e_2 + \Delta x \, \hat e_3, \Vec{x}+ \Delta x \, \hat e_3)$,
those along the $zx$-plane with vertices at
 $(\Vec{x}, \Vec{x}+\Delta x \, \hat e_3, \Vec{x}+\Delta x \, \hat e_1 + \Delta x \, \hat e_3, \Vec{x}+ \Delta x \, \hat e_1)$
and those along the $xy$-plane
 $(\Vec{x}, \Vec{x}+\Delta x \, \hat e_1, \Vec{x}+\Delta x \, \hat e_1 + \Delta x \, \hat e_2, \Vec{x}+ \Delta x \, \hat e_2)$.
The quantity 
\begin{eqnarray}
v_{\Vec{x}}^{1} &=& \frac{1}{2\pi} 
\bigg( 
[ \theta_{\Vec{x} +\Delta x \, \hat e_2} - \theta_{\Vec{x}}  ]_{2\pi}
+
[ \theta_{\Vec{x} +\Delta x \, \hat e_2 + \Delta x \, \hat e_3} - \theta_{\Vec{x} +\Delta x \, \hat e_2} ]_{2\pi}
\nonumber\\
&& 
\qquad 
+
[ \theta_{\Vec{x} +\Delta x \, \hat e_3} - \theta_{\Vec{x} +\Delta x \, \hat e_2 + \Delta x \, \hat e_3}  ]_{2\pi}
+
[\theta_{\Vec{x}} - \theta_{\Vec{x} +\Delta x \, \hat e_3 }]_{2\pi}
\bigg)
\end{eqnarray}
measures the vorticity of the plaquette
$(\Vec{x}, \Vec{x}+\Delta x \, \hat e_2, \Vec{x}+\Delta x \, \hat e_2 + \Delta x \, \hat e_3, \Vec{x}+ \Delta x \, \hat e_3)$.
The $\theta$'s are the phases of the field $\psi$ at the corners of the plaquette and $[\alpha]_{2\pi}$ is the angle 
$\alpha$ modulo $2\pi$, i.e.
$[\alpha]_{2\pi} = \alpha + 2 \pi n$ with $n$ an integer such that $[\alpha]_{2\pi} \in (-\pi, \pi]$.
In this way, a dual oriented linear object is assigned to each  plaquette with 
$v=1$ or $-1$. These oriented line elements join  to form closed vortex loops.
In practice, we decide whether a vortex pierces the plaquette by calculating the flux or winding number
\begin{align}
v_{\Vec{x}}^{1} \equiv \ & \
\ \frac{1}{2 \pi} \ \mathrm{Im} \bigg[ \log \bigg( \frac{\psi_{\Vec{x}+\Delta x \, \hat e_2}}{\psi_{\Vec{x}}} \bigg) + 
\log \bigg( \frac{\psi_{\Vec{x}+\Delta x \, \hat e_2+\Delta x \, \hat e_3}}{\psi_{\Vec{x}+\Delta x \, \hat e_2}} \bigg) 
\nonumber\\
& 
\qquad\qquad
+ \log \bigg( \frac{\psi_{\Vec{x}+\Delta x \, \hat e_3}}{\psi_{\Vec{x}+\Delta x \, \hat e_2+\Delta x \, \hat e_3}} \bigg) + 
\log \bigg( \frac{\psi_{\Vec{x}}}{\psi_{\Vec{x}+\Delta x \, \hat e_3}} \bigg) \bigg]
\label{eq:plaquette-flux}
\end{align}
where $\mathrm{Im}[\log(\psi_B / \psi_A)]$ gives the phase difference $\theta_{AB} \equiv \theta_B - \theta_A + F_{AB}$ 
for the two complex values $\psi_{A,B} \equiv |\psi_{A,B}| e^{i \theta_{A,B}}$.
The phases $\theta_{A,B}$ and the phase difference $\theta_{AB}$ are defined in the range $(-\pi,\pi]$ and the function $F_{AB}$ is given by
\begin{align}
F_{AB} = 
\left\{ \begin{array}{clrcl}
2 \pi & \text{for } & - 2 \pi &< \theta_{B} - \theta_{A} \leq & - \pi \\
0 & \text{for } & - \pi  & < \theta_{B} - \theta_{A} \leq & \pi \\
- 2 \pi & \text{for } & \pi & < \theta_{B} - \theta_{A} < & 2 \pi
\end{array} \right. 
\label{eq:phase-difference}
\end{align}
The flux $v_{\Vec{x}}^1$ takes the form of $v_{\Vec{x}}^1 = (\theta_{AB} + \theta_{BC} + \theta_{CD} + \theta_{DA}) / (2 \pi)$, 
and the phase difference $\theta_{DA}$ in the range $(-\pi,\pi]$ becomes
\begin{align}
\theta_{DA} = \left\{ \begin{array}{llrcl}
- 2 \pi - (\theta_{AB} + \theta_{BC} + \theta_{CD}) & \text{for } & - 3 \pi & < \theta_{AB} + \theta_{BC} + \theta_{CD} & < - \pi \\
- (\theta_{AB} + \theta_{BC} + \theta_{CD}) & \text{for } & - \pi & \leq \theta_{AB} + \theta_{BC} + \theta_{CD} & < \pi \\
2 \pi - (\theta_{AB} + \theta_{BC} + \theta_{CD}) & \text{for } & \pi & \leq \theta_{AB} + \theta_{BC} + \theta_{CD} & < 3 \pi \\
4 \pi - (\theta_{AB} + \theta_{BC} + \theta_{CD}) & \text{for } & \pi &  = \theta_{AB} = \theta_{BC} = \theta_{CD} &
\end{array} \right. 
 \label{eq:flux-condition}
\end{align}
The flux $v_{\Vec{x}}^1$ equals $-1$, $0$, $1$, and $2$ for the first, second, third, and fourth lines in 
Eq.~\eqref{eq:flux-condition}, respectively.
The flux with $v_{\Vec{x}}^1 = 2$ quite rarely arises because three phase differences 
$\theta_{AB}$, $\theta_{BC}$, and $\theta_{CD}$ should be equal to $\pi$ for this to occur,
as shown in Eq.~\eqref{eq:flux-condition}. In the cubic lattice geometry, it is impossible 
to have more than two unit fluxes threading a plaquette.
In other rare cases, the flux $v_{\Vec{x}}^1$ can take fractional values $0 < |v_{\Vec{x}}^1| < 1$ when 
vortex cores just touch one of the four vertices or the sides of plaquettes.
We have never encountered the values $v_{\Vec{x}}^1 = 2$ and $0 < |v_{\Vec{x}}^1| < 1$
in our simulations.
In the same way, we define the fluxes $v_{\Vec{x}}^{2}$ and $v_{\Vec{x}}^{3}$ for the plaquettes
 $(\Vec{x}, \Vec{x}+\Delta x \, \hat e_3, \Vec{x}+\Delta x \, \hat e_1 + \Delta x \, \hat e_3, \Vec{x}+ \Delta x \, \hat e_1)$
 and
 $(\Vec{x}, \Vec{x}+\Delta x \, \hat e_1, \Vec{x}+\Delta x \, \hat e_1 + \Delta x \, \hat e_2, \Vec{x}+ \Delta x \, \hat e_2)$,
respectively.

When $v_{\Vec{x}}^{a}$ takes the value 1, a vortex element with length $\Delta x$ along the 
$\hat{e}_a$-direction pierces the centre of the plaquette from
$(x+\Delta x/2, y+\Delta x/2, z+\Delta x/2) - \Delta x \hat{e}_a$ to $(x+\Delta x/2, y+\Delta x/2, z+\Delta x/2)$.
The direction of the vortex line is reversed in the case $v_{\Vec{x}}^{a}=-1$.

\subsection{Averaged vortex density}

The total vortex length in the system is, therefore, proportional to the total number of plaquettes with non-vanishing flux,  
$\sum_{j,k,l} \sum_{a=1}^3 |v_{\Vec{x}}^a| \neq 0$, and the averaged vortex density $\rho\sub{vortex}$ is defined as
\begin{align}
\rho\sub{vortex} \equiv 
\frac{1}{3L^3} \Bigg\langle \sum_{j,k,l} \sum_{a=1}^3 |v_{\Vec{x}}^a| \Bigg\rangle\sub{stat} . \label{eq:vortex-density}
\end{align}
(Note that this quantity depends on the size of the space discretisation used, $\Delta x$, see App.~\ref{app:mesh} 
and~\cite{Hindmarsch,Kajantie}, for example.)

\vspace{0.25cm}

\begin{figure}[tbh]
\centering
\begin{minipage}{0.49\linewidth}
\centering
(a) \\
\includegraphics[width=0.95\linewidth]{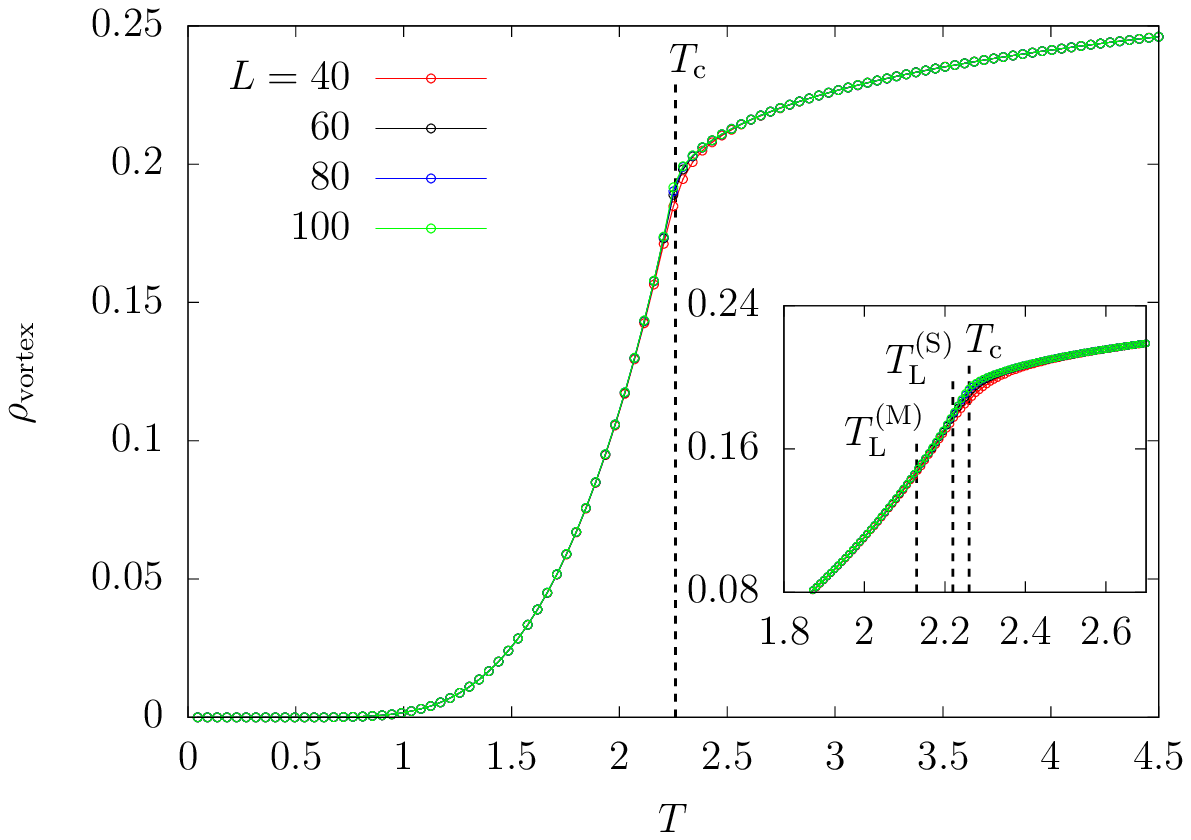}
\end{minipage}
\begin{minipage}{0.49\linewidth}
\centering
(b) \\
\includegraphics[width=0.95\linewidth]{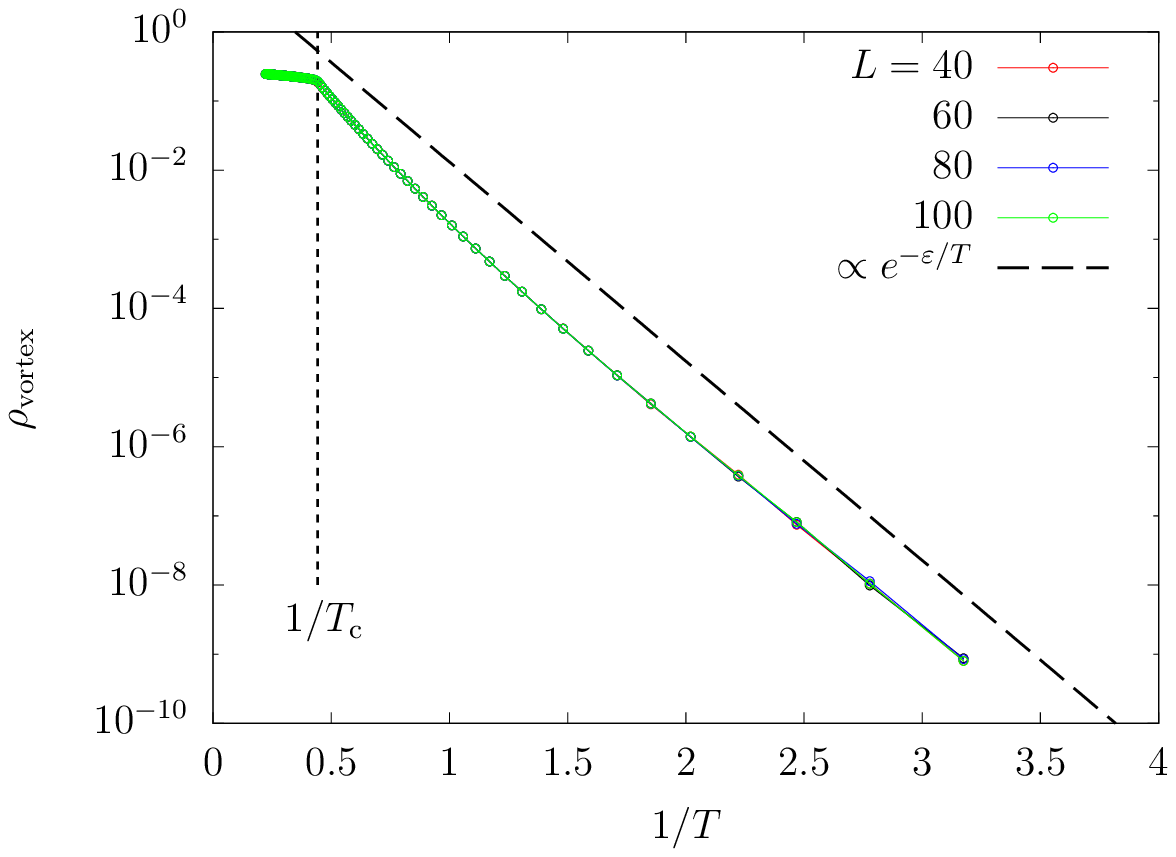}
\end{minipage}
\caption{\label{fig:vortex-all} 
(Color online.)
Dependence of the averaged vortex density $\rho\sub{vortex}$ on 
(a)  temperature and (b) inverse temperature for different system sizes $L$. The critical temperature $T\sub{c}$ 
is indicated with a vertical dotted line in both panels. In the inset to panel (a) we zoom over a temperature interval 
around $T\sub{c}$. The meaning of the temperatures $T\sub{L}\up{(M)}$ and $T\sub{L}\up{(S)}$ will be 
explained below.
The fitting parameter $\varepsilon \simeq 6.63$ for $\rho\sub{vortex} \propto e^{- \varepsilon / T}$ is the 
thermal activation energy for small vortex rings that describes $\rho\sub{vortex}$ at small $T$, shown with a dashed
line in panel~(b).
The equilibrium configurations used in this and all other figures in this subsection
are generated with the under-damped Langevin equation~\eqref{eq:under-damped-Langevin}.
}
\end{figure}

Figure~\ref{fig:vortex-all} shows the dependence of the averaged vortex density 
$\rho\sub{vortex}$ on (a) temperature and 
(b) inverse temperature in linear and linear-log scales, respectively.
$\rho\sub{vortex}$ monotonically increases as a function of temperature. 
From panel (a) one could argue that $\rho\sub{vortex}(T)$ changes concavity at $T\sub{c}$. 
(We have checked that this feature does not change with a different value of $\Delta x$,
although the values of the critical temperature, vortex density and activation energy do change, for example, 
for $\Delta x =2$,  
$T\sub{c} \simeq 2.67$, $\rho\sub{vortex}(T\sub{c}) \simeq 0.18$, $\varepsilon \simeq 8.85$.)
The value  $\rho\sub{vortex}(T_c) \simeq 0.2$ is 
close to the value $0.16$ measured in~\cite{Kajantie} for the $3d$ XY model at its thermodynamic
instability.  We expect the averaged vortex density to approach $\rho\sub{vortex} \to 1/3$ 
in the  infinite temperature limit, $T \to \infty$, at which the phase of the complex field $\psi$ is completely random in time and space, 
see App.~\ref{sec:inf-temp-rho}.
We checked this claim numerically 
obtaining $\rho\sub{vortex} \simeq 0.3332$ for $\Delta x = 1$ and $\rho\sub{vortex} \simeq 0.3333$ for $\Delta x = 2$ 
at  $T = 20 \ T\sub{c}$. 
(Vachaspati and Vilenkin~\cite{Vachaspati} find a different vortex density, $\rho\sub{vortex}=0.29$, for a random configuration
since they use a ``clock" model in which the phase takes only three values and are assigned at random on each 
lattice site.) We observe that 
$\rho\sub{vortex}$   depends very little on the system size $L$. At low temperature, the behaviour is activated, with $\rho\sub{vortex}\propto
e^{-\epsilon/T}$ and $\epsilon\simeq 6.63$, see panel (b).

\vspace{0.25cm}

\begin{figure}[tbh]
\centering
\includegraphics[width=0.75\linewidth]{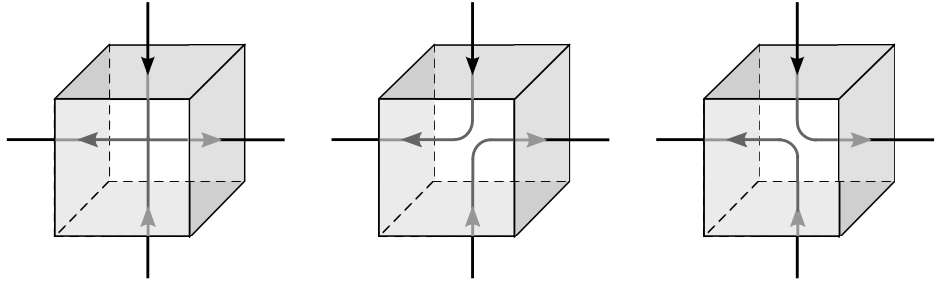}
\caption{\label{fig:vortex-visualization} 
(Color online.)
Example of a unit cube comprising four plaquettes with (shaded) and two plaquettes 
without (not shaded) non-zero flux, respectively.
The vortex elements are shown with arrows and their directions indicate the sign of the fluxes. 
When four vortex elements pierce a unit cube (left), we face an ambiguity in the two ways of connecting them (mid and right images).
Two ways of resolving this ambiguity are explained in the text and in Fig.~\ref{fig:maximal-criterium}.}
\end{figure}

\subsection{The vortex line lengths}

\begin{figure}[tbh]
\vspace{0.75cm}
\centering
\includegraphics[width=0.75\linewidth]{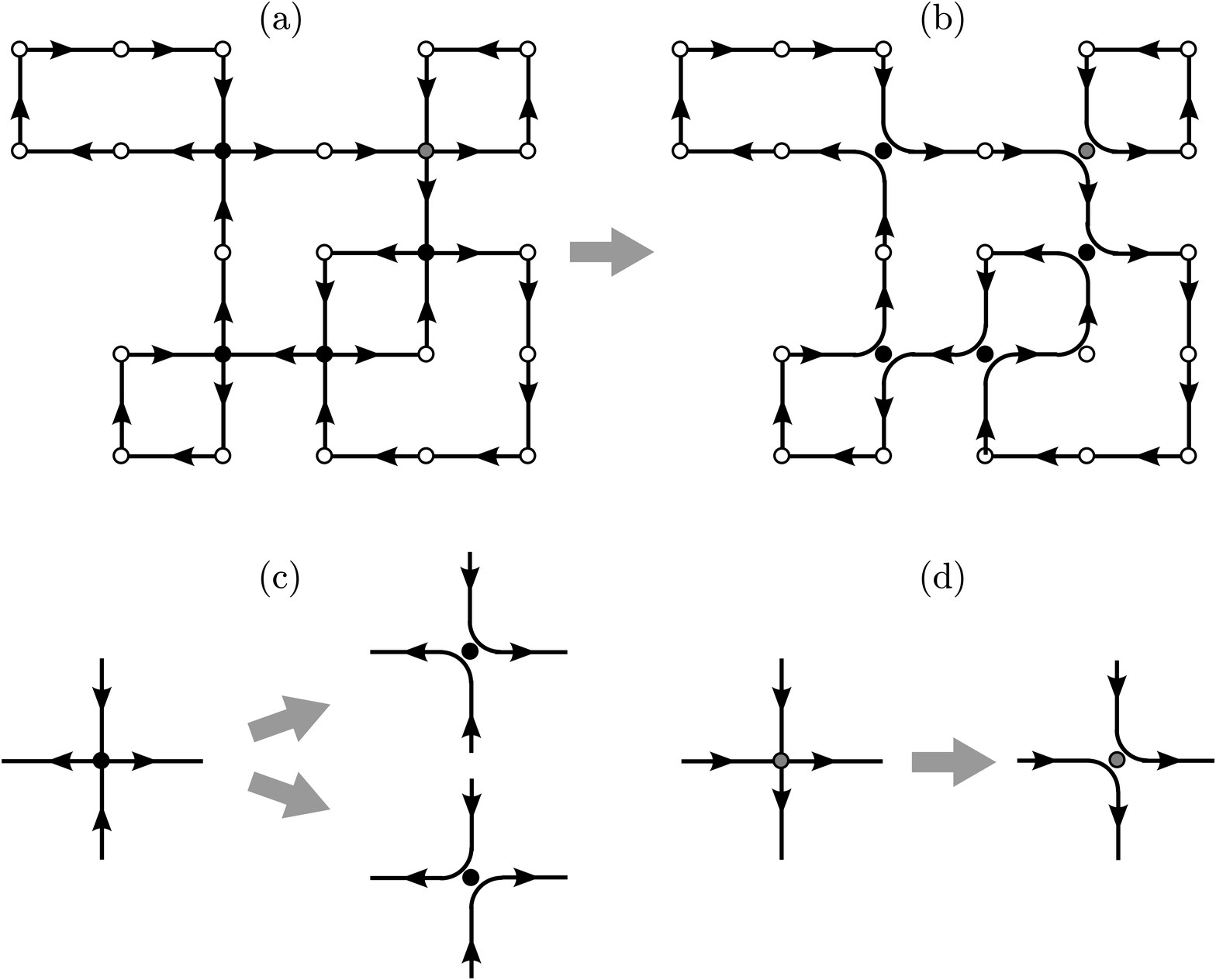} \\
\caption{\label{fig:maximal-criterium} 
(Color online.)
Reconnection of  vortex elements using the maximal criterium.
Black, and grey coloured circles show the centers of the unit cubes with a pair of ingoing and outgoing vortex elements while the white
circles show those with just one ingoing and one outgoing arrow.
In the case of the black circles there are two possible ways of connecting the two ingoing and two outgoing
vortex elements (see Fig.~\ref{fig:vortex-visualization} and panel (c)), whereas in the case of the grey circles the connection is unique
since crossing vortex lines is not allowed (see panel (d)). In practice, 
we first draw all vortex line elements passing through the centres of the cubes (see panel (a)),
and we then select the connection of elements at all black circles in such a way that the length of the total vortex loop is maximised (panel (b)).
At the grey vertices there is no choice (see panel (d)) and the connection may lead to the  separation into two loops, as shown in the 
example in panels (a) and (b).  
}
\end{figure}

We now consider the length $l$ of {\it vortex loops}.
As we discussed above, we place straight vortex line elements at the centres of all plaquettes 
with non-zero flux $|v_{\Vec{x}}^a| = 1$ and we 
connect them with the constraint of not crossing the lines. 
The length of each loop is even in units of $\Delta x$ and the minimal length is $4 \Delta x$.
When four or more plaquettes in one unit cube have non-zero flux, i.e., four or more vortex elements pierce the cube, 
we have to decide how to connect the vortex elements. This is shown in Fig.~\ref{fig:vortex-visualization}.
Several criteria to connect vortex elements are discussed in Refs.~\cite{Kajantie,Bittner}. We adopt 
the maximal and stochastic ones.
The connection is uniquely done so that the vortex loops are joined as much as possible to form a long vortex loop in the maximal criterion, 
see Fig.~\ref{fig:maximal-criterium}, while the connection is done at random with equal 
probability among all possible ways to connect them in the stochastic criterion.
These crossings give rise to vortex recombination.
We verified that all vortices take the form of a closed loop as we expect from the topological prospect for vortices.

\begin{figure}[tbh]
\vspace{0.75cm}
\centering
\begin{minipage}{0.24\linewidth}
\centering
(a) \\
\includegraphics[width=0.95\linewidth]{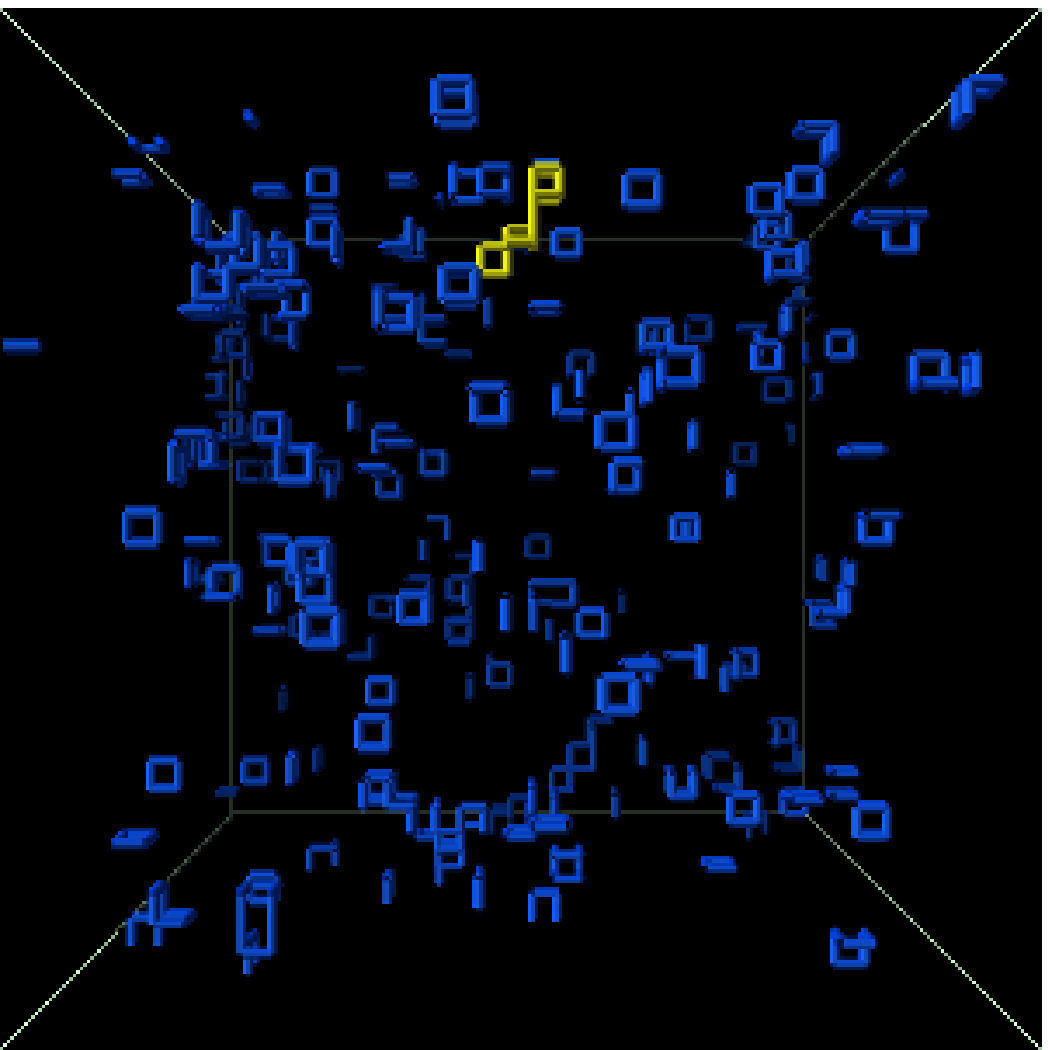} \\
\includegraphics[width=0.95\linewidth]{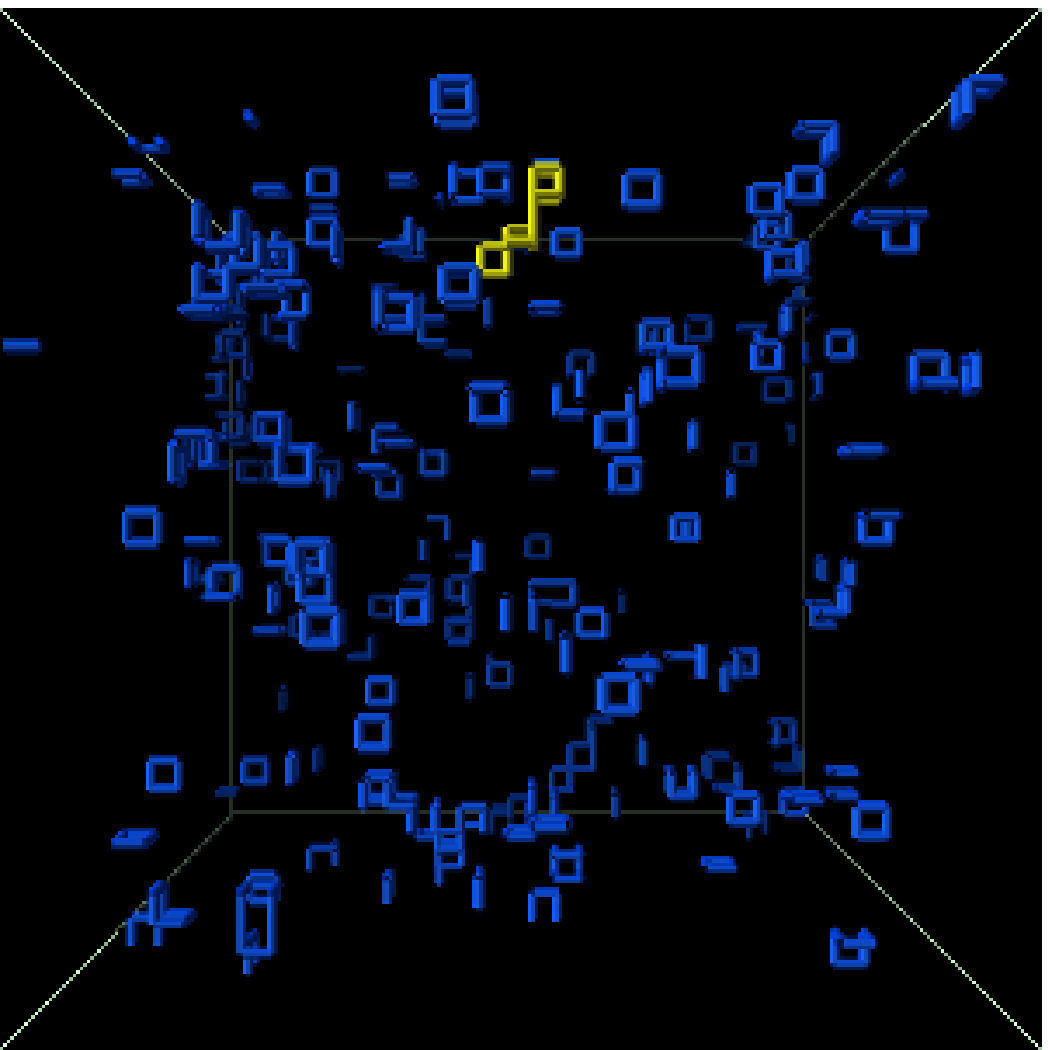}
\end{minipage}
\begin{minipage}{0.24\linewidth}
\centering
(b) \\
\includegraphics[width=0.95\linewidth]{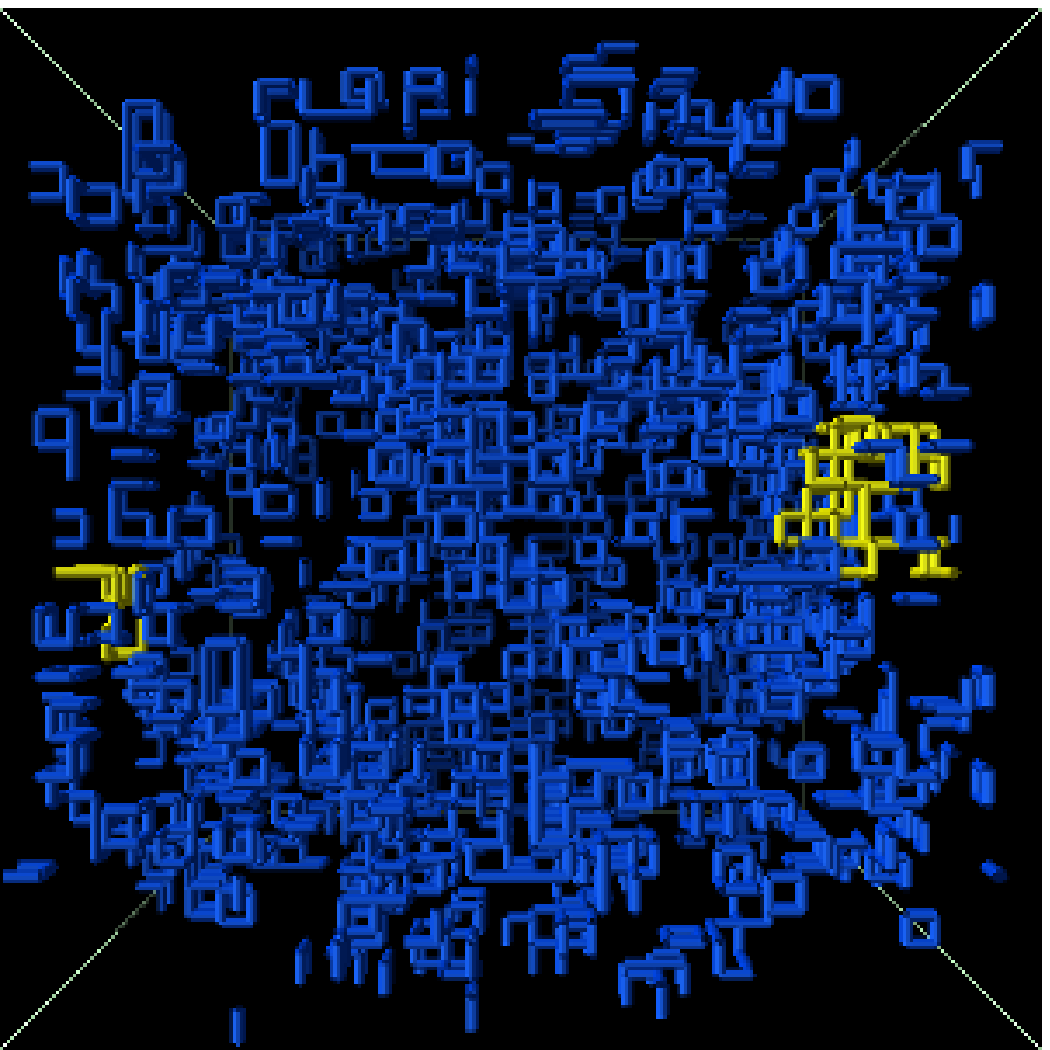} \\
\includegraphics[width=0.95\linewidth]{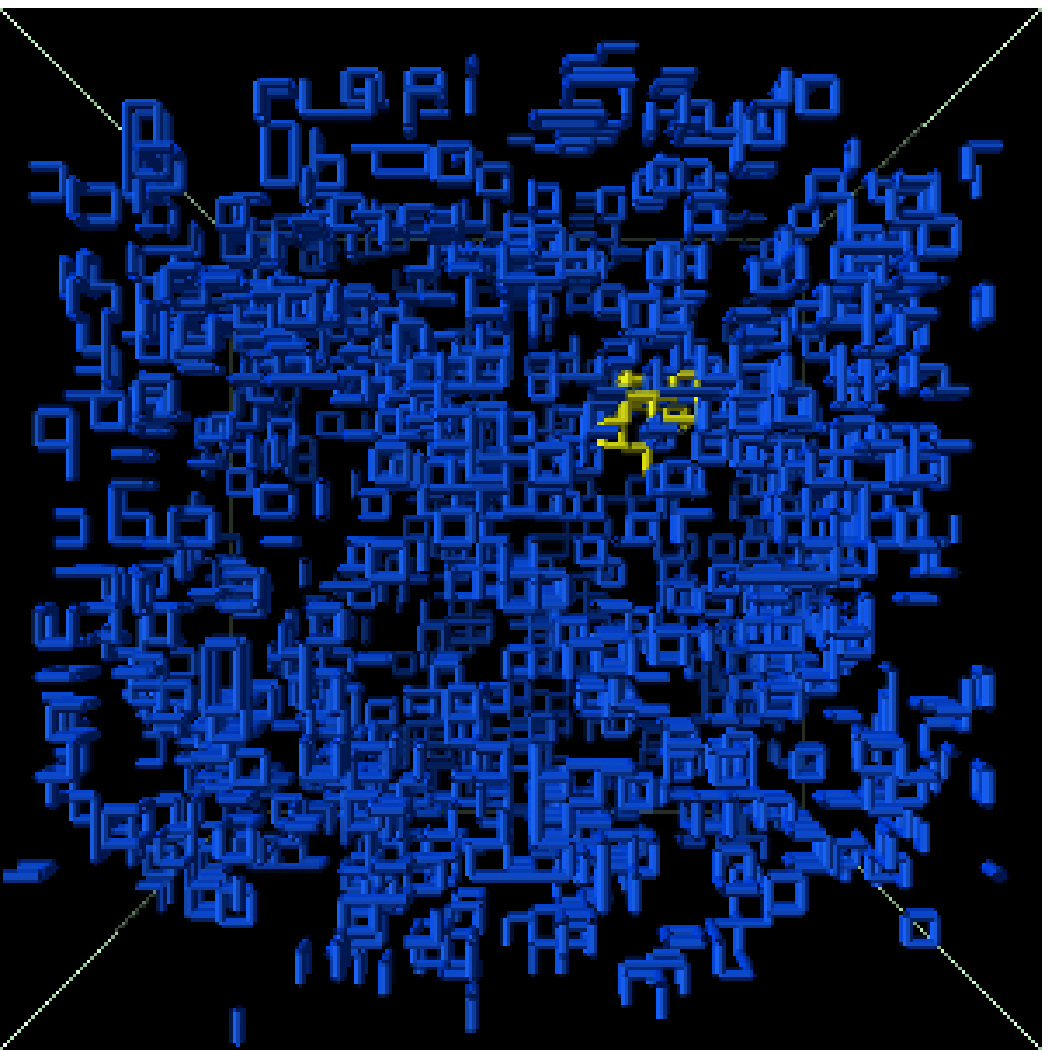}
\end{minipage}
\begin{minipage}{0.24\linewidth}
\centering
(c) \\
\includegraphics[width=0.95\linewidth]{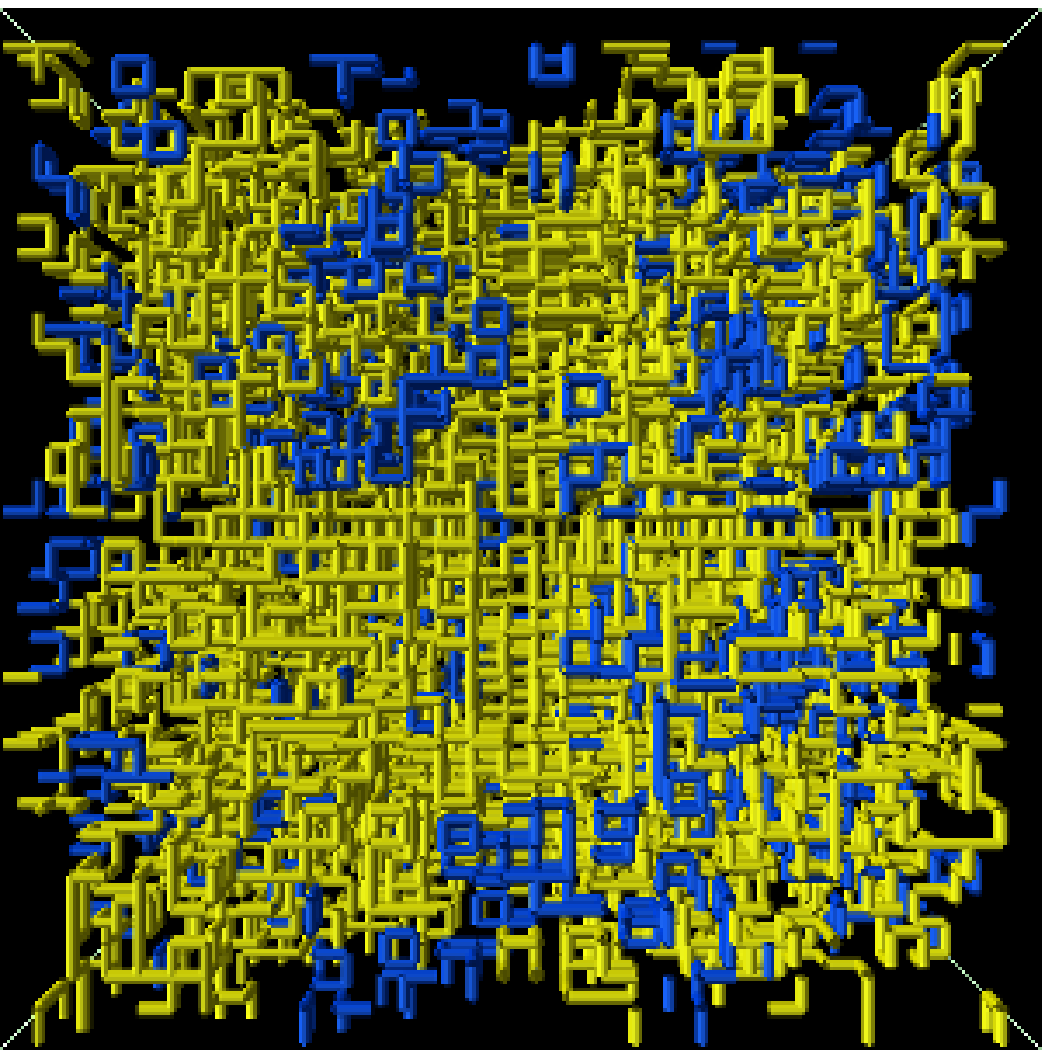} \\
\includegraphics[width=0.95\linewidth]{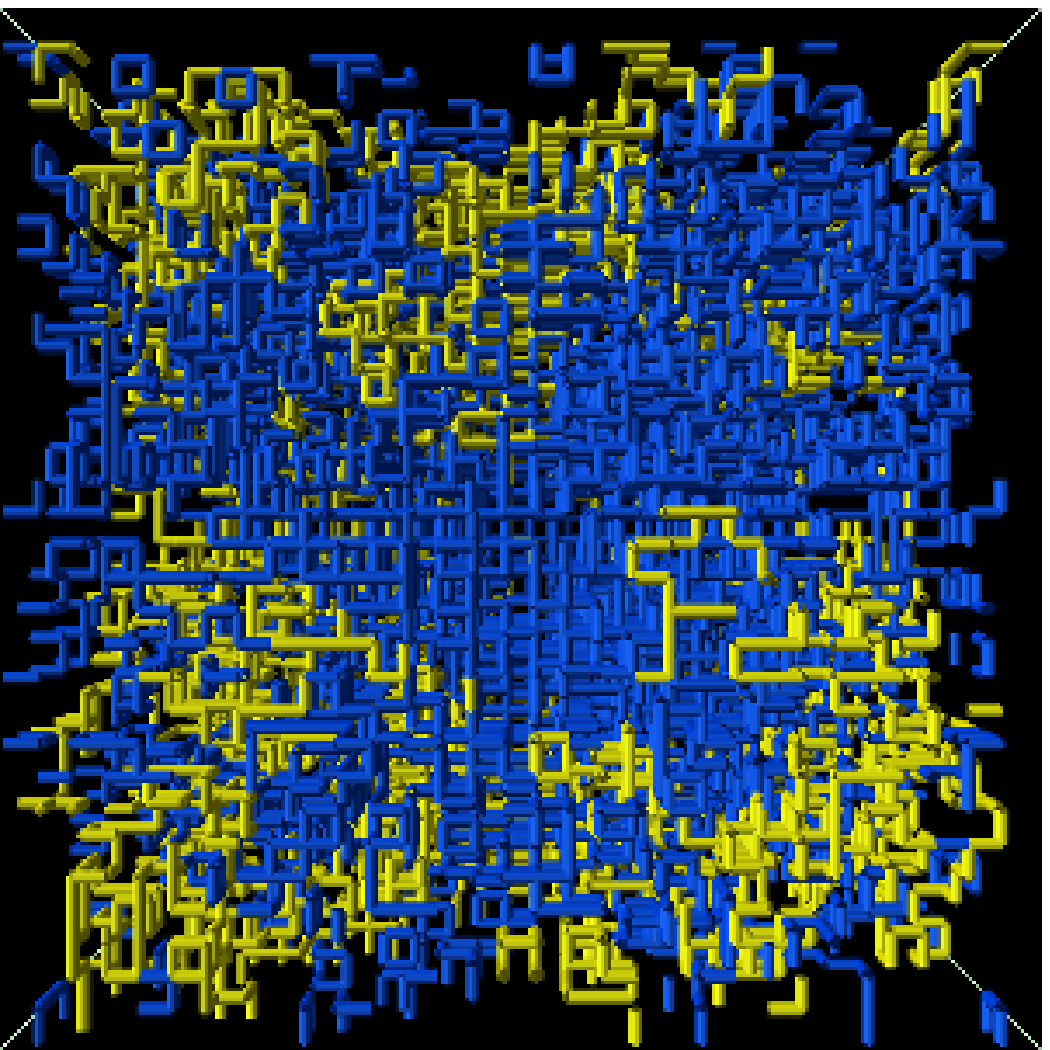}
\end{minipage}
\begin{minipage}{0.24\linewidth}
\centering
(d) \\
\includegraphics[width=0.95\linewidth]{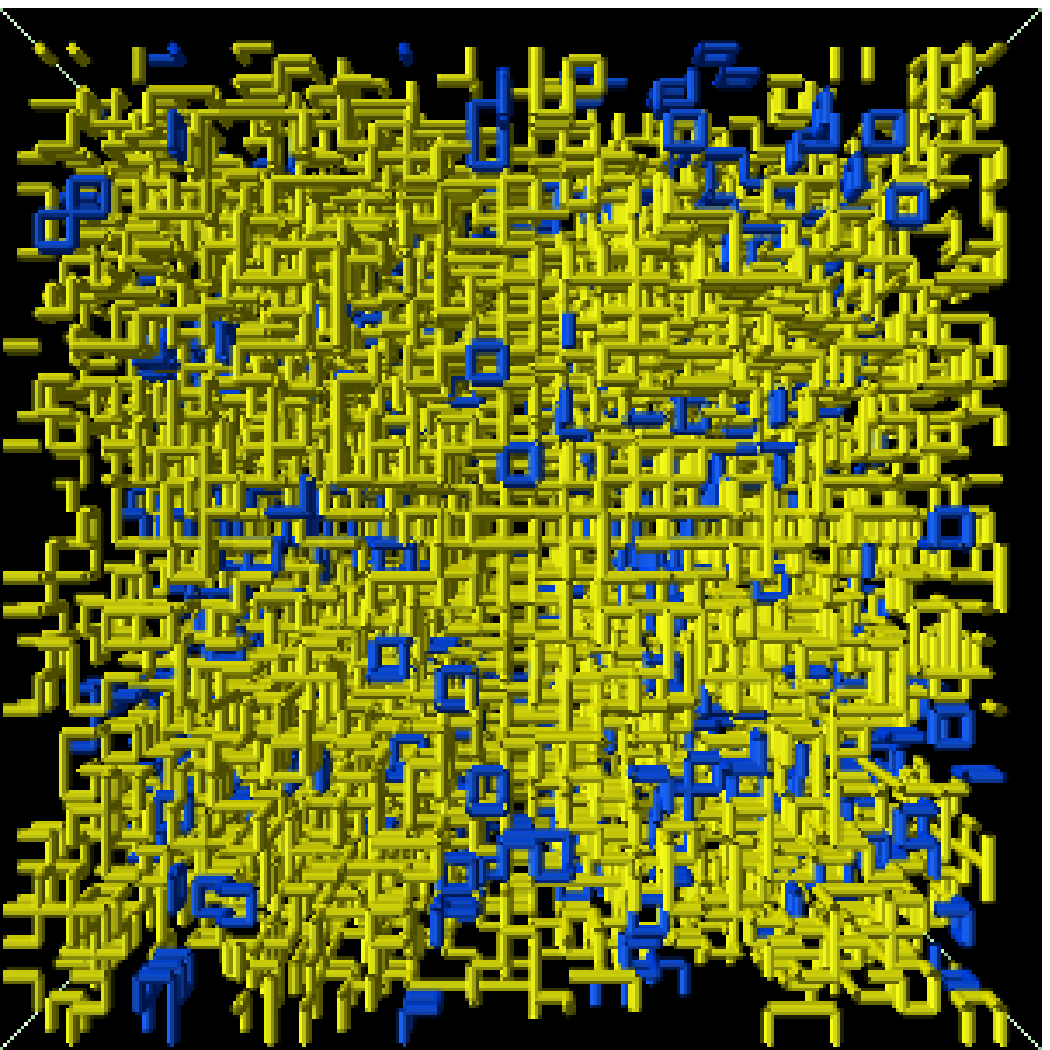} \\
\includegraphics[width=0.95\linewidth]{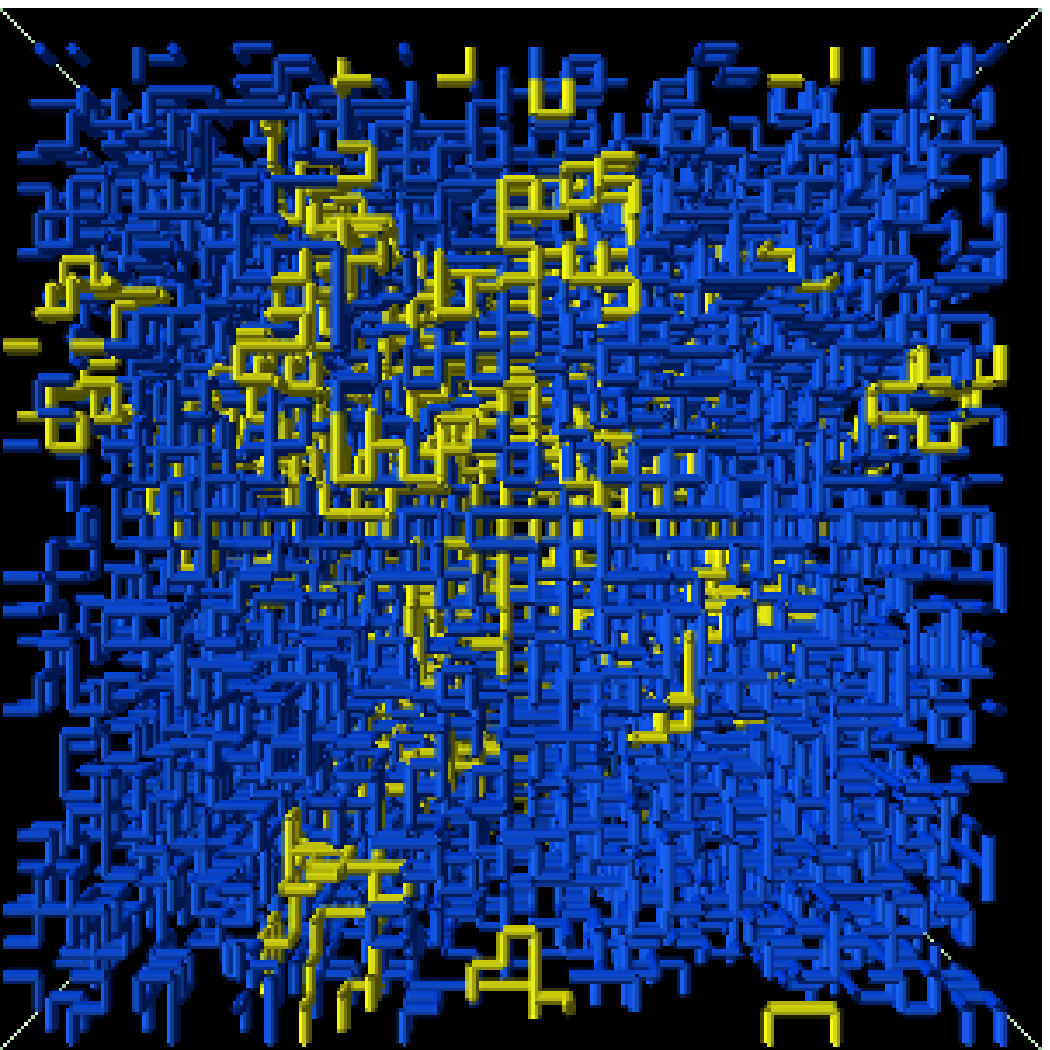}
\end{minipage}
\vspace{0.5cm}
\caption{\label{fig:vortex-location} 
(Color online.)
Equilibrium snapshots of the system configurations at temperatures (a) 
$T = 0.6 \ T\sub{c}$, (b) $T = 0.8 \ T\sub{c}$, (c) $T = T\sub{c}$, and (d) $T = 1.2 \ T\sub{c}$ 
in a system with  size $L = 40$. The vortex line elements are connected with the maximal criterion (upper panels) and 
the stochastic criterion (lower panels) and
they are shown in grey (blue) in the black background. The longest vortex lines in each image are highlighted (in yellow).
}
\end{figure}

Figures~\ref{fig:vortex-location} (a)-(d) show snapshots of equilibrium system configurations where the vortex line elements at the 
centre of the plaquettes with non-zero flux are highlighted. The temperatures of the different snapshots are 
 $T = 0.6 \ T\sub{c}$, $T = 0.8 \ T\sub{c}$, $T = T\sub{c}$, and $T = 1.2 \ T\sub{c}$, from left to right.
Upper and lower panels show the same configurations with the vortex elements connected with the maximal criterion (upper panels)
and the stochastic criterion (lower panels).
At low temperatures, the way in which the elements are connected is irrelevant as the vortex rings are very short,  as shown in 
Fig.~\ref{fig:vortex-location}~(a). We checked that these vortices are rapidly created by thermal fluctuations as small vortex rings
and they are soon annihilated.
It should be hard to experimentally observe such vortices due to the fact that their dynamics occur in
very short time scales.  Accordingly, 
$\rho\sub{vortex}$ is well-fitted by the Arrhenius law $\rho\sub{vortex} \propto e^{- \varepsilon / T}$ 
with the activation energy $\varepsilon \simeq 6.63$ as shown in Fig.~\ref{fig:vortex-all}~(b).

\begin{figure}[tbh]
\vspace{0.75cm}
\centering
\begin{minipage}{0.49\linewidth}
\centering
(a)\\
\includegraphics[width=0.95\linewidth]{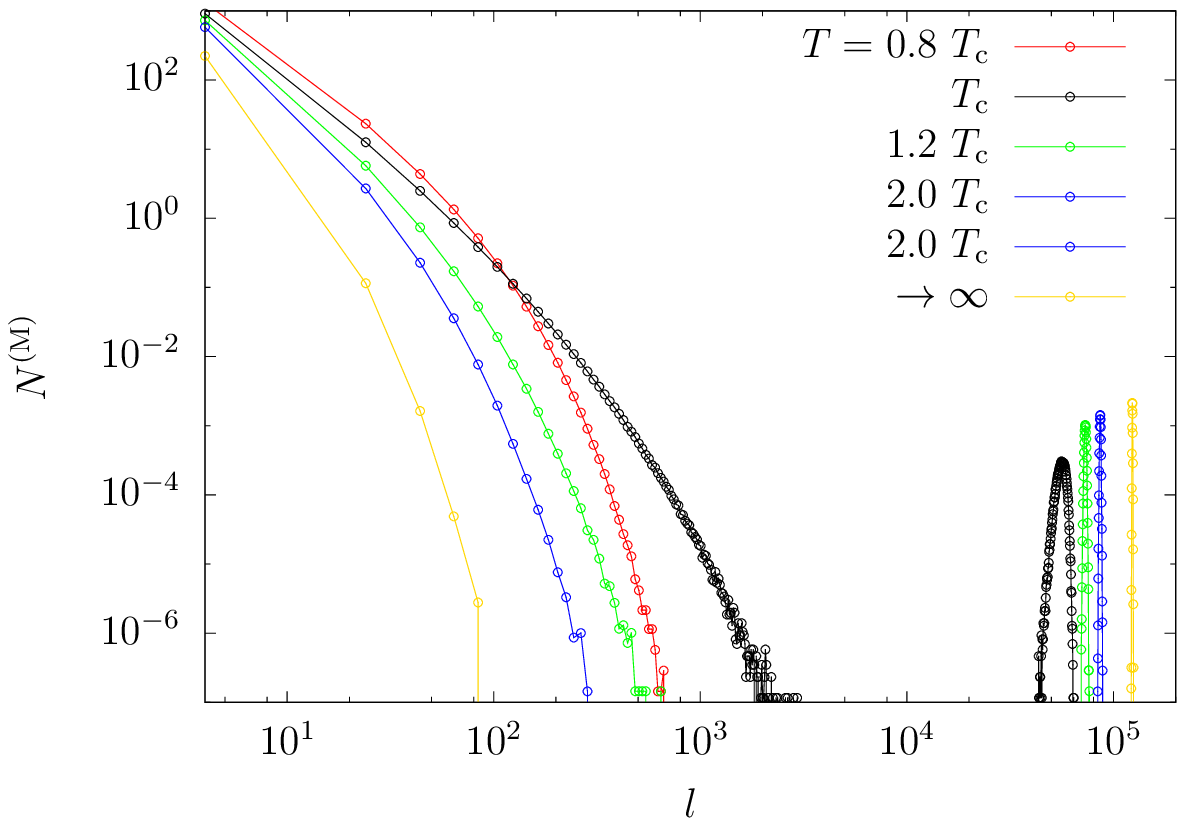} 
(c) \\
\includegraphics[width=0.95\linewidth]{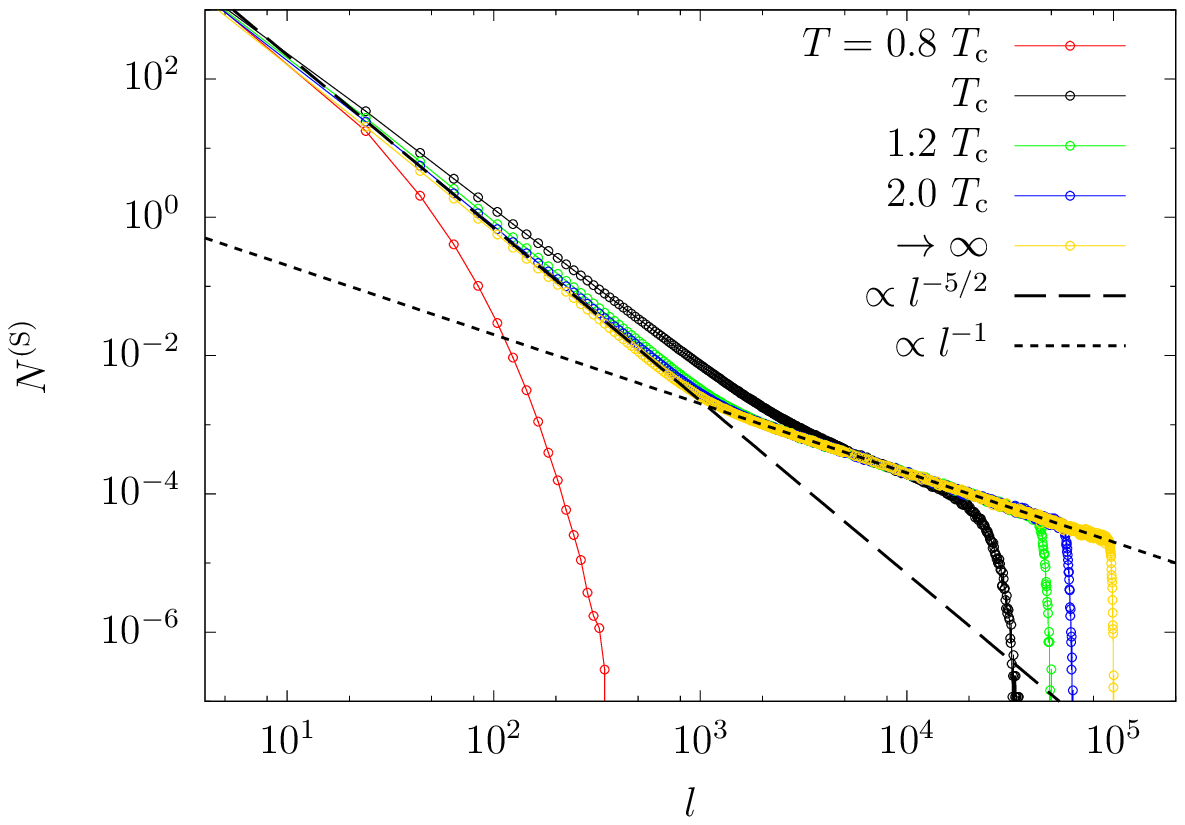}
\end{minipage}
\begin{minipage}{0.49\linewidth}
\centering
(b)\\
\includegraphics[width=0.95\linewidth]{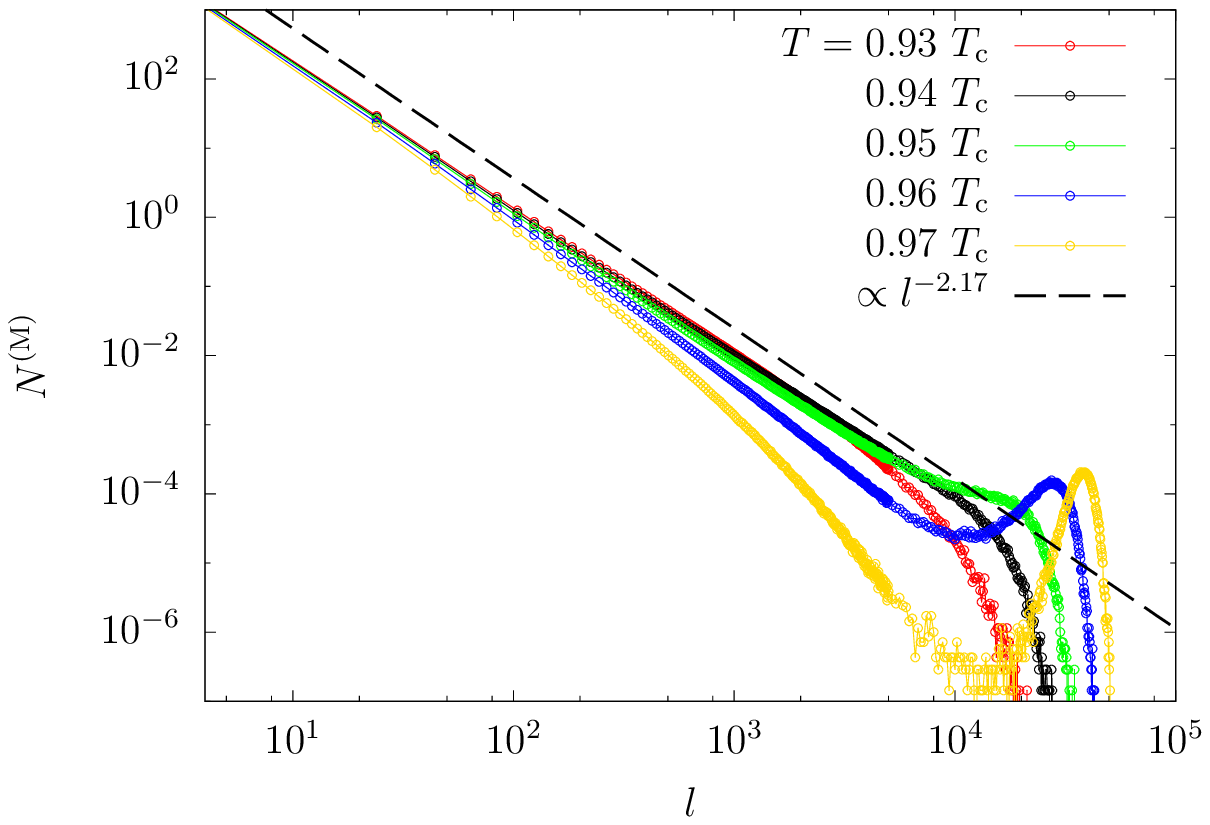} 
(d)\\
\includegraphics[width=0.95\linewidth]{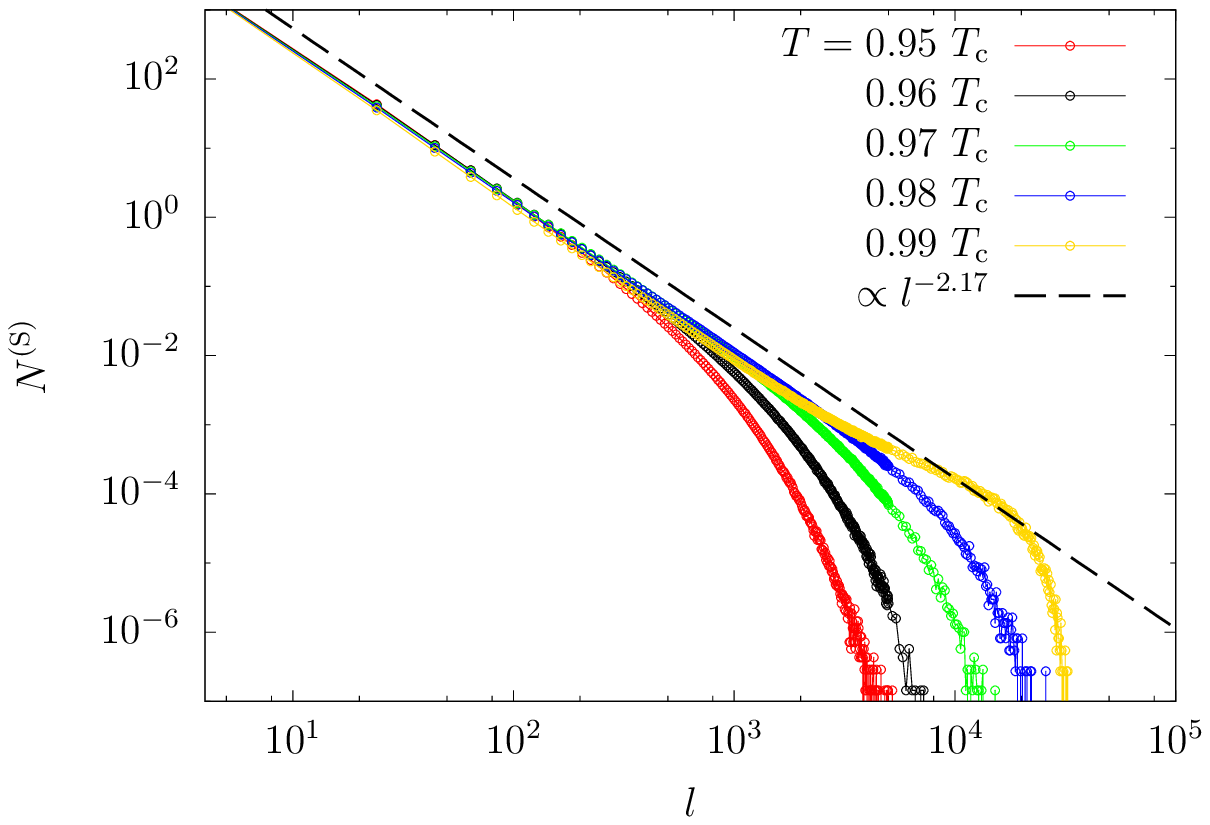}
\end{minipage}
\caption{\label{fig:vortex-size} 
(Color online.)
Equilibrium vortex length number
at various temperatures below and above $T\sub{c}$ in (a) and ($\mbox{c}$) and approaching $T\sub{c}$ from below in (b) and (d), 
for a system with linear size $L=100$.
Upper (lower) panels account for the maximal (stochastic) criteria for connecting vortex elements.
We take a mean over 100 noise realisations and we further average over 1000 different times for each dynamical run.
The dashed line in panels (b) and (d) represent the algebraic decay $l^{-2.17}$.
The two power laws in (c) are $l^{-5/2}$, and $l^{-1}$ as indicated in the key.
}
\end{figure}

At higher temperature, the vortex loops are longer and the method used to connect the 
vortex elements becomes important. The comparison between the upper and lower 
snapshots in Fig.~\ref{fig:vortex-location} (b)-(d) demonstrate that the longest vortex loop
is much longer with the maximal than with the stochastic connection rule.

\begin{figure}[tbh]
\centering
\begin{minipage}{0.49\linewidth}
\centering
(a)\\
\includegraphics[width=0.95\linewidth]{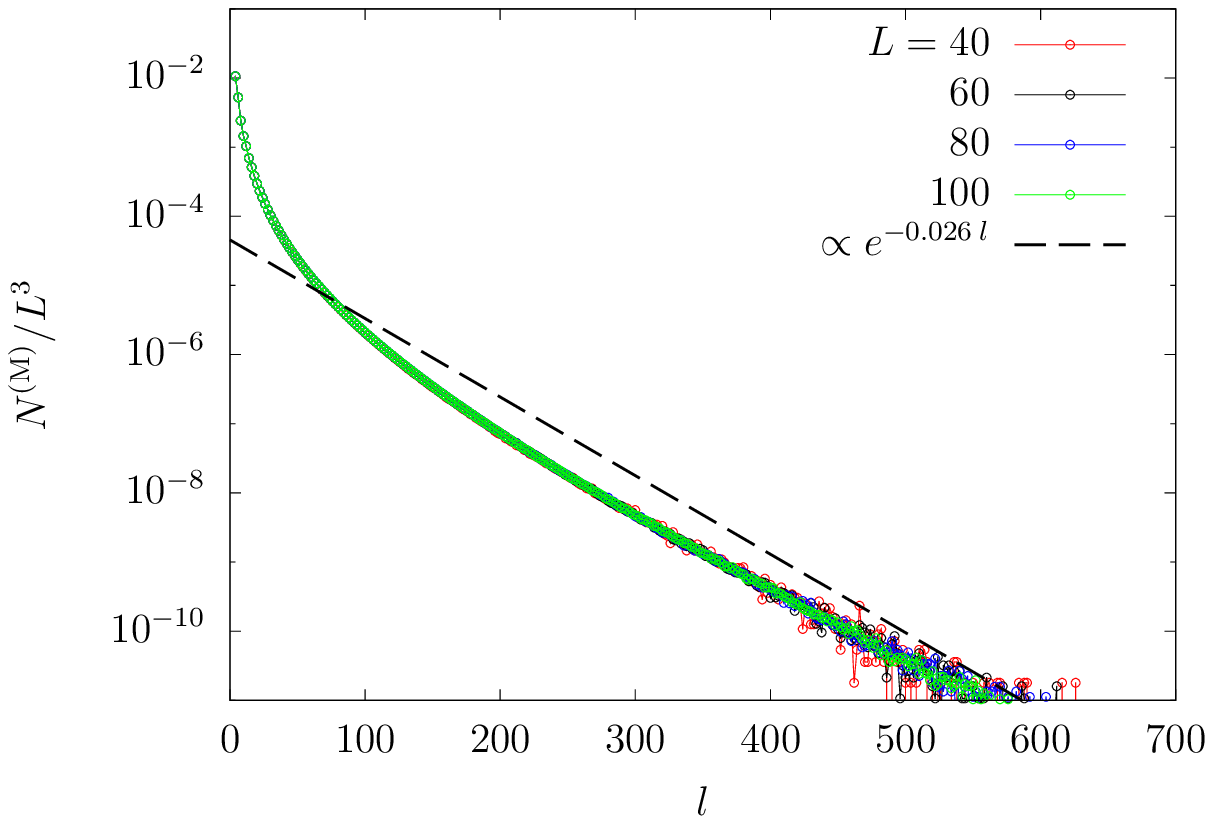} \\
(c)
\includegraphics[width=0.95\linewidth]{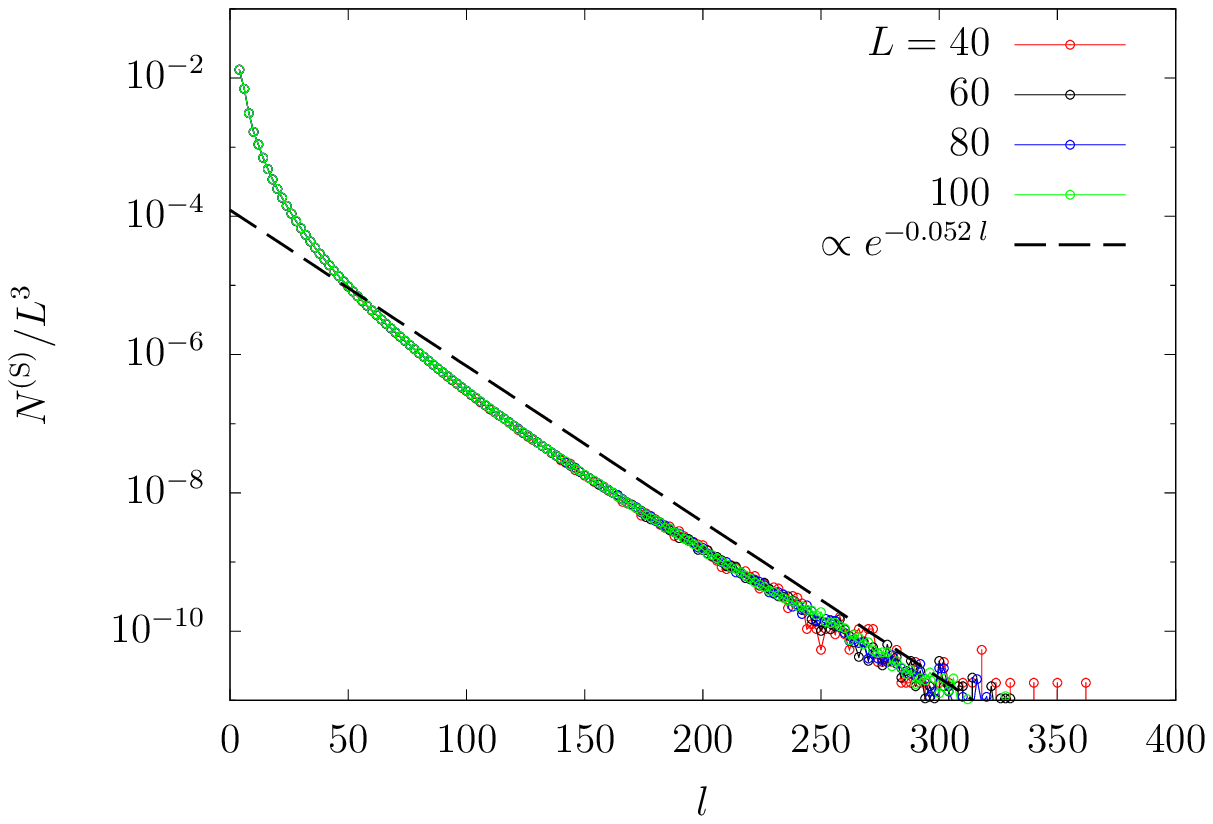}
\end{minipage}
\begin{minipage}{0.49\linewidth}
\centering
(b)\\
\includegraphics[width=0.95\linewidth]{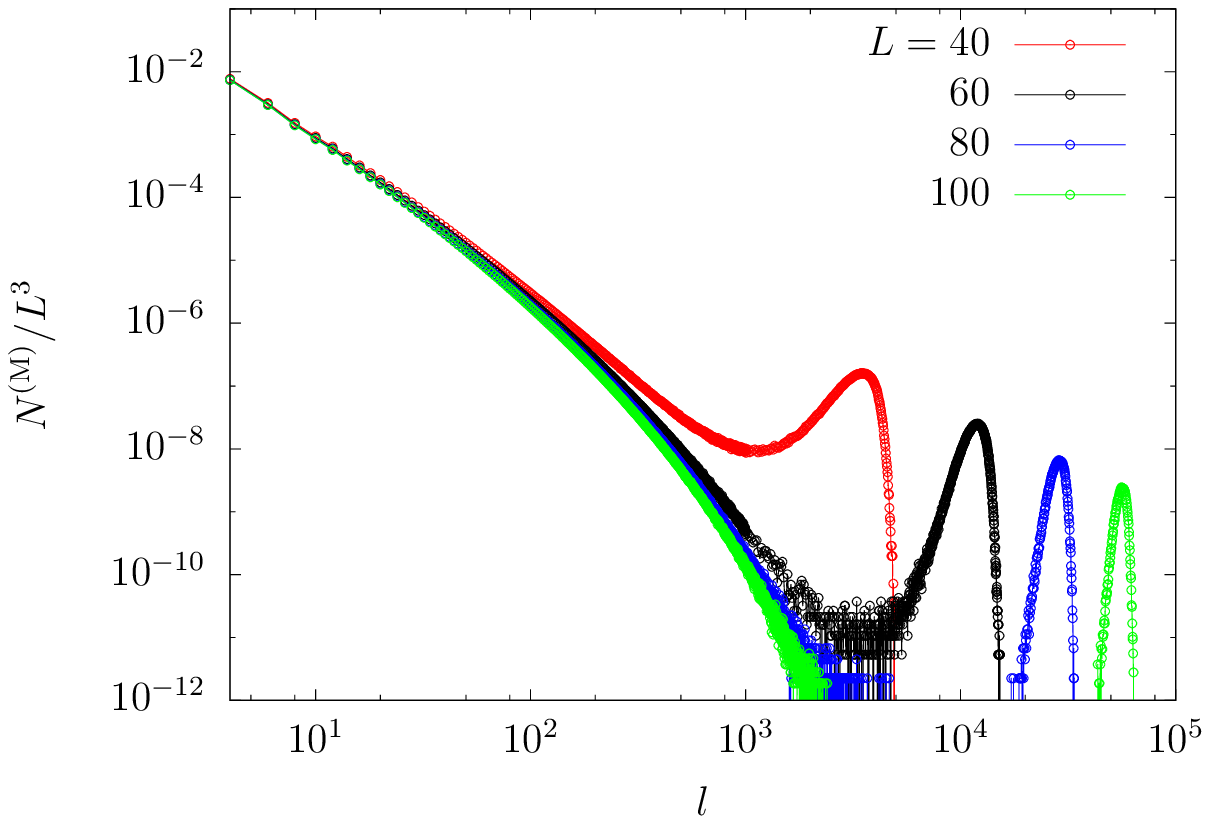} \\
(d)
\includegraphics[width=0.95\linewidth]{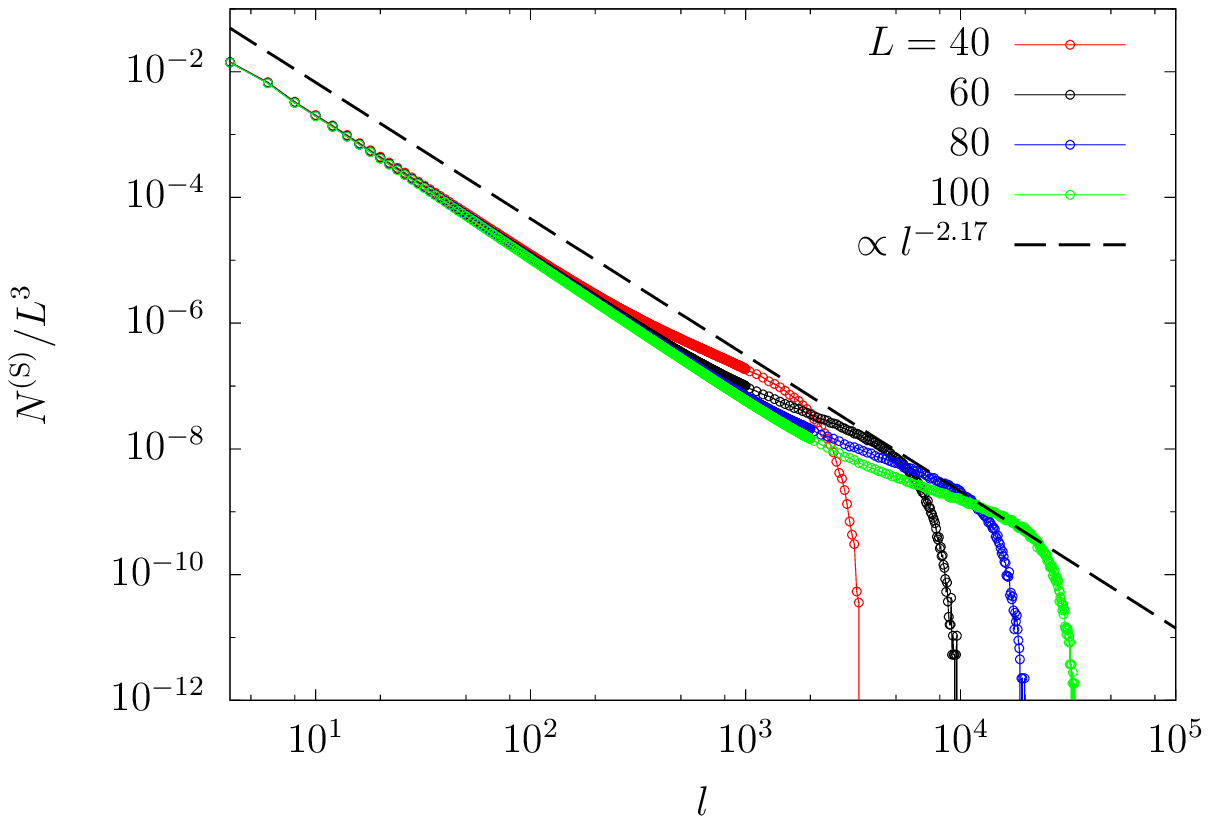}
\end{minipage}
\caption{\label{fig:vortex-size-system} 
(Color online.)
Equilibrium vortex length distribution per unit volume $N(l) / L^3$
at  $T = 0.8 \ T\sub{c}$ in (a) and (c), and $T = \ T\sub{c}$  in (b) and (d) for systems with linear sizes $L = 40$, $60$, $80$, and $100$.
Upper (lower) panels account for the maximal (stochastic) criteria for connecting vortex elements.
In panels  (a) and (c), the dashed straight line represents the exponential decay $e^{-0.026\: l}$ and $e^{-0.052\: l}$, respectively.  
Although the data in panel (d) may suggest that the system is at its
line percolation threshold, at $T\sub{c}$ it is already beyond it, see the text for a discussion.  
}
\end{figure}

The fact that the vortex loops are longer at higher temperature 
 can also be seen from Fig.~\ref{fig:vortex-size}.
Panels~(a) and ($\mbox{c}$) show the number of vortex loops with length $l$, $N\up{(M)}(l)$ calculated with the maximal criterion 
(upper panel) and $N\up{(S)}(l)$ with the stochastic criterion (lower panel) for connecting vortex lines, 
at four temperatures around the critical one, $T = 0.8 \ T_c, \ T_c, 1.2 \ T_c, 2 \ T\sub{c}$.
At $T = 0.8 \ T\sub{c}$ the number density decays exponentially, see also panel (a) in Fig.~\ref{fig:vortex-size-system}, 
irrespectively of the reconnection method used. (This quantity can be turned into a probability 
distribution with its normalisation by the total number of loops in the system, $\int dl \: N\up{(M,S)}(l) = N\up{(M,S)}\sub{loop}$.
As in the dynamic study we will see that this quantity depends on time, we avoid imposing this normalisation.)

As temperature increases from $0.8 \ T\sub{c}$,
longer loops appear and the size of the longest loop increases, as
shown by the fact that the support of $N\up{(M,S)}(l)$ extends further away on the horizontal axis.
With the maximal rule, $N\up{(M)}(l)$ gets close to a power-law,  
$N\up{(M)}(l) \sim l^{-2.17}$ at $T\simeq 0.95 \ T\sub{c}$ and a 
sharp peak at very large value of $l$ starts developing at this temperature (see the solid line in panel (b)
where data for more values of $T$ approaching $T\sub{c}$ from below are shown). 
This bump suggests the existence of very long vortex rings that could wrap around the system many times, 
see Fig.~\ref{fig:vortex-size-system} (b) where the system 
size dependence of the bump is shown explicitly (we will address this issue in detail below). 
At still higher temperature $T > 0.95 \ T\sub{c}$ the weight of the finite size loops decreases but the 
bump remains and gets thinner as less loops with length of the order of the system size exist but their length fluctuates less.
It may become possible to observe such large-scale vortices as a macroscopic fluctuation of the fluid vorticity. 

We also stress the difference between $N\up{(M)}(l)$ and $N\up{(S)}(l)$.
At temperatures far below $T\sub{c}$ ($T = 0.8 \ T\sub{c}$ in Fig.~\ref{fig:vortex-size} (a)), 
there is basically no difference between the data for the two reconnection rules. 
However, the statistics of the strings strongly depends on the reconnection rule
at temperatures close and above $T\sub{c}$. 
The power law $l^{-2.17}$ 
is close to the data for finite loops approaching $T\sub{c}$ from below for both reconnection rules
(panels (b) and (d)) but the behaviour of the distribution at larger scales are totally different.
A bump structure in $N\up{(M)}(l)$ is sharp and clearly seen (panels (a) and (b)),
whereas the statistics of long closed strings at high temperature as obtained with the stochastic criterium crosses over between two 
power-law decays. At $T=2 \ T\sub{c}$, for $\Delta x \ll l \ll L^2$ the chains are Gaussian and $N\up{(S)}(l) \simeq l^{-5/2}$ 
while for $l \gg L^2$ the fact that the loops can wrap around the 
cubic box changes this decay and makes it be $N\up{(S)}(l) \simeq l^{-1}$. 
These two powers are shown with a dashed and dashed lines in panes (c) and (d) where the second power law regime is just incipient 
at $T = 0.99 \ T\sub{c}$. The first power law was also observed in the random phase
clock model studied in~\cite{Vachaspati,Strobl} 
and it is well-known in the field of polymer science~\cite{deGennes}.
The cross-over to the second decay was observed and explained in~\cite{Jaubert} where a fully-packed loop model 
arising in the ice phase of a frustrated magnetic system on the pyrochlore lattice was studied and, in more general 
terms, in~\cite{Nahum,Nahum-book}. Although our system is not fully-packed with loops, the density of
vortex elements is very high at high temperature (e.g. $\rho\sub{vortex} \simeq 0.25$ at $T = 2 \ T\sub{c}$)
and the behaviour is quite similar.

The qualitative change of the vortex line length distribution
and its dependence on the connecting criteria  can already be seen in 
Figs.~\ref{fig:vortex-location} (a)-(d) where the longest vortex line is highlighted (in yellow). On the one hand, 
in panels (a) and (b), at temperatures well below $T\sub{c}$, the longest vortex loop is very short compared to the system size.
On the other hand, in panels (c) and (d), at temperatures at and above $T\sub{c}$, respectively,  
most vortex line elements belong to 
the longest vortex loop, the spatially dominating scale of which is comparable to the system size. 
The longest vortex determined by the maximal criterion is much longer than the one obtained with the stochastic criterion.
Indeed, the longest vortex loop obtained with the maximal criterion contains almost all vortex line elements, and contributes 
to the sharp bump in $N\up{(M)}(l)$ at large $l$.
With the stochastic convention, instead, there are many long vortex loops besides the longest one, 
making $N\up{(S)}(l)$ broader. What is the fraction of vortex mass in an infinite loop is a question of interest in cosmology~\cite{Strobl}.

\vspace{0.5cm}

We now compare the length number density per unit volume $N\up{(M,S)}(l)/L^3$ in systems with different size.
Figures~\ref{fig:vortex-size-system} shows this quantity at temperatures $T = 0.8 \ T\sub{c}$ (a) and (c), and $T = T\sub{c}$ (b) and (d).
At $T = 0.8 \ T\sub{c}$, there is no finite size dependence and there are no long vortices with size comparable to the system size.
At $T = T\sub{c}$, on the other hand, the weight of the number density clearly depends on the system size,
suggesting the existence of very long vortex loops with lengths comparable and increasing with the system size. With the 
maximal convention the tail of the
number density, before the bump, bends down and, clearly,  it is not algebraic. 
With the stochastic one, the data at $T\sub{c}$ suggest a smooth crossover from $l^{-2.17}$ at short length scales
to a different behaviour at long length scales; we will discuss this issue in the next paragraph where we will study the 
percolation phenomenon in detail and we will find that the percolation threshold with the stochastic 
convention although very close to $T\sub{c}$ is not at $T\sub{c}$.

\subsection{The randomly reconnected data in the infinite temperature limit}

We consider now the infinite temperature data and the string length derived with the 
stochastic criterium in more detail and we compare it to predictions
for fully-packed loop models of different kind. 

It was shown in~\cite{Jaubert,Nahum} that the number density of loop lengths in quite generic 
fully-packed loop models behaves as 
\begin{eqnarray}
\frac{l N(l)}{L^3} \simeq 
\left\{
\begin{array}{rcl}
\displaystyle l^{-d/2} & \qquad \mbox{for} & \qquad \Delta x \ll l \ll L^2
\\
\displaystyle L^{-3} & \qquad \mbox{for} & \qquad L^2 \ll l \ll L^3
\label{eq:decay}
\end{array}
\right.
\label{eq:limits}
\end{eqnarray}
as the fractal dimension of the loops at the largest scale is $3$ in our case. 
(Corrections to the power in the second line should be taken into account
for $l\simeq L^3$ and these depend on the model~\cite{Nahum}.) Gaussian statistics for $l \ll L^2$ 
was also found numerically in the nodal statistics of $3d$ complex random wave fields~\cite{Taylor08,Taylor14}.

\begin{figure}[tbh]
\vspace{0.75cm}
\centering
\begin{minipage}{0.49\linewidth}
\centering
(a)\\
\includegraphics[width=0.95\linewidth]{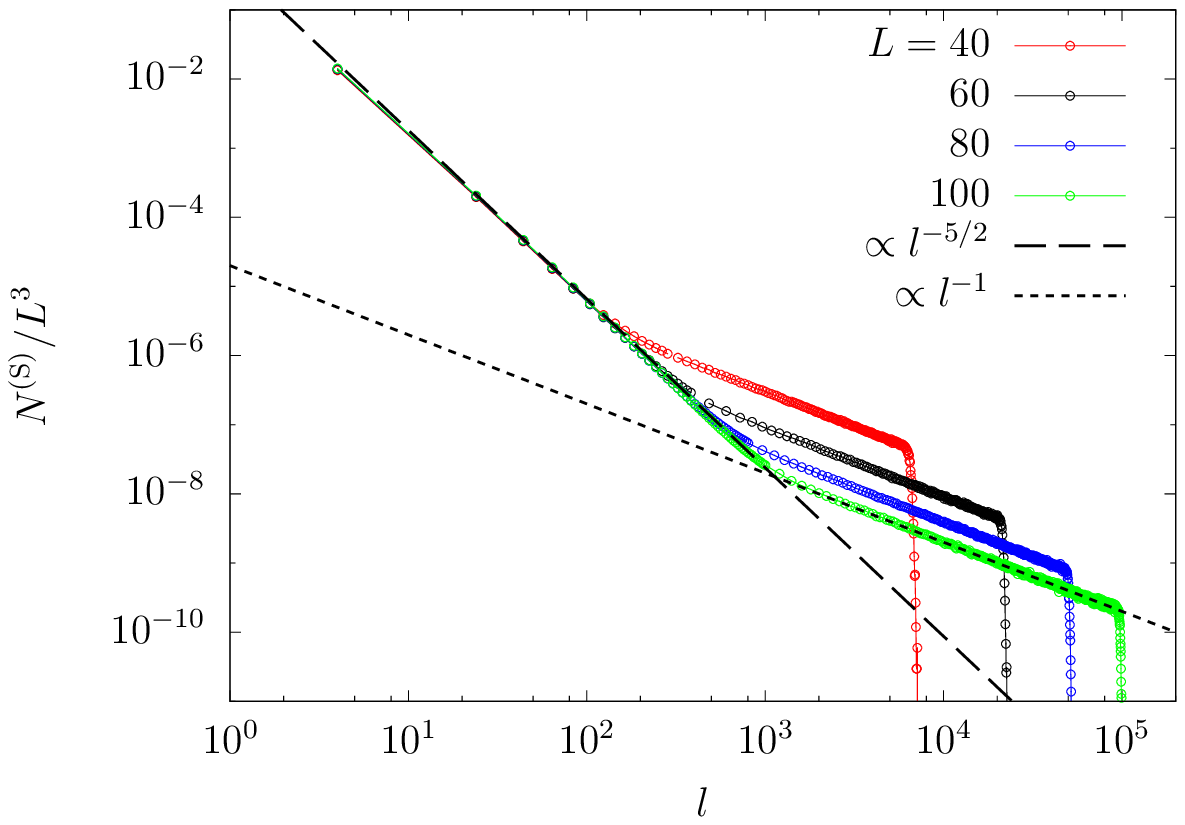}
\end{minipage}
\begin{minipage}{0.49\linewidth}
\centering
(b)\\
\includegraphics[width=0.95\linewidth]{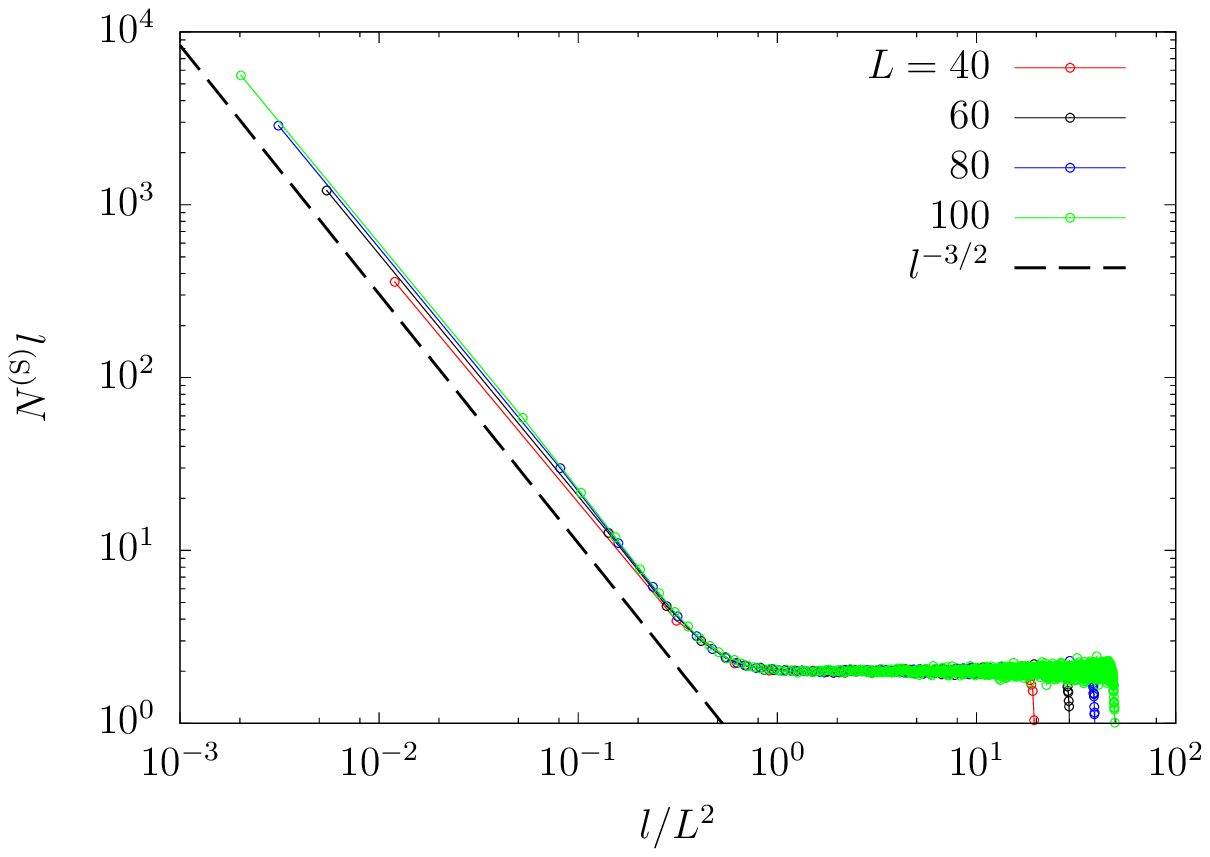}
\end{minipage}
\caption{\label{fig:size-system-scale-inf}
(Color online.)
Stochastically reconnected loop length number density at infinite temperature against 
$l$ (a) and $N\up{(S)} l$ against $l/L^2$ (b).
}
\end{figure}

In Fig.~\ref{fig:size-system-scale-inf} we show $N\up{(S)}(l)$ against $l$  (a)
and $N\up{(S)}(l)$ against $l/L^2$ (b). Data for different system sizes are gathered in the two panels.
We see in (a) that the data do not depend on $L$ for lengths that 
are shorter than $L^2$ while they do for longer scales. In (b) the data 
for $l\ll L^2$ keep the Gaussian statistics (dashed lines) and what remains scales well 
with the proposed scaling variable. The dotted curve is the expected $l^{-1}$ decay in Eq.~(\ref{eq:decay}).
The scaling of the second tail data with the fractal dimension $3$ is good, and
an analogue between $N\up{(S)}(l)$ in the $T\to \infty$ limit and loop soups is thus
confirmed.

\subsection{Vanishing line-tension characteristic temperature}

There are several ways to discuss line percolation in this kind of systems
and they do not yield the same threshold~\cite{Kajantie,Bittner}. 
For this reason, we will be specially careful here.

We adopted the method based on the number density $N\up{(M,S)}(l)$.
Below and close to {\it its} threshold, in the infinite system size limit, 
$N\up{(M,L)}(l)$ should behave as in Eq.~(\ref{eq:dist-threshold})
\begin{align}
N\up{(M,S)}(l) \propto l^{- \alpha\up{(M,S)}\sub{L}} \ e^{- l m\up{(M,S)}\sub{L}} 
\label{eq:pdf-loop-lengths}
\end{align}
with the ``Fisher" exponent $\alpha\up{(M,S)}\sub{L}=1+d/D\up{(M,S)}\sub{L}$ 
being related to the fractal dimension, $D\up{(M,S)}\sub{L}$, of the vortex lines, 
and the ``mass" $m\up{(M,S)}\sub{L}$ that vanishes at the threshold. 
Following Ref.~\cite{Kajantie}, we call this temperature the line-tension point $T\up{(M,S)}\sub{L}$.

\vspace{0.25cm}

\begin{figure}[tbh]
\centering
\begin{minipage}{0.49\linewidth}
\centering
(a)\\
\includegraphics[width=0.95\linewidth]{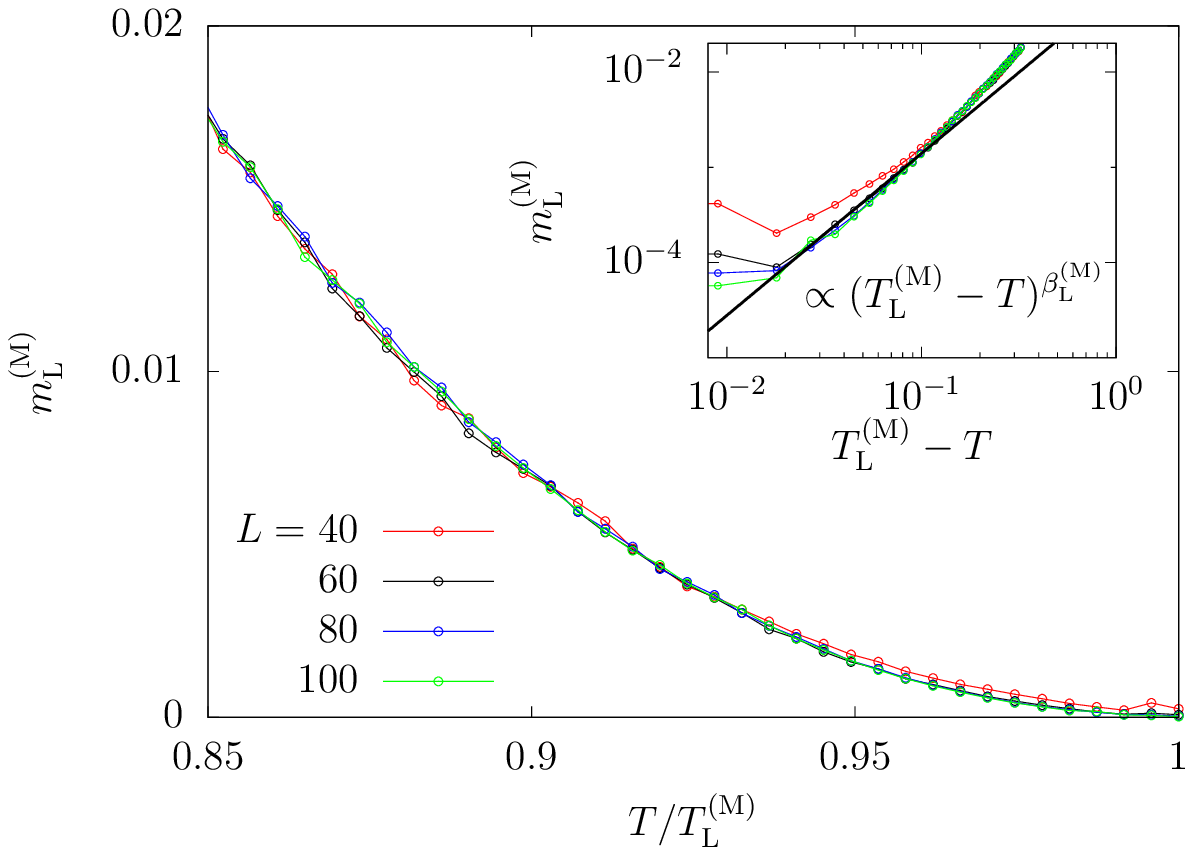} \\
\includegraphics[width=0.95\linewidth]{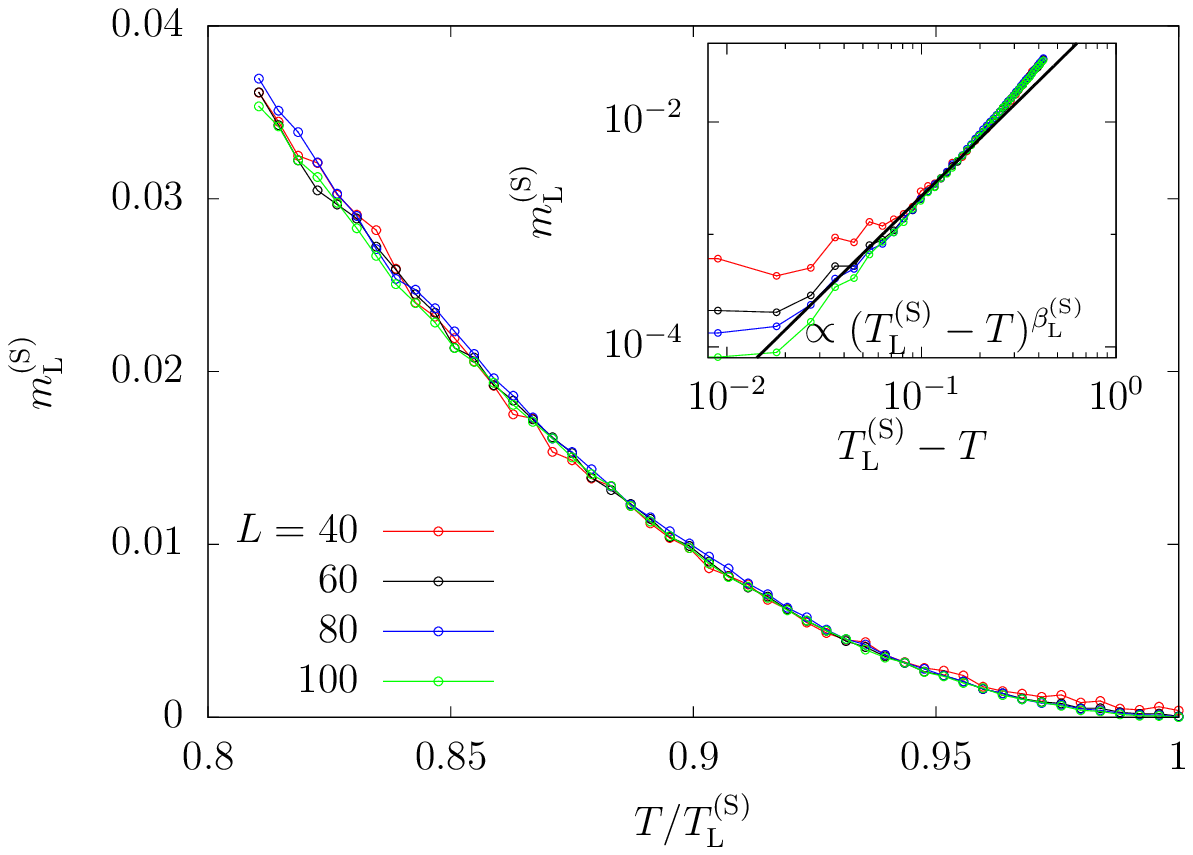}
\end{minipage}
\begin{minipage}{0.49\linewidth}
\centering
(b)\\
\includegraphics[width=0.95\linewidth]{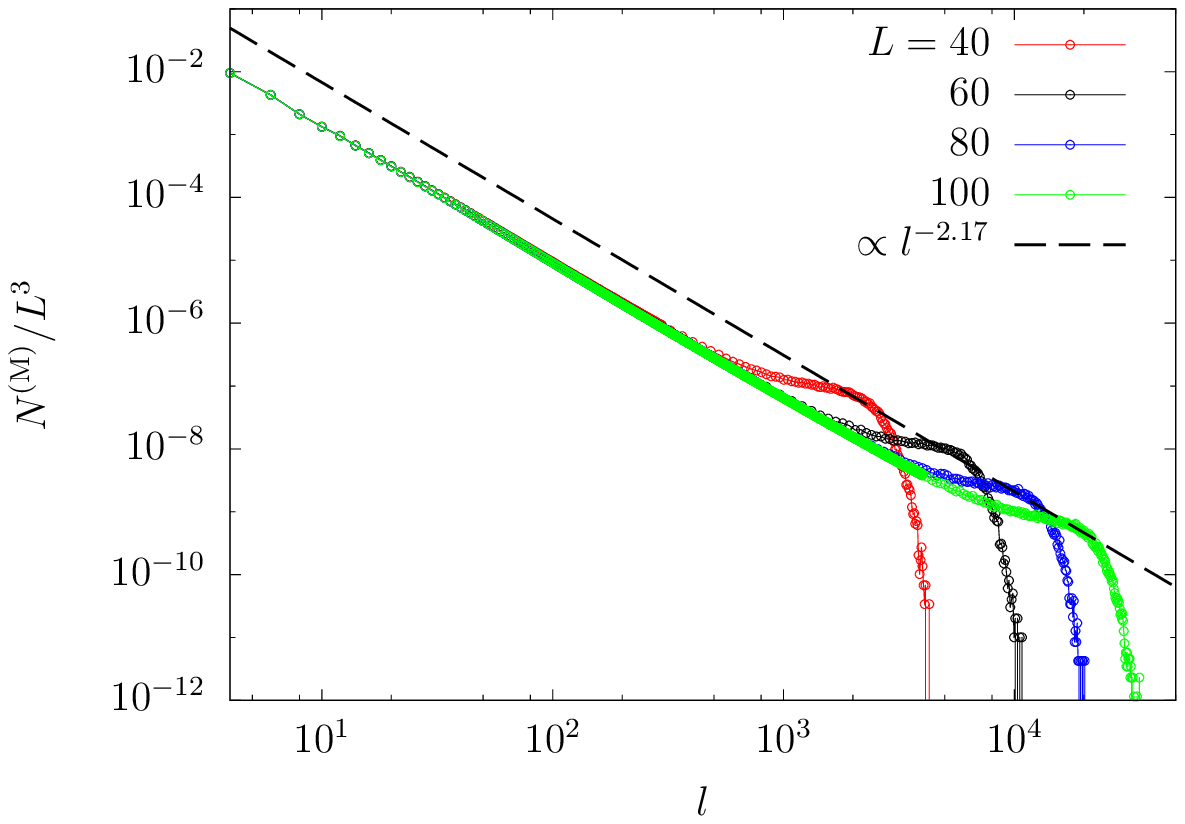} \\[5pt]
\includegraphics[width=0.95\linewidth]{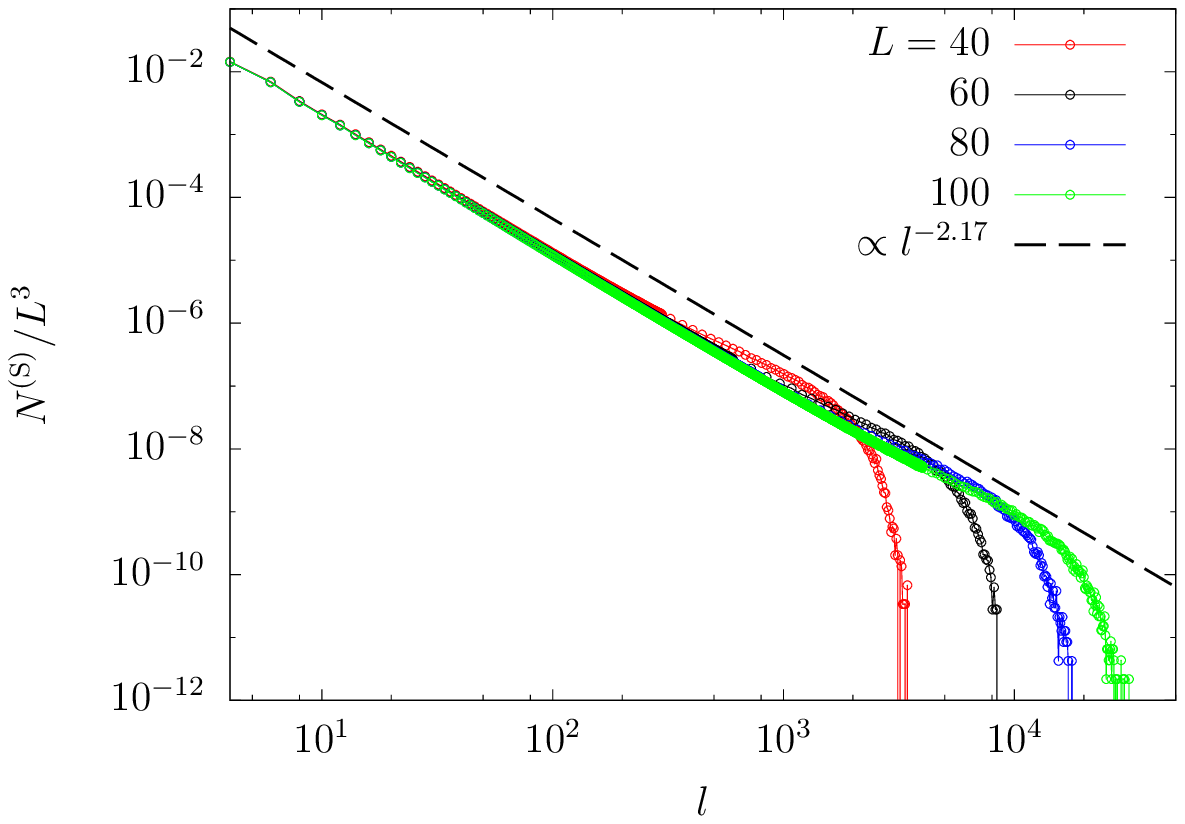}
\end{minipage}
\caption{\label{fig:porder} 
(Color online.)
(a) Temperature dependence of the mass
parameter $m\up{(M,S)}\sub{L}$ in the fit (\ref{eq:pdf-loop-lengths}) to the 
vortex length number densities. The
insets show $m\up{(M,S)}\sub{L}$ as a function of $T\up{(M,S)}\sub{L} - T$ with $T\up{(M)}\sub{L} = 0.94 \ T\sub{c}$ and
$T\up{(S)}\sub{L} = 0.98 \ T\sub{c}$ in double logarithmic scale together with an algebraic dependence with 
$\beta\up{(M,S)}\sub{L}=1.7$.
(b) Equilibrium vortex length number per unit volume $N\up{(M,S)}(l) / L^3$ at the line-tension point $T\up{(M,S)}\sub{L}$.
Upper (lower) panels account for the maximal (stochastic) criteria for connecting vortex elements.
}
\end{figure}

In Fig. \ref{fig:vortex-size} (b) we show the length number density
$N\up{(M)}(l)$ at $T = 0.93 \ T\sub{c}$, $0.94 \ T\sub{c}$, $0.95 \ T\sub{c}$, $0.96 \ T\sub{c}$, and $0.97 \ T\sub{c}$
with the maximal line-reconnection criterion (upper panel), and $N\up{(S)}(l)$ at
$T = 0.95 \ T\sub{c}$, $0.96 \ T\sub{c}$, $0.97 \ T\sub{c}$, $0.98 \ T\sub{c}$, and $0.99 \ T\sub{c}$
with the stochastic line-reconnection criterion (lower panel) for the largest system size that we simulated, $L=100$.
At $0.94 \ T\sub{c}$ ($0.98 \ T\sub{c}$) the data are close to algebraic with an incipient bump 
at the largest scales in the upper (lower) panel. The short length-scale, say $10^2 \lesssim l \lesssim 10^3$, behaviour of 
$N\up{(M,S)}(l)$ is rather well fitted by
\begin{align}
N\up{(M,S)}(l) \propto l^{- \alpha\up{(M,S)}\sub{L}}
\qquad \mbox{with} \qquad
\alpha\up{(M,S)}\sub{L} \simeq 2.17 \; .
\label{eq:complete-loop-length} 
\end{align}

Figure \ref{fig:porder} (a) shows the mass
$m\up{(M,S)}\sub{L}$  extracted from fits with the full function \eqref{eq:pdf-loop-lengths} with the exponent $\alpha\up{(M,S)}\sub{L}$ fixed to 
$\alpha\up{(M,S)}\sub{L} = 2.17$.
From the data fits the mass $m\up{(M)}\sub{L}$ ($m\up{(S)}\sub{L}$) vanishes at 
$T \simeq 2.13 \simeq 0.94\ T\sub{c}$ ($T \simeq 2.22 \simeq 0.98\ T\sub{c}$).
We therefore estimate the temperature at which $N\up{(M,S)}(l)$ is purely algebraic as
\begin{align}
\begin{split}
T\up{(M)}\sub{L} \simeq 2.13 \simeq 0.94 \ T\sub{c},\qquad T\up{(S)}\sub{L} \simeq 2.22 \simeq 0.98 \ T\sub{c},
\end{split}
\end{align} 
and they do not coincide with the one for the
thermodynamic instability $T\sub{c}$, see the analysis in Sec.~\ref{sec:equilibrium}.
In the temperature range $0.05 \leq T\up{(M,S)}\sub{L} - T \leq 0.1$, 
the masses $m\up{(M,S)}\sub{L}$ for the system sizes $L = 80$ and $L=100$ 
are rather well fitted by 
\begin{equation}
m\up{(M,S)}\sub{L} \propto (T\up{(M,S)}\sub{L} - T)^{\beta\up{(M,S)}\sub{L}}
\qquad \mbox{with} \qquad
\beta\up{(M,S)}\sub{L} \simeq 1.7 \; .
\end{equation}
This value hardly depends on the criteria for connecting vortices and is 
consistent with the results obtained with Monte Carlo simulations of the $3d$ $XY$-model~\cite{Kajantie} and this model~\cite{Bittner}.
Figure~\ref{fig:porder} (b) shows the length number density per unit volume $N\up{(M,S)}(l) / L^3$ for systems with  different linear sizes at the vortex line-tension point $T\up{(M,S)}\sub{L}$.
Except for the bump structure in the region of long $l$, $N\up{(M,S)}(l) / L^3$ at $T = T\up{(M,S)}\sub{L}$ does not depend upon $L$.

Finally, we note that simulations with different $\Delta x$ 
give different values for $T\sub{c}$, $T\sub{L}\up{(M)}$, and $T\sub{L}\up{(S)}$, but the algebraic behaviour 
at the vortex line-tension point hardly depends on $\Delta x$ with the same exponents 
$\alpha\sub{L}\up{(M,S)} \simeq 2.17$ and $\beta\sub{L}\up{(M,S)} \simeq 1.7$ within our numerical accuracy. 

\subsection{The bump}

We consider now the bump structure in the number density $N\up{(M,S)}(l)$.
A bump in the number density $N\up{(M,S)}(l)$ at large value of $l$ starts to develop at $T\up{(M,S)}\sub{L}$ 
and is due to the finite size of the system.
To describe it one should write a finite system 
size additive correction~\cite{Stauffer} to the vortex length distribution $N\up{(M,S)}(l)$ in \eqref{eq:pdf-loop-lengths}:
\begin{align}
N\up{(M,S)}\sub{finite\ size\ corr}(l) 
= N\up{(M,S)}\sub{L}\big(\tilde{l}\up{(M,S)}\big),\qquad 
\tilde{l}\up{(M,S)} \equiv l / L^{D\up{(M,S)}\sub{L}}
\label{eq:pdf-loop-lengths-finite-size}
\end{align}
with $N\up{(M,S)}\sub{L}\big(0\big)=0$.
Equation \eqref{eq:pdf-loop-lengths-finite-size} states that the finite size correction to the number density 
$N\up{(M,S)}(l)$ should be a universal function of $\tilde{l}\up{(M,S)} 
\equiv l / L^{D\up{(M,S)}\sub{L}}$ with $D\up{(M,S)}\sub{L} = d / (\alpha\up{(M,S)}\sub{L} - 1) \simeq 2.56$ the fractal dimension of the lines and $d = 3$ the dimension of space.

\begin{figure}[tbh]
\vspace{0.75cm}
\centering
\begin{minipage}{0.49\linewidth}
\centering
(a) \\
\includegraphics[width=0.95\linewidth]{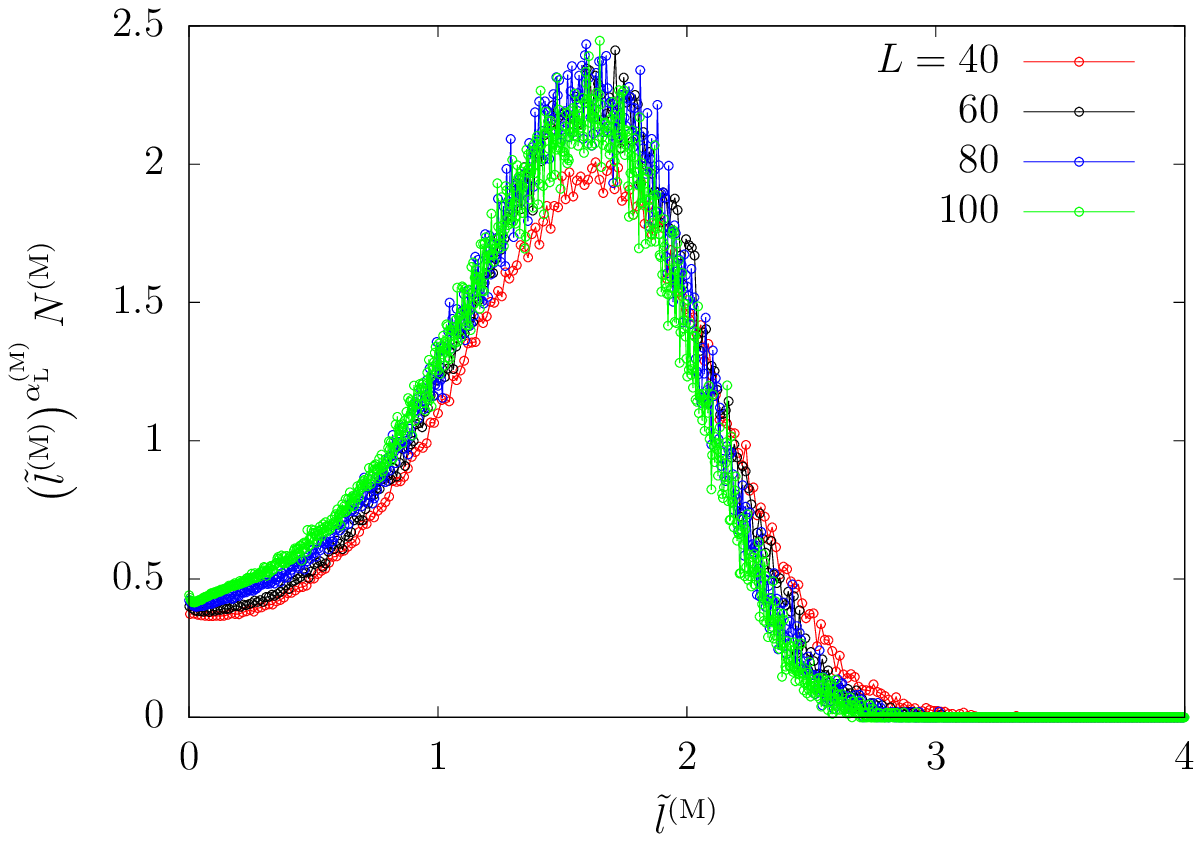}\\
\includegraphics[width=0.95\linewidth]{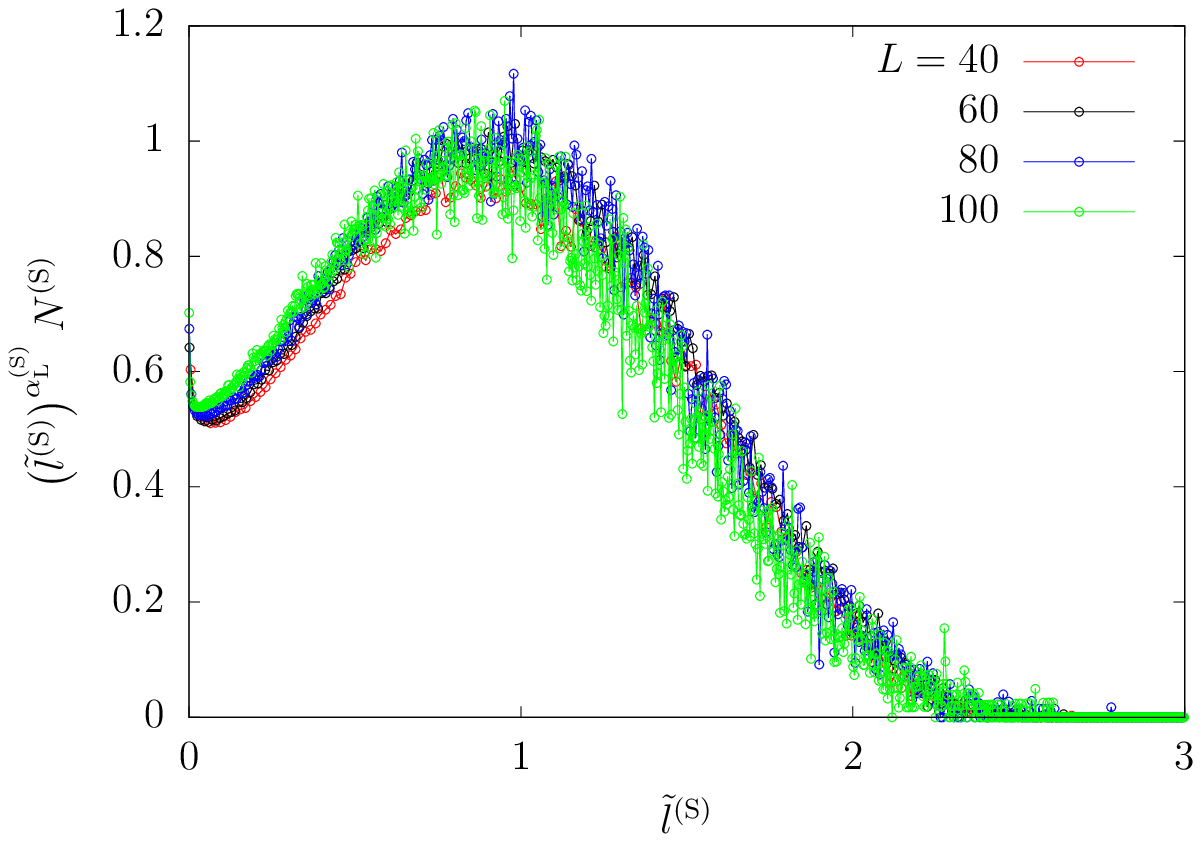}
\end{minipage}
\begin{minipage}{0.49\linewidth}
\centering
(b) \\
\includegraphics[width=0.95\linewidth]{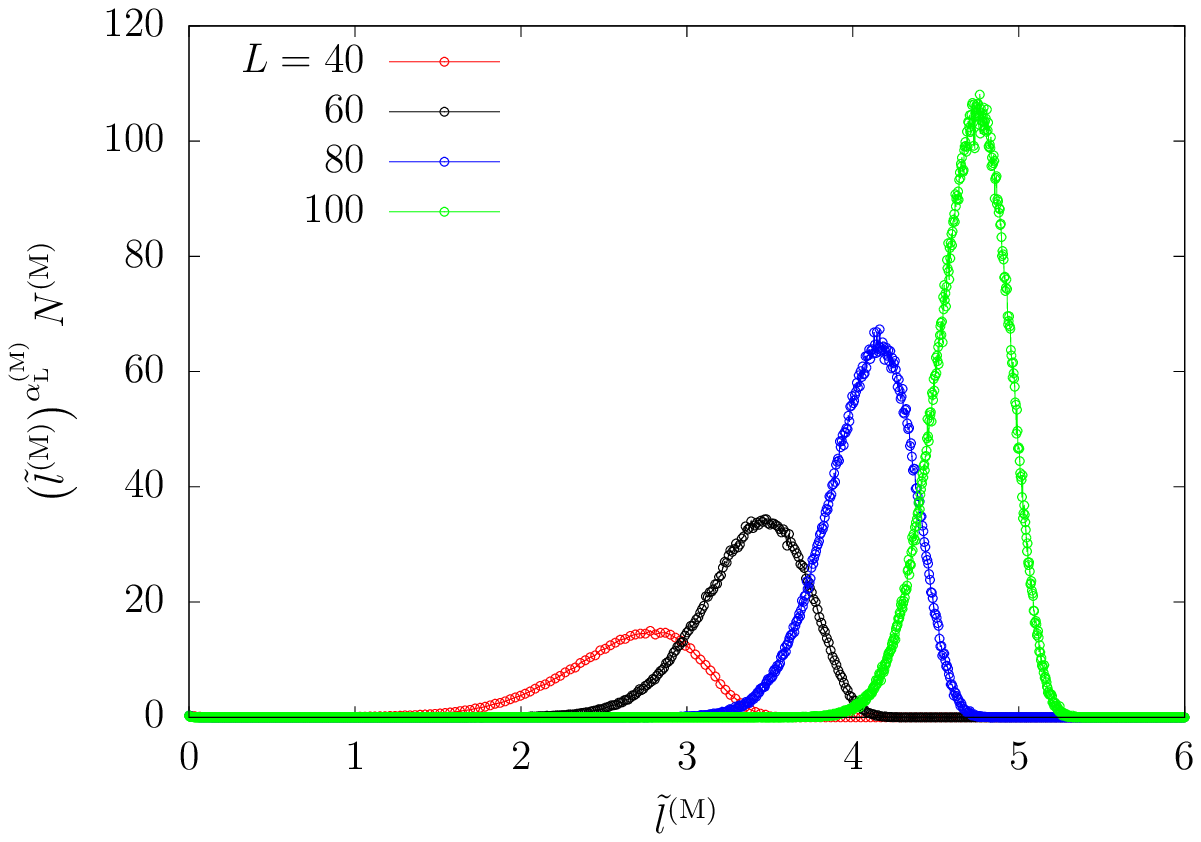}\\
\includegraphics[width=0.95\linewidth]{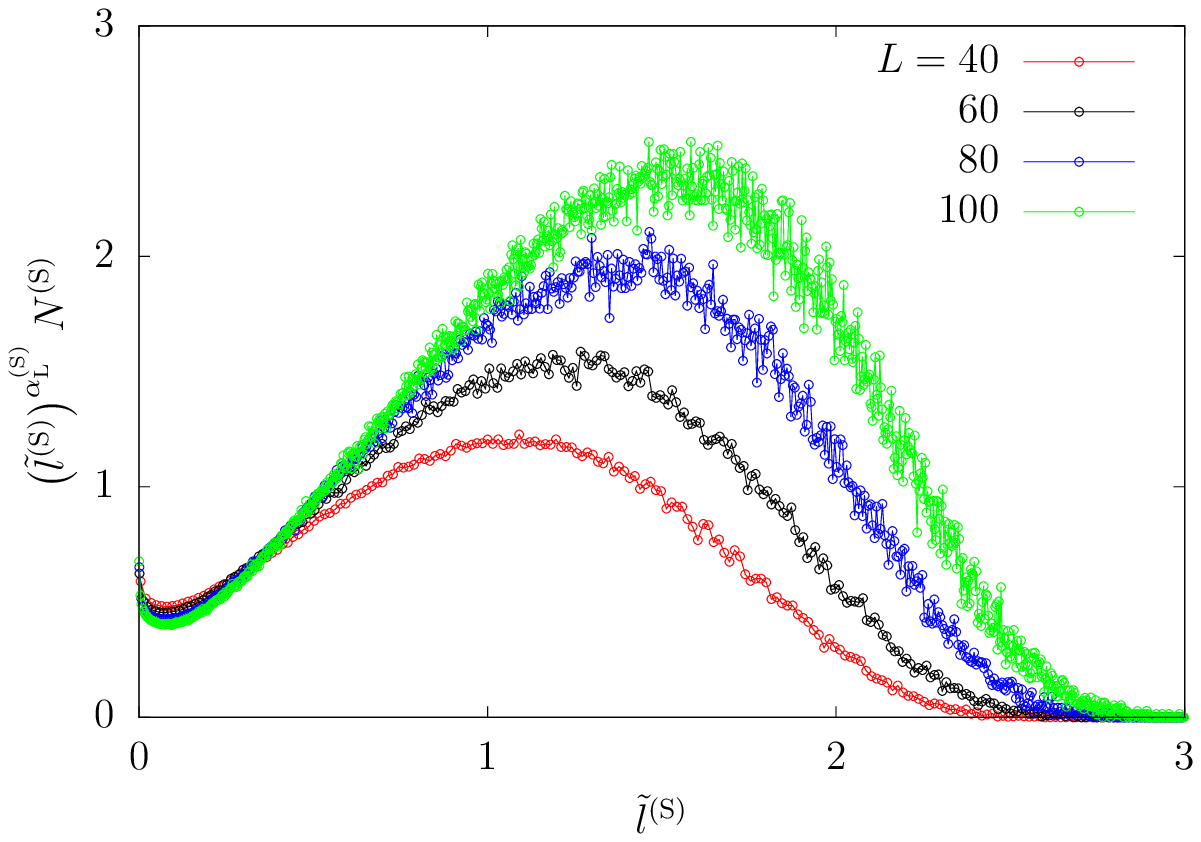}
\end{minipage}
\caption{\label{fig:size-system-scale2} 
(Color online.)
Finite-size scaling of the bump structure in 
$N\up{(M,S)}\sub{L}$
at (a) the vortex line-tension point $T\up{(M,S)}\sub{L}$ and (b) the critical temperature $T\sub{c}$.
Upper (lower) panels account for the maximal (stochastic) criterium for connecting vortex elements.
The scaling variables are ${\tilde l}\up{(M,S)} = l / L^{D\up{(M,S)}\sub{L}}$ and the 
fractal dimensions are fixed to $D\up{(M,S)}\sub{L} = 2.56$, see the text for a discussion.
Their is data collapse in (a) but not in (b).
}
\end{figure}

Figures \ref{fig:size-system-scale2} (a) and (b) show $\big(l\up{(M,S)}\big)^{\alpha\sub{L}\up{(M,S)}}\: N\up{(M,S)}(l)$ 
as a function of $\tilde{l}\up{(M,S)}$, at $T\up{(M,S)}\sub{L}$ and $T\sub{c}$, respectively,  and for the system sizes $L = 40$, $60$, $80$, $100$.
After multiplying by $\big(l\up{(M,S)}\big)^{\alpha\sub{L}\up{(M,S)}}$ the finite length contribution should become just an irrelevant additive 
constant and all the variation is due to the finite system-size correction.
The universal behaviour of the bump structure as shown in Eq.~\eqref{eq:pdf-loop-lengths-finite-size} holds at 
$T\up{(M,S)}\sub{L}$ and it does not at $T\sub{c}$, confirming the fact that line percolation occurs at $T\sub{L}\up{(M,S)}$.

\subsection{Mean number of vortex loops}

\begin{figure}[tbh]
\centering
\begin{minipage}{0.45\linewidth}
\centering
(a) \\
\includegraphics[width=1\linewidth]{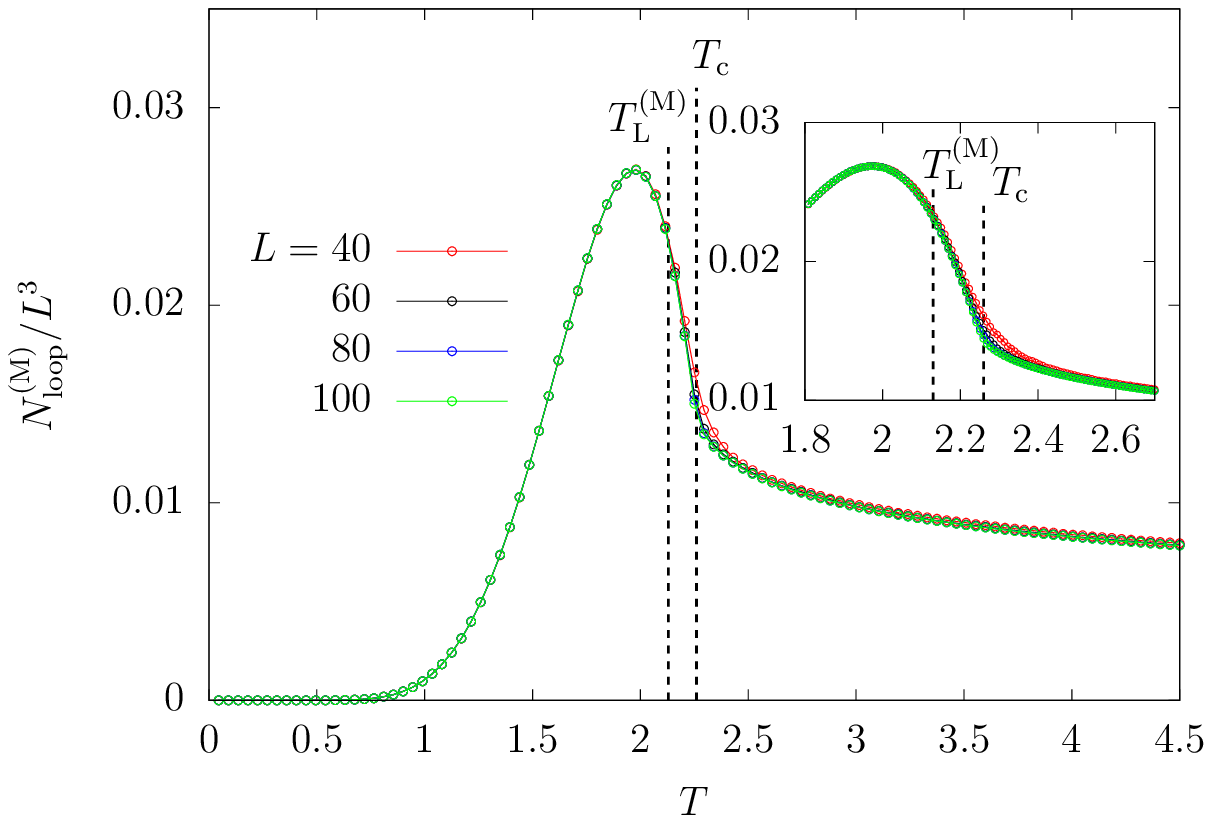}
\end{minipage}
\begin{minipage}{0.45\linewidth}
\centering
(b) \\
\includegraphics[width=1\linewidth]{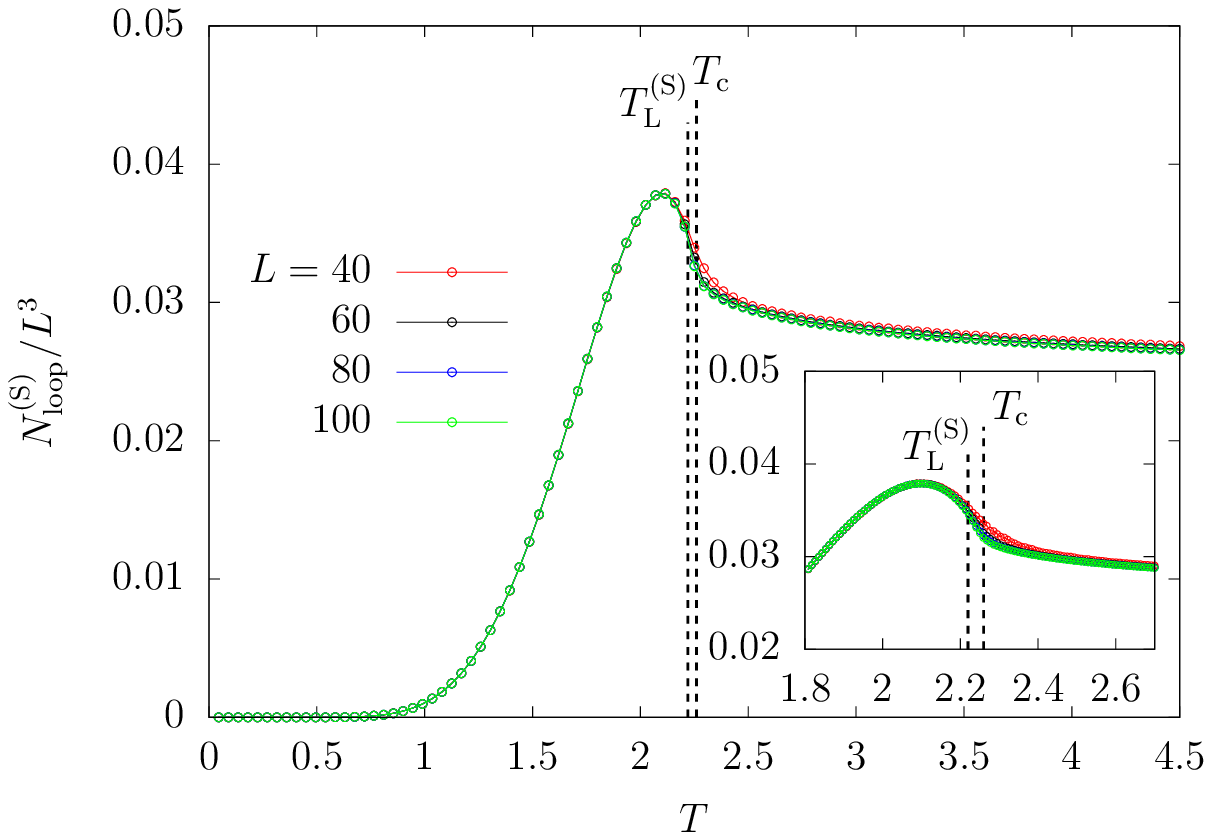}
\end{minipage}
\caption{\label{fig:loop} 
(Color online.)
Temperature dependence of the mean number of vortex loops 
connected with the maximal rule $N\up{(M)}\sub{loop}$ (a) and with the stochastic rule $N\up{(S)}\sub{loop}$ (b)
in both cases  normalized by the system size. Several system sizes were used and 
are given in the key. In the insets, zooms over the peaks.
}
\end{figure}

Figure~\ref{fig:loop} shows the temperature dependence of the mean number of vortex loops 
\begin{equation}
N\up{(M,S)}\sub{loop} \equiv \langle \text{total number of vortex loops} \rangle\sub{stat}
\end{equation}
normalized by the size of the simulation box $L^3$.
$N\up{(M)}\sub{loop}$ is consistently smaller than $N\up{(S)}\sub{loop}$. Both figures show no $L$ dependence.
At low temperatures, $N\up{(M,S)}\sub{loop}$ is an increasing function of temperature.
Above a temperature that is slightly lower than $T\sub{L}\up{(M,S)}$, $N\up{(M,S)}\sub{loop}$ reaches a maximum 
and next decreases with increasing temperature, 
suggesting that many small vortex rings merge to form longer loops, as $\rho\sub{vortex}$ is still increasing with temperature.
The fact that  $N\up{(M)}\sub{loop}$ decreases faster than $N\up{(S)}\sub{loop}$ with temperature is due to the fact that 
more vortex elements are joined to the longest vortex loop with the maximal than with the stochastic 
rule. Notably, the curvature of the curves changes at $T\sub{c}$ but we do not see any special feature at
$T\sub{L}\up{(M,S)}$. 

\vspace{0.5cm}

\subsection{Wrapping vs. contractible loops}

As discussed above, the scaling of the bump in $N\up{(M,S)}(l)$, and the peak and tail in $N\up{(M)}(l)$ and 
$N\up{(S)}(l)$ at high temperature, suggest the existence of very long vortex loops 
with length of the order of, or even much longer than, $L$.
In order to distinguish loops that wrap around the system from long but 
contractible loops, we define and calculate two quantities.

\begin{figure}[b]
\centering
\includegraphics[width=0.75\linewidth]{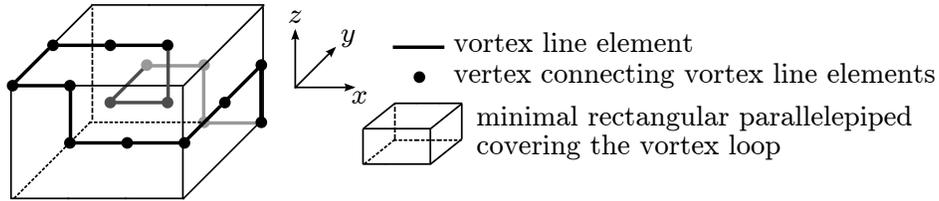}
\caption{\label{fig:loop-size}
(Color online.)
The size of a vortex loop explained with an example.
A vortex loop is shown with a broken solid line made of straight vortex line elements.
The minimal rectangular parallelepiped that covers the loop in the $x$, $y$, and $z$-directions is also shown.
The size of the vortex loop is defined as the maximal linear length of the faces of the covering parallelepiped. In the 
case in the figure, the three linear sizes of the rectangular parallelepiped are $3$, $2$, and $2$, and the size of the vortex loop is 
$\mathrm{max}(3,2,2) = 3$.}
\end{figure}

The first observable just focuses on the size of the vortices, that we define as the maximal side of the rectangular parallelepiped 
covering the vortex loop in the $x$, $y$, and $z$-directions (see Fig.~\ref{fig:loop-size}). The size is the 
length the string would have after smoothing out all small scale irregularities (and it yields a length scale
similar to $R$ in Eq.~(\ref{eq:def-fractal})).

\vspace{0.5cm}

\begin{figure}[tbh]
\centering
\begin{minipage}{0.49\linewidth}
\centering
(a)\\
\includegraphics[width=0.95\linewidth]{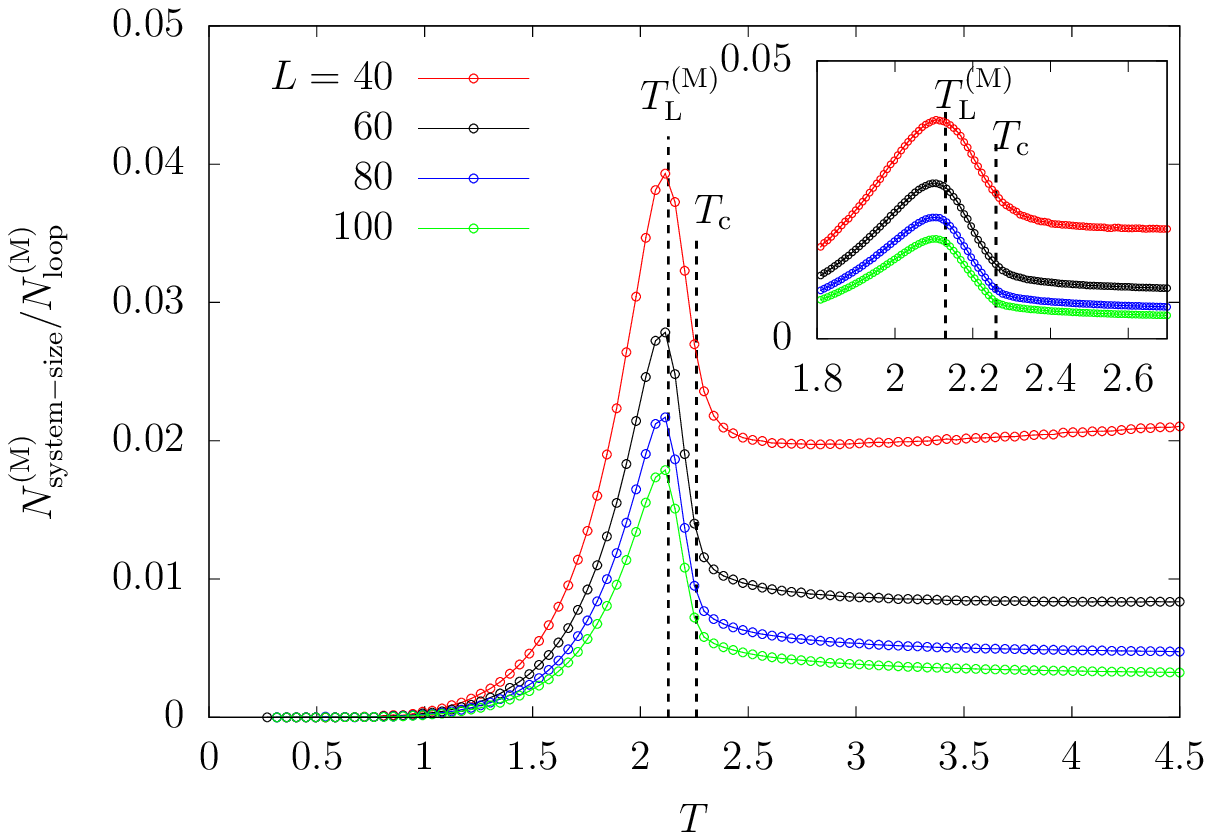} 
(c) \\
\includegraphics[width=0.95\linewidth]{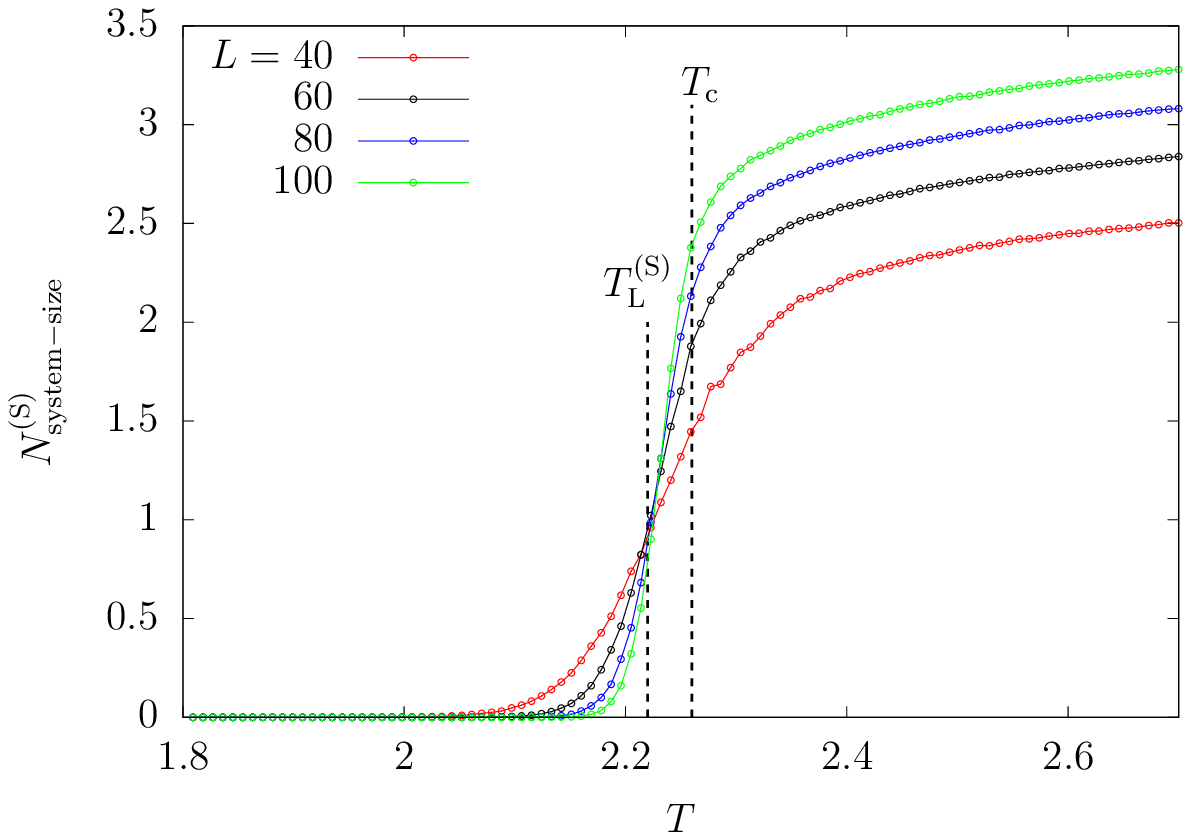}
\end{minipage}
\begin{minipage}{0.49\linewidth}
\centering
(b)\\
\includegraphics[width=0.95\linewidth]{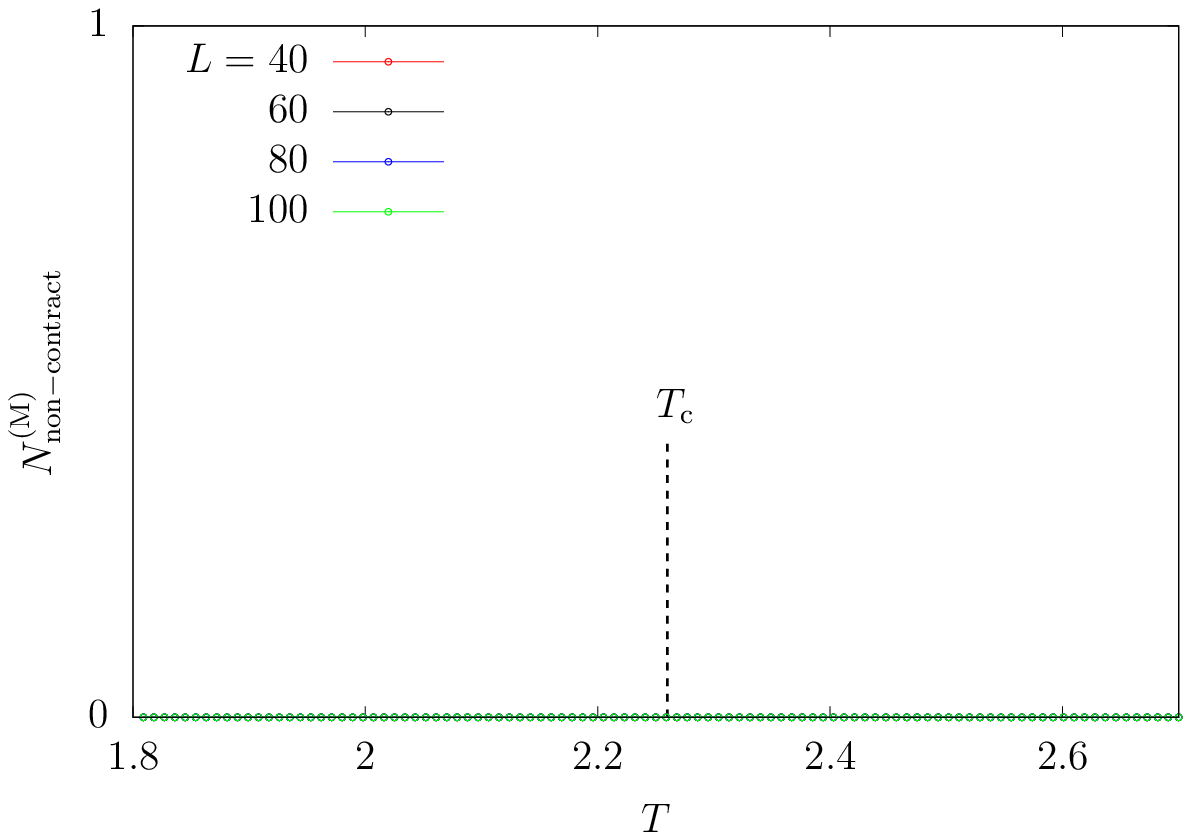} 
(d) \\
\includegraphics[width=0.95\linewidth]{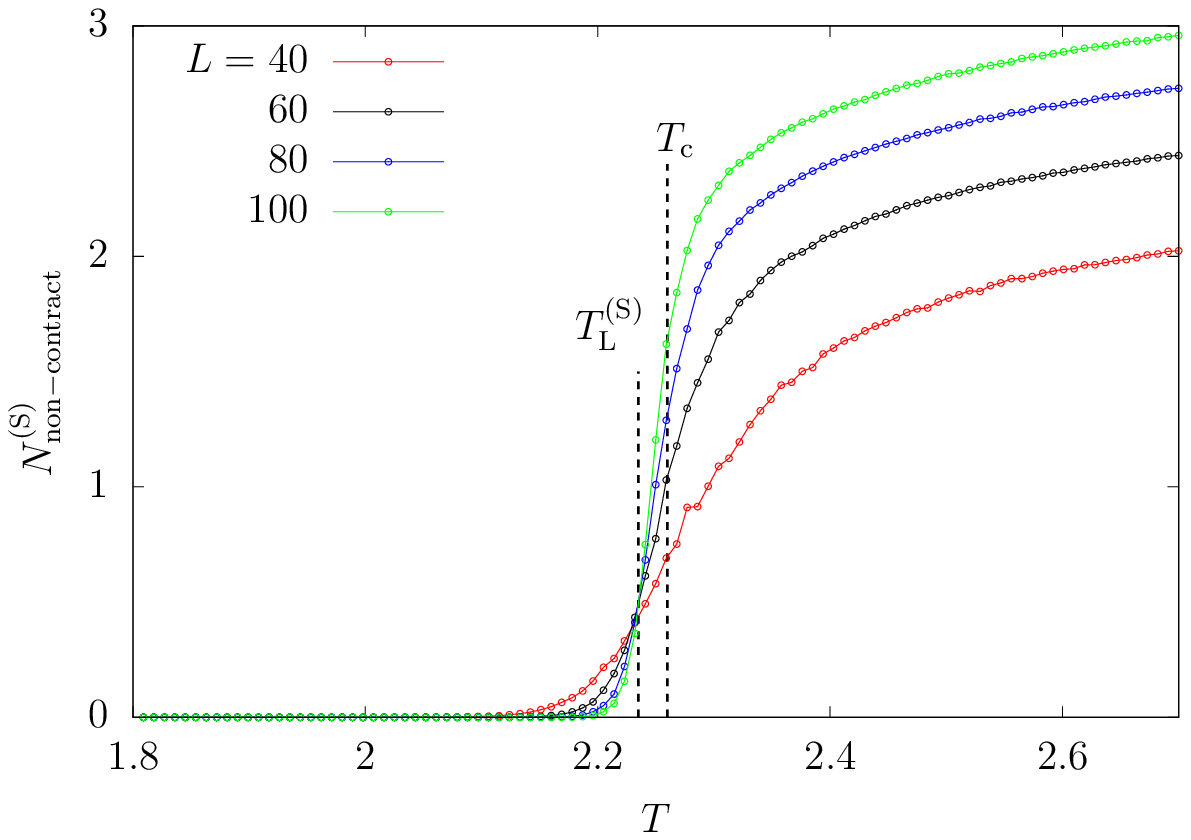}
\end{minipage}
\caption{\label{fig:noncontract} 
(Color online.)
Temperature dependence of the number of vortices satisfying
different conditions. 
(a)  Ratio between the number of vortex loops 
larger than the system size $N\up{(M)}\sub{system-size}$ and the total number of vortex loops $N\up{(M)}\sub{loop}$.
In the inset, a zoom over the peak.
(b) Averaged number of non-contractible vortex loops $N\up{(M)}\sub{non-contract}$.
(c) Number of vortex loops larger than the system size $N\up{(S)}\sub{system-size}$.
(d) Number of non-contractible loops.
In (a) and (b) the maximal criterium is used. In (c) and (d) the stochastic one is used.
The system sizes are given in the keys.
}
\end{figure}

We then count the number of vortex loops, the size of which is larger than the system size $L$, and we calculate the 
statistical average:
\begin{align}
\begin{split}
& N\up{(M,S)}\sub{system-size} \equiv \langle \text{Number of vortex loops, the size of which is larger than $L$}  \rangle\sub{stat}.
\end{split}
\end{align}
Figure \ref{fig:noncontract} (a) shows the temperature dependence of the fraction $N\up{(M)}\sub{system-size} / N\up{(M)}\sub{loop}$.
It detaches from zero at $T \simeq 0.59 \simeq 0.26 \ T\sub{c}$ 
(while $\rho\sub{vortex}$ detaches from zero at $T \simeq 0.32 \simeq 0.14 \ T\sub{c}$) 
and has a peak at a temperature that is very close to the value of $T\sub{L}\up{(M)}$ found with the analysis of 
$N\up{(M)}(l)$ in the infinite system size limit.

The behaviour of $N\sub{system-size}\up{(S)}$ shown in Fig.~\ref{fig:noncontract} (c) is quantitatively different from the one of $N\sub{system-size}\up{(M)}$: 
it does not have a peak and it detaches from zero at a temperature slightly lower than $T\up{(S)}\sub{L}$ 
found with the analysis of $N\up{(M)}(l)$.
At a temperature very close to $T\up{(S)}\sub{L}$, $N\sub{system-size}\up{(S)}$ 
loses its size dependence and $N\sub{system-size}\up{(S)} \simeq 1.07$.
In the limit of infinite system size $L \to \infty$, one may expect a sharp transition from 
$N\sub{system-size}\up{(S)} = 0$ to $N\sub{system-size}\up{(S)} >1$ at~$T\up{(S)}\sub{L}$.

With the second method we count only non-contractible loops that are topologically distinct from contractible ones 
due to the periodic boundary condition.
We can check whether a vortex loop is non-contractible or not in the following way.
We set the winding numbers along the $x$, $y$, and $z$-directions to zero, $w_x = w_y = w_z = 0$.
We then start from a point on the loop and we follow the loop path.
When the loop jumps from $(0,y,z)$ ($(L,y,z)$) to $(L,y,z)$ $((0,y,z))$, we change $w_x \to w_x + 1$ ($w_x \to w_x - 1$).
In the same manner we update  $w_y$ and $w_z$ when going across the system's ``boundary'' in the $y$ and 
$z$ directions. After going back to the starting point, at least one of the three winding numbers 
$w_x$, $w_y$, and $w_z$ take non-zero value when the loop is non-contractible. 

\begin{figure}[tbh]
\vspace{0.75cm}
\centering
\includegraphics[width=0.55\linewidth]{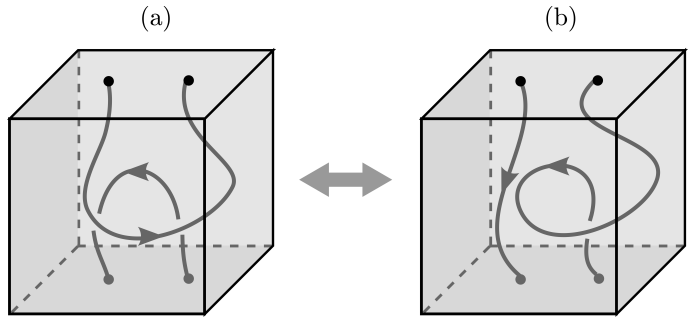} 
\caption{\label{fig:contract-example}
(Color online.)
Examples of contractible and non-contractible vortex loops.
In panel (a) there is one contractible vortex loop, the size of which is larger than the system size.
In panel (b) there are two non-contractible vortex loops.
The two vortex configurations in panels (a) and (b) can be continuously transformed one into the other 
through the reconnection of two vortex elements.
}
\end{figure}
We define $N\sub{non-contract}$ as
\begin{align}
\begin{split}
& N\sub{non-contract} \equiv \langle \text{Number of non-contractible vortex loops}  \rangle\sub{stat}.
\end{split}
\end{align}
We should note that the summation of winding numbers for all vortex loops vanishes identically:
\begin{align}
\sum\sub{loops} (w_x, w_y, w_z) = (0,0,0)
\end{align}
showing that there is no net rotational flow.
Another property is that $N\up{(M,S)}\sub{system-size} \geq N\up{(M,S)}\sub{non-contract}$.
Figure~\ref{fig:contract-example} shows examples of contractible and non-contractible vortex loops.
In Fig.~\ref{fig:contract-example} (a), there is one long contractible vortex loop, the size of which is larger than the system linear 
size. Through the reconnection of two vortex elements in the loop, the contractible vortex loop splits into two
non-contractible vortex loops as shown in the panel (b) in the same figure.

\begin{figure}[tbh]
\vspace{0.75cm}
\centering
\begin{minipage}{0.24\linewidth}
\centering
(a) \\
\includegraphics[width=0.95\linewidth]{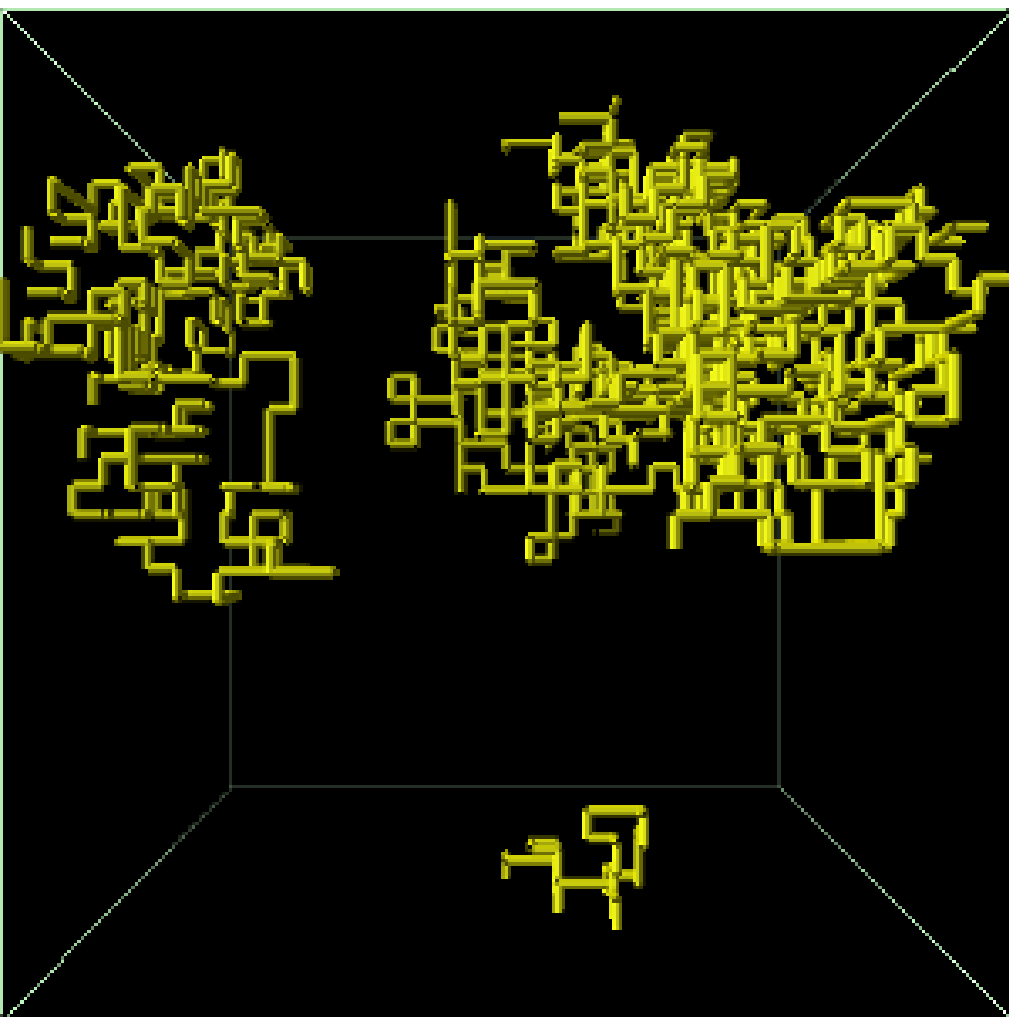}
\end{minipage}
\begin{minipage}{0.24\linewidth}
\centering
(b) \\
\includegraphics[width=0.95\linewidth]{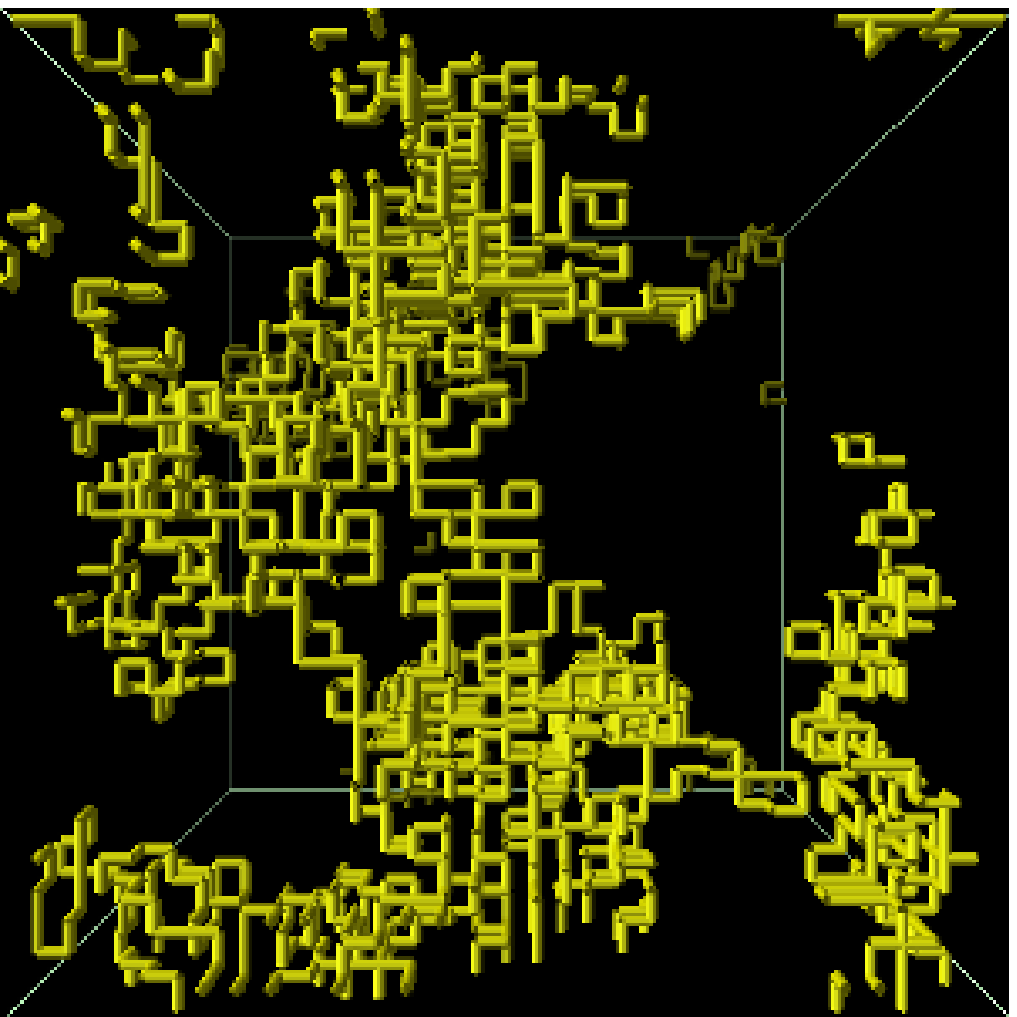}
\end{minipage}
\begin{minipage}{0.24\linewidth}
\centering
(c) \\
\includegraphics[width=0.95\linewidth]{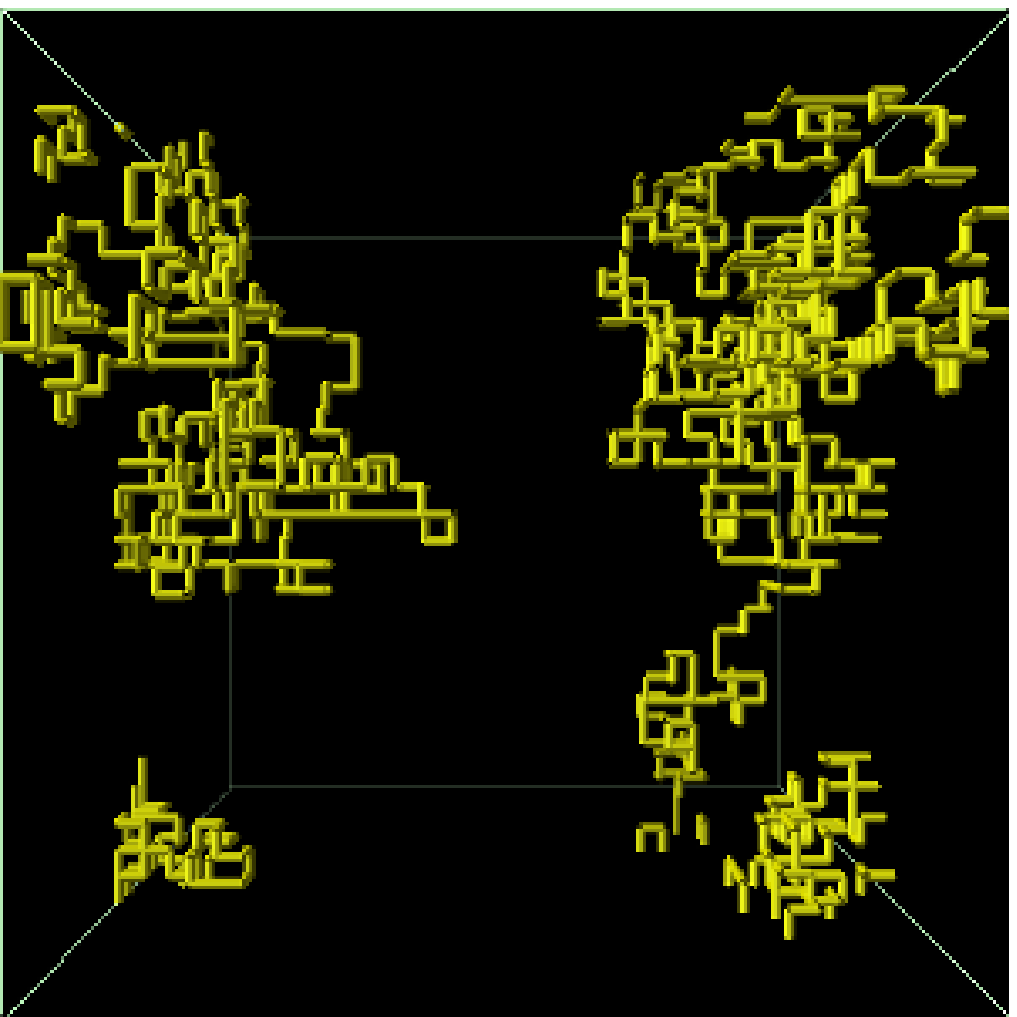}
\end{minipage}
\vspace{0.5cm}
\caption{\label{fig:vortex-noncont} 
(Color online.)
Three equilibrium snapshots of the system at $T = T\up{(S)}\sub{L}$.
We show the longest vortex loop obtained with the stochastic criterion for the connection of vortex elements.
In panel (a) the size of the vortex loop is smaller than the system size $L$.
In panel (b) the size of the vortex loop is longer than the system size $L$ both in the horizontal and vertical directions, but it is contractible.
In panel (c) the vortex loop is non-contractible in the vertical direction.
}
\end{figure}

Figure~\ref{fig:vortex-noncont} (a)-(c) shows the longest vortex loops in three equilibrium configurations at $T = T\up{(S)}\sub{L}$.
The vortex elements were connected using the stochastic criterion.
In panel (a), the size the vortex loop is smaller than the system size $L$.
Although the sizes of vortex loops are larger than the system size in panels (b) and (c), the vortex loop in panel (b) is contractible while the
 one in panel (c) is non-contractible in the vertical direction.

Figure~\ref{fig:noncontract} (b) shows the temperature dependence of $N\up{(M)}\sub{non-contract}$ (upper panel) 
and $N\up{(S)}\sub{non-contract}$ (lower panel). With the maximal criterium for connecting vortex elements, 
we have $N\up{(M)}\sub{non-contract} = 0$ and there are no non-contractible vortex loops.
This  {\it a priori} surprising results is due to the fact that with the maximal criterium all non-contractible vortex loops get  connected to neighboring
ones to form a large contractible vortex loop (see Fig.~\ref{fig:contract-example}: the configuration in panel (a) 
is preferred because the length of the single vortex loop is longer than the one of the two non-contractible vortices in panel (b)).
With the stochastic criterium for connecting vortex elements, we have a finite number of non-contractible vortex loops 
$N\up{(S)}\sub{non-contract} >0$.
As well as $N\up{(S)}\sub{system-size}$ in (c), $N\up{(S)}\sub{non-contract}$ in (d) loses its size dependence at 
$T\up{(S)}\sub{L}$ (within our numerical accuracy) and $N\up{(S)}\sub{non-contract} \simeq 0.532$.
In the limit of infinite system size $L\to \infty$, we expect a 
sharp transition from $N\sub{non-contract} = 0$ to $N\sub{non-contract} > 0 $ at 
a temperature close to $T\up{(S)}\sub{L}$, which suggests that one vortex loop larger than the system size $L$ at 
$T = T\up{(S)}\sub{L}$ is non-contractible with a probability close to $0.532$. 

\begin{figure}[tbh]
\centering
\begin{minipage}{0.49\linewidth}
\centering
(a)\\
\includegraphics[width=0.95\linewidth]{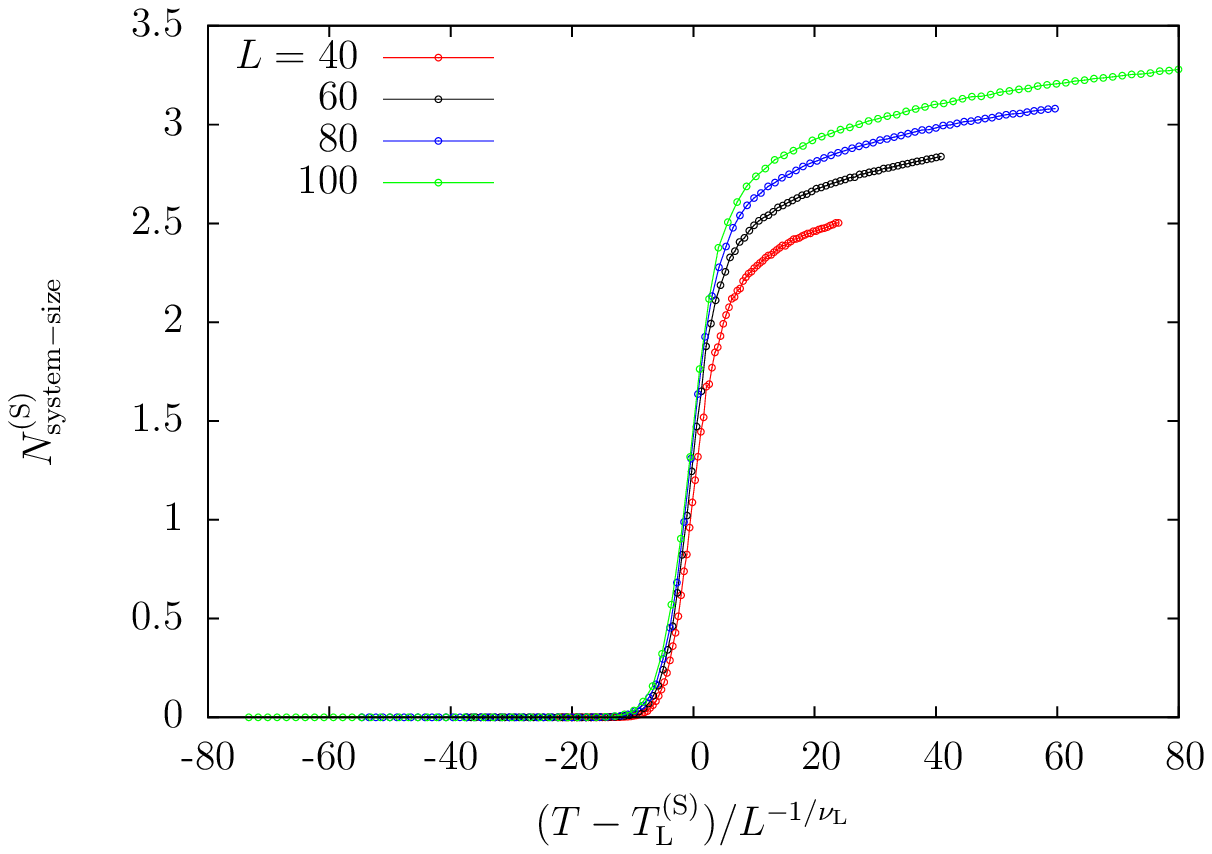} \\
(c)\\
\includegraphics[width=0.95\linewidth]{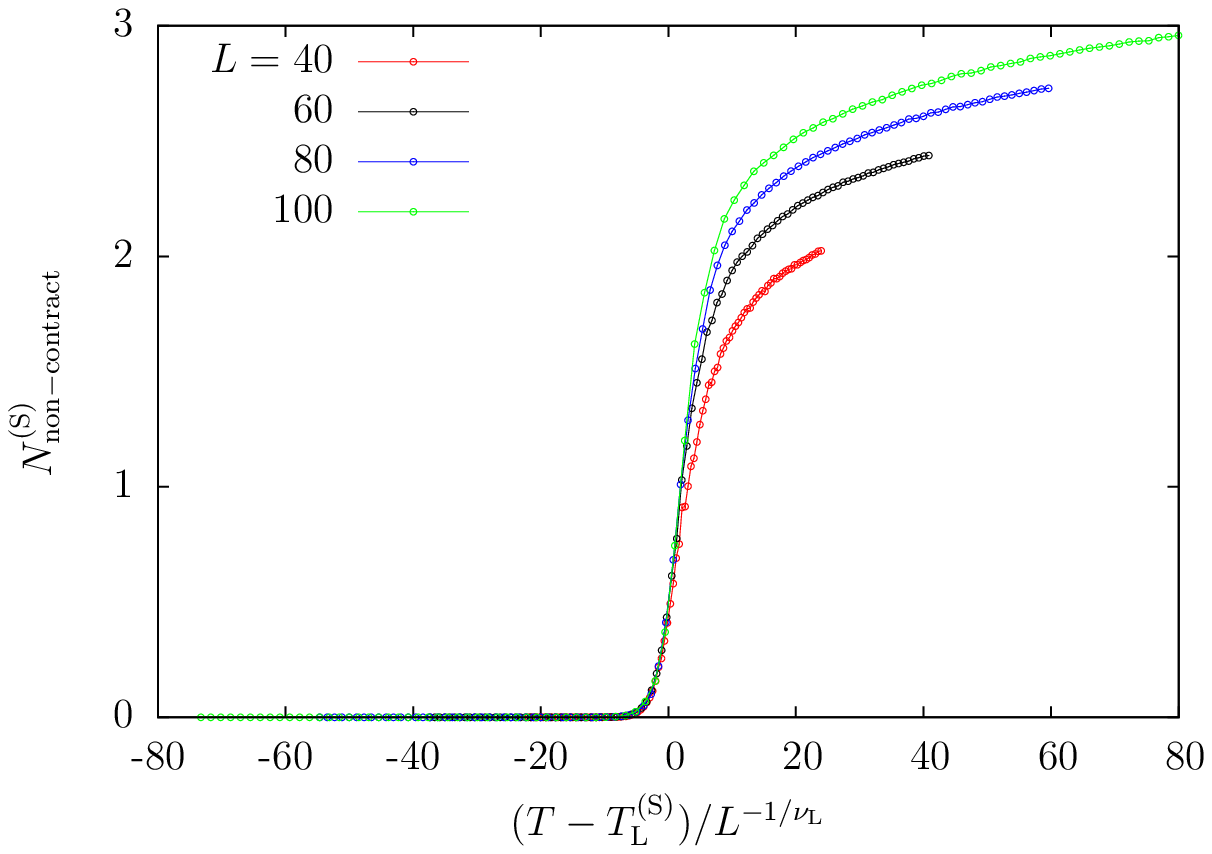} \\
\end{minipage}
\begin{minipage}{0.49\linewidth}
\centering
(b)
\includegraphics[width=0.95\linewidth]{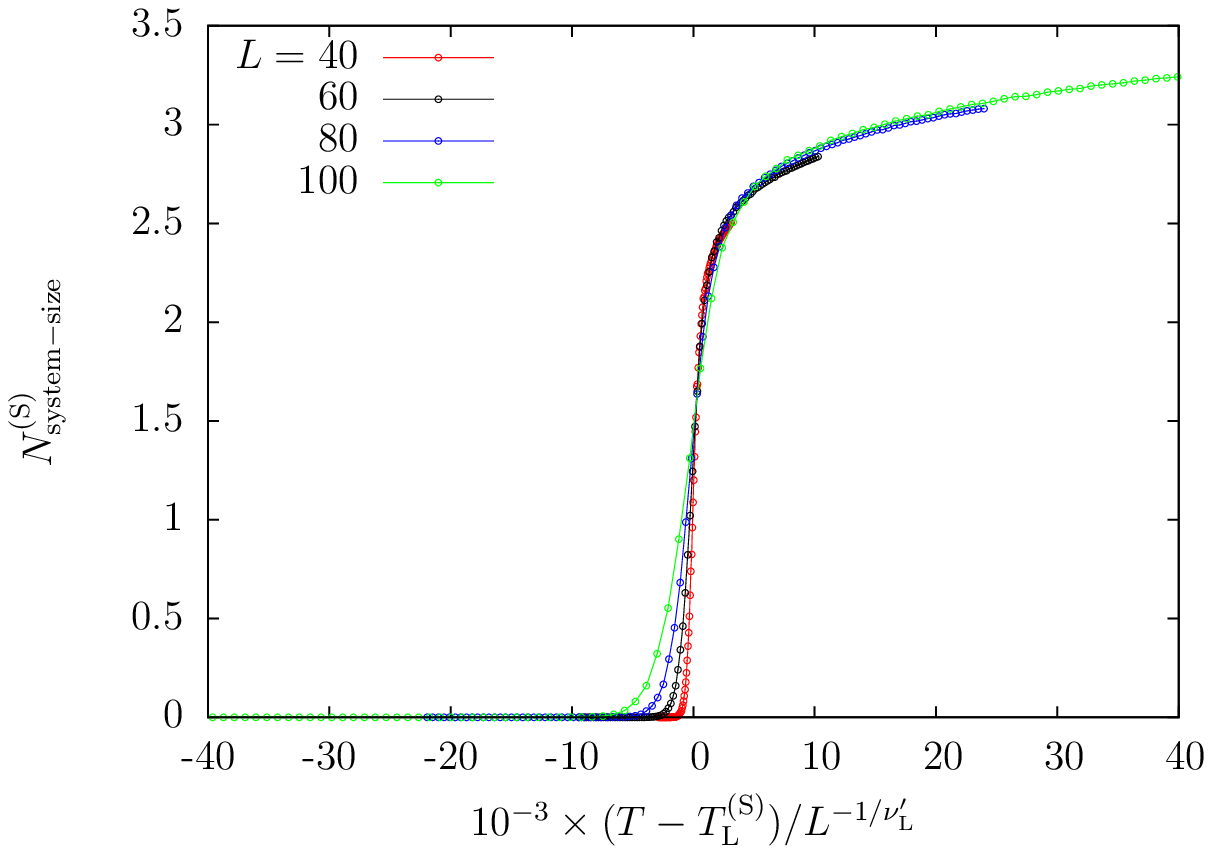} \\
(d)
\includegraphics[width=0.95\linewidth]{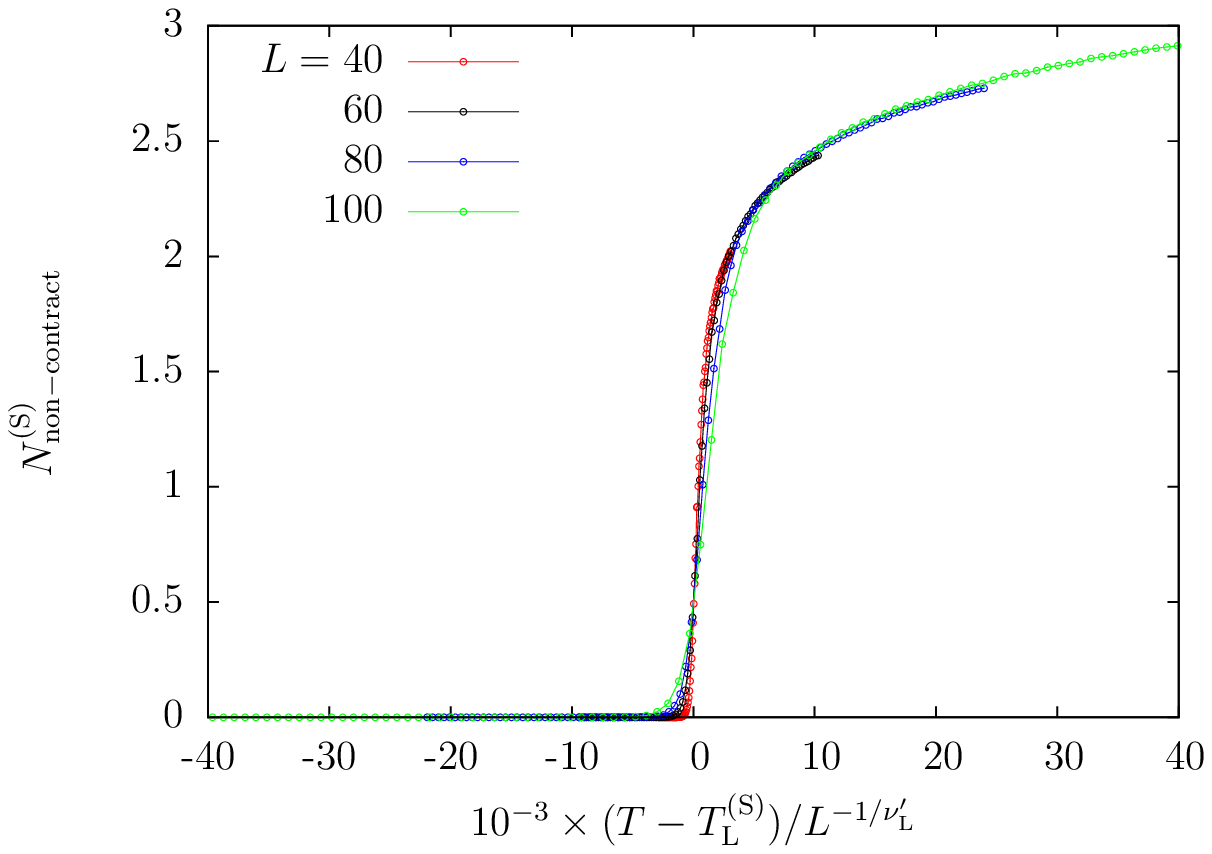} \\
\end{minipage}
\vspace{0.5cm}
\caption{\label{fig:noncontract-scale}
(Color online.)
In panels (a) and (b) the finite-size scaling of the data in Fig.~\ref{fig:noncontract} (c) for the 
number of vortex loops larger than the system size $N\up{(S)}\sub{system-size}$. 
In panels (c) and (d) the finite-size scaling of the data in Fig.~\ref{fig:noncontract} (d) for the 
number of non-contractible vortex loops $N\up{(S)}\sub{non-contract}$. 
The stochastic rule for connecting vortex elements was used.
The system sizes are given in the keys.
From the scaling analysis, we obtain the exponent $\nu\sub{L} = 0.76$ for $T < T\sub{L}\up{(S)}$ 
in panels (a) and (c), and $\nu\sub{L}^\prime = 0.34$ for
$T > T\sub{L}\up{(S)}$ in panels (b) and (d).
}
\end{figure}

\begin{figure}[tbh]
\centering
\includegraphics[width=0.49\linewidth]{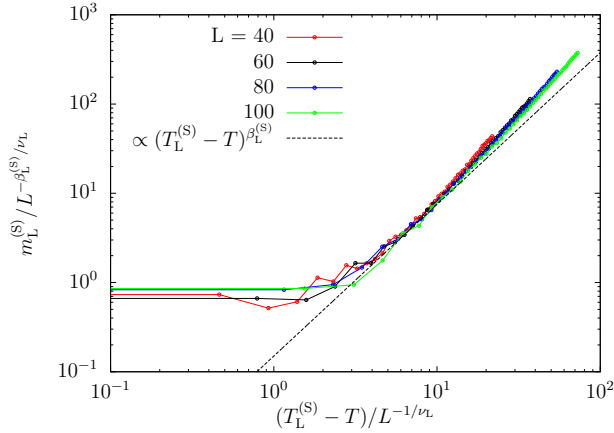}
\caption{\label{fig:porder-scale2}
(Color online.)
Finite-size scaling of the mass parameter $m\up{(S)}\sub{L}$. 
The sizes used are given in the key and the dotted (blue) line represents the 
critical behaviour close to the transition.
}
\end{figure}

From the fact that the number of vortex loops larger than the system size $N\up{(S)}\sub{system-size}$
and the number of non-contractible loops $N\up{(S)}\sub{non-contract}$ obtained with the stochastic criterium  
(see Figs.~\ref{fig:noncontract} (c) and (d)) are size independent at the vortex line-tension point $T = T\sub{L}\up{(S)}$,
we can expect them to be universal functions of $(T\sub{L}\up{(S)} - T) / L^{- \nu\sub{L}}$ ($(T - T\sub{L}\up{(S)}) / L^{- \nu^\prime\sub{L}}$)
with some exponents $\nu\sub{L}$ ($\nu\sub{L}^\prime$) at temperatures $T < T\sub{L}\up{(S)}$ ($T > T\sub{L}\up{(S)}$).
Figures~\ref{fig:noncontract-scale} (a) ((b)) and (c) ((d)) show $N\up{(S)}\sub{system-size}$ and $N\up{(S)}\sub{non-contract}$ as 
functions of $(T\sub{L}\up{(S)} - T) / L^{- \nu\sub{L}}$ ($(T - T\sub{L}\up{(S)}) / L^{- \nu^\prime\sub{L}}$) with $\nu\sub{L} = 0.76$ 
($\nu\sub{L}^\prime = 0.34$). The data  show good collapse on both sides of the line-tension point $T\sub{L}$.
In Fig.~\ref{fig:porder-scale2} we show the finite size scaling behaviour of $m\up{(S)}\sub{L}$
using  two exponents $\nu\sub{L}$ and $\beta\up{(S)}\sub{L}$ for the mass parameter
$m\up{(S)}\sub{L}$ (see the inset in Fig.~\ref{fig:porder} (a)).

\subsection{Discussion}

In this Section we analysed the statistical properties of the vortex tangle in equilibrium. 

The full vortex configuration is independent of the reconnection method and boundary conditions 
in the low temperature regime. The distribution of vortex lengths is simply exponential.

Different percolation thresholds can be identified by working with different vortex-related observables~\cite{Kajantie,Bittner}.
A natural characterisation of the loop ensemble is given by their length distribution, from which a critical point 
is identified as the temperature at which the mass parameter vanishes, the so-called line-tension point.
The result $T\up{(M,S)}\sub{L} < T\sub{c}$ shows that the spontaneous breaking of the U(1) symmetry 
is not directly connected to the percolation of vortex lines, which is consistent with previous work for 
the $3d$ XY model~\cite{Kajantie} and the O(2) model~\cite{Bittner}, and contrary to claims 
in~\cite{Kohring,AntunesBettencourt,Schakel,Nguyen,Camarda}. The fact 
that $T\sub{L}\up{(M)} < T\sub{L}\up{(S)}$ is reasonable since strings are longer in the 
former than in the latter case. The critical properties of finite loops remain
independent of the reconnection rule and the size of the mesh used to discretize 
space (within numerical accuracy).

At $T\up{(M,S)}\sub{L}$ we find the Fisher exponent $\alpha\up{(M,S)}\sub{L} \simeq 2.17$, and from this value 
we deduce $D\up{(M,S)}\sub{L} \sim 2.56$. Moreover,  $\alpha\up{(M,S)}\sub{L} < 2.5$ 
suggests that vortices at the line tension point behave as a self-seeking random walk. Bittner~{\it et al}~\cite{Bittner}
found $\alpha\sub{L} \simeq 2.26 - 2.27$ in the continuous O(2) field theory and
Kajantie {\it et al}~\cite{Kajantie} $\alpha\sub{L} \simeq 2.11$ in Monte Carlo simulations  of the $3d$ XY model, 
both with the stochastic reconnection rule and at the percolation point. Similarly, 
Ortu\~no {\it et al.} computed $\alpha\up{(M,S)}\sub{L} = 2.184(3)$ at the critical point of a network model for the 
disorder-induced localisation transition.
We recall that~\cite{Kajantie} also showed that the percolation-observable critical properties may also depend upon the 
reconnection rule adopted. 

We do not see any special feature in the density of vortex elements or the mean number of vortex 
lines at the line tension point. However, signatures of the vanishing line tension point are seen in other quantities.
We see a maximum in the ratio between the number of loops that are longer than the system size and the total number of loops 
when the maximal criterium is used (though the height of the maximum 
decreases with $L$ increasing and we cannot exclude that this effect disappears in the thermodynamic limit). 
(This feature is not shared by the numerator in this ratio.) 
The number of vortex loops that are longer than the system size and that are non-contractible 
constructed with the stochastic criterium behave similarly to an order parameter for the 
geometric transition. No quantity 
of this kind for the maximal rule behaves as an order parameter.
Interestingly enough, the thermodynamic threshold $T\sub{c}$ seems to appear as the 
temperature at which $\rho\sub{vortex}$ and $N\sub{loop}\up{(M,S)}$ change concavity.

At high temperature the influence of the reconnection method becomes very 
important as there are loops with length of the order of the linear size of the 
system or longer. The boundary conditions also become important. As in the 
quench dynamic analysis we will use
the equilibrium state at high temperature as the initial configuration, 
it is specially important to characterise the vortex tangle at
very high temperature. With the maximal criterium we found that one line carries
most of the vortex mass in the sample at very high temperature. With the stochastic 
criterium we found that the statistics of loops with length $l \ll L^2$ is Gaussian 
while even longer loops exist and their number density falls-off as $(l/L^3)^{-1}$. 

Vachaspati \& Vilenkin~\cite{Vachaspati} used a simple $Z_3-symmetric$ 
model to generate the putative initial conditions 
of the field theory that should describe the state of the universe before undergoing 
a phase transition. This is a
clock model with three phase values attributed at random with equal probability 
on each vertex of a regular cubic lattice with {\it open boundary conditions}. They used the stochastic rule 
to reconnect the vortex elements on a cell.
Strobl \& Hindmarsch increase the number of discrete angles  from 3 to 255 in a formally infinite lattice~\cite{Strobl}.
They both found the statistics of a Gaussian random walk ($D=2$) as for 
a dense polymer network~\cite{deGennes}. 
At very high temperature the statistics of our loop ensemble, when treated with the stochastic reconnection 
rule, and for length scales such that $l \ll L^2$, also approaches this result. 
Instead, the statistics is very different with the maximal reconnection criterium or 
beyond the crossover at $L^2$. 

We note that the behaviour of  vortex loops in the three-dimensional model is quite 
different from the one of the topological defects in the Kosterlitz-Thouless transition 
of the two-dimensional system. In the latter, the phase transition occurs at 
the same temperature at which the vortex pairs  unbind. In the former, percolation occurs at a different
temperature from $T\sub{c}$. This is similar to what happens in the Ising model of magnetism: in two-dimensions
the percolation of geometric clusters occurs at the  critical temperature while in three-dimensions 
this is not the case. The fact that percolation of geometric objects does not always occur 
at the thermodynamic critical phenomenon has been known since the work in~\cite{Muller}.

\section{Fast quench dynamics}
\label{sec:dynamics}

In this section, we consider the stochastic dynamics following an instantaneous quench 
from equilibrium at $T = 2 \ T\sub{c}$ to $T=0$. The analysis in the previous section 
allowed us to characterise the vortex configurations at the 
initial state at high temperature  in full detail. Here we will be particularly concerned with the evolution
of these states after an infinitely fast deep quench. We will show that during the low temperature dynamics
vortex lengths with statistics and fractal dimension numerically identical to the one at the 
percolation threshold $T\sub{L}\up{(M,S)}$ will be relevant, although the quench protocol 
 does not spend any time at  nor even close to it.  These features exist for all microscopic 
 dynamic rules.

\subsection{The initial state}

Whether the initial state has an order parameter $\overline{\psi}$ that vanishes or not, 
can have a highly non-trivial influence on the subsequent dynamics. This fact was derived by Toyoki and Honda~\cite{Toyoki-analytic}
and later confirmed numerically~\cite{Toyoki,Mondello}. Here, we use equilibrium initial states such that the vortex
configuration, see Fig.~\ref{fig:fast-vortex-ini} (a), is characterised by the density $\rho\sub{vortex} \simeq 0.25$ in Fig.~\ref{fig:vortex-all} 
and the distribution of vortex loop lengths shown  in Fig.~\ref{fig:vortex-size}~(a). 
At $T = 2 \ T\sub{c}$, the order parameter suffers from finite size corrections 
and we measure $\overline{\psi} \simeq 0.030$ for 
$L = 20$, $\overline{\psi} \simeq 0.016$ for $L = 30$, $\overline{\psi} \simeq 0.010$ for $L = 40$, and 
$\overline{\psi} \simeq 0.0073$ for $L = 50$.
These values are small enough for the dynamics to be regarded as subsequent to a 
zero average field initial condition, i.e., 
$\overline{\psi} \sim 0$. This is also confirmed by the fact that the scaling regime is reached 
 independently of the system size $L$, see  Figs.~\ref{fig:fast-xid} (a)-(d) and Figs.~\ref{fig:fast-rhov}~(a)-(d).
(In some references, e.g.~\cite{Mondello}, such quenches are named ``critical''. In the statistical physics context a ``critical quench" 
is a quench to the critical temperature $T\sub{c}$, so we rather not use this terminology here.)

The statistical and geometrical properties of the vortex loops at high temperatures were characterised in 
detail in Sec.~\ref{sec:equilibrium-vortex} and we will use this information here.

\subsection{The initial stage of evolution}

\subsubsection{Instability}

Let us consider the dynamics in the initial stage of evolution within the mean-field framework.
By approximating the initial high temperature state as $\psi(t = 0) \simeq 0$ 
and the time-dependent field as $\psi = \delta \psi = u e^{i (\Vec{k} \cdot \Vec{x} - \omega t)} + v^\ast e^{- i (\Vec{k} \cdot \Vec{x} - \omega^\ast t)}$,
the Bogoliubov-de Gennes equation becomes 
\begin{align}
\begin{split}
\{ \omega^2 / c^2 - k^2 + (2 \mu + i \gamma\sub{L}) \omega + g \rho \} u + O(\delta \psi^2) &= 0, 
\\
\{ \omega^2 / c^2 - k^2 - (2 \mu - i \gamma\sub{L}) \omega + g \rho \} v + O(\delta \psi^2) &= 0.
\end{split}
\end{align}
The solution to the linear set of equations is
\begin{align}
\begin{split}
& \omega_1^\pm(k) = \frac{- i c^2 (\gamma\sub{L} - 2 i \mu) \pm c \sqrt{4 k^2 - c^2 (\gamma\sub{L} - 2 i \mu)^2 - 4 g \rho}}{2}, \\
& \omega_2^\pm(k) = \frac{- i c^2 (\gamma\sub{L} + 2 i \mu) \pm c \sqrt{4 k^2 - c^2 (\gamma\sub{L} + 2 i \mu)^2 - 4 g \rho}}{2}.
\end{split}
\end{align}
$\omega_{1,2}^-(k)$ are rapidly decaying modes for all $k$ and vanish in the non-relativistic limit $c\to\infty$, 
while $\omega_{1,2}^+(k)$ are slowly growing modes for $k \lesssim \sqrt{g \rho}$ and decaying modes for $k \gtrsim \sqrt{g \rho}$.
In the first stage of the ordering process, $\omega_{1,2}^+(k \lesssim \sqrt{g \rho})$ are the most important modes.
The time scale of the growth is $t_0 \sim (\mathrm{Im}[\omega_{1,2}^+(k)])^{-1} \sim (g \rho - k^2)^{-1/2}$
and the mean-field approximation breaks down beyond it.

\subsubsection{Irrelevance of the reconnection rule}

\begin{figure}[tbh]
\centering
\begin{minipage}{0.24\linewidth}
\centering
(a) \\
\includegraphics[width=0.95\linewidth]{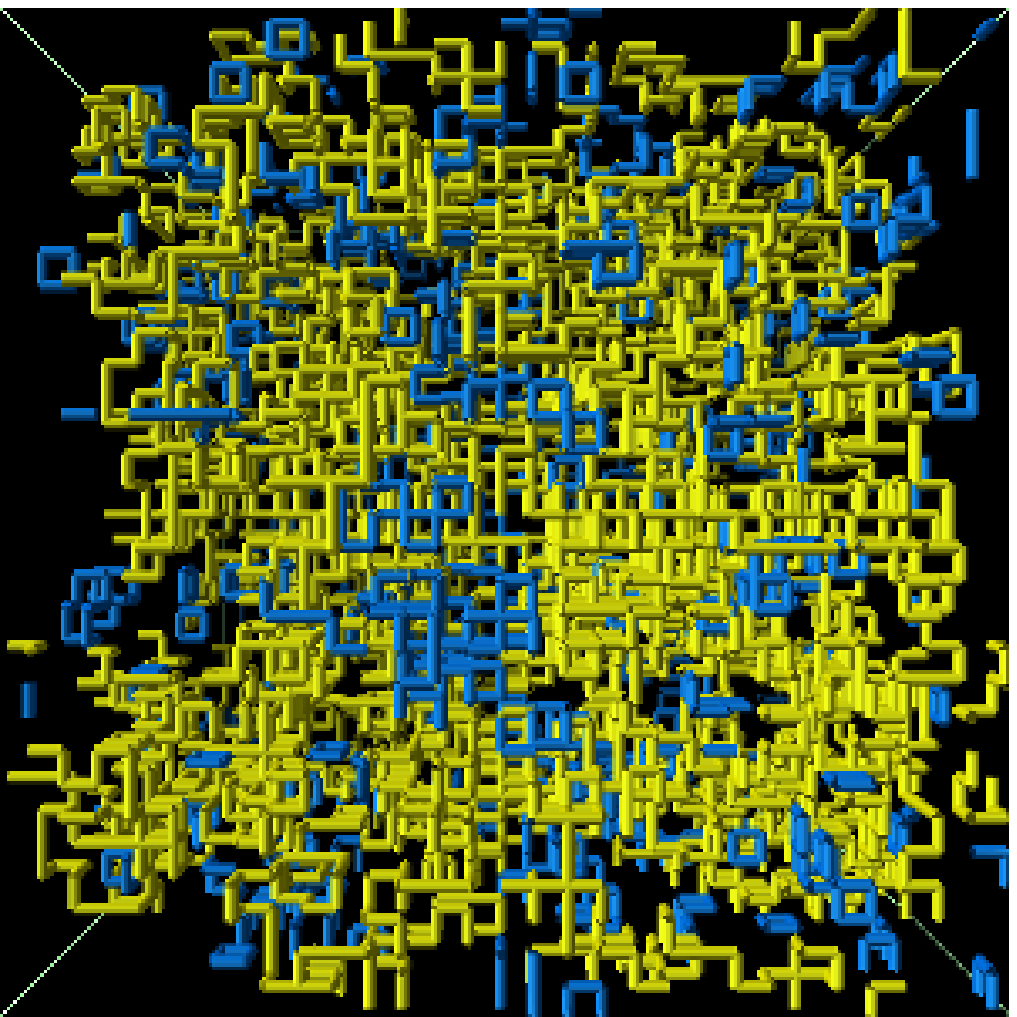} \\
\includegraphics[width=0.95\linewidth]{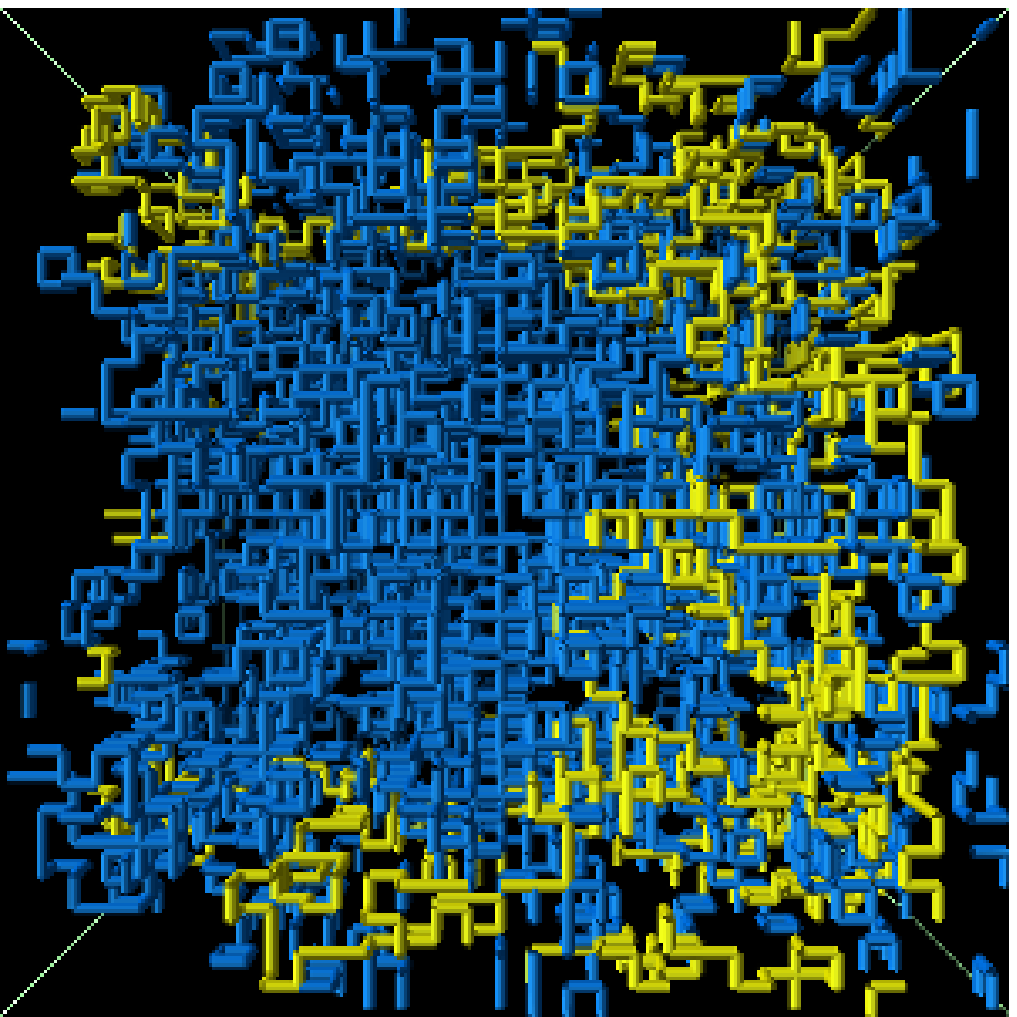} \\
\end{minipage}
\begin{minipage}{0.24\linewidth}
\centering
(b) \\
\includegraphics[width=0.95\linewidth]{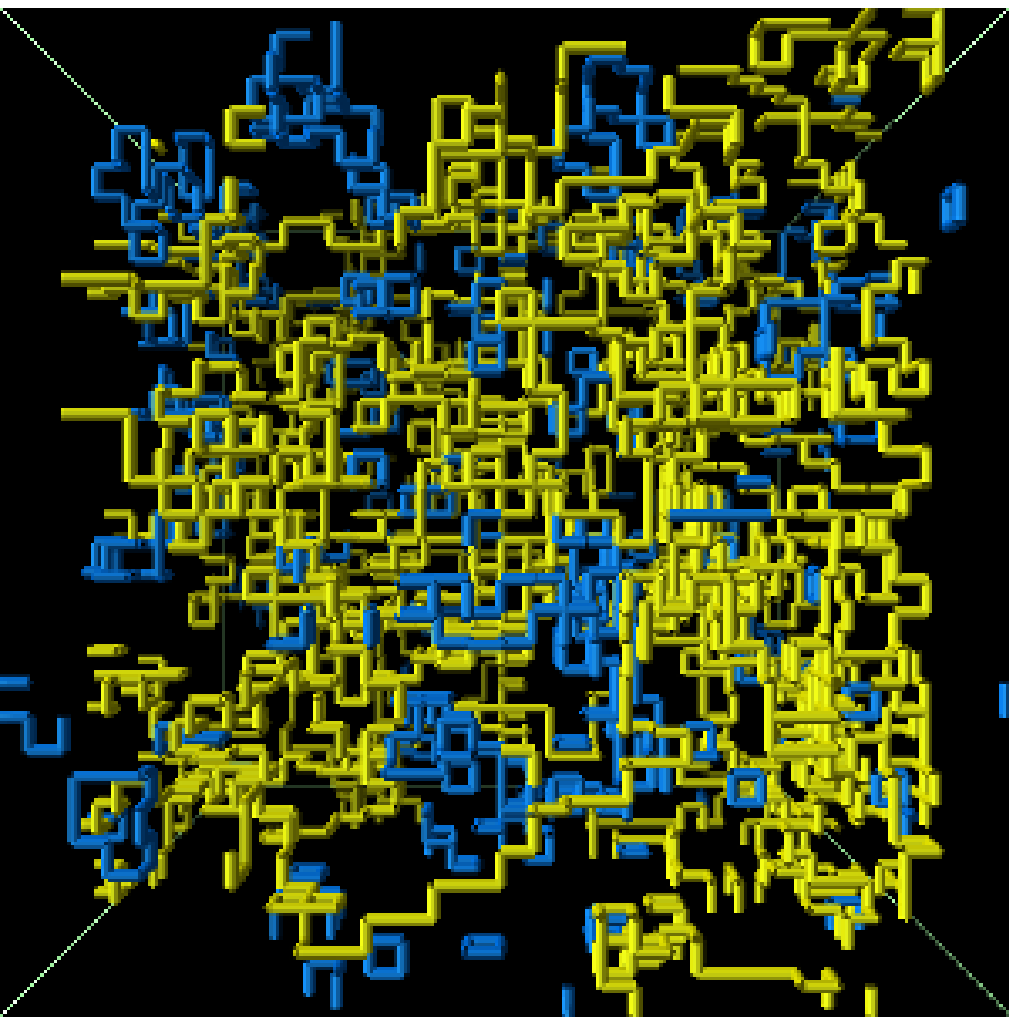} \\
\includegraphics[width=0.95\linewidth]{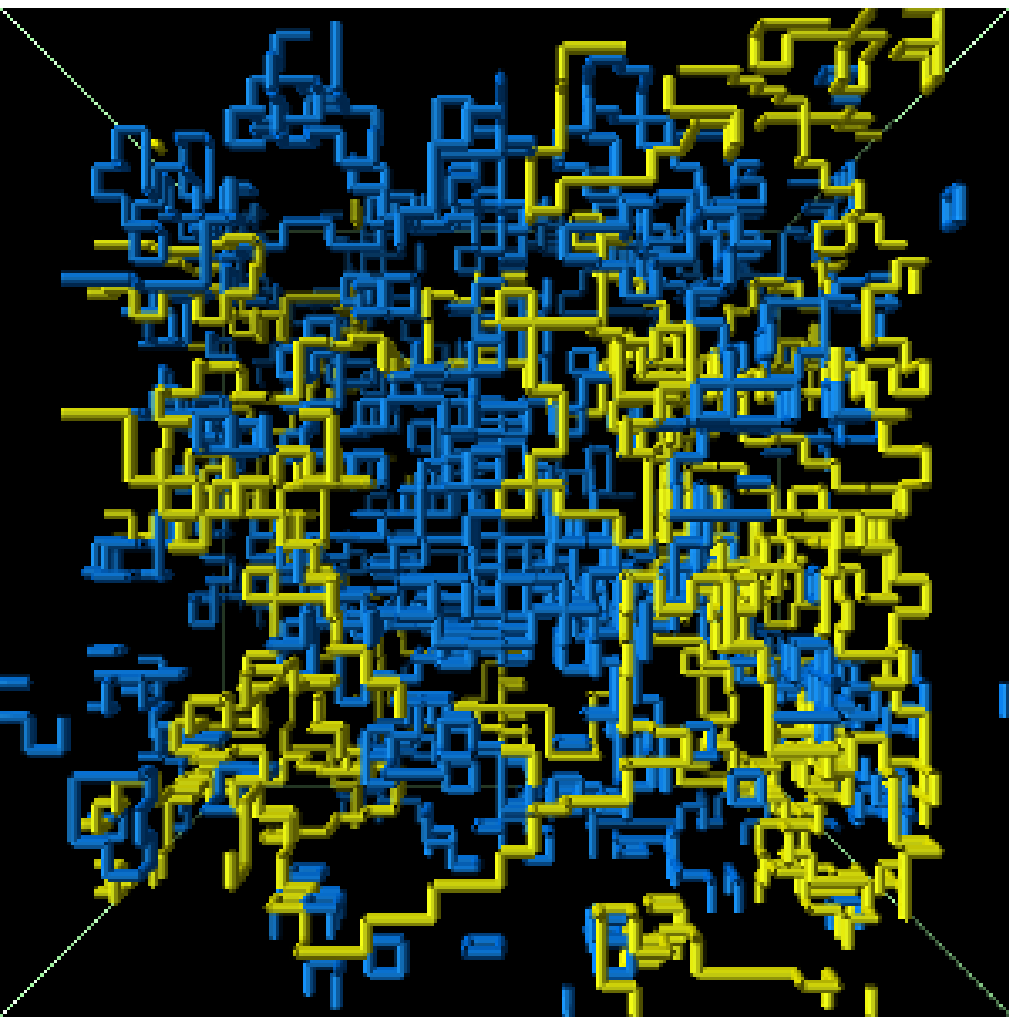} \\
\end{minipage}
\begin{minipage}{0.24\linewidth}
\centering
(c) \\
\includegraphics[width=0.95\linewidth]{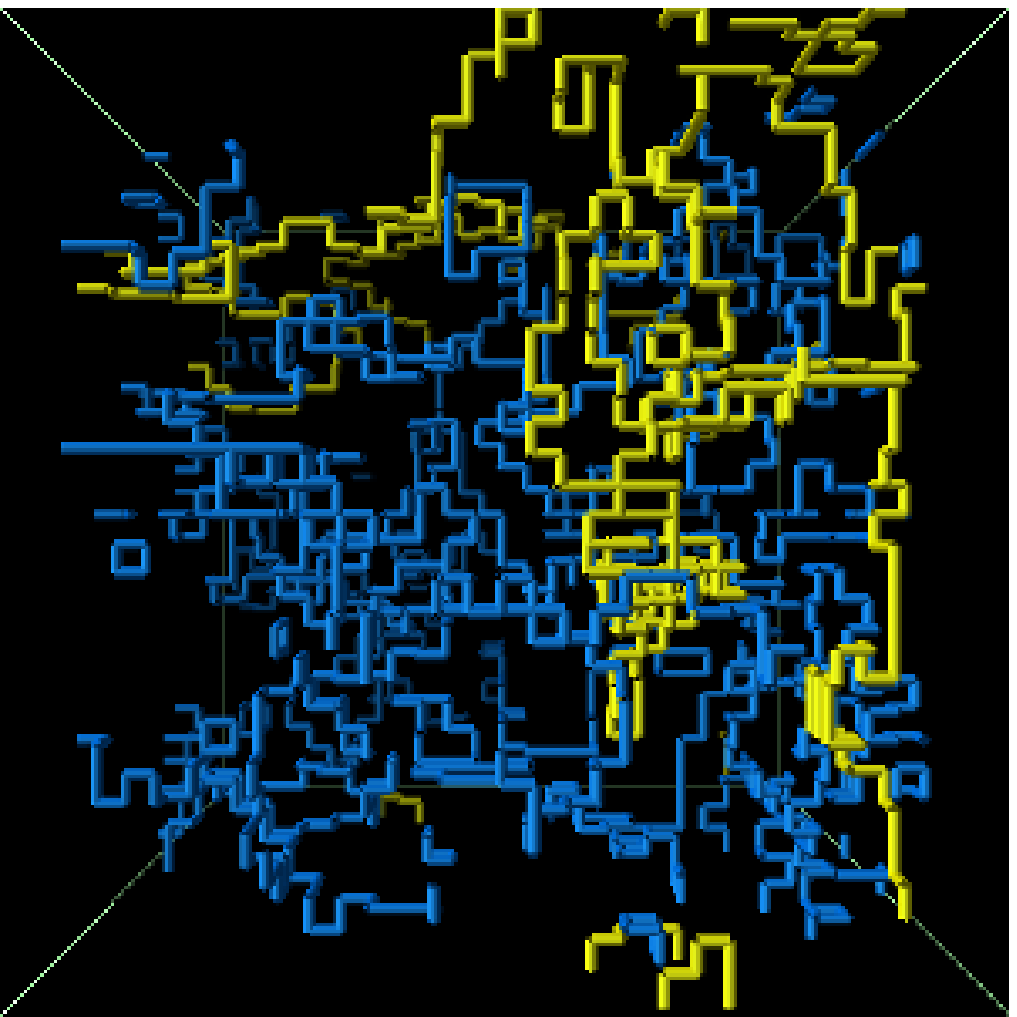} \\
\includegraphics[width=0.95\linewidth]{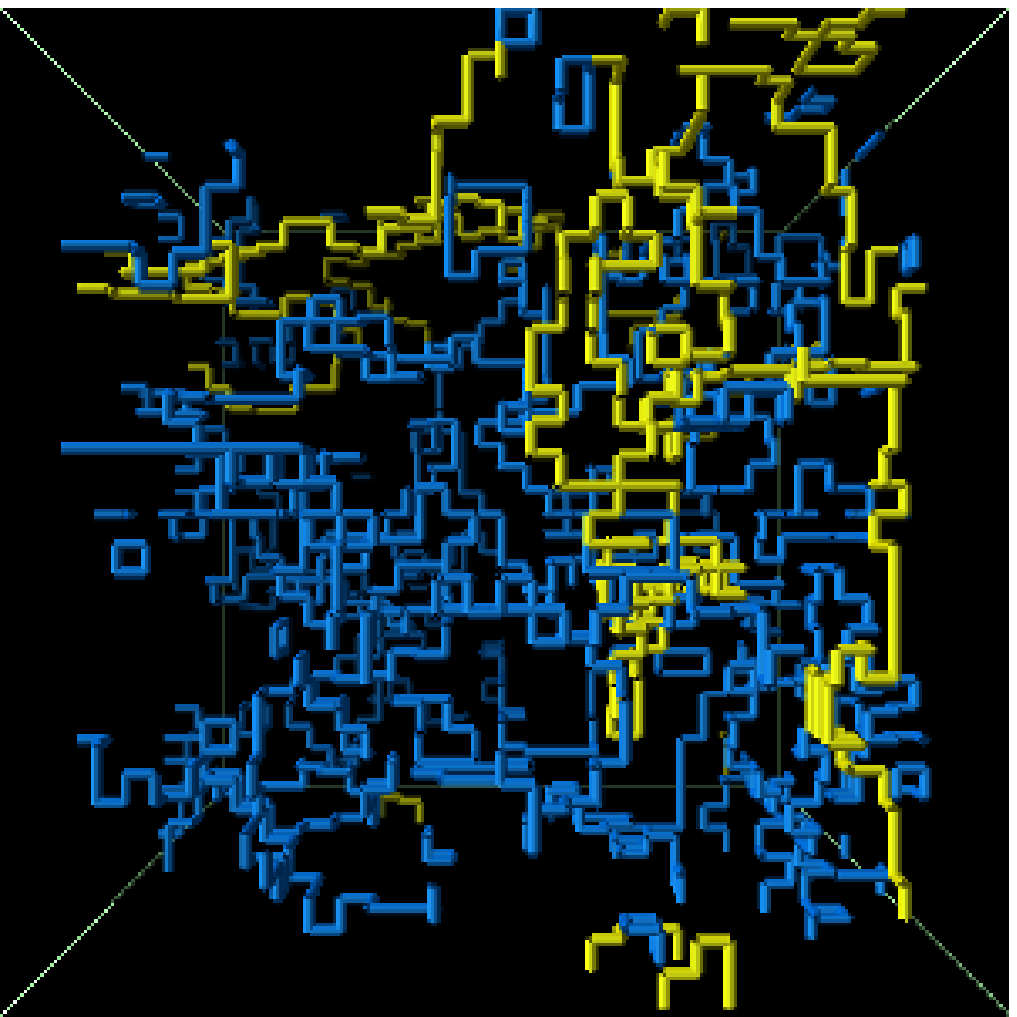} \\
\end{minipage}
\begin{minipage}{0.24\linewidth}
\centering
(d) \\
\includegraphics[width=0.95\linewidth]{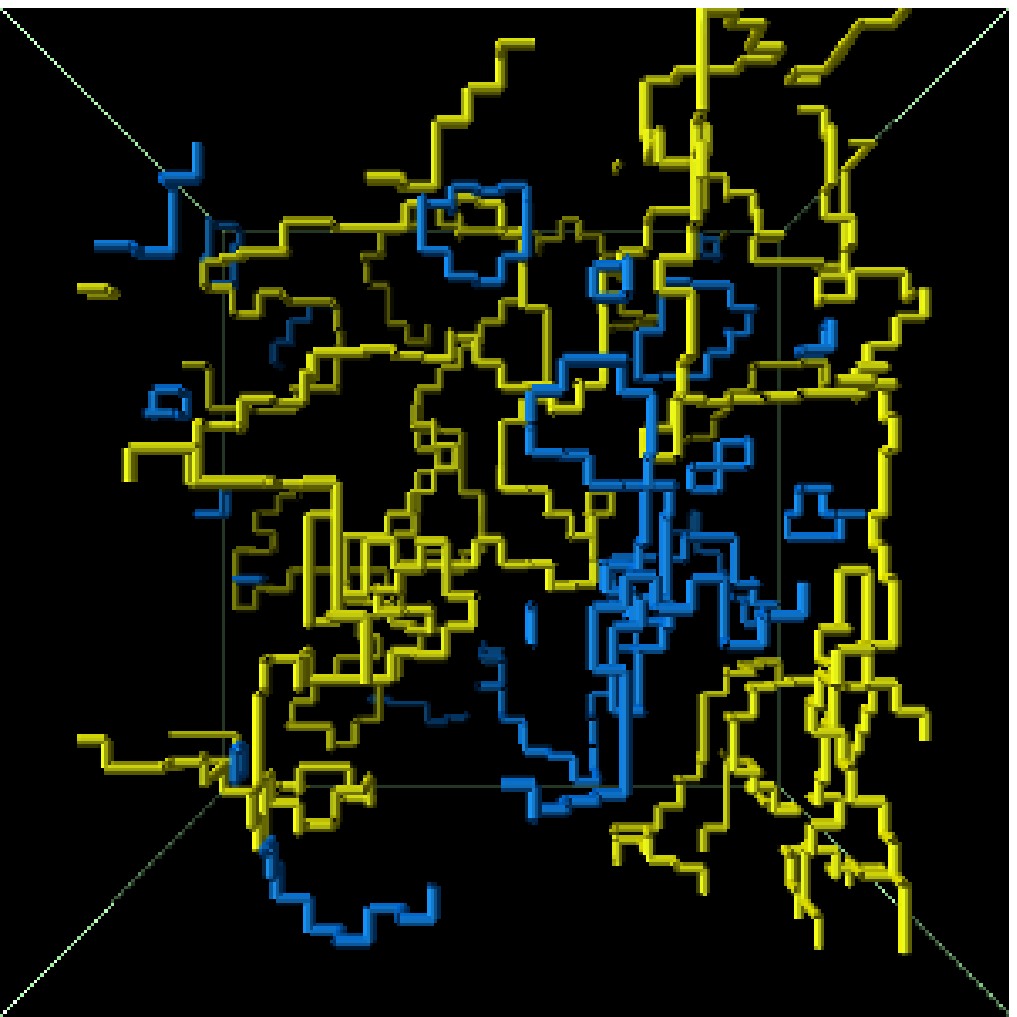} \\
\includegraphics[width=0.95\linewidth]{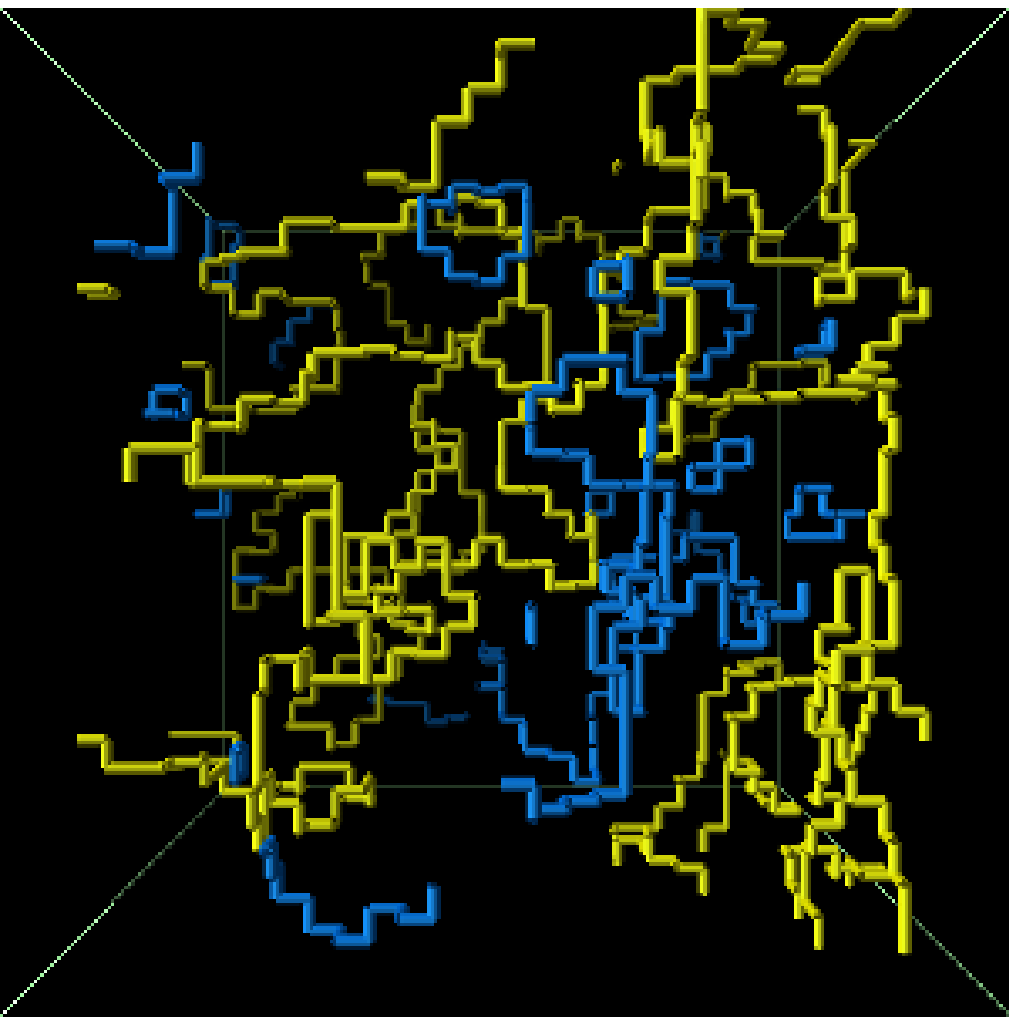} \\
\end{minipage}
\vspace{0.5cm}
\caption{\label{fig:fast-vortex-ini} 
(Color online.)
Snapshots of the vortex configurations at (a) 
$t = 2$, (b) $t = 3$, (c) $t = 4$, and (d) $t = 5$, after an instantaneous quench 
at $t=0$ from equilibrium at $2 \ T\sub{c}$. 
We plot all vortex line elements at the centers of plaquettes with non-zero flux (the total system linear size is $L = 40$).
The vortex line elements are shown in grey (blue) in the black background and the longest vortex 
lines in each image are highlighted (in yellow).
The configurations are generated with the under-damped Langevin 
equation~\eqref{eq:under-damped-Langevin} running at $T = 0$.
The maximal and stochastic line reconnection criteria  were used in the upper and lower panels, respectively. We 
note that while they influence the vortex tangle at the initial state, the rule used 
to link the vortex elements becomes irrelevant after a very short time scale, $t\simeq 4$.
}
\end{figure}

Figure~\ref{fig:fast-vortex-ini} shows the vortex loop configurations at four instants soon after the 
quench. One sees from the pictures that the reconnection rule used to build the vortex loops 
becomes irrelevant relatively soon, as the configurations in the upper and lower panels 
in the column (c) are very similar and in (d) are identical.

\subsection{The dynamic correlation length}

After a transient, the system is expected to enter a dynamic scaling regime~\cite{Bray} 
characterised by a growing length scale, $\xi\sub{d}(t) \simeq t^{1/z\sub{d}}$. How soon or not this is achieve 
will be discussed in the following subsections; for the moment we assume the scaling 
regime established and we study global correlation functions and observables within its framework.

The dynamic exponent in the low temperature phase, $z\sub{d}$, is different from the one found in quenches to the critical point that, in 
turn, coincides with the equilibrium critical one $z\sub{eq}$ discussed in Sec.~\ref{sec:equilibrium-relax-time}. We will now 
determine $z\sub{d}$.

The dynamic growing length, $\xi\sub{d}(t)$, can be measured in different ways by exploiting the 
dynamic scaling hypothesis~\cite{Bray}. Under this assumption, in the infinite size limit, the space-time correlation function and the 
dynamic structure factor after a quench to very low-temperatures should scale as
\begin{equation}
S(k,t) \simeq \xi\sub{d}^d(t) \ \Phi(k \xi\sub{d}(t))
\qquad\qquad
C(r,t) \simeq f\left(\frac{r}{\xi\sub{d}(t)} \right) \; .
\end{equation}
with $\Phi$ and $f$ two scaling functions. 

\vspace{0.25cm}

\begin{figure}[tbh]
\centering
\begin{minipage}{0.49\linewidth}
\centering
(a)\\
\includegraphics[width=0.95\linewidth]{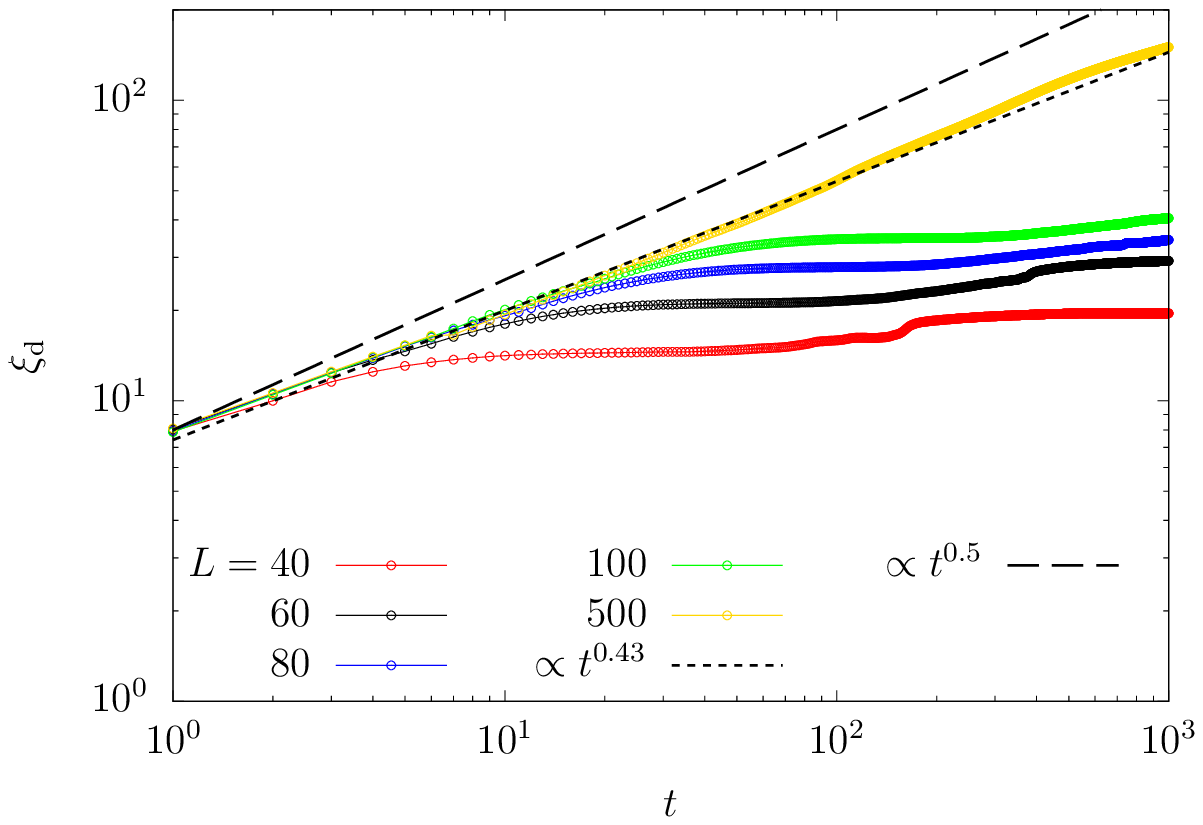}
\end{minipage}
\begin{minipage}{0.49\linewidth}
\centering
(b)\\
\includegraphics[width=0.95\linewidth]{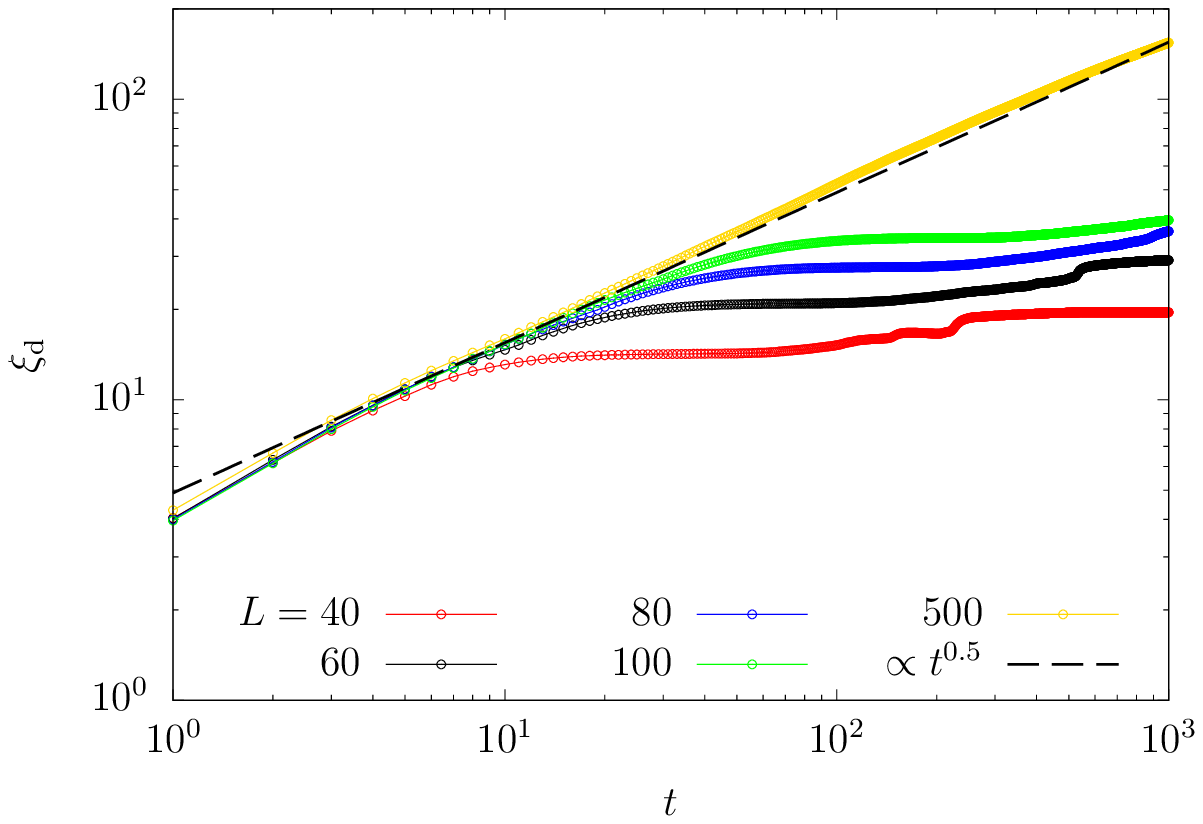}
\end{minipage}
\begin{minipage}{0.49\linewidth}
\centering
(c)\\
\includegraphics[width=0.95\linewidth]{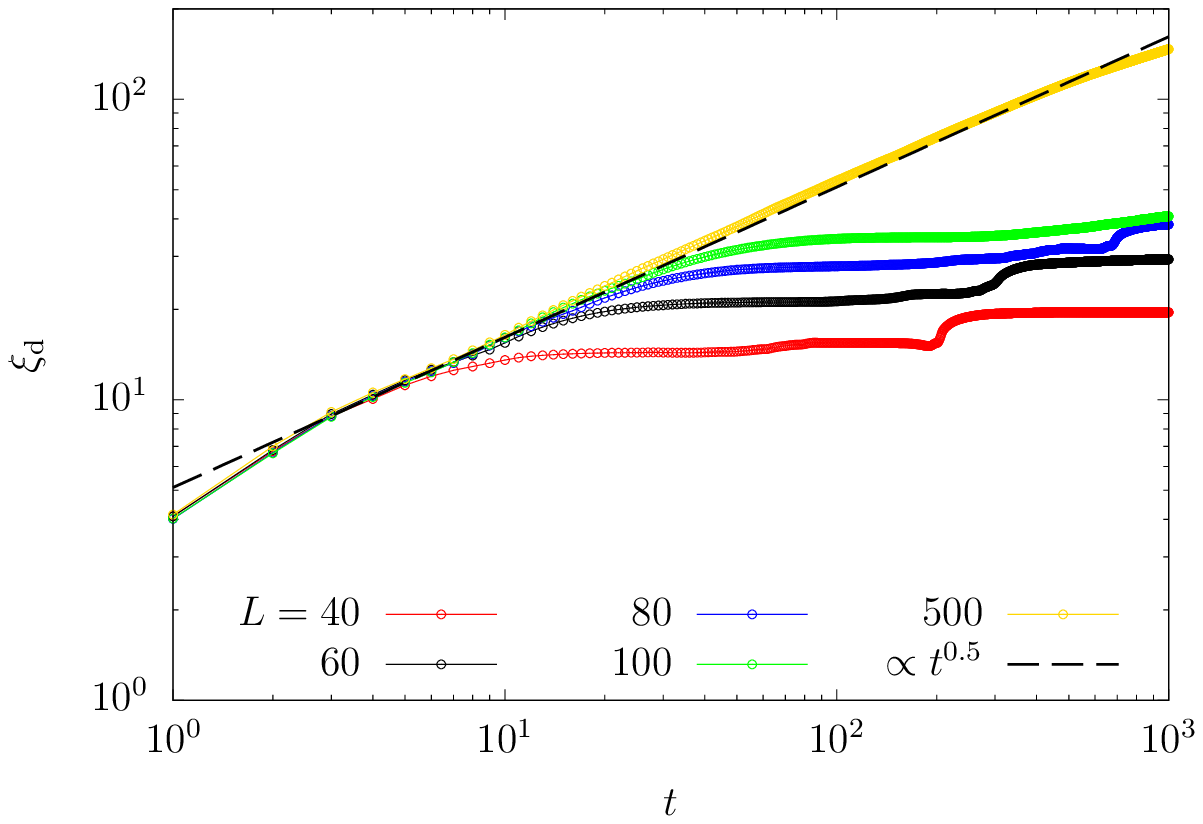}
\end{minipage}
\begin{minipage}{0.49\linewidth}
\centering
(d)\\
\includegraphics[width=0.95\linewidth]{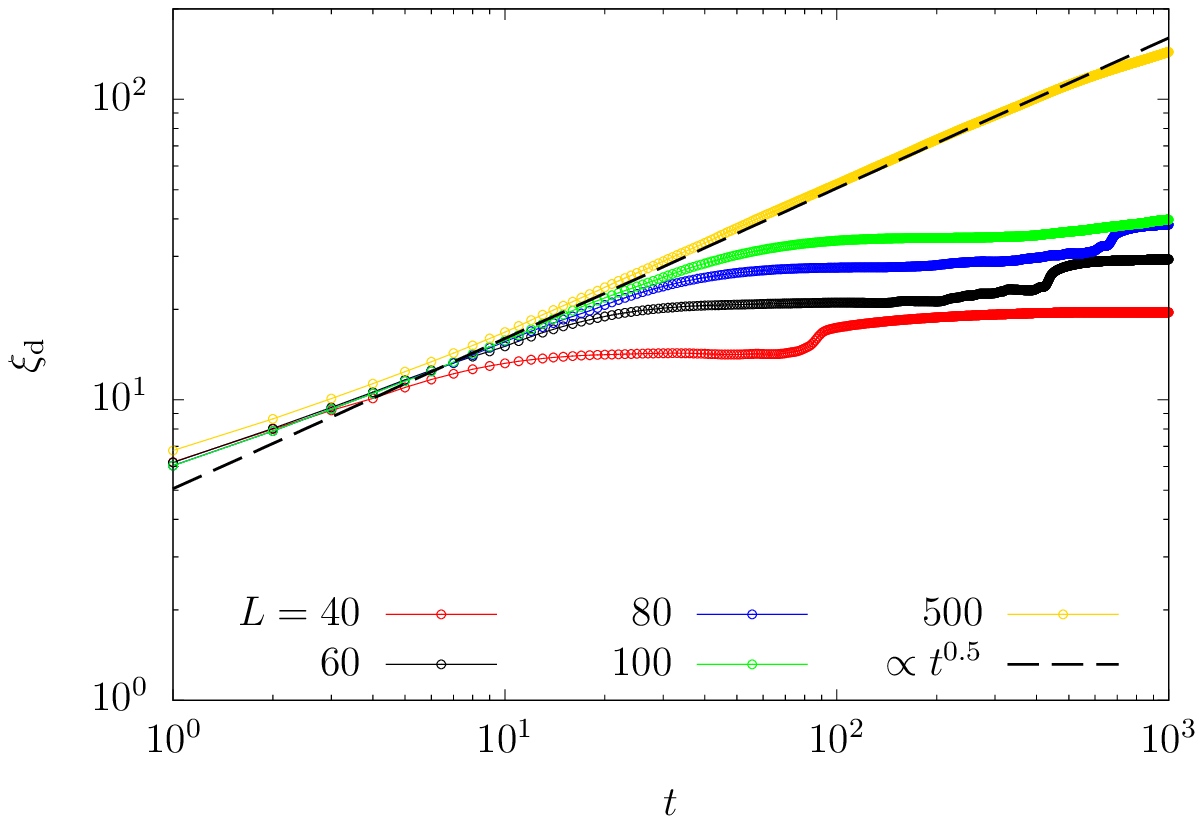}
\end{minipage}
\caption{\label{fig:fast-xid} 
(Color online.)
The dynamical correlation length $\xi\sub{d}$ against time 
$t$ obtained from the over-damped Langevin equation \eqref{eq:over-damped-Langevin} (a),
the under-damped Langevin equation \eqref{eq:under-damped-Langevin} (b), 
the ultra-relativistic limit of the under-damped Langevin equation \eqref{eq:under-damped-ultra-relativistic} (c)  
and the  non-relativistic limit of under-damped Langevin equation \eqref{eq:under-damped-non-relativistic} (d)
in systems with different system sizes given in the key.}
\end{figure}

Figures \ref{fig:fast-xid} (a)-(d) show the dynamical correlation length $\xi\sub{d}$ obtained from
\begin{align}
\frac{S(k = 2 \pi / \xi\sub{d}(t))}{S(k \to 0)} = 10^{-1}, \quad
S(k \to 0) \equiv 2 S(\Delta k) - S(2 \Delta k), 
\end{align}
a similar definition to the one used for the equilibrium correlation length $\xi\sub{eq}$ 
in Eqs.~\eqref{eq:correlation-length}. 
In the algebraic regime, $\xi\sub{d}(t) \simeq t^{1/z\sub{d}}$, the 
estimated exponents, $z\sub{d}$, are 0.43 for the over-damped Langevin equation \eqref{eq:over-damped-Langevin}, 0.50 for the 
under-damped Langevin equation \eqref{eq:under-damped-Langevin}, 0.50 for the ultra-relativistic limit of the 
under-damped Langevin equation \eqref{eq:under-damped-ultra-relativistic}, and 0.50 for the non-relativistic limit of the under-damped 
Langevin equation \eqref{eq:under-damped-non-relativistic}. 
We have not observed an appreciable change in the value of $z\sub{d}$ by varying the 
increments in space and time $\Delta x$ and $\Delta t$ in our algorithm for the over-damped 
evolution.

As we will show in Sec.~\ref{subsec:time-dep-vortex}, 
the exponents characterising the vortex density $\rho\sub{vortex}$ decay, and the growth of 
the dynamical correlation length $\xi\sub{d}$, are well related by 
$\xi\sub{d} \propto 1 / \sqrt{\rho\sub{vortex}}$ within numerical accuracy. Again, we obtain a weak discrepancy between our 
result for the over-damped Langevin equation and the prediction; $\xi\sub{d} \propto t^{1/2}$~\cite{BrayHumayun}, 
and good agreement for the other three under-damped Langevin equations. The slight disagreement with theory in the numerical data for the over-damped dynamics was also observed in~\cite{Toyoki,Mondello}. We will give a possible reason for it in Sec.~\ref{subsec:time-dep-vortex}.

At sufficient long times and for finite system sizes the growing length saturates.
Saturation is 
observed in the curves for $L\leq 100$ when the curves depart from the power law and reach a 
plateau. For $L=500$ the saturation is pushed beyond the numerical 
time window. We postpone the finite-size scaling analysis of the dynamic correlation length in 
Sec.~\ref{subsec:finite-size}.

\subsection{Time-dependent vortex density}
\label{subsec:time-dep-vortex}

We now examine the phase ordering process from the point of view of the vortex dynamics.
In Figs.~\ref{fig:fast-vortex-ini}~(a)-(d), we show snapshots of the vortex elements in the initial stage of 
evolution; $0 \leq t \leq 5$. The two rows compare the loop configurations for the two 
reconnection conventions for the same field configurations. We have already noted 
that while the loop configurations are different 
at very short times, $t=2, \ 3$, they are the same at the two latest times, $t=4, \ 5$.
In both cases, the longest vortex present in the initial configuration breaks up generating shorter 
vortex loops, the shortest 
vortex rings on the scale of the grid are rapidly annihilated, and loops of finite but long length are
still present during the evolution. 

\vspace{0.25cm}

\begin{figure}[tbh]
\centering
\begin{minipage}{0.49\linewidth}
\centering
(a)\\
\includegraphics[width=0.95\linewidth]{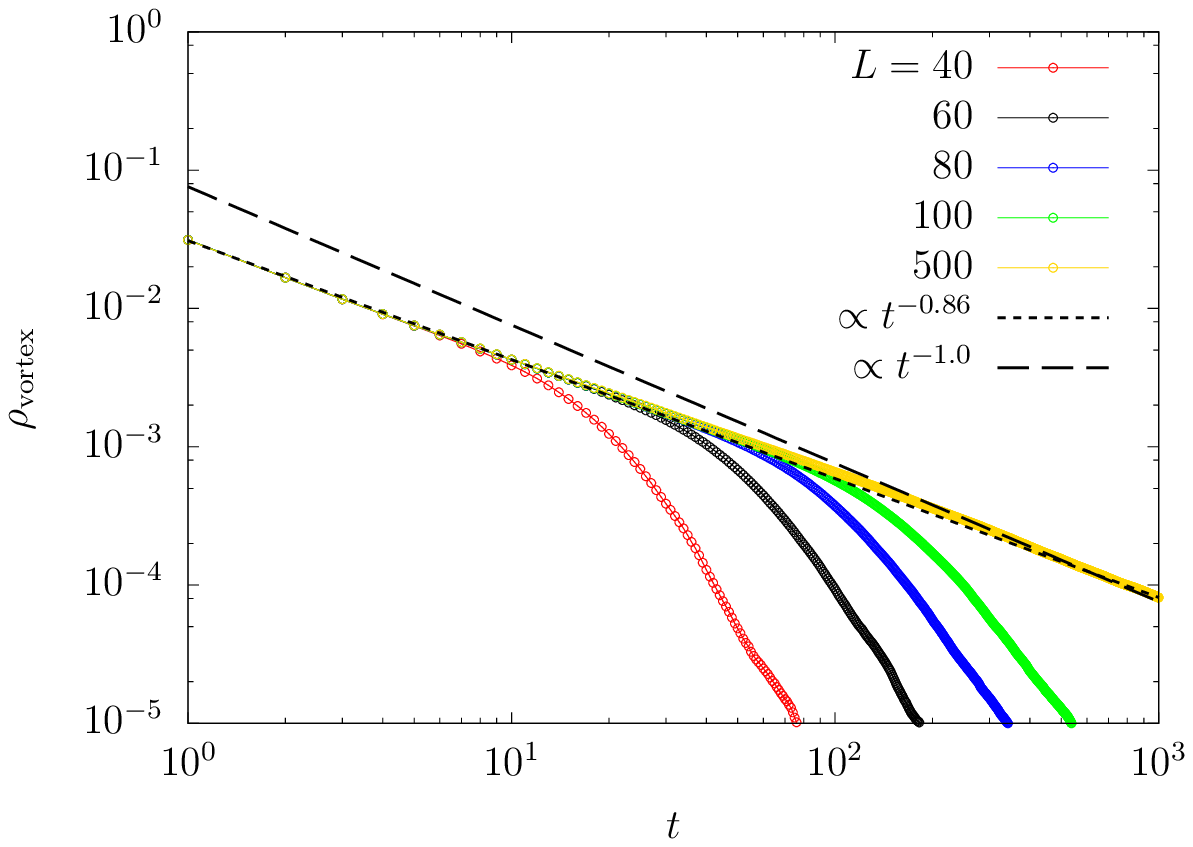}
\end{minipage}
\begin{minipage}{0.49\linewidth}
\centering
(b)\\
\includegraphics[width=0.95\linewidth]{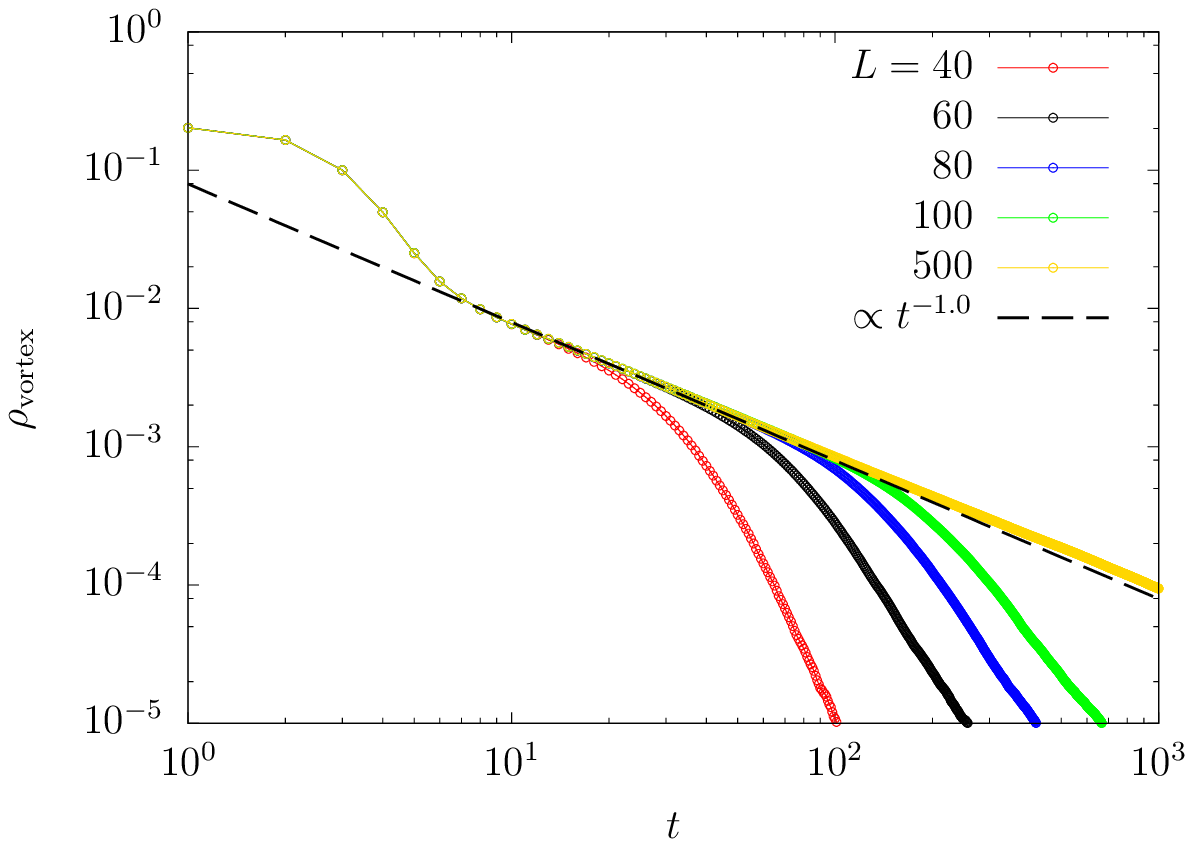}
\end{minipage}
\begin{minipage}{0.49\linewidth}
\centering
(c)\\
\includegraphics[width=0.95\linewidth]{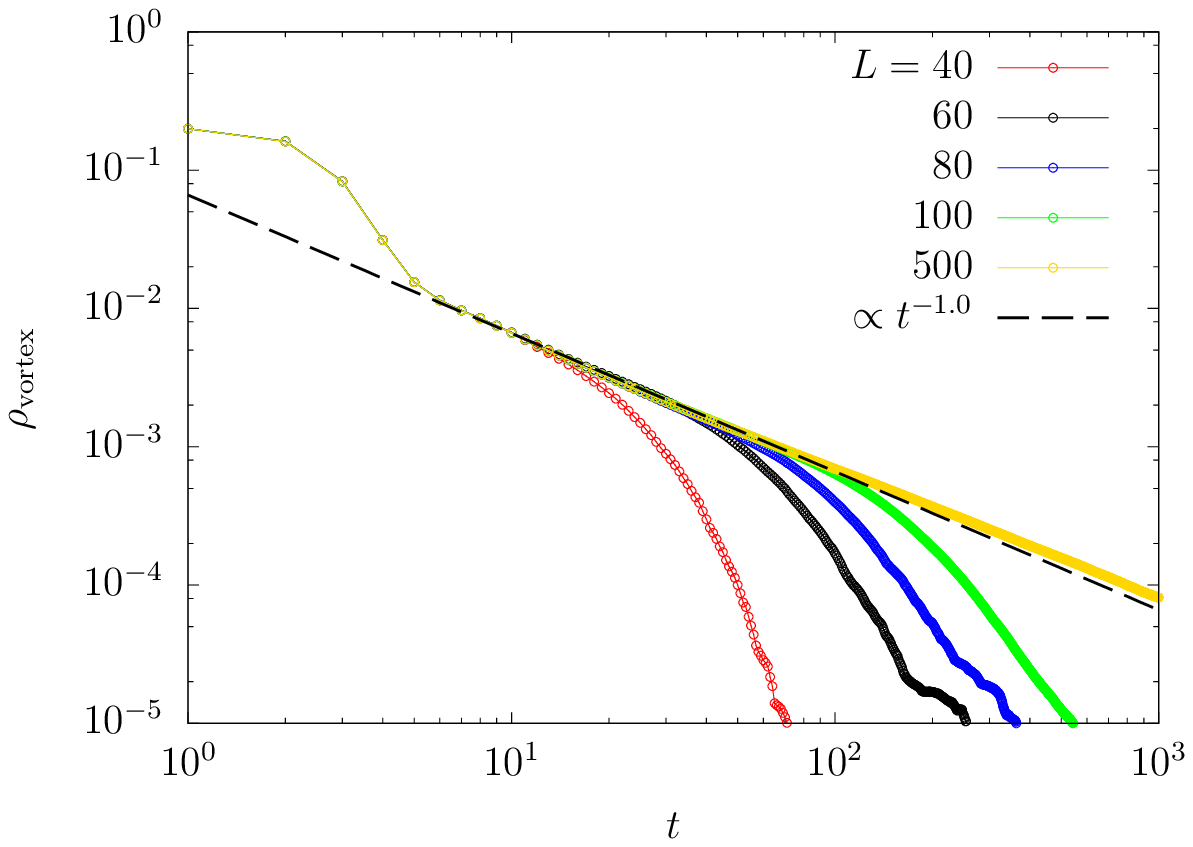}
\end{minipage}
\begin{minipage}{0.49\linewidth}
\centering
(d)\\
\includegraphics[width=0.95\linewidth]{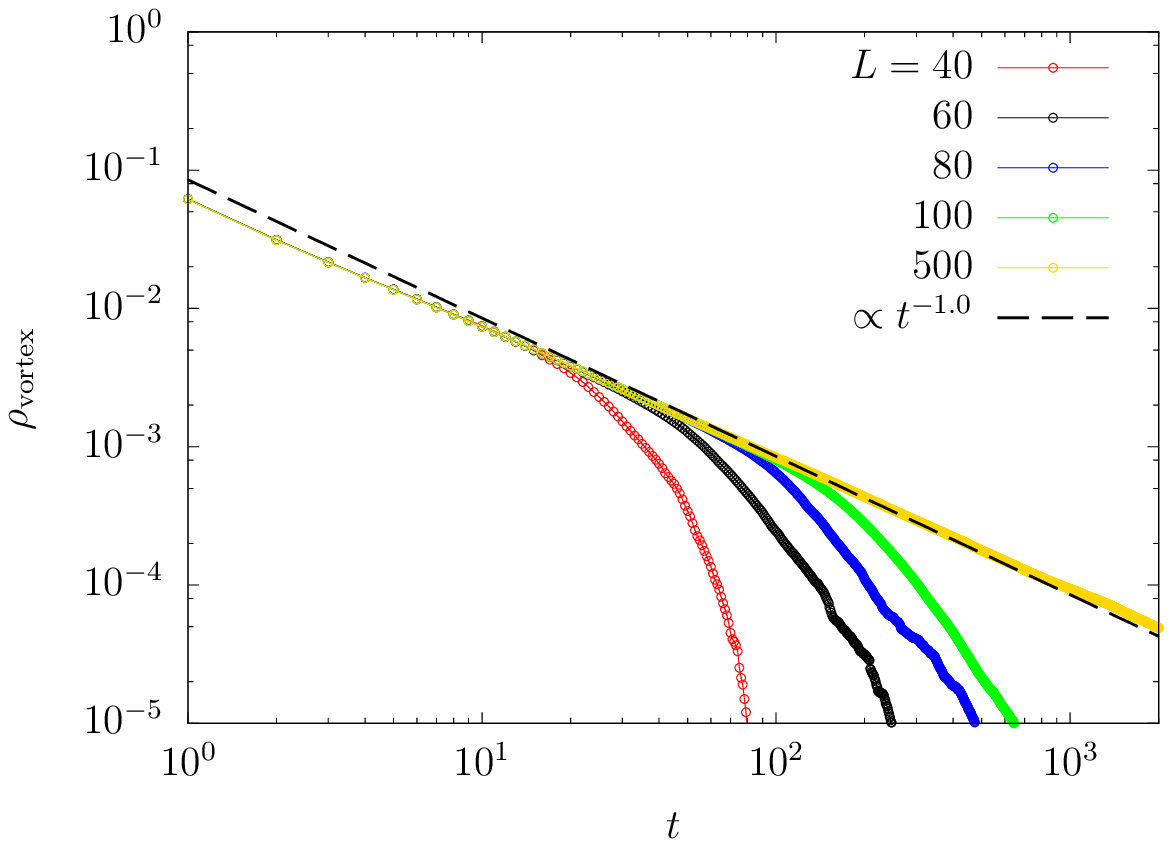}
\end{minipage}
\caption{\label{fig:fast-rhov} 
(Color online.)
The time-dependent vortex density $\rho\sub{vortex}(t)$.  (a) 
over-damped Langevin equation \eqref{eq:over-damped-Langevin}, (b) under-damped dynamics \eqref{eq:under-damped-Langevin}, 
(c) ultra-relativistic limit of the under-damped Langevin equation \eqref{eq:under-damped-ultra-relativistic}, and (d) non-relativistic limit of 
under-damped Langevin equation \eqref{eq:under-damped-non-relativistic}. The different curves in each panel are for 
different system  and mesh sizes.
}
\end{figure}

Our next task is to examine how do the vortex dynamics depend on the evolution equation.
Figures~\ref{fig:fast-rhov} (a)-(d) show the $t$-dependence of the averaged vortex density $\rho\sub{vortex}$ 
as obtained from the Langevin equations \eqref{eq:over-damped-Langevin}, \eqref{eq:under-damped-Langevin}, 
\eqref{eq:under-damped-ultra-relativistic}, and \eqref{eq:under-damped-non-relativistic}, respectively. 
We calculate $\rho\sub{vortex}(t)$ from Eq.~\eqref{eq:vortex-density} by replacing the ensemble average 
$\langle \cdots \rangle\sub{stat}$ by an average over 1000 independent initial states in equilibrium at
$T = 2 \ T\sub{c}$.

In the initial stage of evolution for $0 \lesssim t \lesssim 5$, inertial effects are apparent in the behaviour of $\rho\sub{vortex}(t)$ 
for the under-damped dynamics and the ultra-relativistic limit independently of the system size, 
as shown in Figs.~\ref{fig:fast-rhov} (b) and ($\mbox{c}$). 
These are absent in Figs.~\ref{fig:fast-rhov} (a) and (d) with $c \to \infty$. 

After a short transient of the order of $t\simeq 5$ for our system sizes, 
the vortex density enters the proper scaling regime in which $\rho\sub{vortex}(t)$ should be 
proportional to $t^{-1}$~\cite{Bray, BrayHumayun}.
The numerical exponents are, however, weakly dependent on the type of Langevin equation; 
we measure $-0.86$ for the over-damped Langevin equation \eqref{eq:over-damped-Langevin} (a) 
(for comparison, the algebraic decay $t^{-1}$ is also shown in this figure), 
$-1.0$ for the under-damped Langevin equation \eqref{eq:under-damped-Langevin} (b), 
$-1.0$ for the ultra-relativistic limit of the under-damped Langevin equation \eqref{eq:under-damped-ultra-relativistic} ($\mbox{c}$), 
and also $-1.0$ for the non-relativistic limit of the under-damped Langevin equation \eqref{eq:under-damped-non-relativistic} (d).
The value $-0.86$ for the over-damped Langevin equation is similar the value $-0.90(2)$ found 
in~\cite{Toyoki,Mondello} using a cell-dynamics integration scheme. 
On the other hand, the under-damped Langevin equation and its ultra-relativistic and non-relativistic limits give values that are much closer to the 
analytic ones. 

\vspace{0.25cm}

\begin{figure}[tbh]
\centering
\begin{minipage}{0.24\linewidth}
\centering
(a) \\
\includegraphics[width=0.95\linewidth]{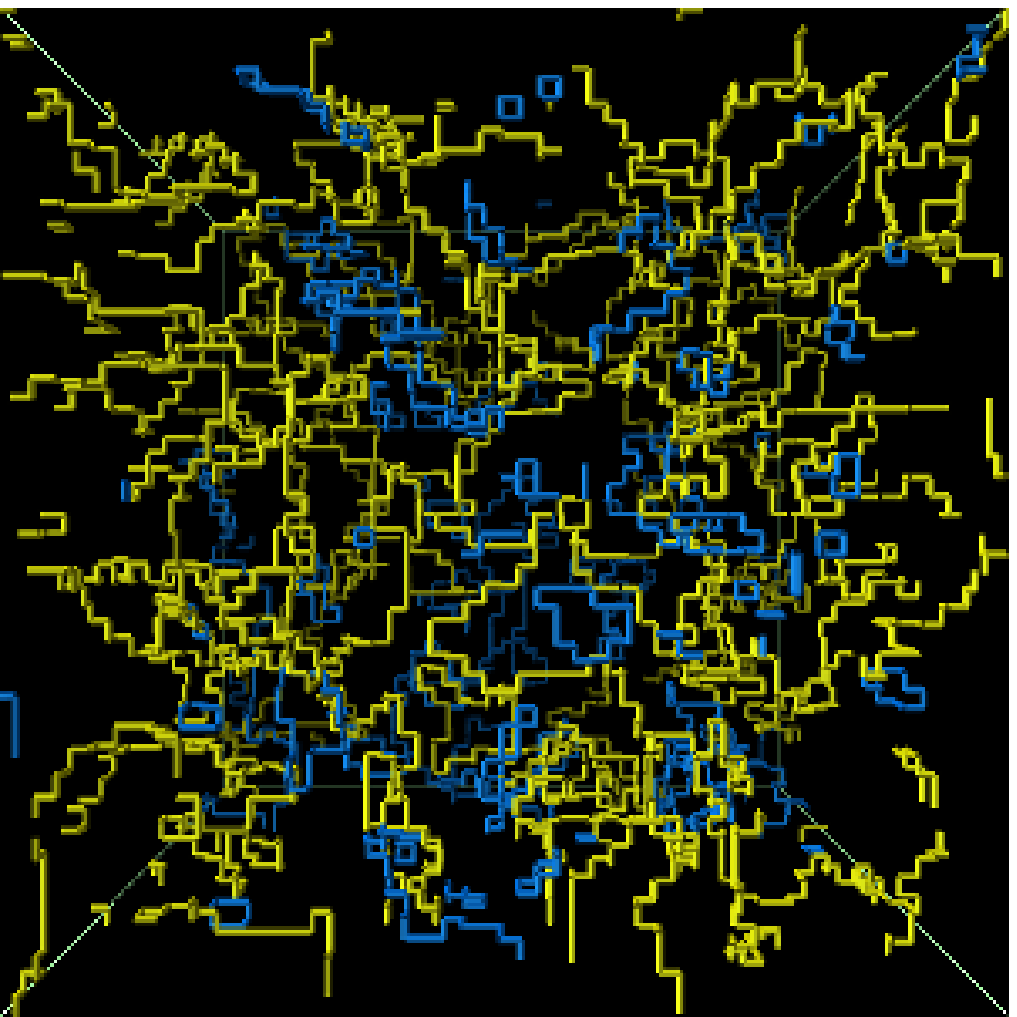}
\end{minipage}
\begin{minipage}{0.24\linewidth}
\centering
(b) \\
\includegraphics[width=0.95\linewidth]{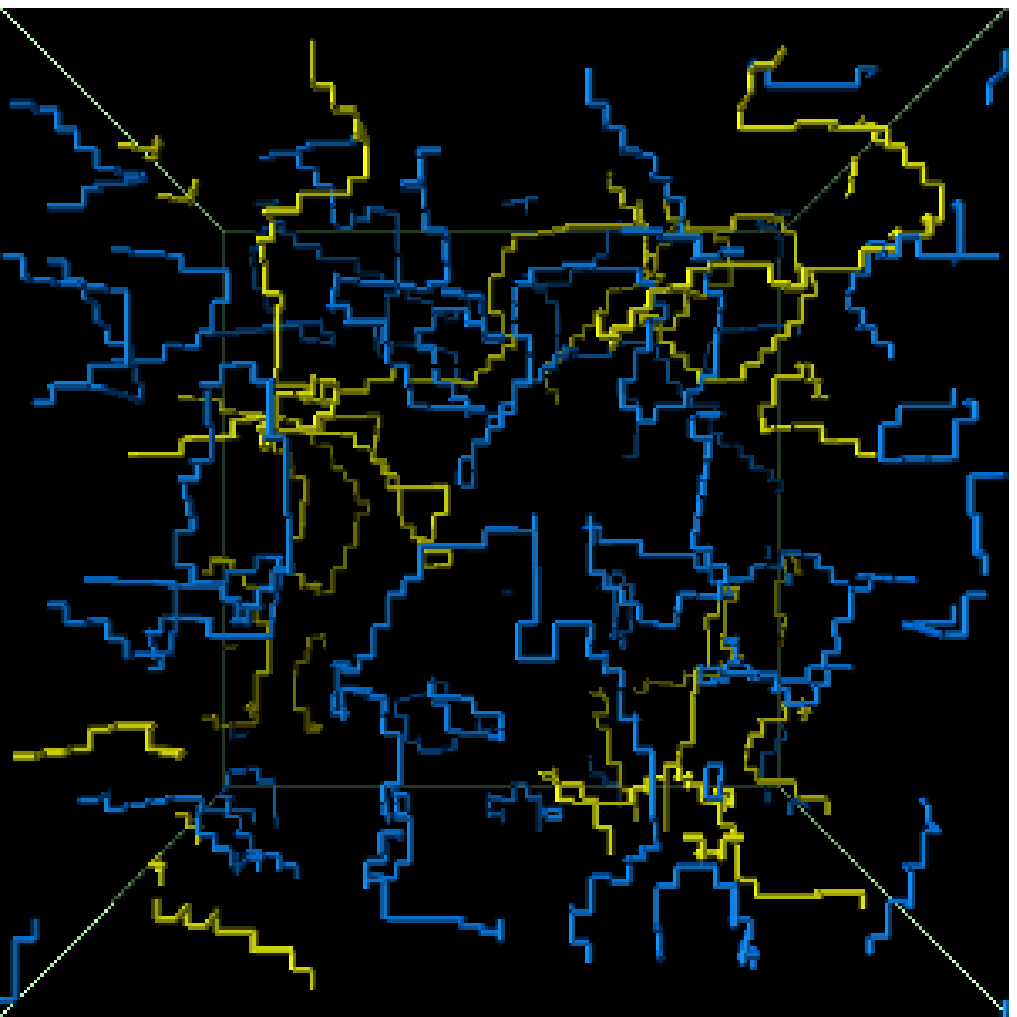}
\end{minipage}
\begin{minipage}{0.24\linewidth}
\centering
(c) \\
\includegraphics[width=0.95\linewidth]{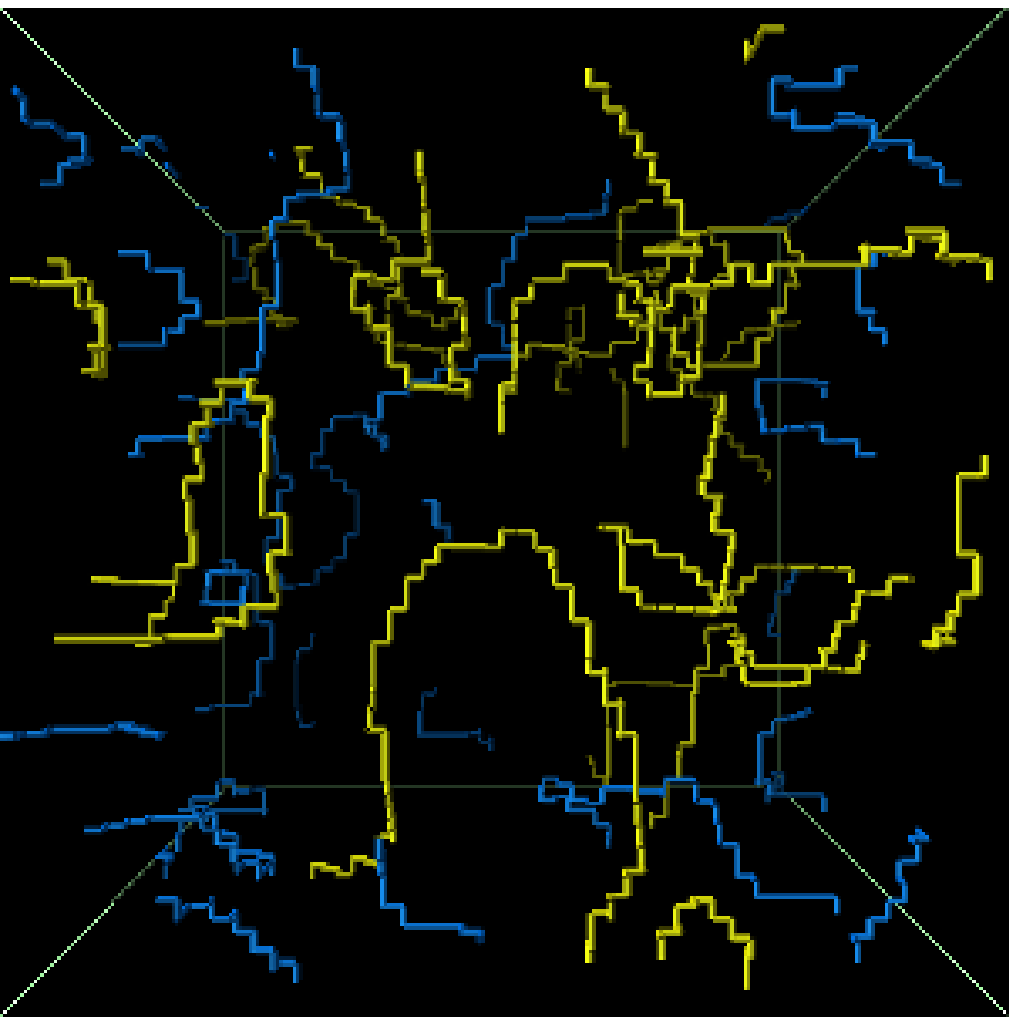}
\end{minipage}
\begin{minipage}{0.24\linewidth}
\centering
(d) \\
\includegraphics[width=0.95\linewidth]{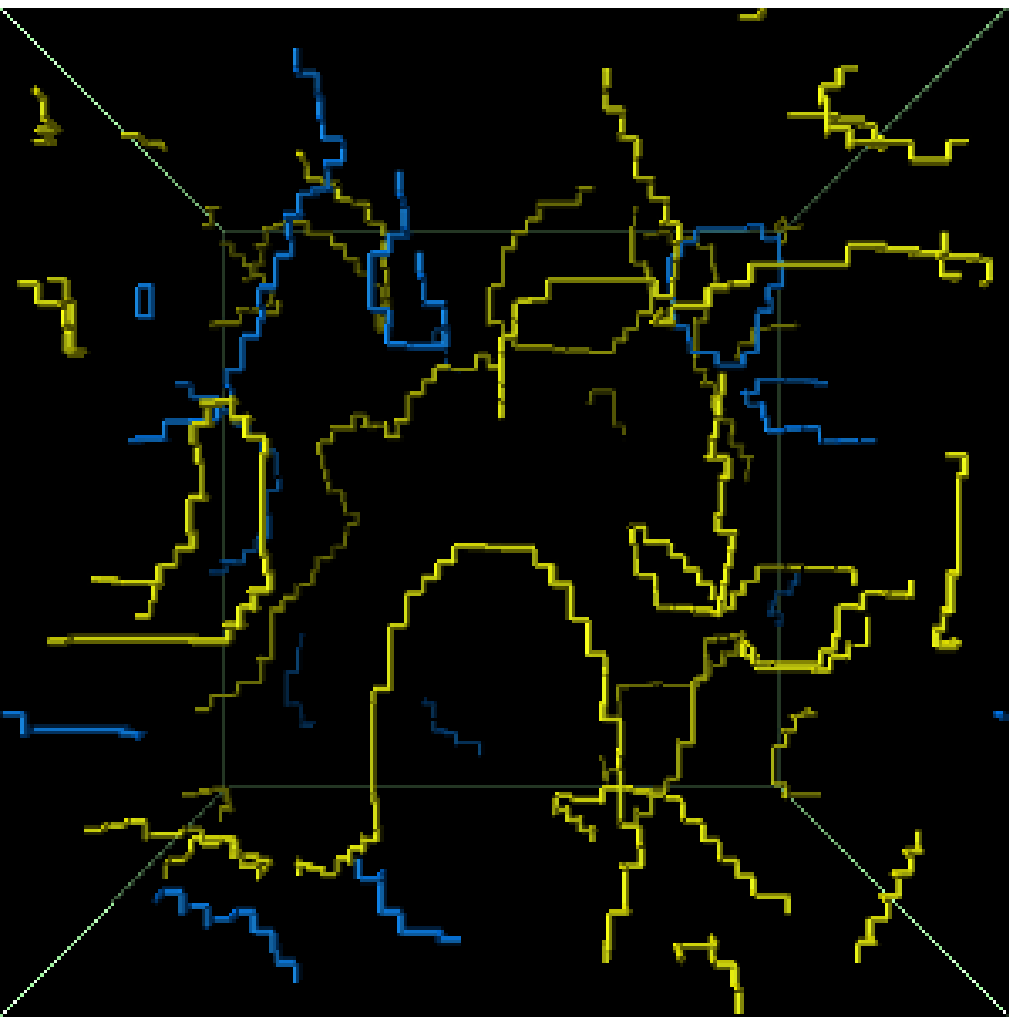}
\end{minipage}
\vspace{0.5cm}
\caption{\label{fig:fast-vortex-power} 
(Color online.)
Dynamic scaling regime.
Snapshots of the vortex configurations at (a) 
$t = 5$, (b) $t = 10$, (c) $t = 15$, and (d) $t = 20$, after an instantaneous quench 
at $t=0$ from equilibrium at $2 \ T\sub{c}$. 
We plot all vortex line elements at the centers of the 
plaquettes with non-zero flux (the total system linear size is $L = 60$).
The vortex line elements are shown in grey (blue) in the black background and the longest vortex 
lines in each image are highlighted (in yellow).
The configurations are generated with the under-damped Langevin 
equation~\eqref{eq:under-damped-Langevin} running at $T = 0$.
The reconnection criterium is not important at this time scale.
}
\end{figure}

\begin{figure}[tbh]
\centering
\begin{minipage}{0.4\linewidth}
\centering
(a) \\
\includegraphics[width=1.17\linewidth]{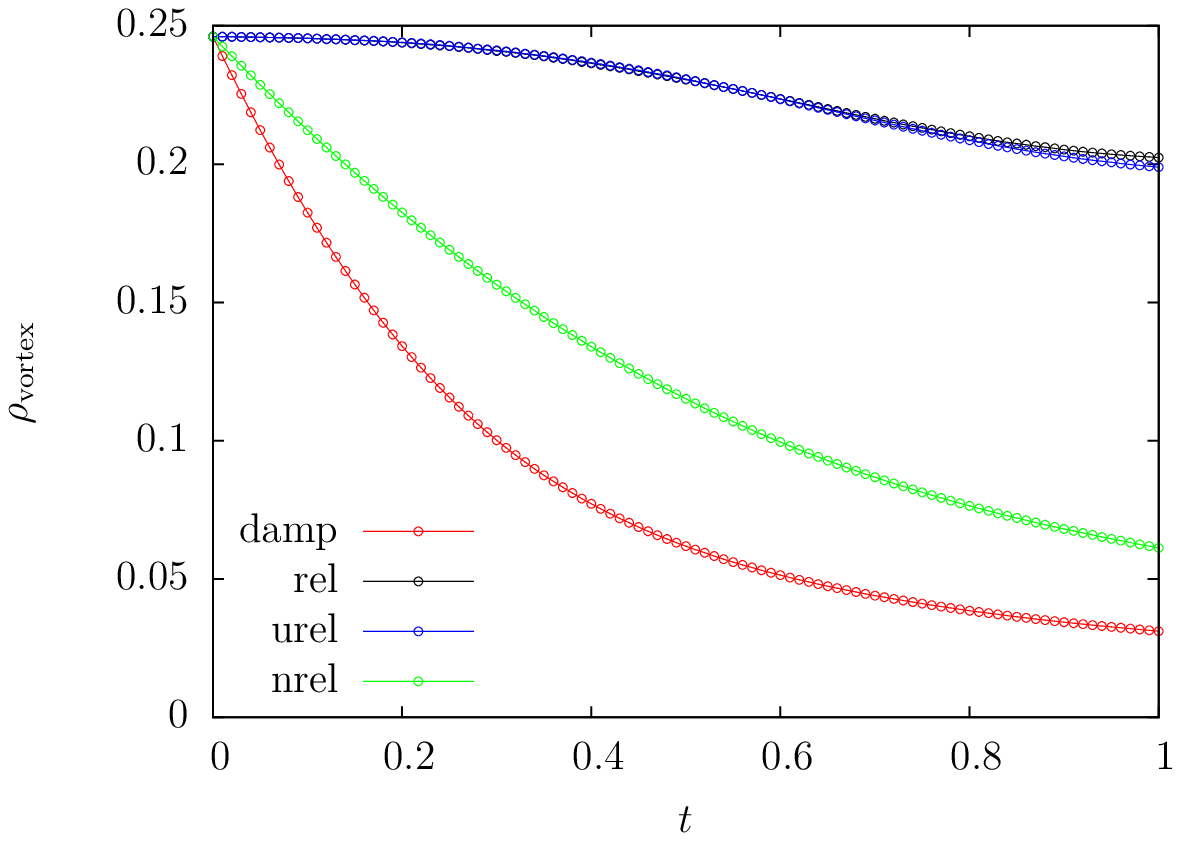}
\end{minipage}
\hspace{1.5cm}
\begin{minipage}{0.4\linewidth}
\centering
(b) \\
\includegraphics[width=1.17\linewidth]{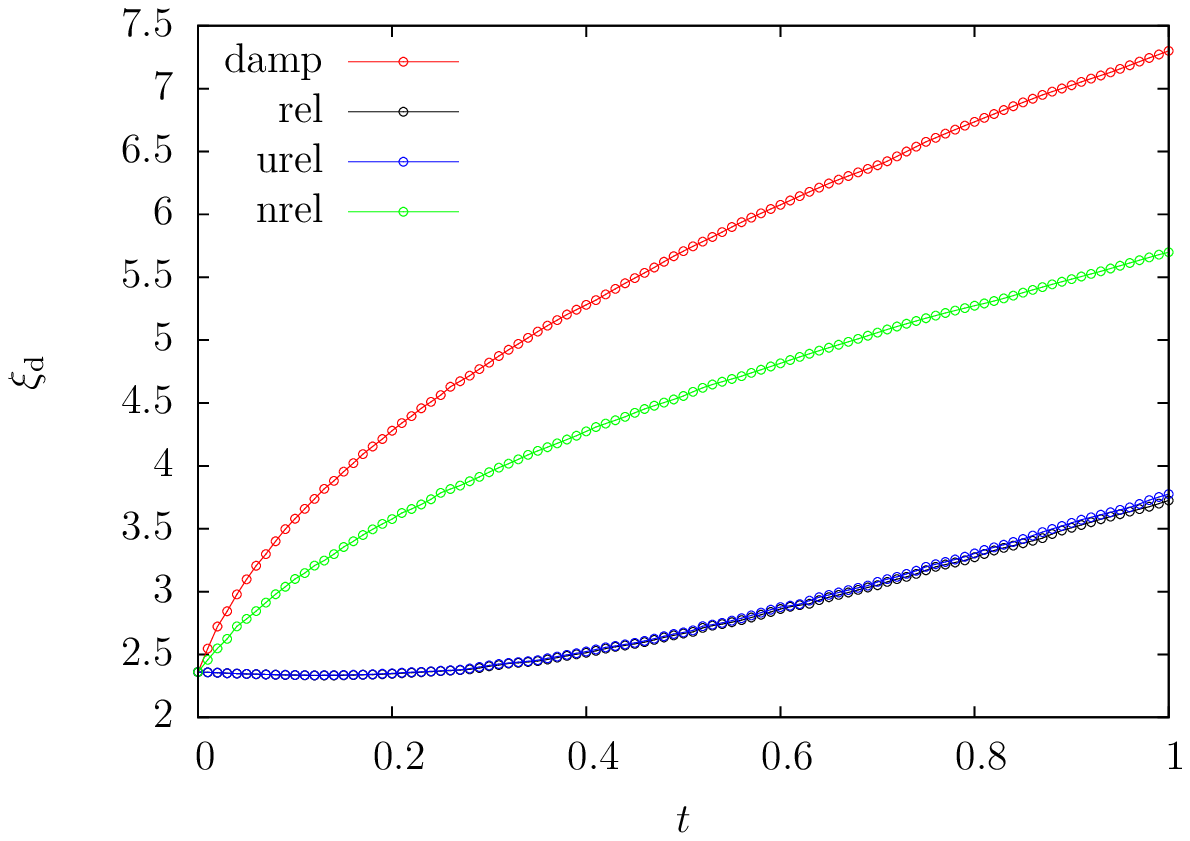}
\end{minipage}
\hspace{1cm}
\vspace{0.5cm}
\caption{\label{fig:very-short-time-scale} 
(Color online.)
Very short time scale dynamics. Differences induced by the various microscopic dynamics in the vortex density and 
dynamic correlation length.
}
\end{figure}

We have calculated the dynamic correlation length and the vortex density using other values of the time and space 
discretisation parameters and we found essentially the same estimates for the exponent $z\sub{d}$ with deviation 
from the expected value $z\sub{d}=2$ for the over-damped dynamics. We may ascribe the origin of this 
difference to the fact that with this kind of dynamics the vortices are very soon diluted in the sample, for times $0 \leq t \lesssim 1$
see Fig.~\ref{fig:very-short-time-scale},
and they cannot properly reach their own scaling regime. 

\vspace{0.25cm}

The various curves in each panel in Fig.~\ref{fig:fast-vortex-power} correspond to different linear system sizes given in the keys. 
The time lapse over which the dynamics remain in the dynamic scaling regime 
is no more than a decade for $L\leq 100$ and  finite size effects are causing the departure of these 
curves from a master
one and their rapid bending down. This effect is pushed beyond the maximal time simulated for the largest 
system size, $L=500$.

The two complementary panels in Fig.~\ref{fig:very-short-time-scale} make 
manifest the differences induced by the dynamic equations in the initial instants. 
The short-time evolution of $\rho\sub{vortex}$ and $\xi\sub{d}$ are the fastest for the over-damped dynamics \eqref{eq:over-damped-Langevin},
intermediate for the non-relativistic limit of the under-damped equation \eqref{eq:under-damped-non-relativistic}  and the slowest for under-damped  
\eqref{eq:under-damped-Langevin}
and ultra-relativistic limit of this same equation \eqref{eq:under-damped-ultra-relativistic} that 
yield undistinguishable curves on these plots.

\begin{figure}[tbh]
\centering
\begin{minipage}{0.49\linewidth}
\centering
(a)\\
\includegraphics[width=0.95\linewidth]{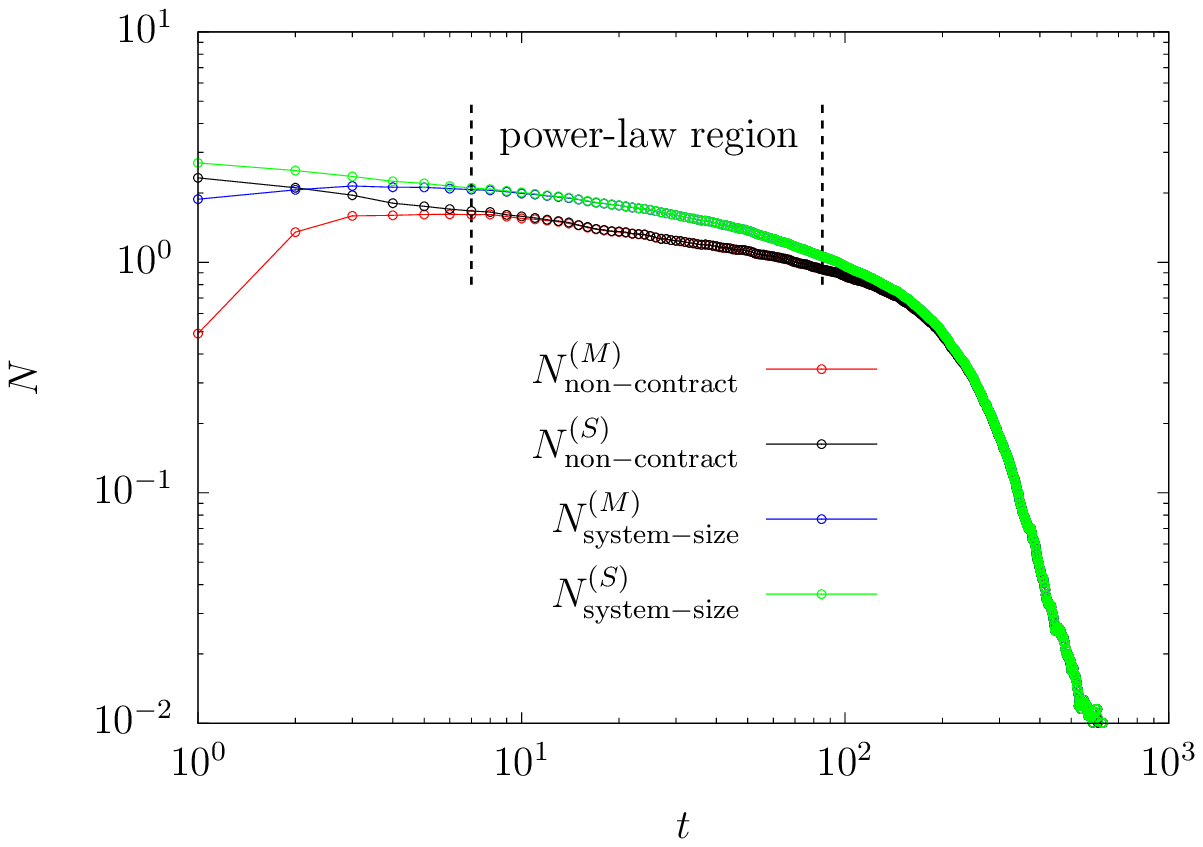}
\end{minipage}
\begin{minipage}{0.49\linewidth}
\centering
(b)\\
\includegraphics[width=0.95\linewidth]{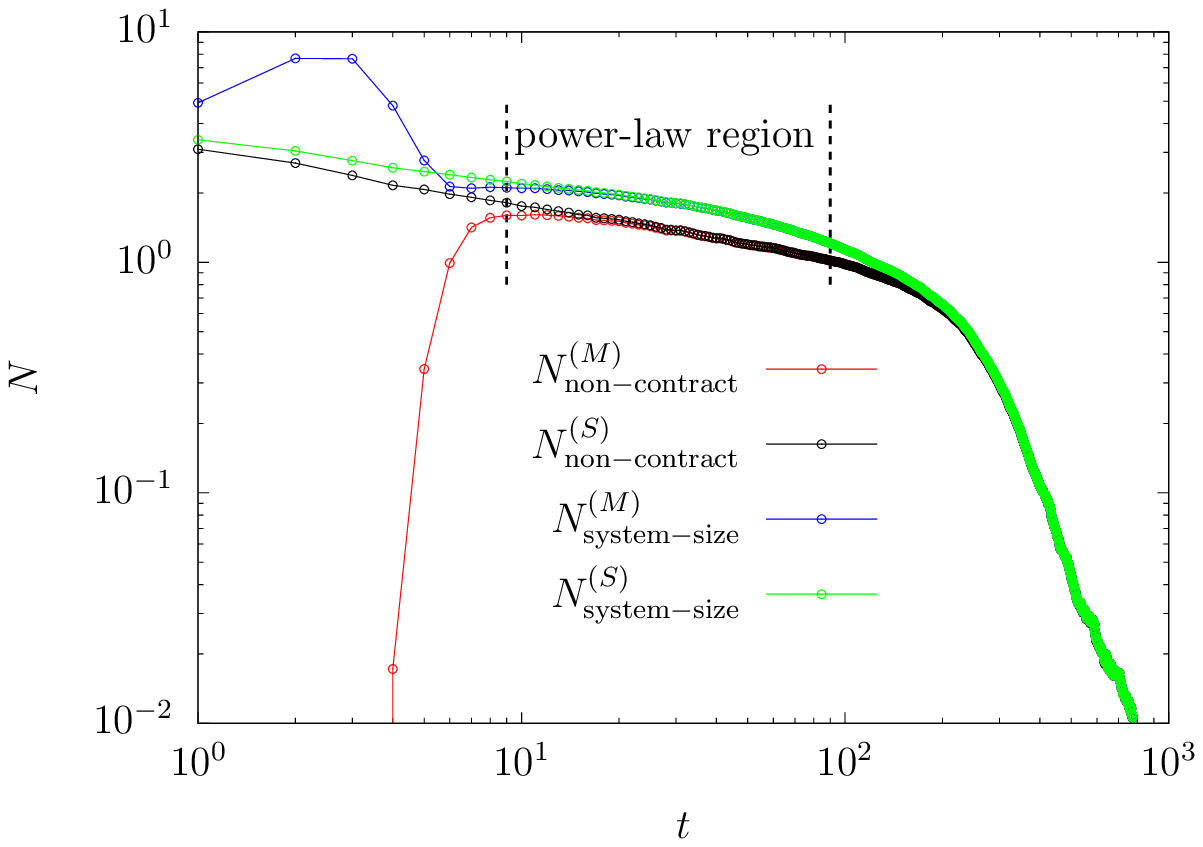}
\end{minipage}
\begin{minipage}{0.49\linewidth}
\centering
(c)\\
\includegraphics[width=0.95\linewidth]{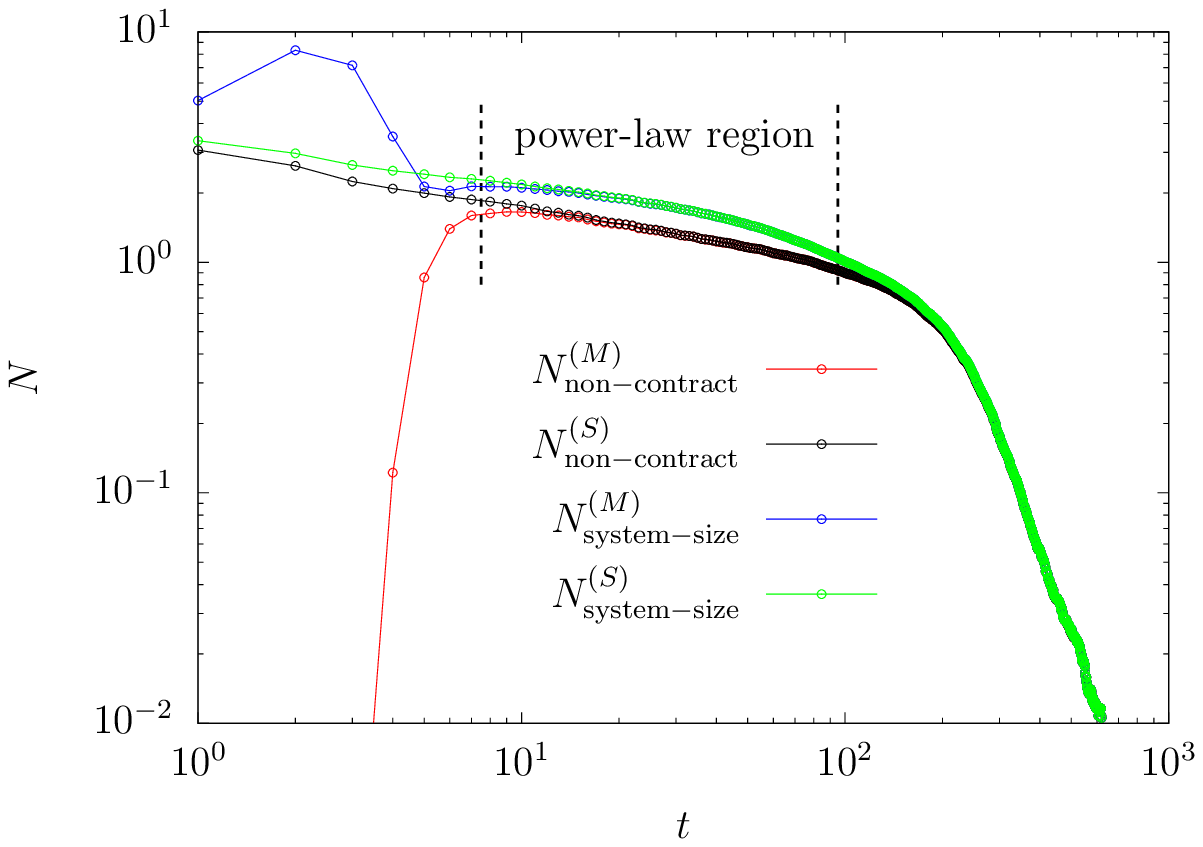}
\end{minipage}
\begin{minipage}{0.49\linewidth}
\centering
(d)\\
\includegraphics[width=0.95\linewidth]{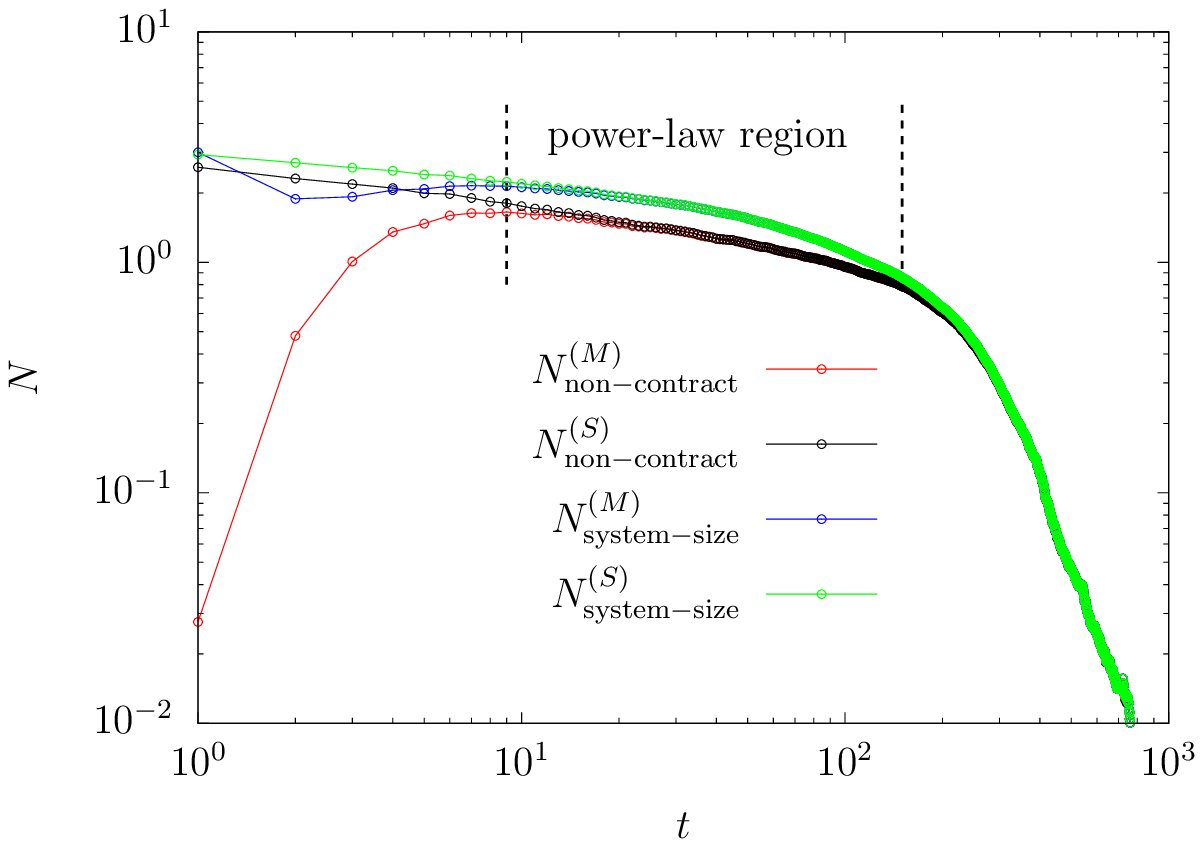}
\end{minipage}
\caption{\label{fig:fast-contract}
(Color online.)
Time-dependent number of vortex loops that are 
larger than the system size $N\sub{system-size}\up{(M,S)}$ and time-dependent number of 
non-contractible vortex loops $N\sub{non-contractible}\up{(M,S)}$. (a) 
over-damped Langevin dynamics \eqref{eq:over-damped-Langevin}, (b) under-damped dynamics \eqref{eq:under-damped-Langevin}, 
(c) ultra-relativistic limit of the under-damped Langevin equation \eqref{eq:under-damped-ultra-relativistic}, and (d) non-relativistic limit of 
the under-damped Langevin equation \eqref{eq:under-damped-non-relativistic}.
The interval over which the algebraic decay of $\rho\sub{vortex}$ is apparent (see Figs.~\ref{fig:fast-rhov}~(a)-(d)) is shown
in each panel.
}
\end{figure}

Figures~\ref{fig:fast-vortex-power}~(a)-(d) display snapshots of the vortex elements in a system with 
linear size $L = 60$
at times $t = 5, \ 10, \ 15, \ 20$, in the early stages of the dynamic scaling regime.
In all panels the longest vortex loop is highlighted.
The percolation across the system of these vortices is confirmed by counting the number of vortex loops
the size of which is larger than the system size $N\sub{system-size}\up{(M,S)}$ and the number of non-contractible loops $N\sub{non-contract}\up{(M,S)}$.
Figures~\ref{fig:fast-contract}~(a)-(d) show $N\sub{system-size}\up{(M,S)}$ and $N\sub{non-contract}\up{(M,S)}$ in a system with linear size
$L = 100$. The power-law behaviour is apparent at $7 \lesssim t \lesssim 85$ for the over-damped dynamics (panel (a)),
$9 \lesssim t \lesssim 90$ for the under-damped dynamics (panel (b)),
$8 \lesssim t \lesssim 95$ for the ultra-relativistic limit of the under-damped dynamics (panel (c)),
and $9 \lesssim t \lesssim 150$ for the non-relativistic limit of the under-damped dynamics (panel (d)).
In these power-law regimes, there is little difference between the results for the
maximal and stochastic criteria for connecting vortex-line elements.
We also see that 1 or 2 vortices contribute to $N\sub{system-size}\up{(M,S)}$ and $N\sub{non-contract}\up{(M,S)}$.

\begin{figure}[tbh]
\vspace{\baselineskip}
\centering
\begin{minipage}{0.24\linewidth}
\centering
(a) \\
\includegraphics[width=0.95\linewidth]{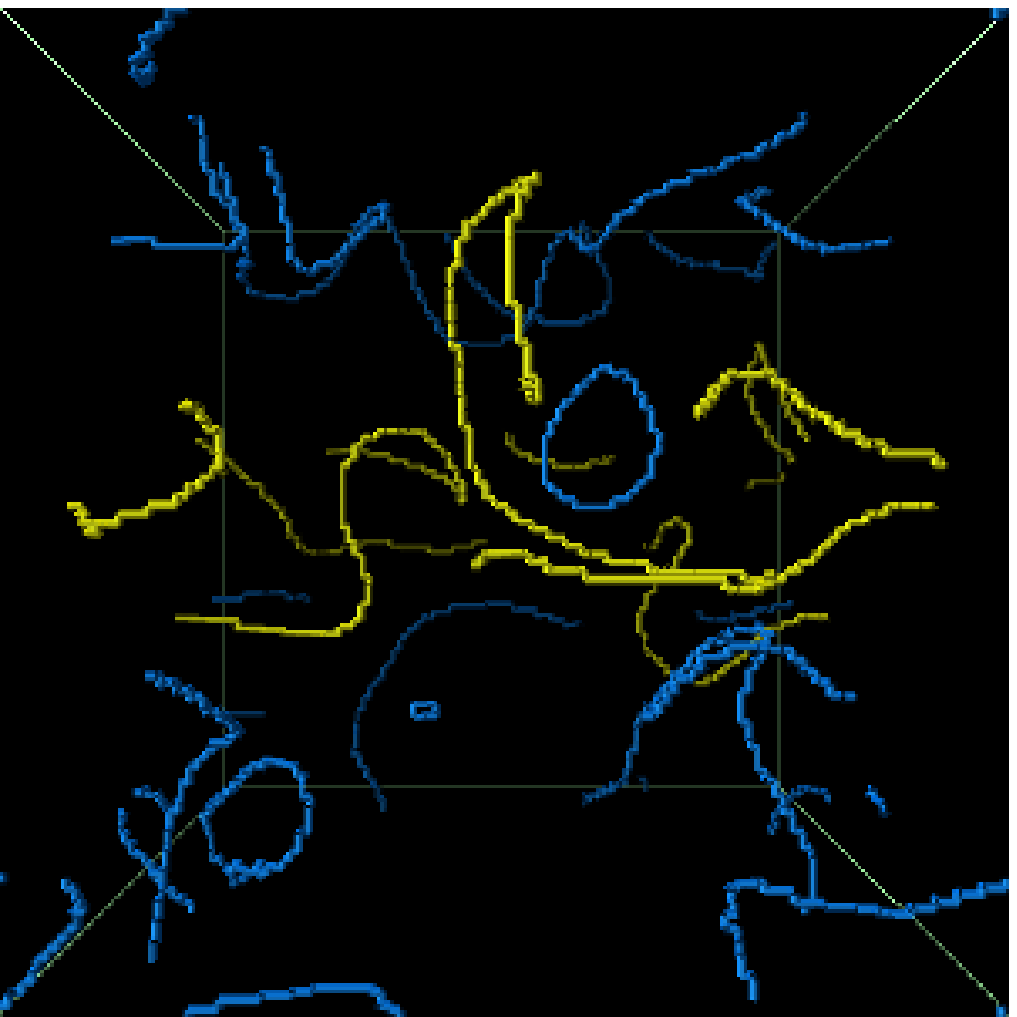}
\end{minipage}
\begin{minipage}{0.24\linewidth}
\centering
(b) \\
\includegraphics[width=0.95\linewidth]{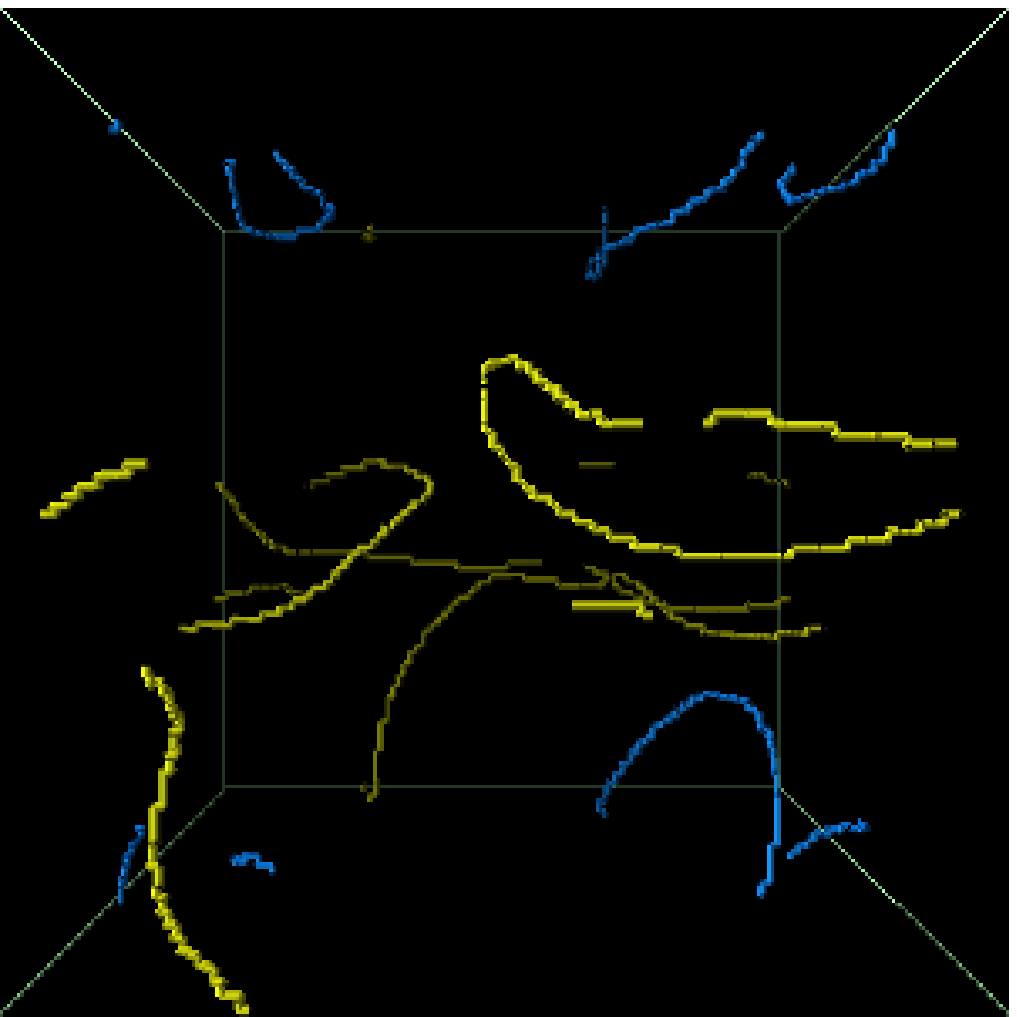}
\end{minipage}
\begin{minipage}{0.24\linewidth}
\centering
(c) \\
\includegraphics[width=0.95\linewidth]{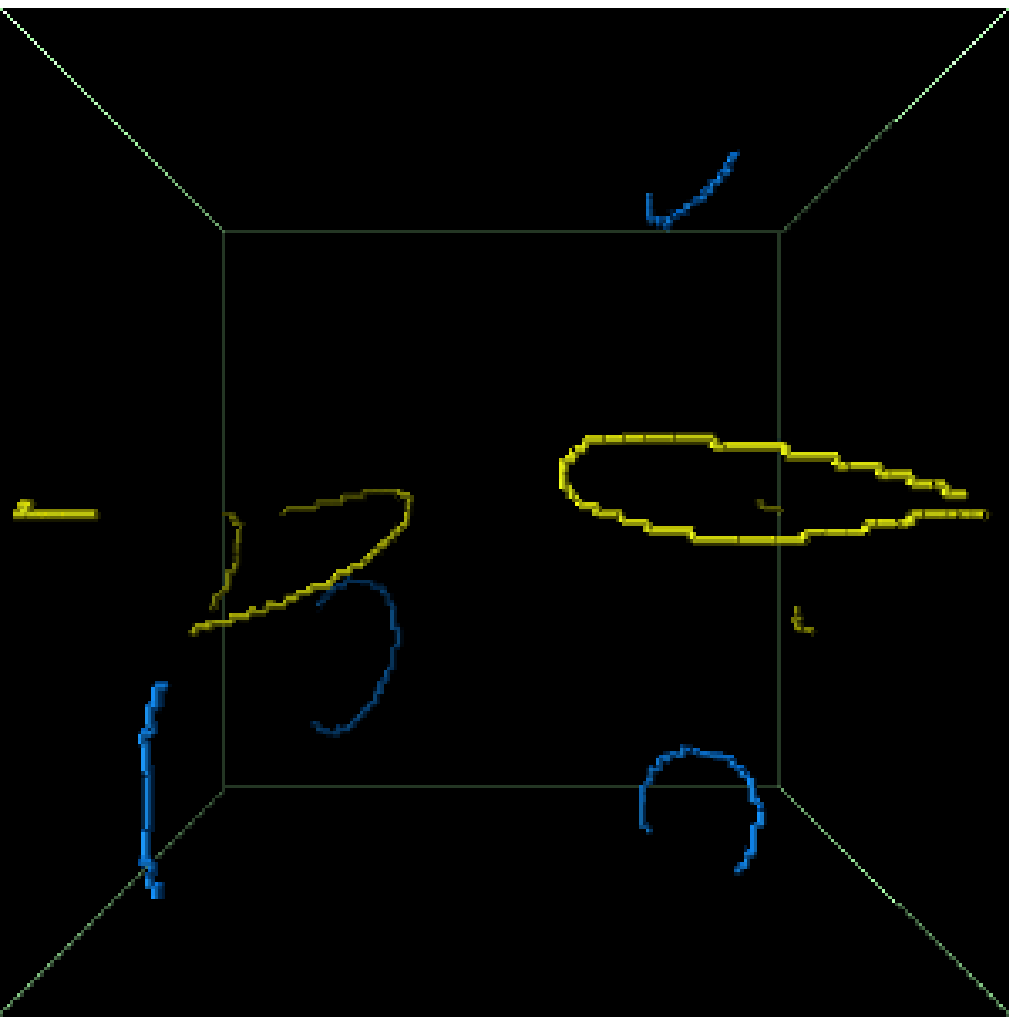}
\end{minipage}
\begin{minipage}{0.24\linewidth}
\centering
(d) \\
\includegraphics[width=0.95\linewidth]{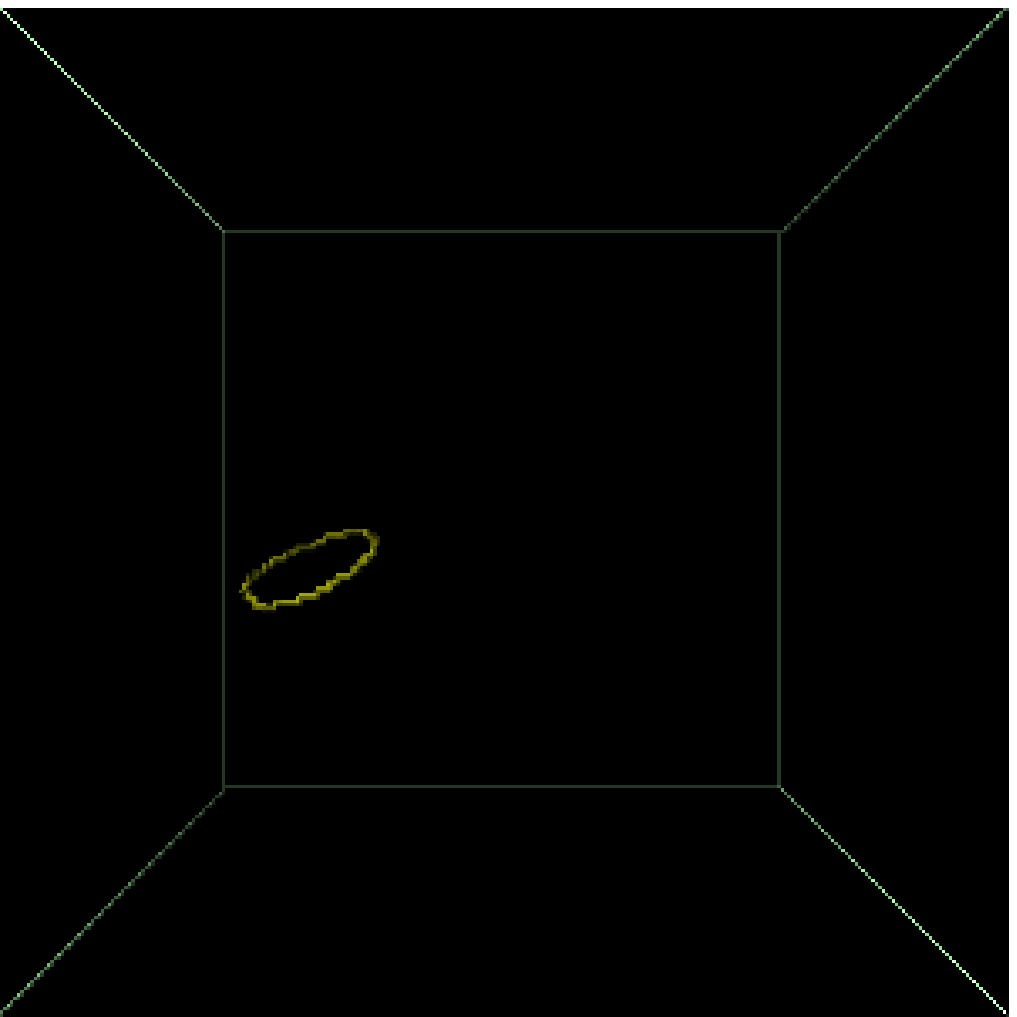}
\end{minipage}
\vspace{0.5cm}
\caption{\label{fig:fast-vortex-one-loop} 
(Color online.)
Late epochs.
Snapshots of the vortex configurations at (a) 
$t = 50$, (b) $t = 100$, (c) $t = 150$, and (d) $t = 200$, after an instantaneous quench 
at $t=0$ from equilibrium at $2 \ T\sub{c}$. 
We plot all vortex line elements at the centers of the plaquettes with non-zero flux.  The  system linear size is $L = 100$.
The vortex line elements are shown in grey (blue) in the black background and the longest vortex 
lines in each image are highlighted (in yellow).
The configurations are generated with the under-damped Langevin 
equation~\eqref{eq:under-damped-Langevin} running at $T = 0$.
At these times the reconnection rule is irrelevant.}
\end{figure}

Figures~\ref{fig:fast-vortex-one-loop} (a)-(d) show snapshots of the vortex elements in a  system with linear size $L = 100$
at four later times in the interval $50 \leq t \leq 200$, that is to say, in the late stages of the dynamic scaling 
regime and the final approach to equilibrium.
At $t = 50$, panel (a), $\rho\sub{vortex}(t)$ enters the power-law $t^{-1}$ regime. At
$t=100$, panel (b), the dynamics exit this scaling regime. 
In panels (a) and (b) the size of the longest vortex loop is larger than the system size.
At $t = 150$ and $200$ (panels (c) and (d)), $\rho\sub{vortex}(t)$ decays faster than $t^{-1}$,
and there are only finite size contractible vortices left, which just shrink via the viscosity.

\subsection{Finite-size scaling of $\xi\sub{d}$ and $\rho\sub{vortex}$}
\label{subsec:finite-size}

Here, we discuss the finite-size scaling properties of the vortex density $\rho\sub{vortex}(t)$ and the dynamic correlation length $\xi\sub{d}(t)$.

\begin{figure}[b]
\centering
\begin{minipage}{0.49\linewidth}
\centering
(a)\\
\includegraphics[width=0.95\linewidth]{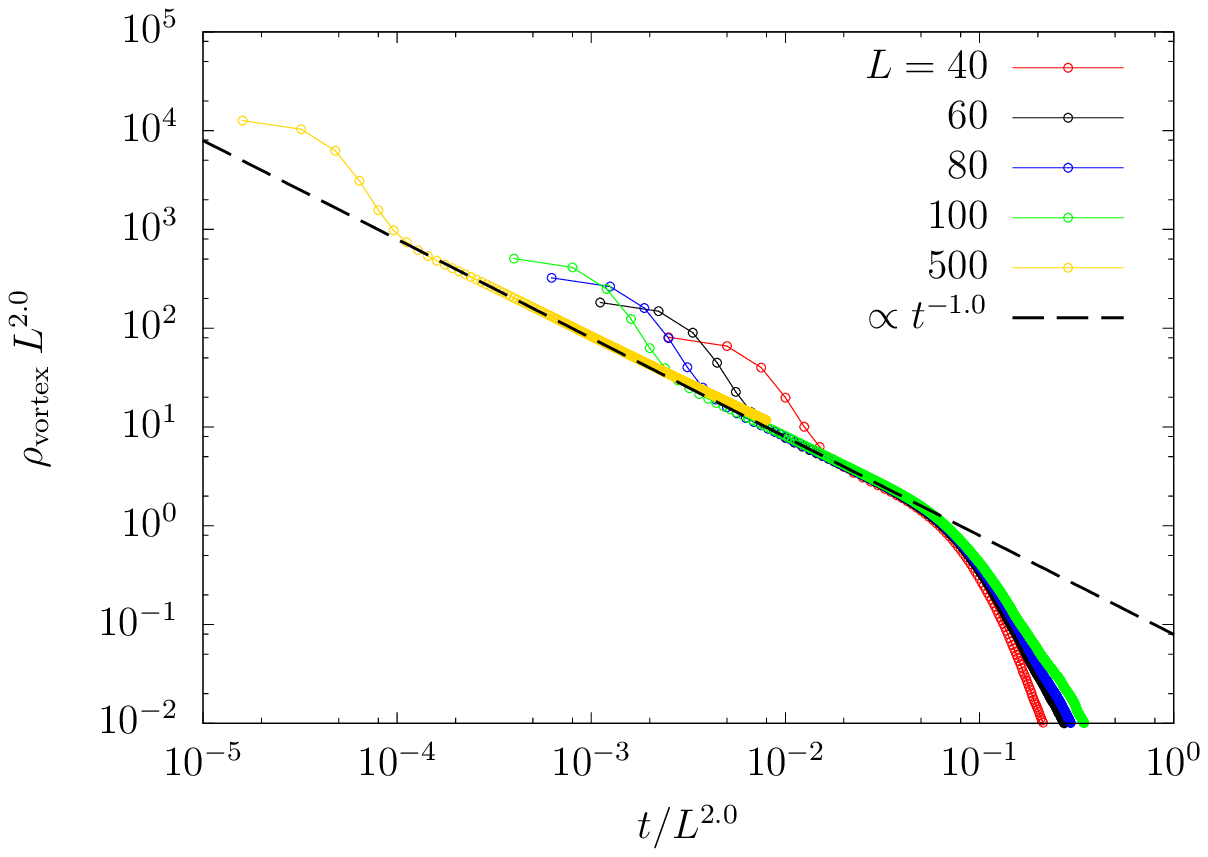}
\end{minipage}
\begin{minipage}{0.49\linewidth}
\centering
(b)\\
\includegraphics[width=0.95\linewidth]{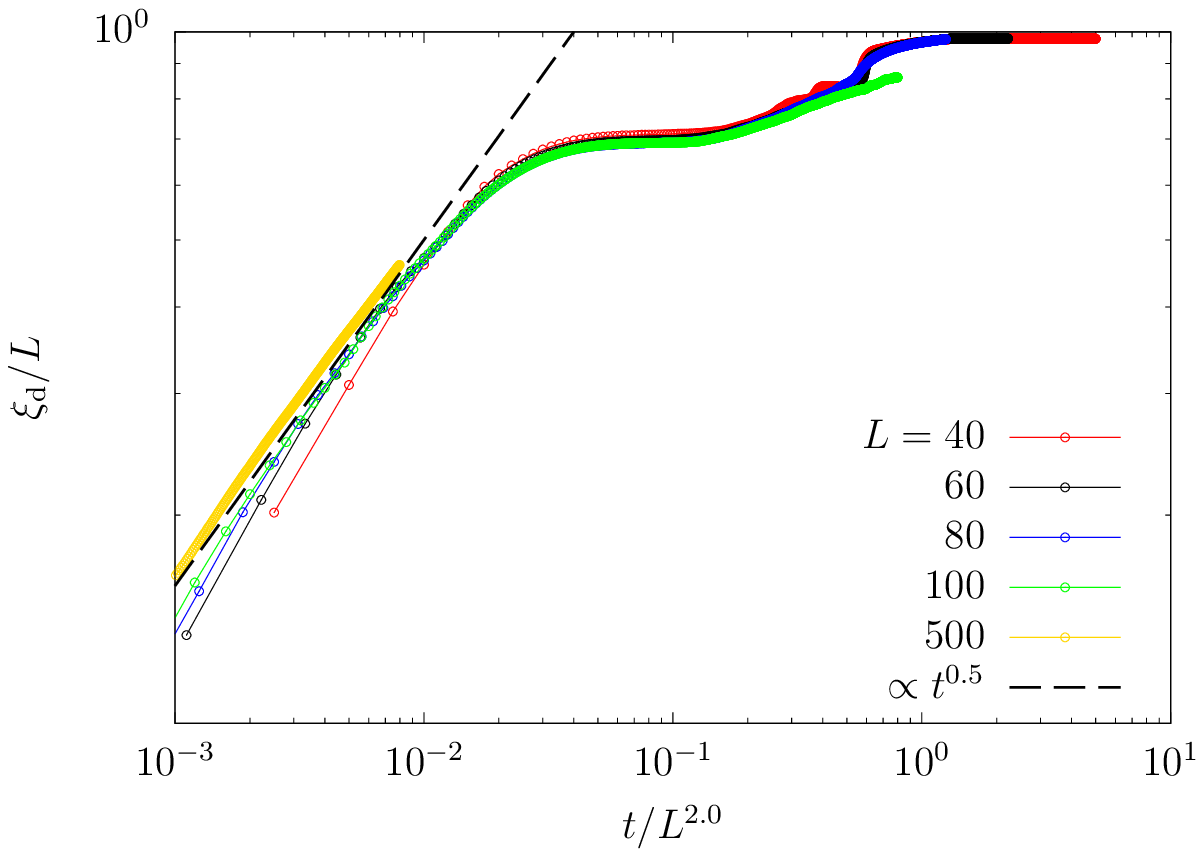}
\end{minipage}
\caption{\label{fig:fast-scale-rel} 
(Color online.)
Finite-size scaling plots for (a) the vortex density 
$\rho\sub{vortex}(t)$ and (b) the dynamic correlation length with the dynamical exponent 
$z\sub{d} = 0.5$ obtained from the under-damped Langevin equation \eqref{eq:under-damped-Langevin} 
at $T = 0$.
}
\end{figure}
Since the dynamic correlation length grows in time as 
$\xi\sub{d}(t) \propto t^{1 / z\sub{d}}$ in  the infinite system size limit, 
$\rho\sub{vortex}(t)$ and $\xi\sub{d}(t)$ are expected to be 
universal functions of $t / L\up{z\sub{d}}$ in the late stages of evolution of finite 
size systems:
\begin{equation}
\frac{\xi\sub{d}}{L} = f_\xi\left( \frac{t}{L^{z\sub{d}}} \right) \qquad\mbox{and}\qquad
\rho\sub{vortex} = f_\rho\left( \frac{t}{L^{z\sub{d}}} \right) 
\; . 
\end{equation}
Figures \ref{fig:fast-scale-rel} (a) and (b) show $\rho\sub{vortex}(t)$ and $\xi\sub{d}(t)$ as  functions of $t / L\up{z\sub{d}}$ with $z\sub{d} = 2$ obtained from the under-damped Langevin equation at $T = 0$.
Except for the initial stage of evolution where another scaling variable characterising the approach to 
a percolating structure may also be necessary~\cite{Blanchard14}, 
the universal behaviour is good.

\begin{figure}[tbh]
\centering
\begin{minipage}{0.49\linewidth}
\centering
(a)\\
\includegraphics[width=0.95\linewidth]{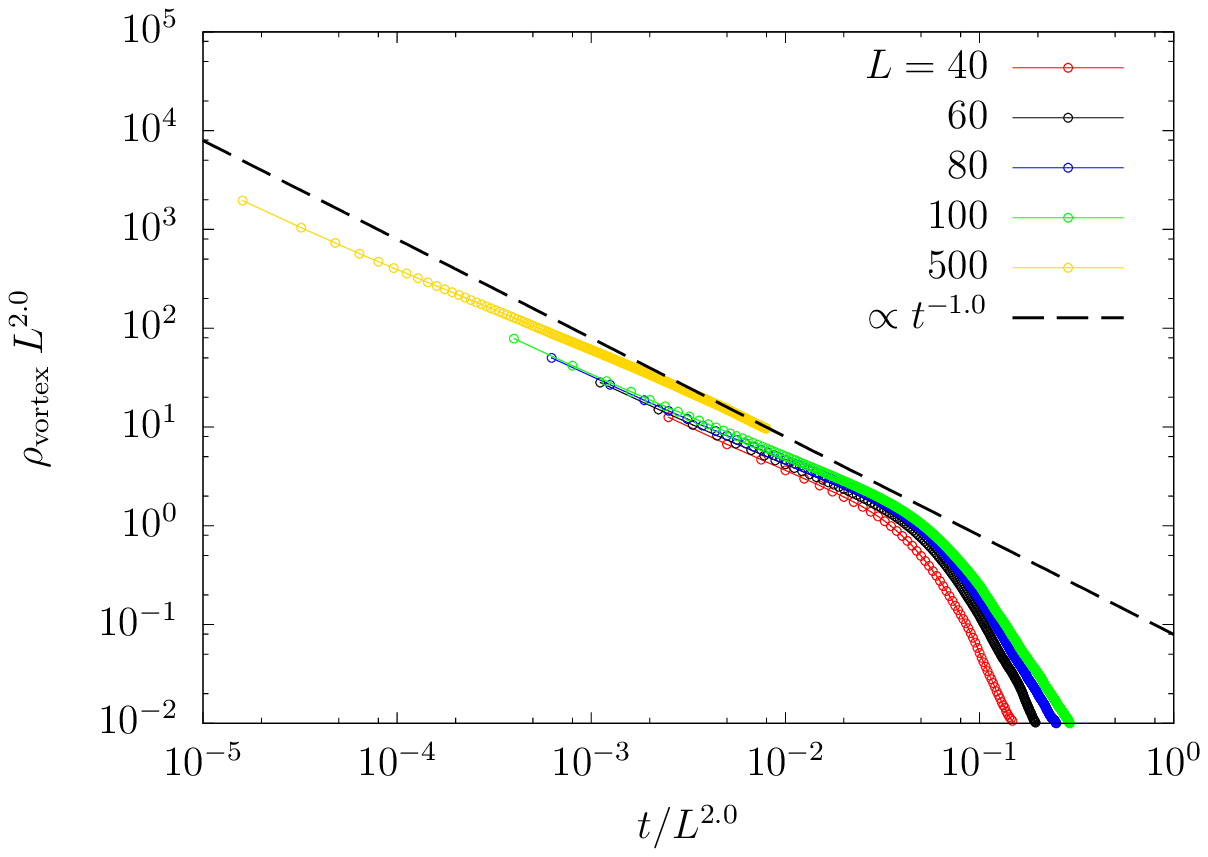}
\end{minipage}
\begin{minipage}{0.49\linewidth}
\centering
(b)\\
\includegraphics[width=0.95\linewidth]{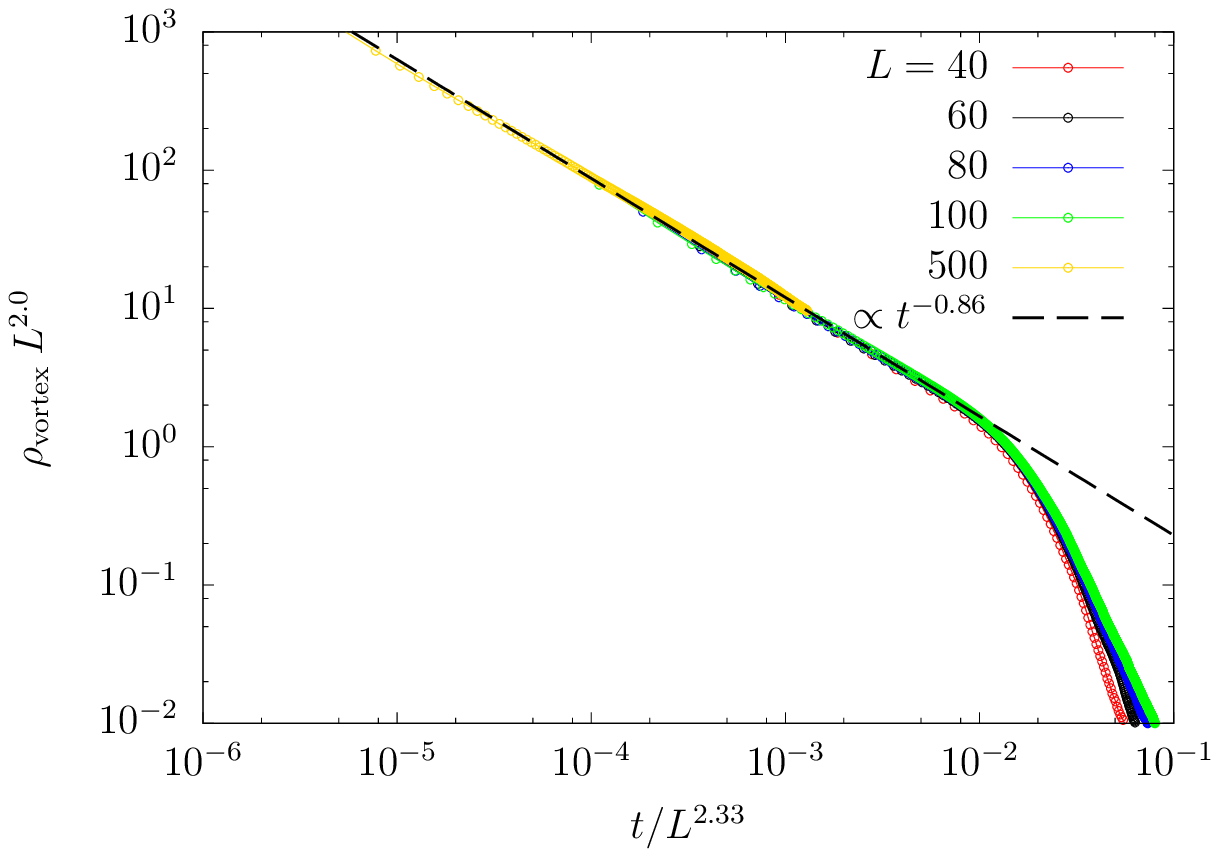}
\end{minipage} \\
\begin{minipage}{0.49\linewidth}
\centering
(c)\\
\includegraphics[width=0.95\linewidth]{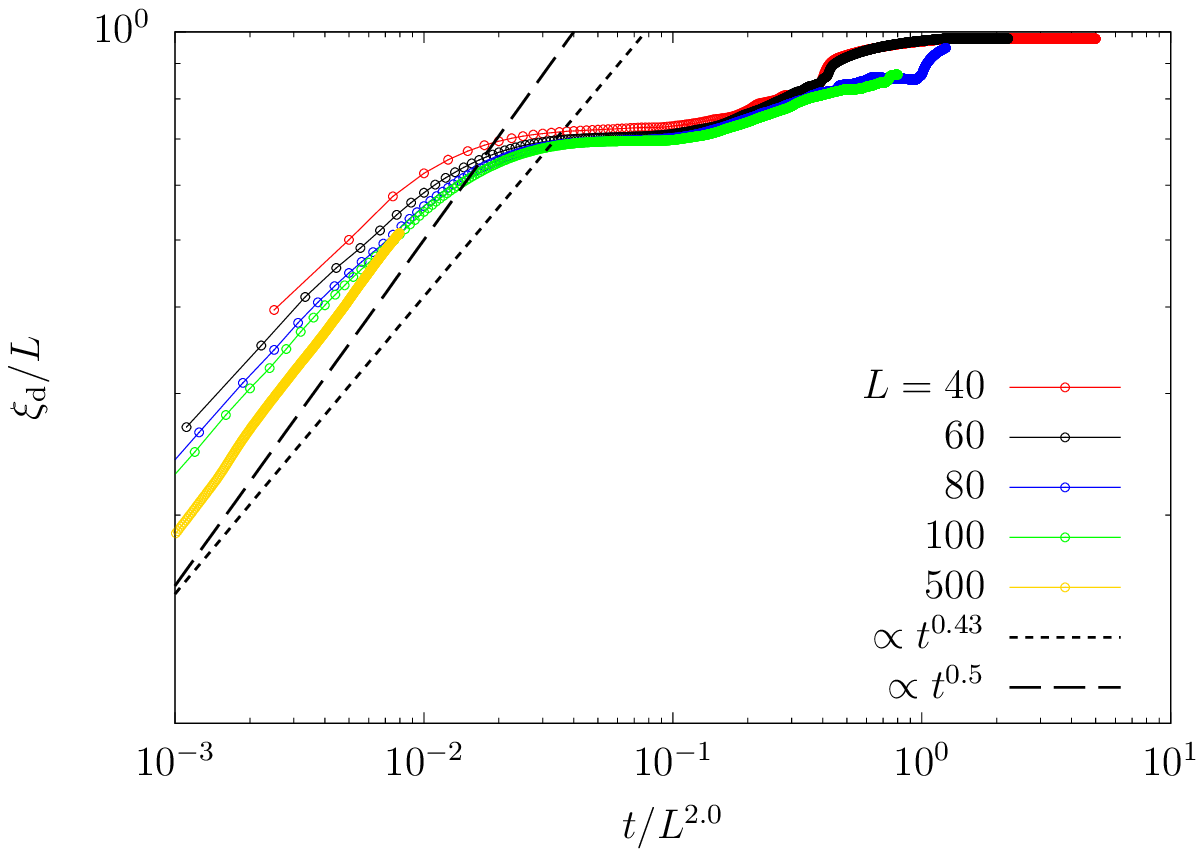}
\end{minipage}
\begin{minipage}{0.49\linewidth}
\centering
(d)\\
\includegraphics[width=0.95\linewidth]{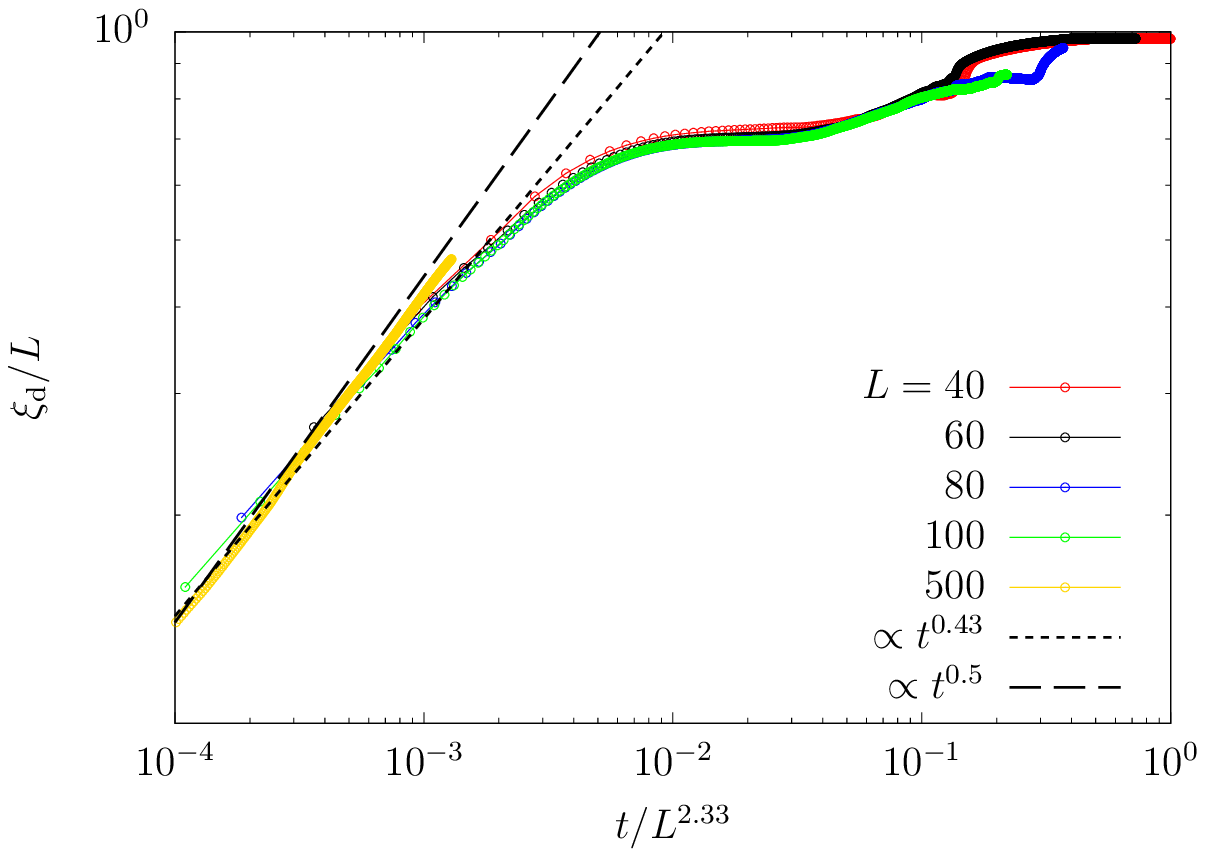}
\end{minipage}
\caption{\label{fig:fast-scale-damp}
(Color online.)
 Finite-size scaling plots of
the vortex density $\rho\sub{vortex}(t)$ (panels (a) and (b)) and the dynamic correlation length (panels (c) and (d)) 
obtained from the over-damped Langevin equation \eqref{eq:under-damped-Langevin} at $T = 0$ 
with the dynamical exponent $1/z\sub{d} = 0.5$ (panels (a) and (c)) and $1/z\sub{d} = 0.43$ (panels (b) and (d)). 
}
\end{figure}

We note that similar good universal properties have been obtained using the 
ultra-relativistic limit of the under-damped Langevin equation \eqref{eq:under-damped-ultra-relativistic}, 
and the non-relativistic limit of the under-damped Langevin equation \eqref{eq:under-damped-non-relativistic} 
at $T = 0$ with the same dynamical critical exponent $z\sub{d}$.

With the over-damped Langevin dynamics \eqref{eq:over-damped-Langevin} at $T = 0$, we measured a different dynamical 
exponent $1/z\sub{d} \simeq 0.43$ in Fig \ref{fig:fast-rhov} (a) for the vortex density and Fig. \ref{fig:fast-xid} (a) for the 
dynamic correlation length.
We then compare the scaling with the two dynamical exponents $1/z\sub{d} = 0.5$ and $0.43$.
Figure \ref{fig:fast-scale-damp} shows $\rho\sub{vortex}(t)$ (panels (a) and (b)) and $\xi\sub{d}(t)$ (panels (c) and (d) as 
functions of $t / L\up{z\sub{d}}$ with $z\sub{d} = 2$ (panels (a) and (c)) and $1/z\sub{d} = 0.43$ (panels (b) and (d))
and these dynamics. As expected, in Figs.~\ref{fig:fast-rhov} (a) and \ref{fig:fast-xid} (a), we find better universal behaviour 
with $1/z\sub{d} = 0.43$, although the analytic expectation for $1/z\sub{d}$ is $0.5$.

\subsection{Number densities of string lengths}

We now analyse the statistics of vortex lengths in the course of time. We have already identified
three time regimes from the study of the growing length and vortex density: transient, dynamic scaling, 
and saturation. We therefore study 
the vortex length statistics in each of these regimes separately.

\subsubsection{Short-time transient}

We first focus on the short-time transient, say $t \leq  7$, 
just before $\rho\sub{vortex}$ enters the scaling regime in which the space-time correlation 
scales with the growing length and the vortex density 
$\rho\sub{vortex}$ relaxes algebraically. As already observed in the analysis of 
$\rho\sub{vortex}$ the reconnection rule and microscopic dynamics affect the 
observations during this transient. Accordingly, we present the data
for the stochastic and maximal criteria separately.

\begin{figure}[tbh]
\centering
\begin{minipage}{0.49\linewidth}
\centering
(a)\\
\includegraphics[width=0.95\linewidth]{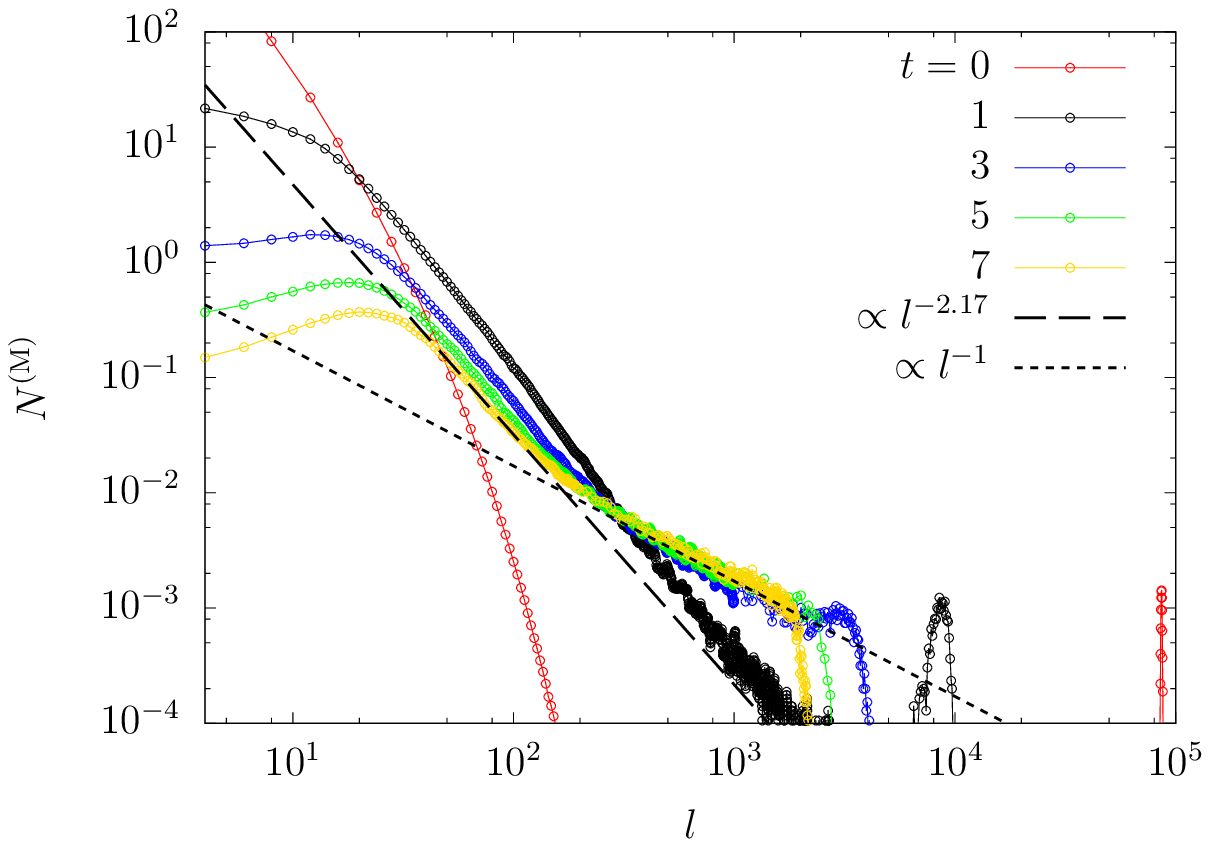}
\end{minipage}
\begin{minipage}{0.49\linewidth}
\centering
(b)\\
\includegraphics[width=0.95\linewidth]{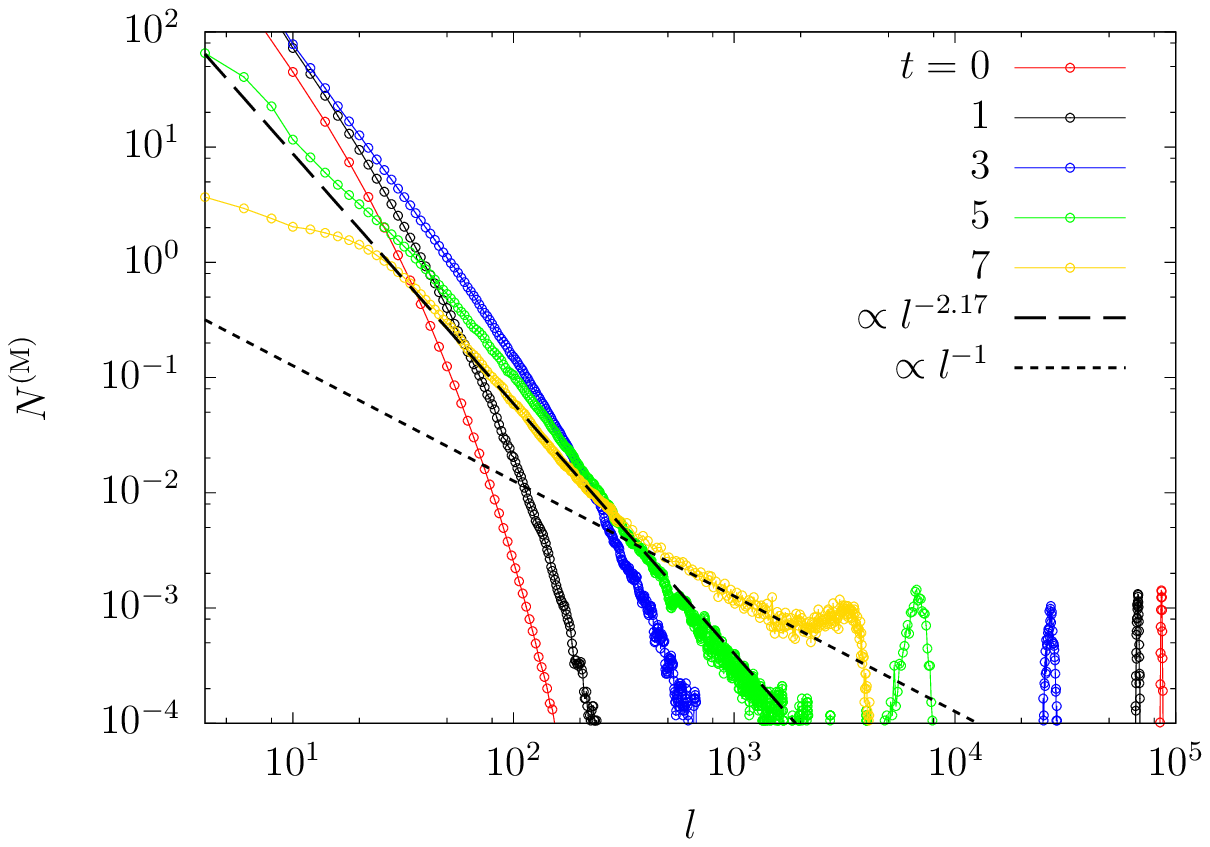}
\end{minipage}
\begin{minipage}{0.49\linewidth}
\centering
(c)\\
\includegraphics[width=0.95\linewidth]{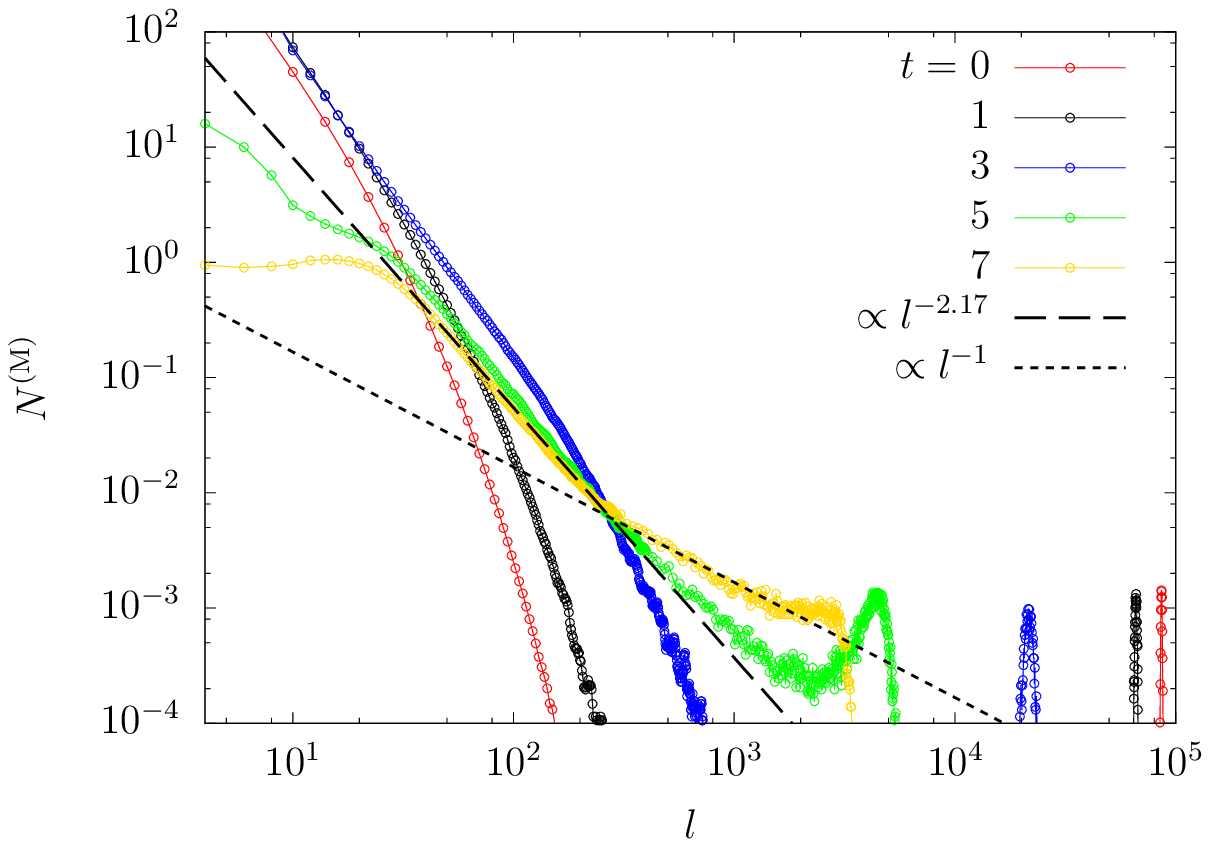}
\end{minipage}
\begin{minipage}{0.49\linewidth}
\centering
(d)\\
\includegraphics[width=0.95\linewidth]{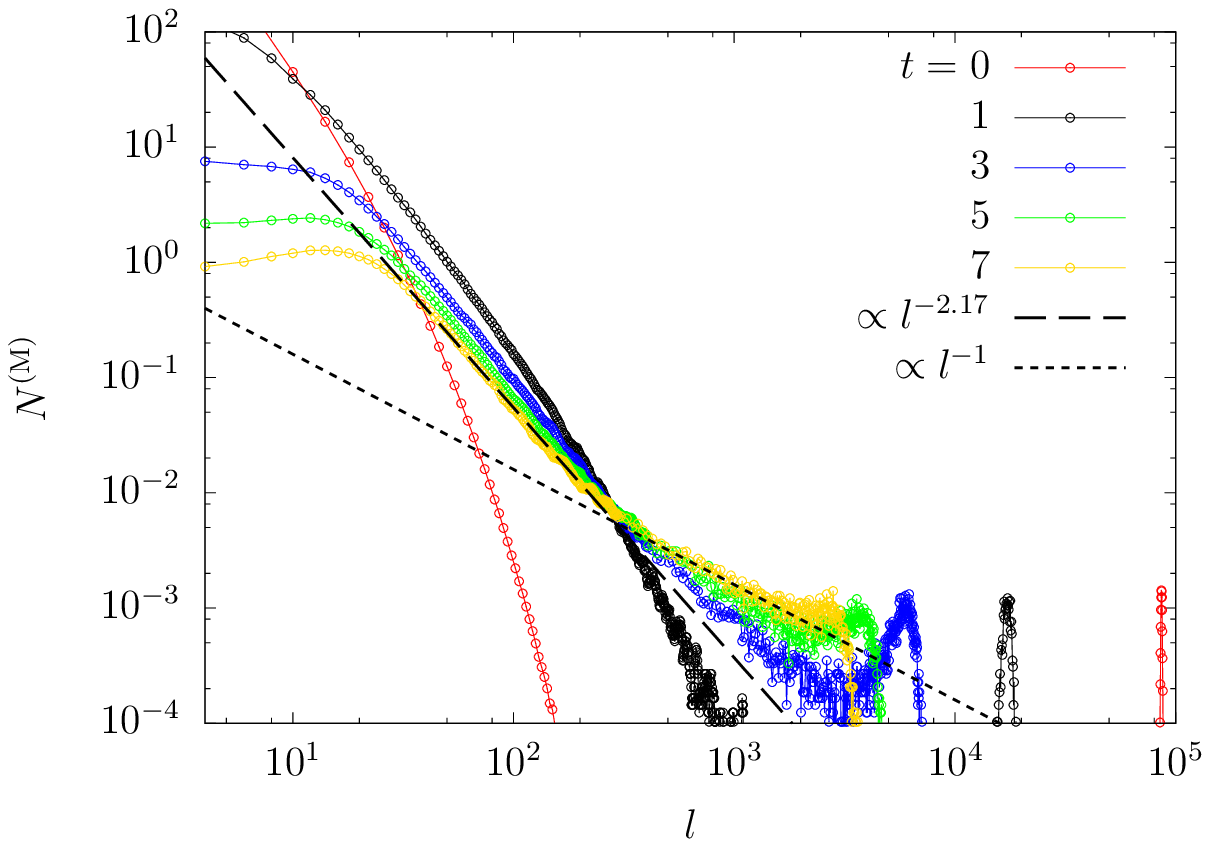}
\end{minipage}
\caption{\label{fig:fast-size-initial}
(Color online.)
Early stages of evolution.
Time dependent length number density $N\up{(M)}(l,t)$  in the 
initial stage of evolution, $t = 1$, $3$, $5$, $7$ of the (a) over-damped Langevin dynamics \eqref{eq:over-damped-Langevin},
(b) under-damped dynamics \eqref{eq:under-damped-Langevin}, (c) ultra-relativistic limit of the under-damped 
Langevin equation \eqref{eq:under-damped-ultra-relativistic},
and (d) non-relativistic limit of under-damped Langevin equation \eqref{eq:under-damped-non-relativistic}.
The maximal reconnection rule was used here to identify the vortex loops. The linear system size is $L = 100$.
The dashed line is the power law $l^{-\alpha\sub{L}\up{(M)}}$ with $\alpha\sub{L}\up{(M)}=2.17$, the dotted line the power $l^{-1}$,  
and the almost vertical peaks at the far right of the plot correspond length scales that diverge with the 
system size.
}
\end{figure}

Figure \ref{fig:fast-size-initial} shows the number of vortex loops with length 
$l$, i.e. $N\up{(M)}(l,t)$, in the 
initial stage of evolution obtained with the maximal criterium for vortex reconnection.
We recall that initially $N\up{(M)}(l)$ is given by the (blue) data in 
Fig.~\ref{fig:vortex-size} (a) with an exponential decay for finite size loops and  
a very sharp peak at $l \simeq L^3$. 

First, we confirm that the dependence on the microscopic dynamics is very strong during this 
initial period but it disappears at around $ t \simeq 7$. 

Second, we can see that the peak at long $l$ is progressively washed out as the very long loops 
break up into smaller ones.

Third, we observe that the curves at $t\simeq 7$ have three distinct
length regimes with smooth crossovers between them:

 - an incipient smooth increase at very short lengths, say $l \lesssim 20$, 

 - an algebraic decay, $\simeq l^{-2.17}$, at $20 \lesssim l \lesssim 200$ and 

 - a slower algebraic decay, $\simeq l^{-1}$, at $ 200 \lesssim l$. 

A very interesting feature of these curves is that the algebraic dependence after $l \gtrsim 20$
for $t=3-7$ strongly resembles the power-law decay of the number densities $N\up{(M,S)}(l)$ at the percolation
temperature $T\sub{L}\up{(M,S)}$ shown in Fig.~\ref{fig:porder}, 
$N\up{(M,S)}(l,t) \propto l^{-2.17}$, see the dashed line included as a guide-to-the-eye in all panels.
This fact suggests that the early dynamics spontaneously takes the system close to 
a percolating state similar to the equilibrium one at the  percolation threshold $T\sub{L}\up{(M)}$.

\begin{figure}[tbh]
\centering
\begin{minipage}{0.49\linewidth}
\centering
(a)\\
\includegraphics[width=0.95\linewidth]{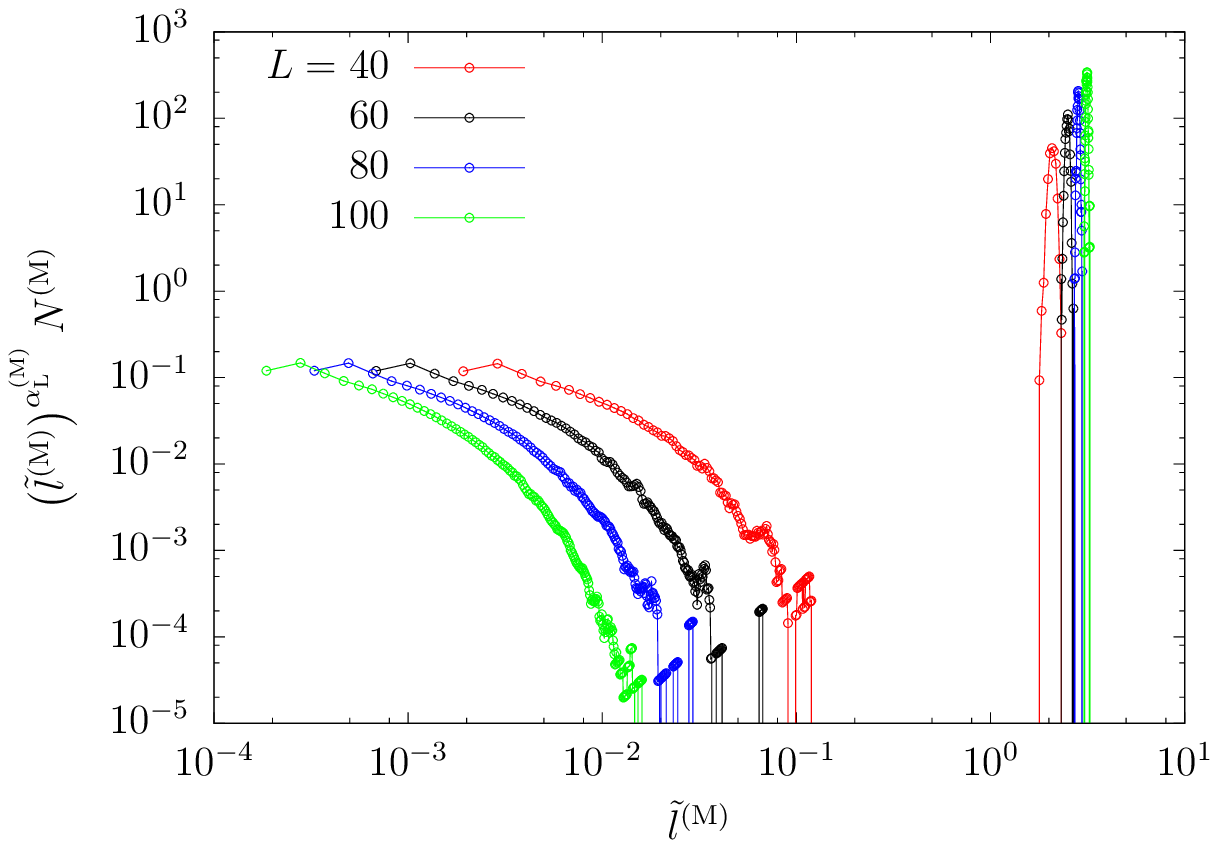}
\end{minipage}
\begin{minipage}{0.49\linewidth}
\centering
(b)\\
\includegraphics[width=0.95\linewidth]{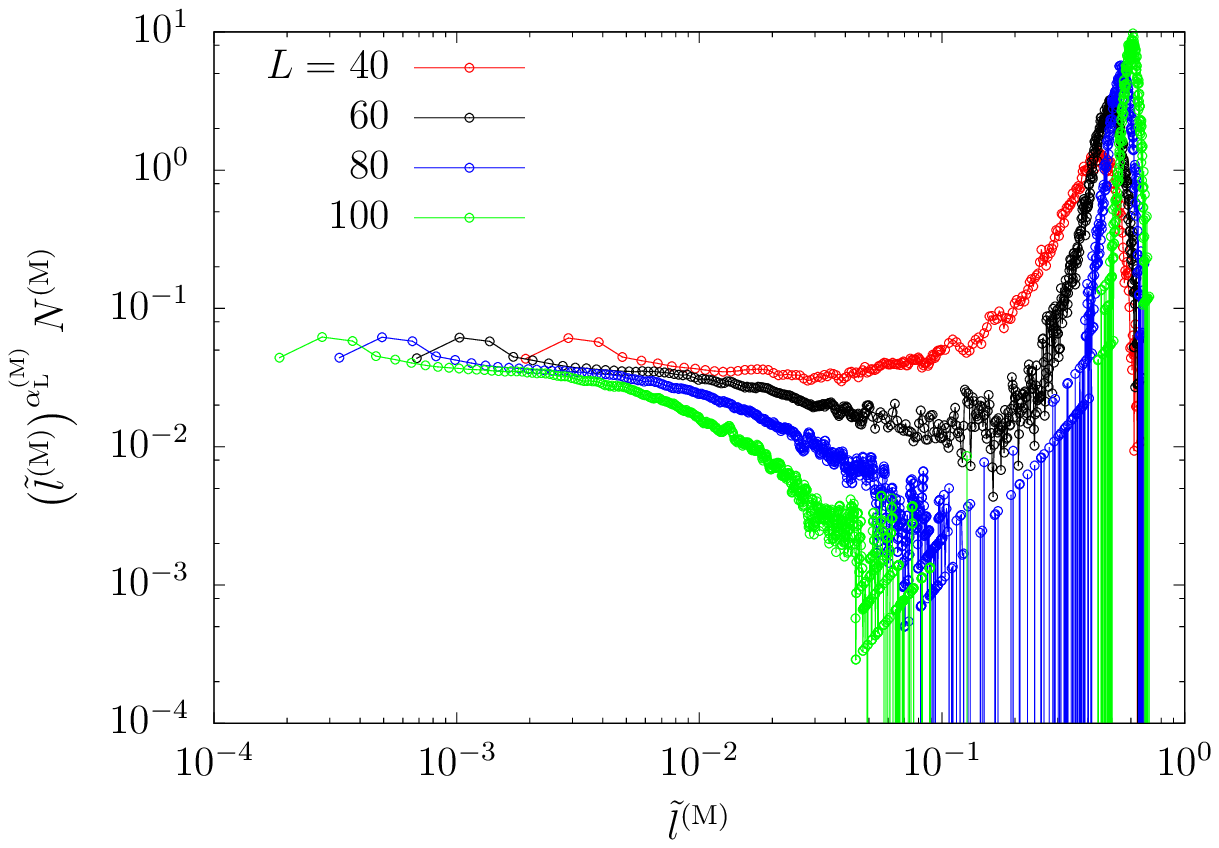}
\end{minipage}
\begin{minipage}{0.49\linewidth}
\centering
(c)\\
\includegraphics[width=0.95\linewidth]{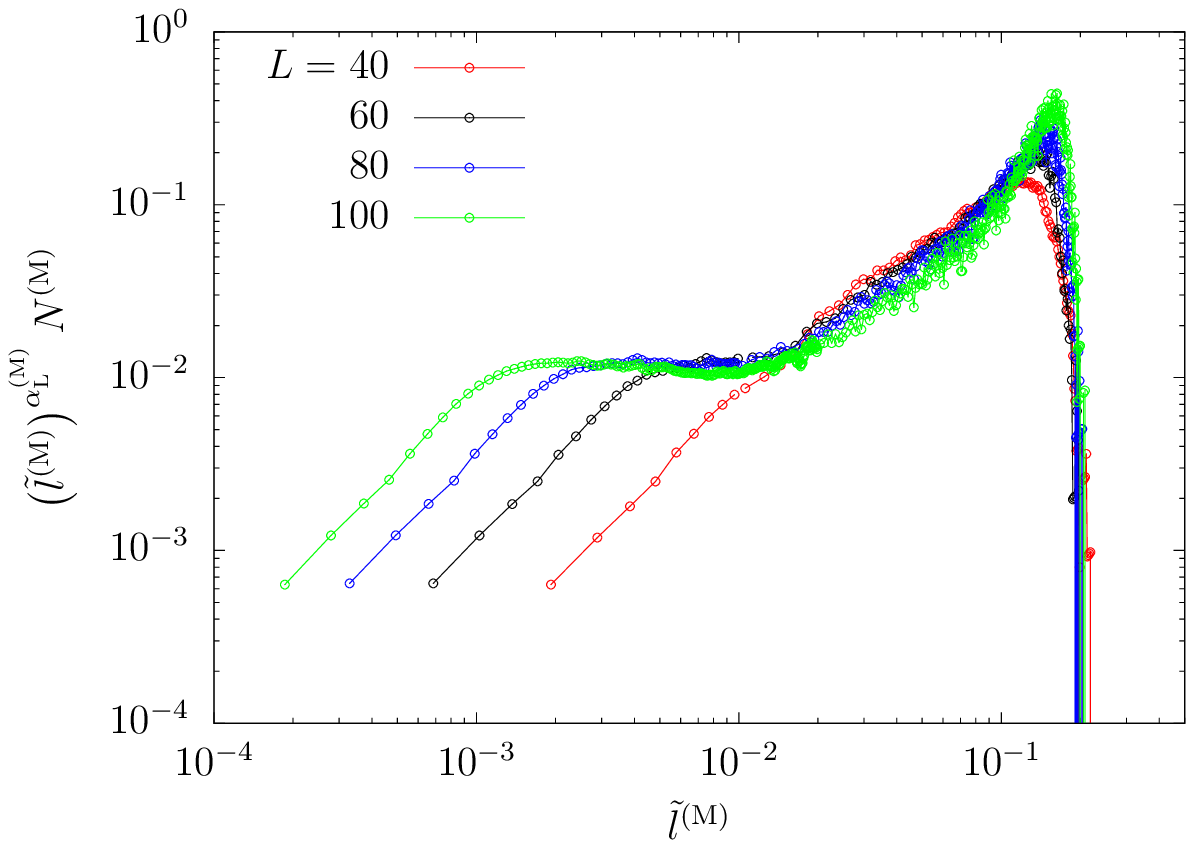}
\end{minipage}
\begin{minipage}{0.49\linewidth}
\centering
(d)\\
\includegraphics[width=0.95\linewidth]{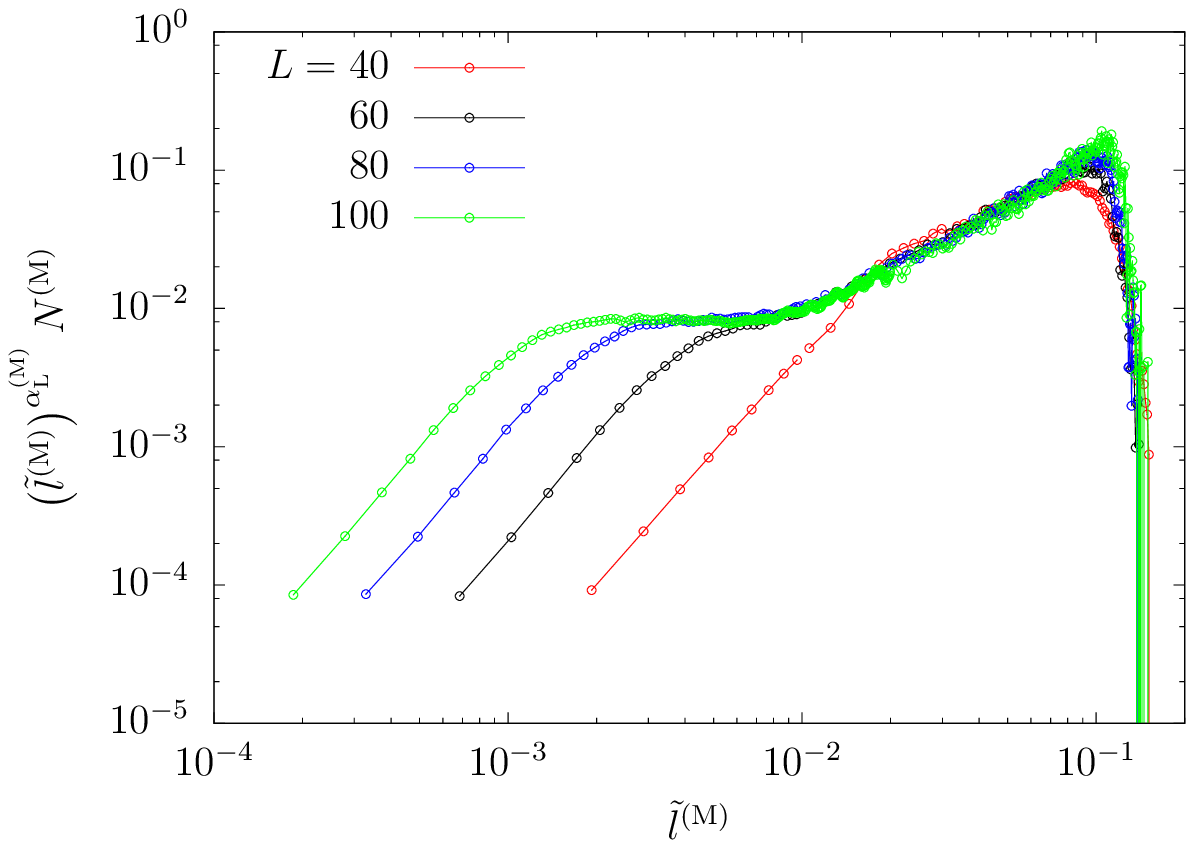}
\end{minipage}
\caption{\label{fig:fast-size-initial-scale}
(Color online.)
Early stages of evolution.
Finite-size scaling of the number density $N\up{(M)}(l,t)$ at (a) $t = 1$, (b) $t = 4$, (c) $t = 7$, and (d) $t = 10$ of the 
under-damped dynamics \eqref{eq:under-damped-Langevin}, with the maximal reconnection rule to identify the vortex loops.
The data are presented in a way that selects the weight at very long $l$. The scaling variable is 
$\tilde{l}\up{(M)} = l / L^{D\up{(M)}\sub{L}}$.
}
\end{figure}

Another fact to remark is the disappearance of the peak at very large $l$ (a feature of the initial condition 
treated with the maximum rule that is absent from the data analysed with the stochastic one)
and the generation of the $l^{-1}$ tail characteristic of fully-packed loop models (that was absent initially for this recombination rule).
 
In order to check the scenario of the spontaneous approach to  the percolating state, we study the
scaling of the large vortex loop weight as done in Fig.~\ref{fig:size-system-scale2} with the same scaling
variable $\tilde{l}\up{(M)} = l / L^{D\up{(M)}\sub{L}}$ and the fractal dimension 
$D\up{(M)}\sub{L} = d / (\alpha\sub{L}\up{(M)} - 1) \simeq 2.56$
(see~\cite{Blanchard14} for a similar analysis of the quench dynamics of the $2d$ Ising model).
Just after the quench, around $t = 1$, the number density $N\up{(M)}(l,t)$ is not universal with strong size-dependence.
As time elapses, the size-dependence gets weaker, and a scaling behaviour at large $l$ establishes at 
$t \simeq 10$ as shown in panel (d).
We can therefore conclude that the system enters the scaling regime around $t \sim 10$, and that 
this value does not  strongly depend on the exact form of the Langevin equation.
We note that the data for linear system size $L = 40$ slightly deviates from the scaling behaviour.

\begin{figure}[tbh]
\centering
\begin{minipage}{0.49\linewidth}
\centering
(a)\\
\includegraphics[width=0.95\linewidth]{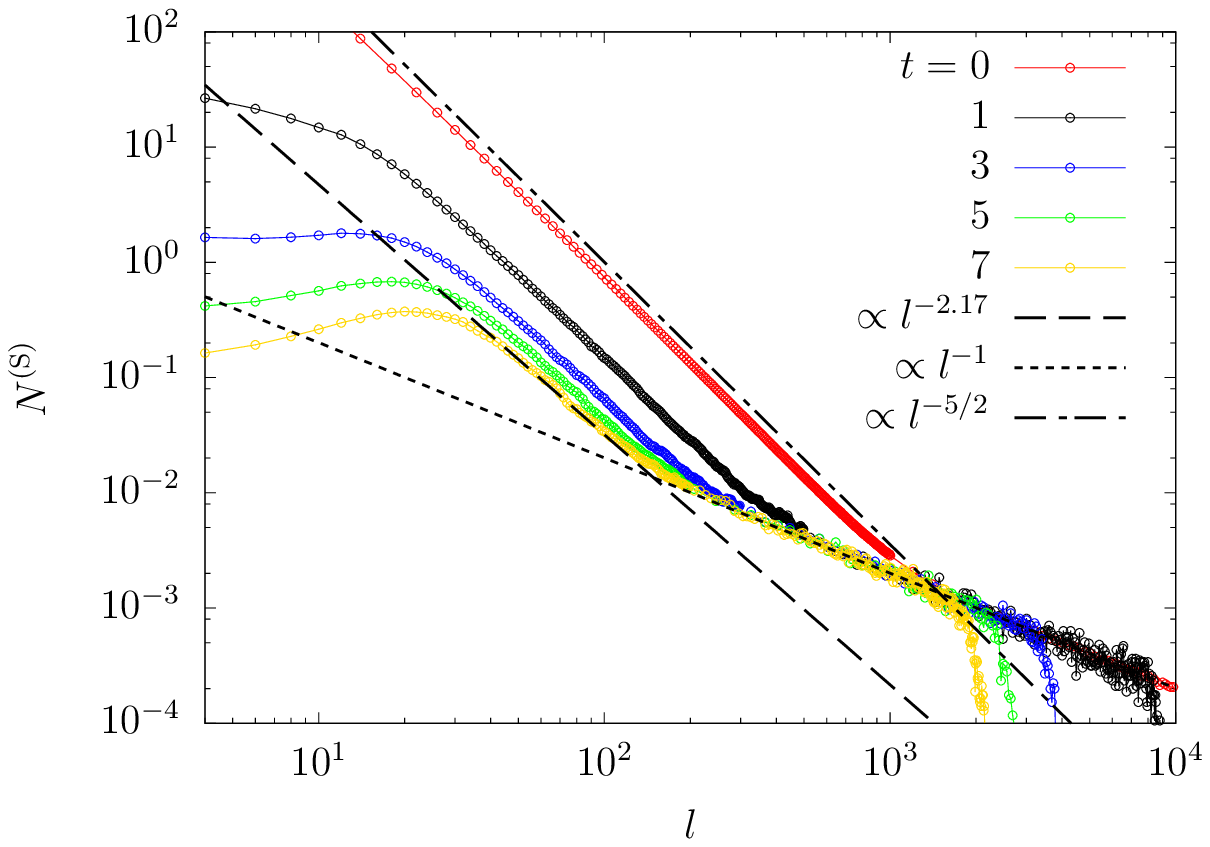}
\end{minipage}
\begin{minipage}{0.49\linewidth}
\centering
(b)\\
\includegraphics[width=0.95\linewidth]{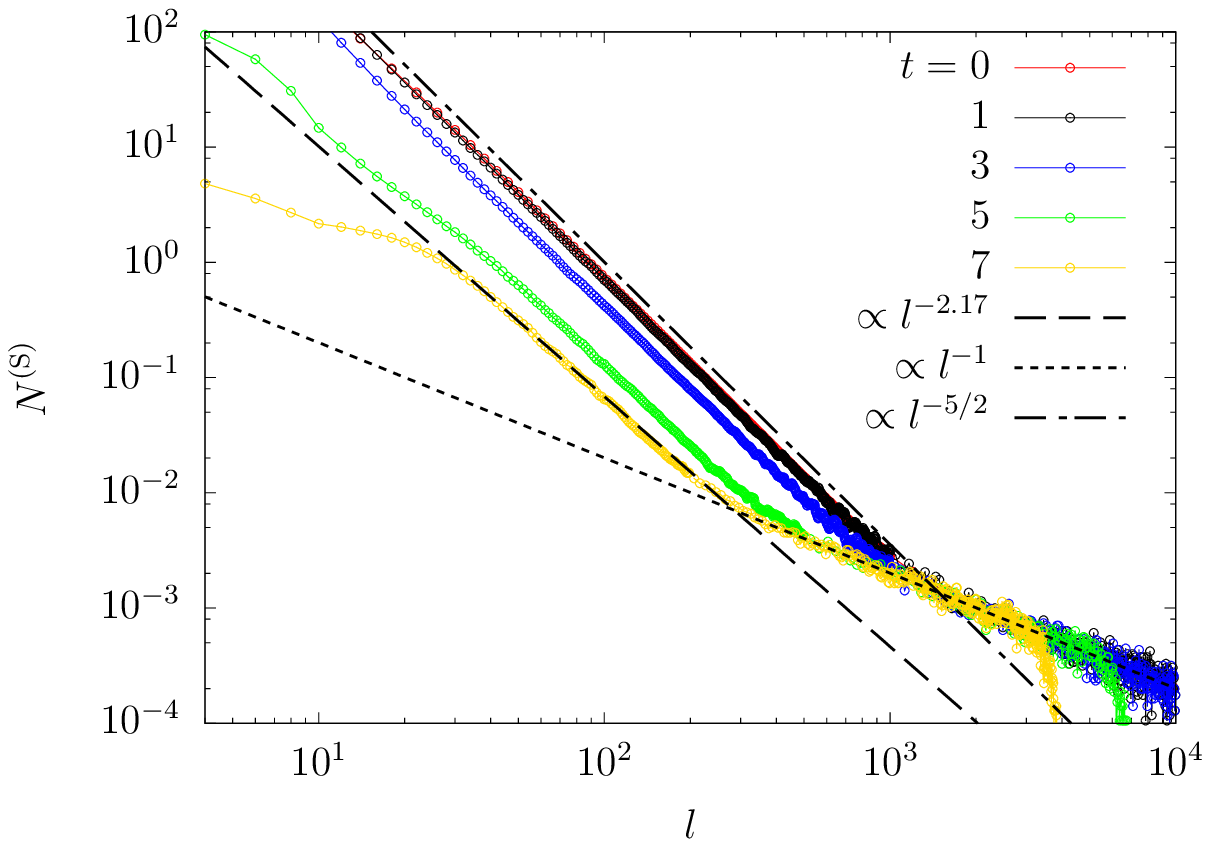}
\end{minipage}
\begin{minipage}{0.49\linewidth}
\centering
(c)\\
\includegraphics[width=0.95\linewidth]{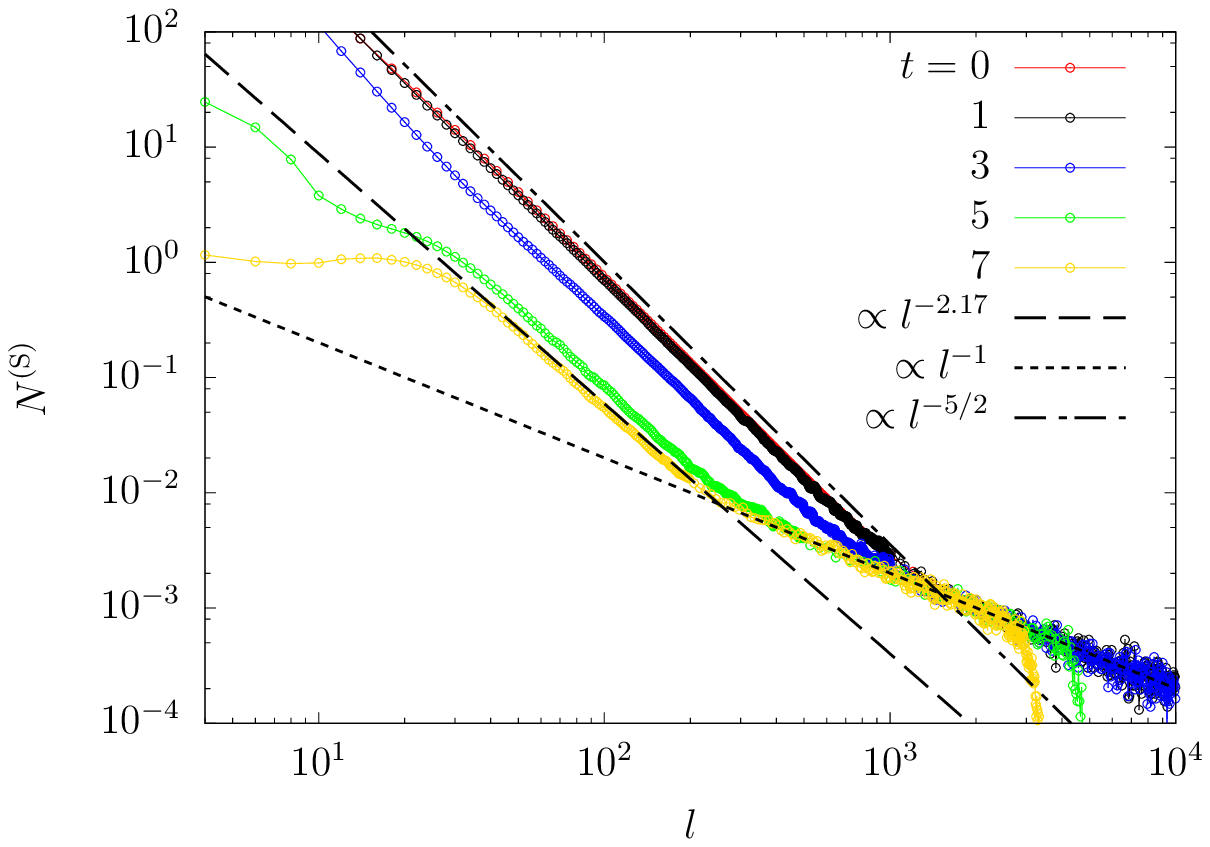}
\end{minipage}
\begin{minipage}{0.49\linewidth}
\centering
(d)\\
\includegraphics[width=0.95\linewidth]{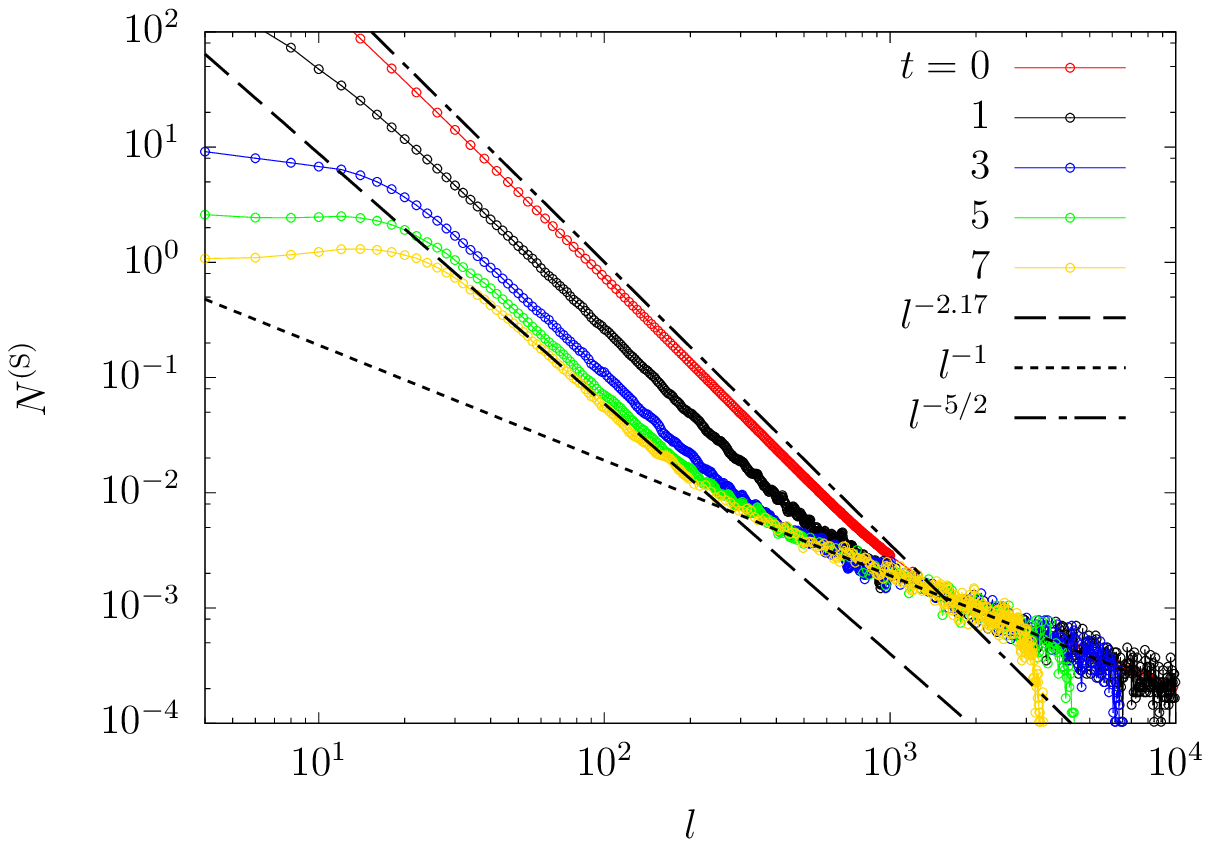}
\end{minipage}
\caption{\label{fig:fast-size-initial-stochastic}
(Color online.)
Early stages of evolution. 
Time dependent length number densities $N\up{(S)}(l,t)$  in the 
initial stage of evolution, $t = 1$, $3$, $5$, $7$ with (a) over-damped Langevin dynamics \eqref{eq:over-damped-Langevin},
(b) under-damped dynamics \eqref{eq:under-damped-Langevin}, (c) the ultra-relativistic limit of the under-damped Langevin equation
 \eqref{eq:under-damped-ultra-relativistic},
and (d) the non-relativistic limit of the under-damped Langevin equation \eqref{eq:under-damped-non-relativistic}. 
In all cases the stochastic reconnection rule was used to identify the vortex loops.
The linear system size is $L = 100$.
The dark dashed line is the power law $l^{-\alpha\sub{L}\up{(S)}}$ with $\alpha\sub{L}\up{(S)}=2.17$, 
the light dotted line is the power law $l^{-1}$ characterising the large scale statistics at high temperatures,
and the dashed-dotted line is the power law $l^{-5/2}$ of the Gaussian random walks that characterise the 
finite size loops at high temperature.
}
\end{figure}

Figure \ref{fig:fast-size-initial-stochastic} shows $N\up{(S)}(l,t)$ calculated with the stochastic 
criterium for vortex reconnections. We recall that initially, $N\up{(S)}$ is given by the (blue) data in 
Fig.~\ref{fig:vortex-size} (c) with a broken algebraic decay with exponents $5/2$ (Gaussian, lengths shorter than $L^2$) 
and $1$ (fully-packed, very long). 
All panels demonstrate the development of three length-scale regimes in the data-sets; again, 
very short lengths, 
$l \lesssim 20$, intermediate lengths, $20 \lesssim l \lesssim 200$, and 
very long lengths, $l \gtrsim 200$, as for the maximal criterium. 
In the course of time, the very long-tail remains proportional to $l^{-1}$, as in the equilibrium 
data at high $T$. The intermediate regime very soon acquires an algebraic decay that is numerically indistinguishable from the
one at the  
critical percolation point $T\sub{L}\up{(S)}$, given by the exponent $\alpha\sub{L} \simeq 2.17$. The 
weight of the number density at short loops is different, it increases with $l$ and decreases with $t$,
as for the maximal criterium.  We reckon that already at $t=1$ the Gaussian statistics of 
long loops with $l \ll L^2$ present in the initial condition has disappeared and the algebraic one has replaced it.

We end the analysis of the early dynamics by stating that, apart from the very specific peak at very long $l$ in 
the initial state with  the maximum criterium
that is soon erased dynamically, the dynamic vortex tangle built with the two rules has the same statistical and 
geometric properties. The quantitative analysis of the system-size dependence of the time needed to achieve the 
percolation structure at the intermediate length scales~\cite{Blanchard14,Tartaglia15} (that we very roughly estimated to be 
a few time units here) is beyond the scope of this paper.

\begin{figure}[tbh]
\centering
\begin{minipage}{0.49\linewidth}
\centering
(a)\\
\includegraphics[width=0.95\linewidth]{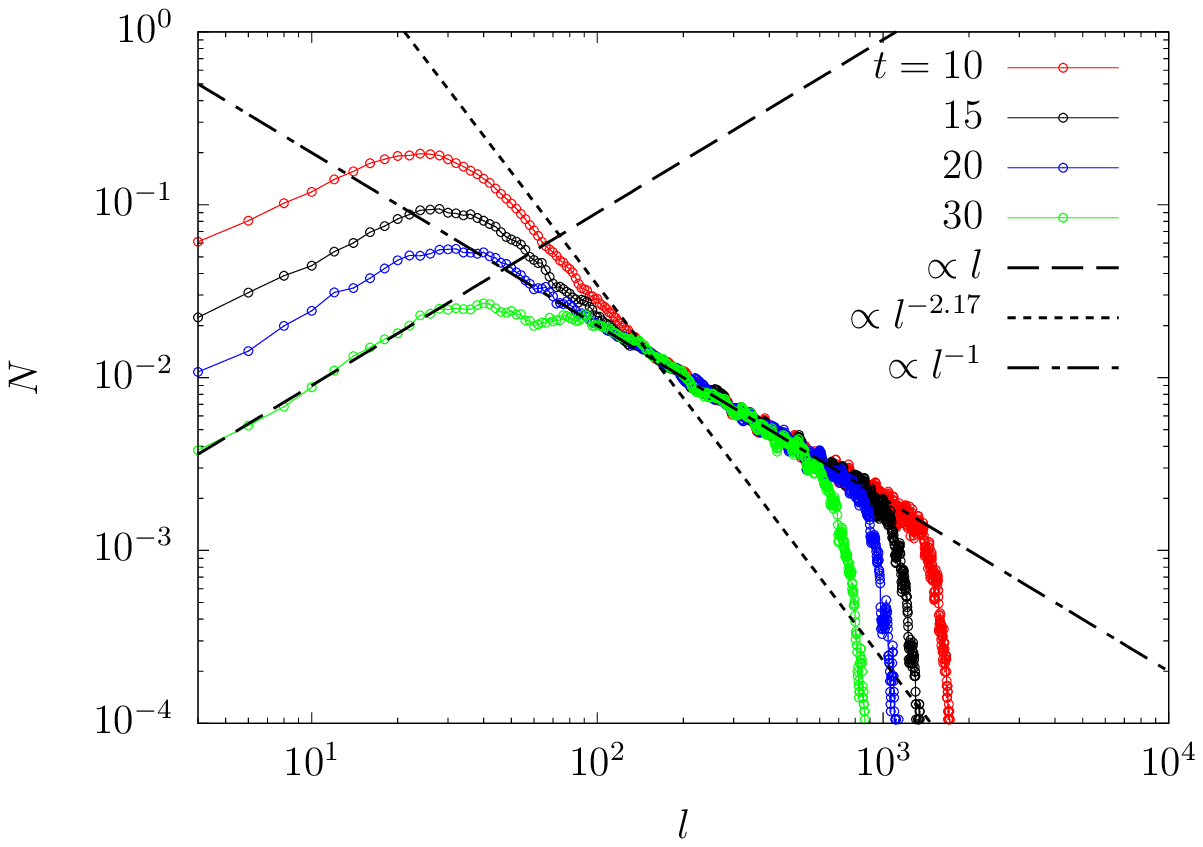}
\end{minipage}
\begin{minipage}{0.49\linewidth}
\centering
(b)\\
\includegraphics[width=0.95\linewidth]{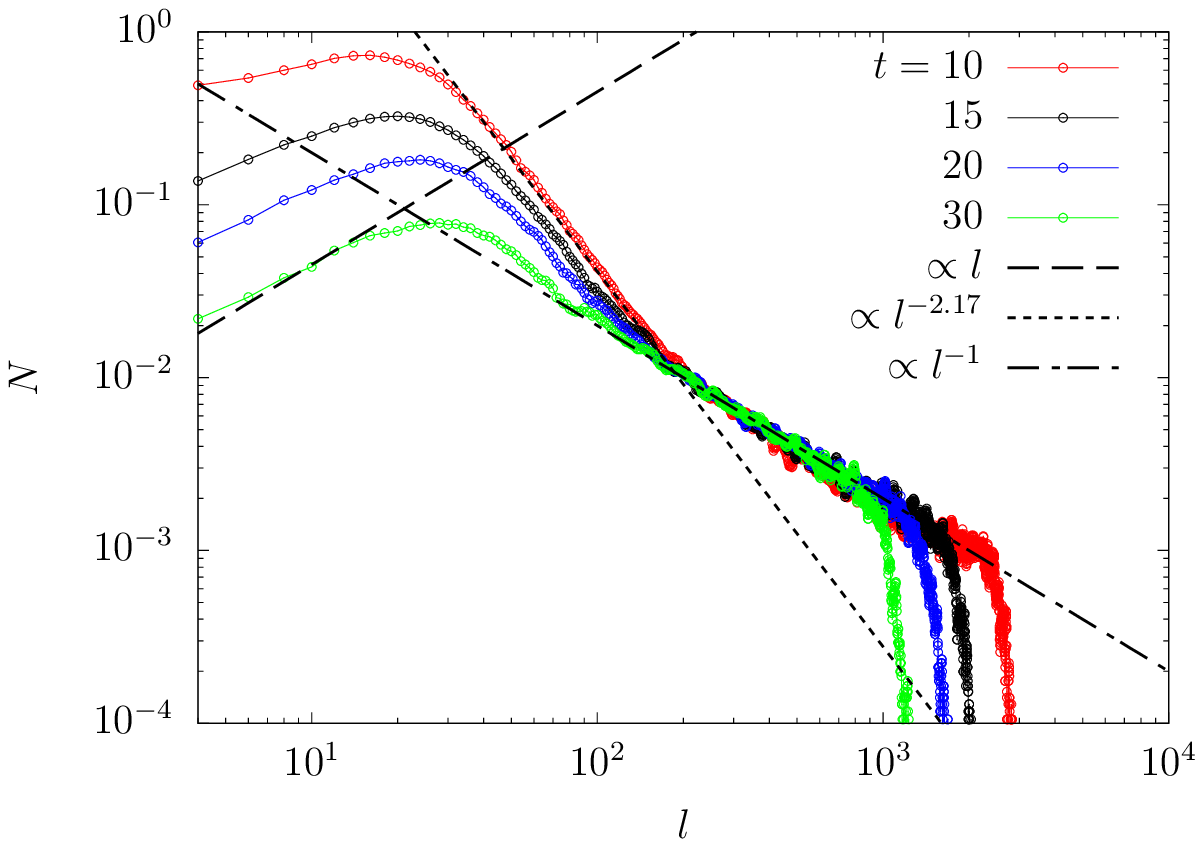}
\end{minipage}
\begin{minipage}{0.49\linewidth}
\centering
(c)\\
\includegraphics[width=0.95\linewidth]{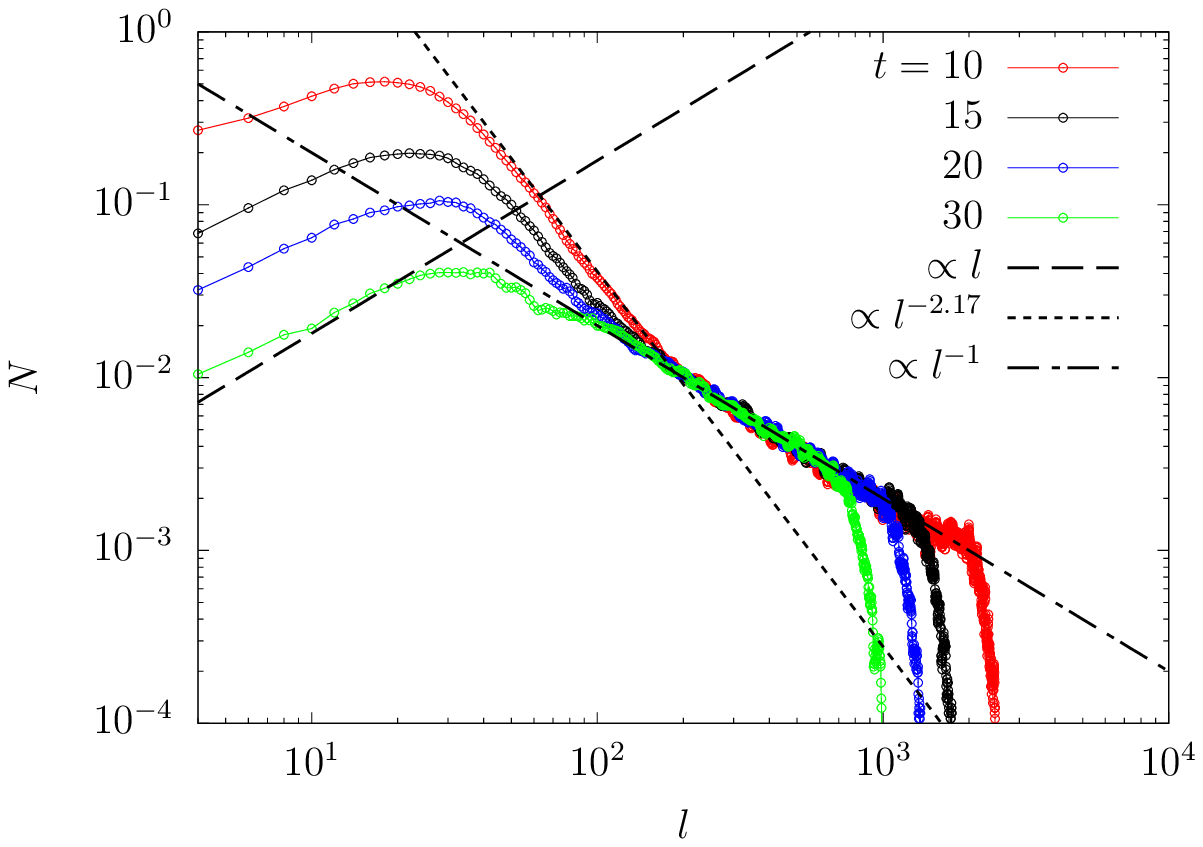}
\end{minipage}
\begin{minipage}{0.49\linewidth}
\centering
(d)\\
\includegraphics[width=0.95\linewidth]{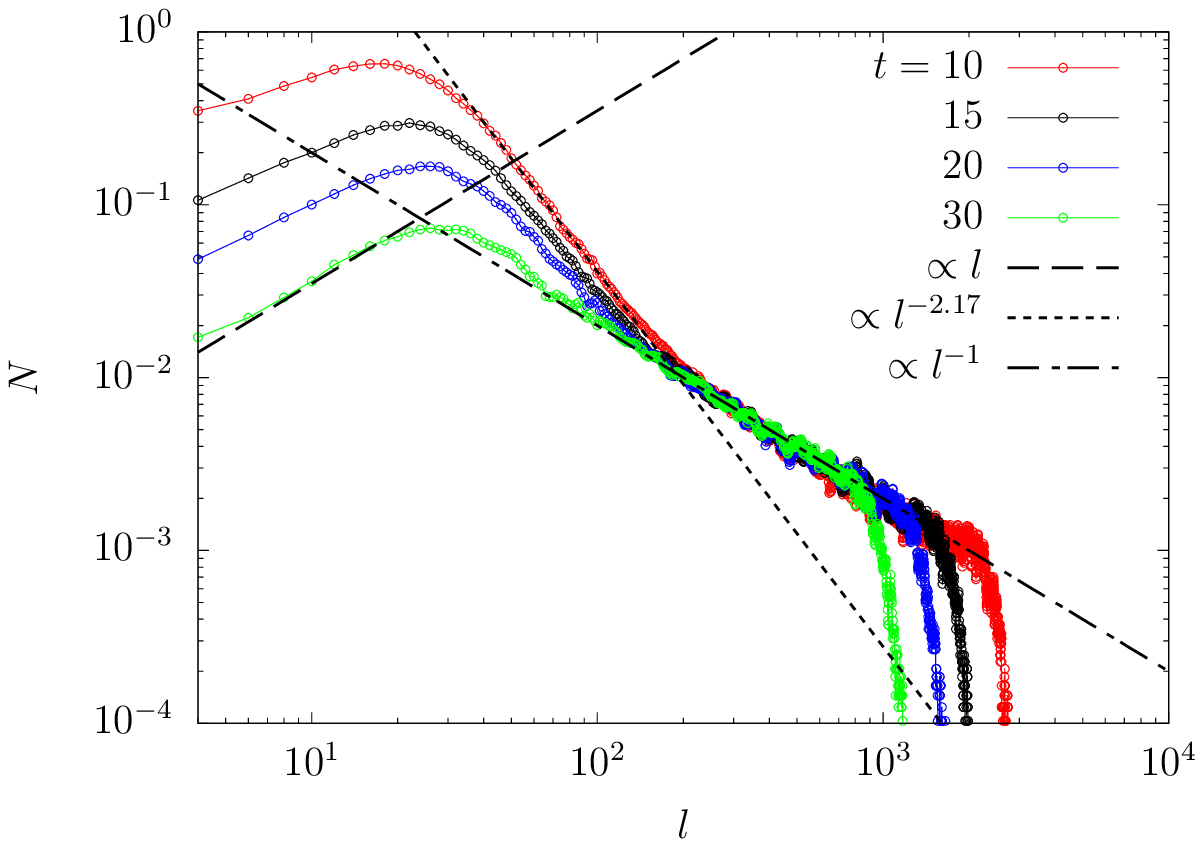}
\end{minipage}
\caption{\label{fig:fast-size-power}
(Color online.)
Dynamic scaling regime.
Time dependent length number densities $N(l,t)$  in the 
early stages of the scaling regime, $t = 10$, $15$, $20$, $30$ of the (a) over-damped Langevin dynamics \eqref{eq:over-damped-Langevin},
(b) under-damped dynamics \eqref{eq:under-damped-Langevin}, (c) ultra-relativistic limit of the under-damped Langevin 
equation \eqref{eq:under-damped-ultra-relativistic},
and (d) non-relativistic limit of under-damped Langevin equation \eqref{eq:under-damped-non-relativistic}. 
We use here the maximal reconnection rule to identify the vortex loops, although this choice is irrelevant in this regime.
The dashed line is the power law $l^{-\alpha\sub{L}}$ with $\alpha\sub{L}=2.17$,
the dotted line is the power law $l^{-1}$, and we also include a line proportional to $l$ for 
the statistics of the shortest loops. 
}
\end{figure}

\subsubsection{Scaling regime}

We now turn to the scaling regime in which the growing length and vortex density 
grow and decay algebraically, respectively. As the recombination rule becomes irrelevant
in this time-regime, we simply omit the upper-scripts (M) or (S).

\begin{figure}[tbh]
\centering
\begin{minipage}{0.49\linewidth}
\centering
(a)\\
\includegraphics[width=0.95\linewidth]{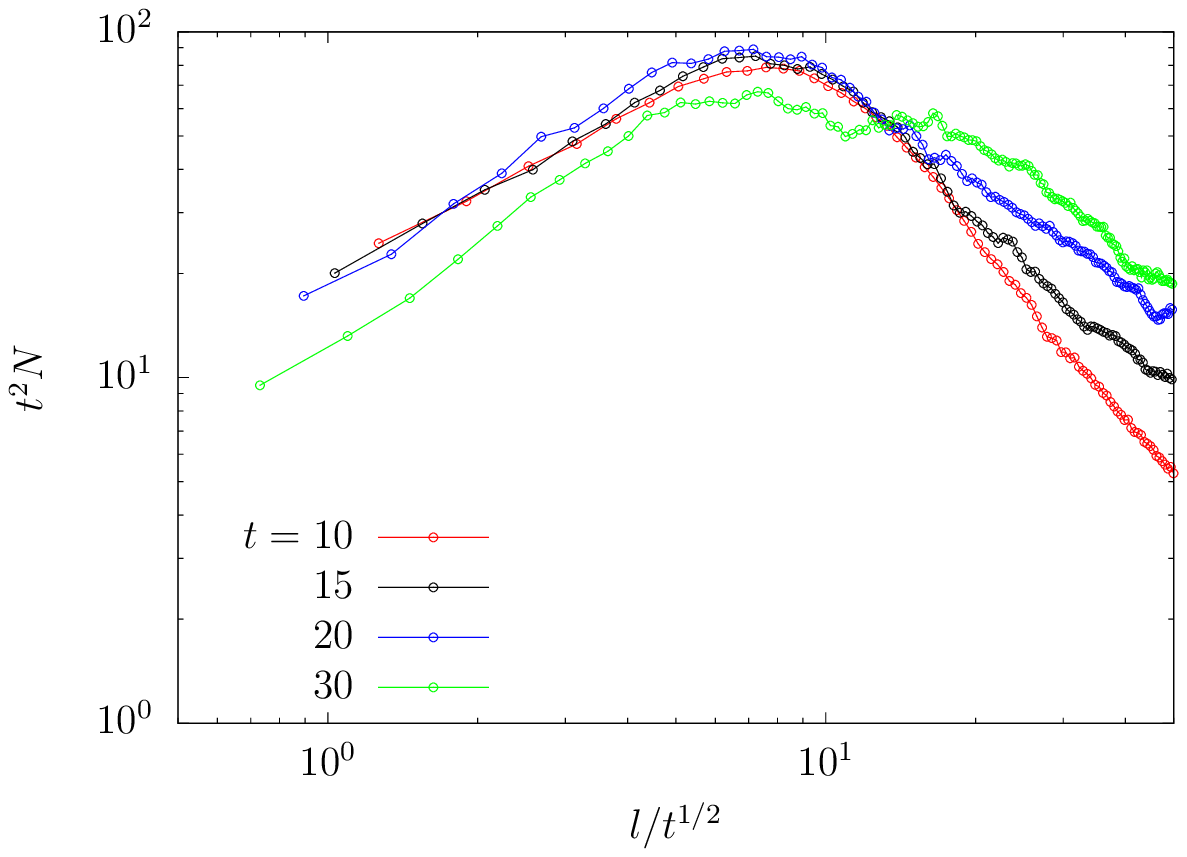}
\end{minipage}
\begin{minipage}{0.49\linewidth}
\centering
(b)\\
\includegraphics[width=0.95\linewidth]{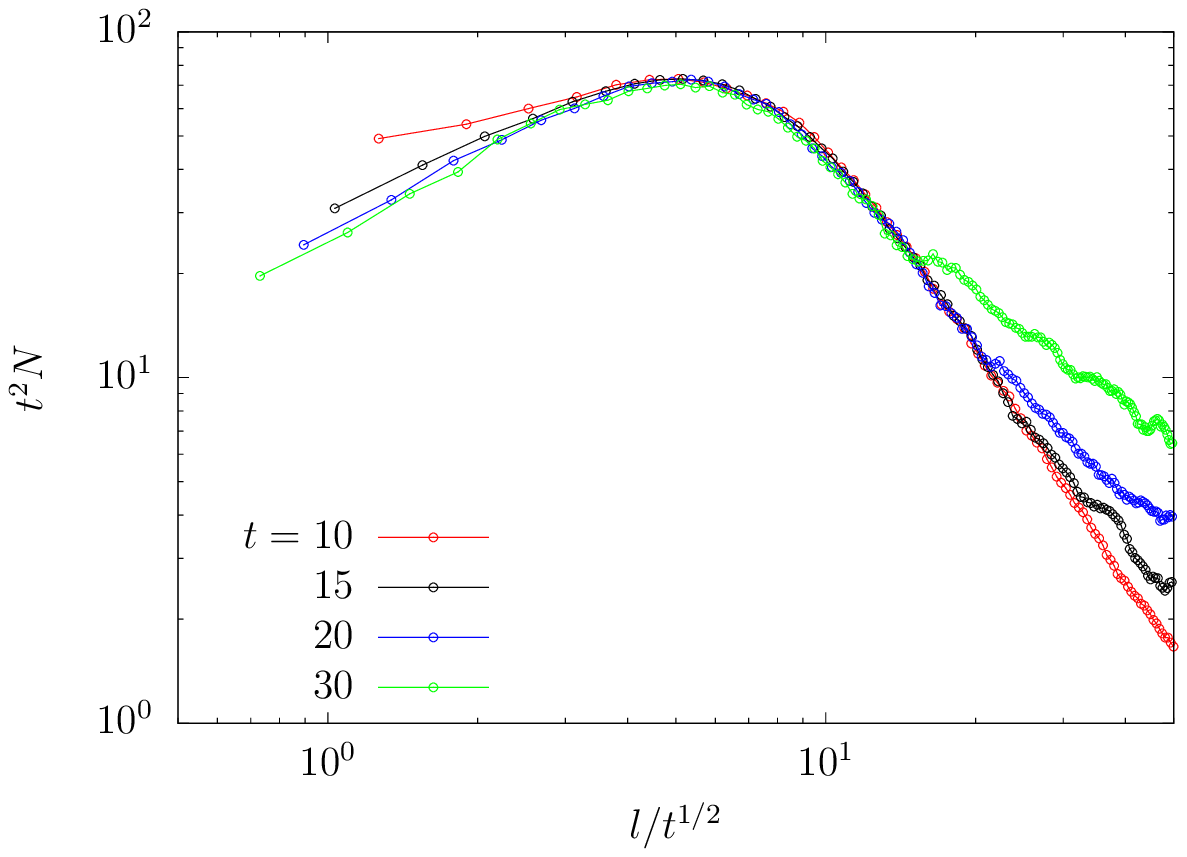}
\end{minipage}
\begin{minipage}{0.49\linewidth}
\centering
(c)\\
\includegraphics[width=0.95\linewidth]{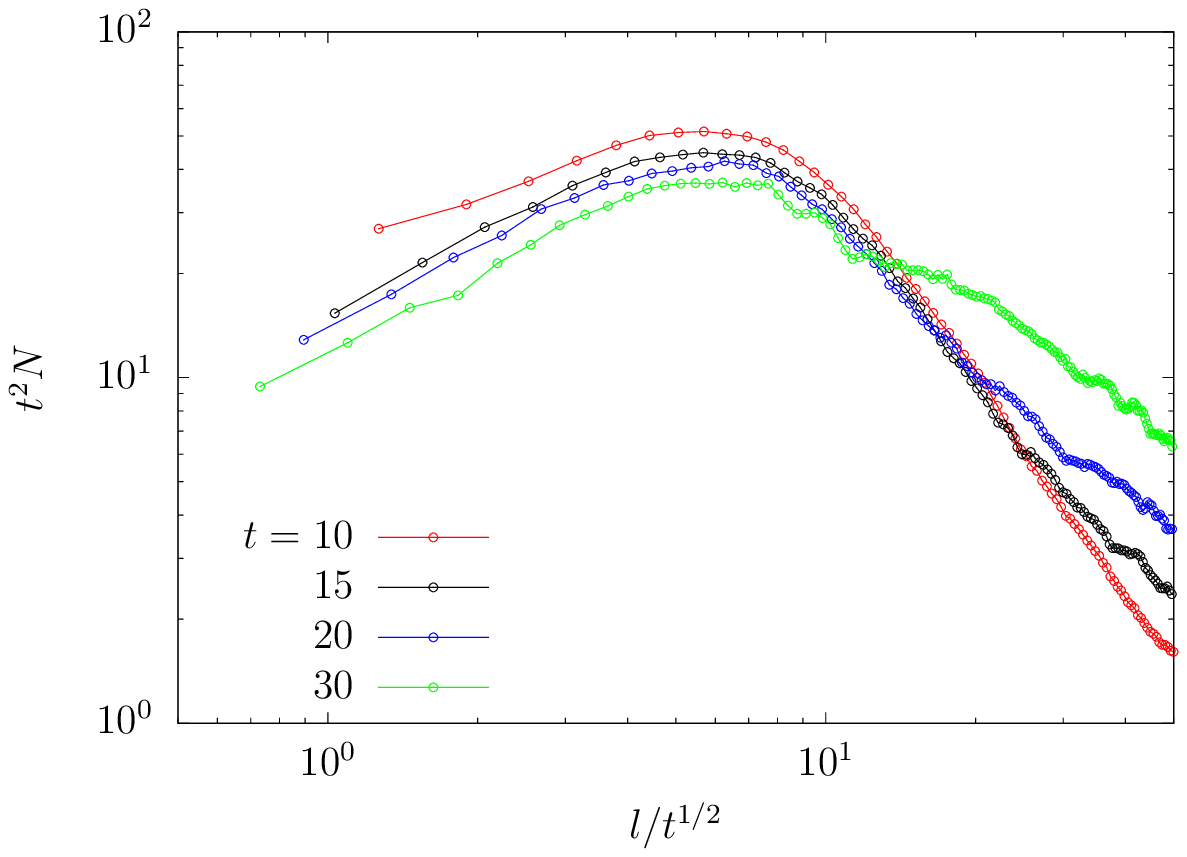}
\end{minipage}
\begin{minipage}{0.49\linewidth}
\centering
(d)\\
\includegraphics[width=0.95\linewidth]{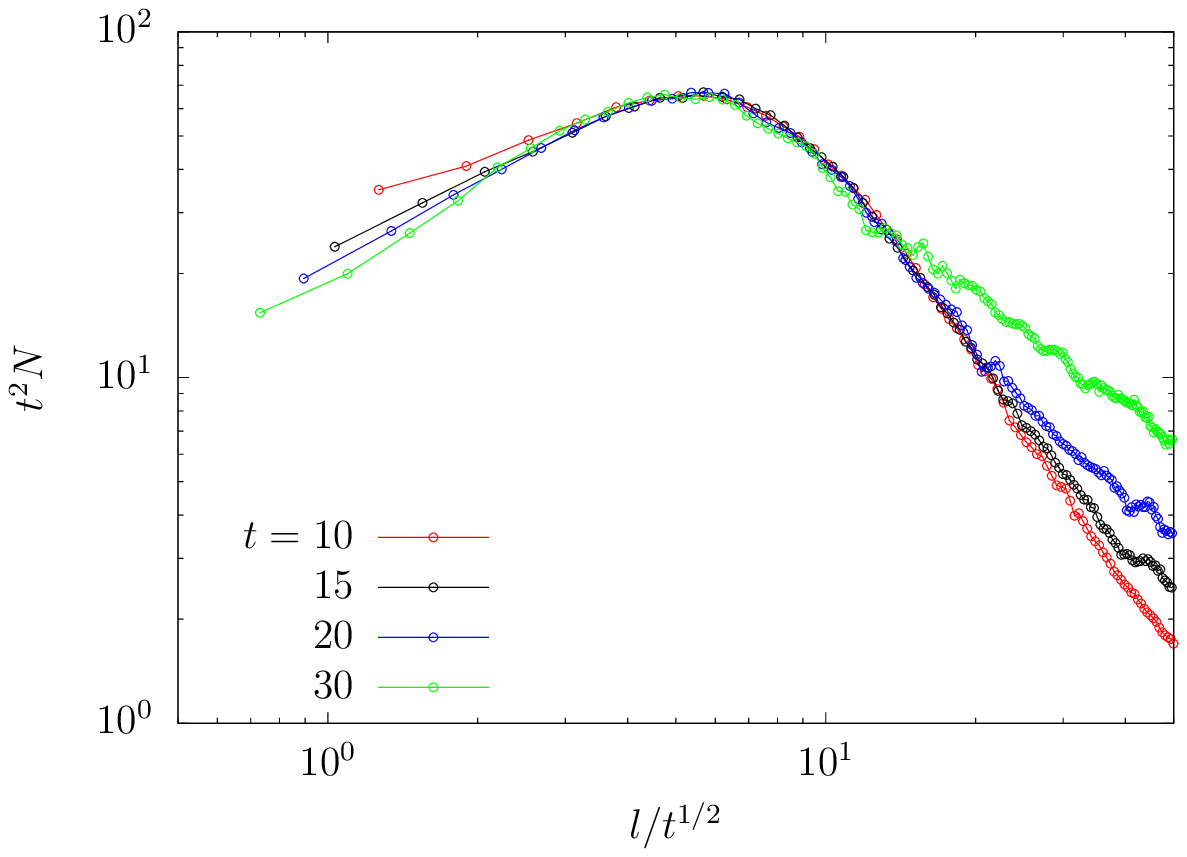}
\end{minipage}
\caption{\label{fig:short-length-scaling}
(Color online.)
Dynamic scaling regime.
Time dependent length number densities scaled as $t^2 N(l,t)$ as a function of $l / t^{1/2}$ in the 
early stages of the scaling regime, $t = 10$, $15$, $20$, $30$ of the (a) over-damped Langevin dynamics \eqref{eq:over-damped-Langevin},
(b) under-damped dynamics \eqref{eq:under-damped-Langevin}, (c) ultra-relativistic limit of the under-damped Langevin 
equation \eqref{eq:under-damped-ultra-relativistic},
and (d) non-relativistic limit of under-damped Langevin equation \eqref{eq:under-damped-non-relativistic}.
(The maximal reconnection rule was used here  although this choice is irrelevant at these times.)
}
\end{figure}

\begin{figure}[h!]
\centering
\begin{minipage}{0.49\linewidth}
\centering
(a)\\
\includegraphics[width=0.95\linewidth]{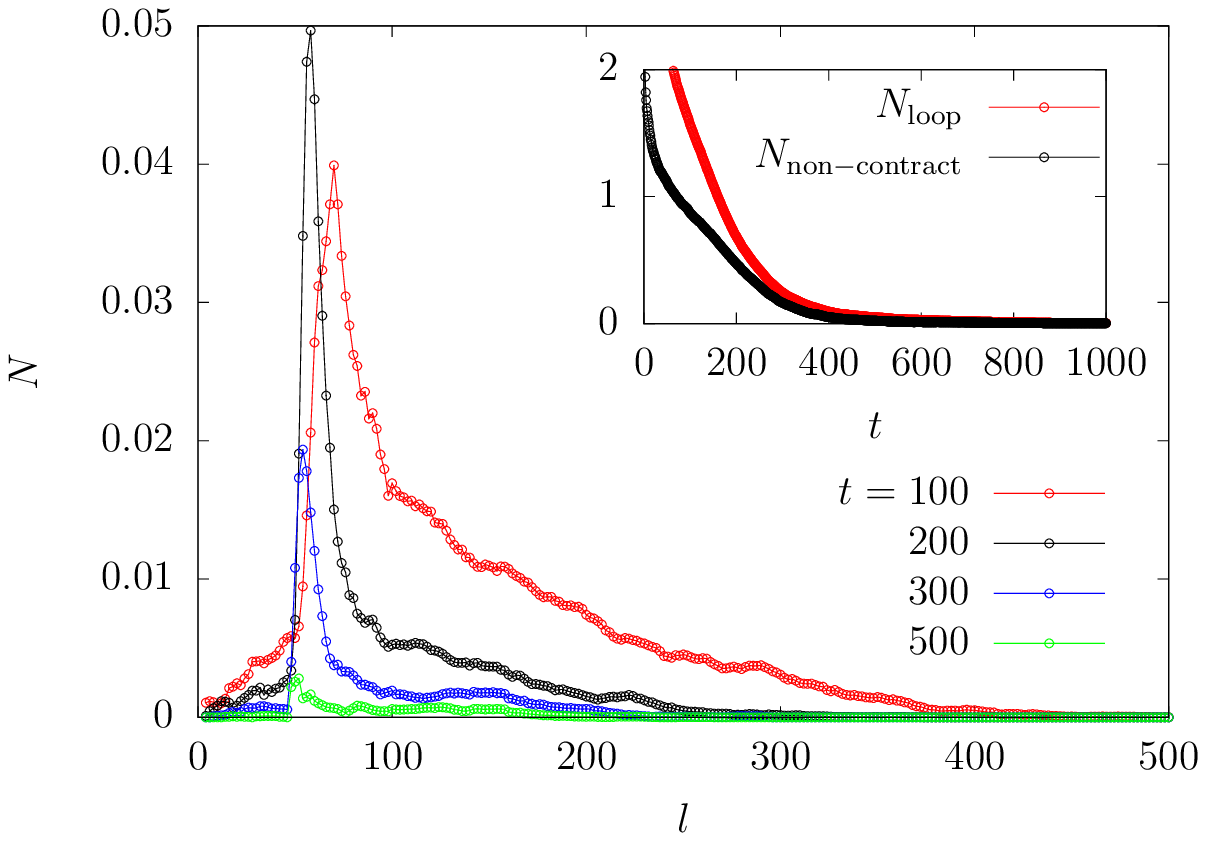}
\end{minipage}
\begin{minipage}{0.49\linewidth}
\centering
(b)\\
\includegraphics[width=0.95\linewidth]{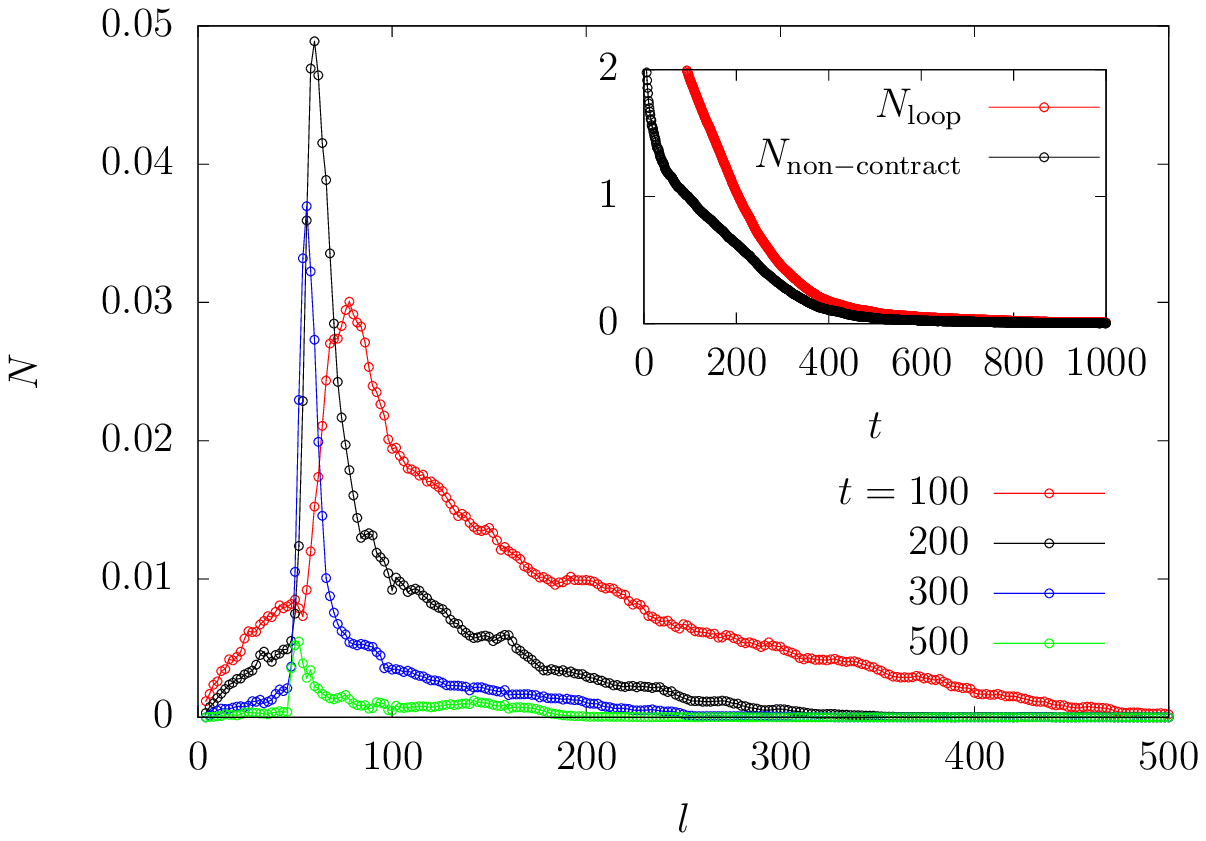}
\end{minipage}
\begin{minipage}{0.49\linewidth}
\centering
(c)\\
\includegraphics[width=0.95\linewidth]{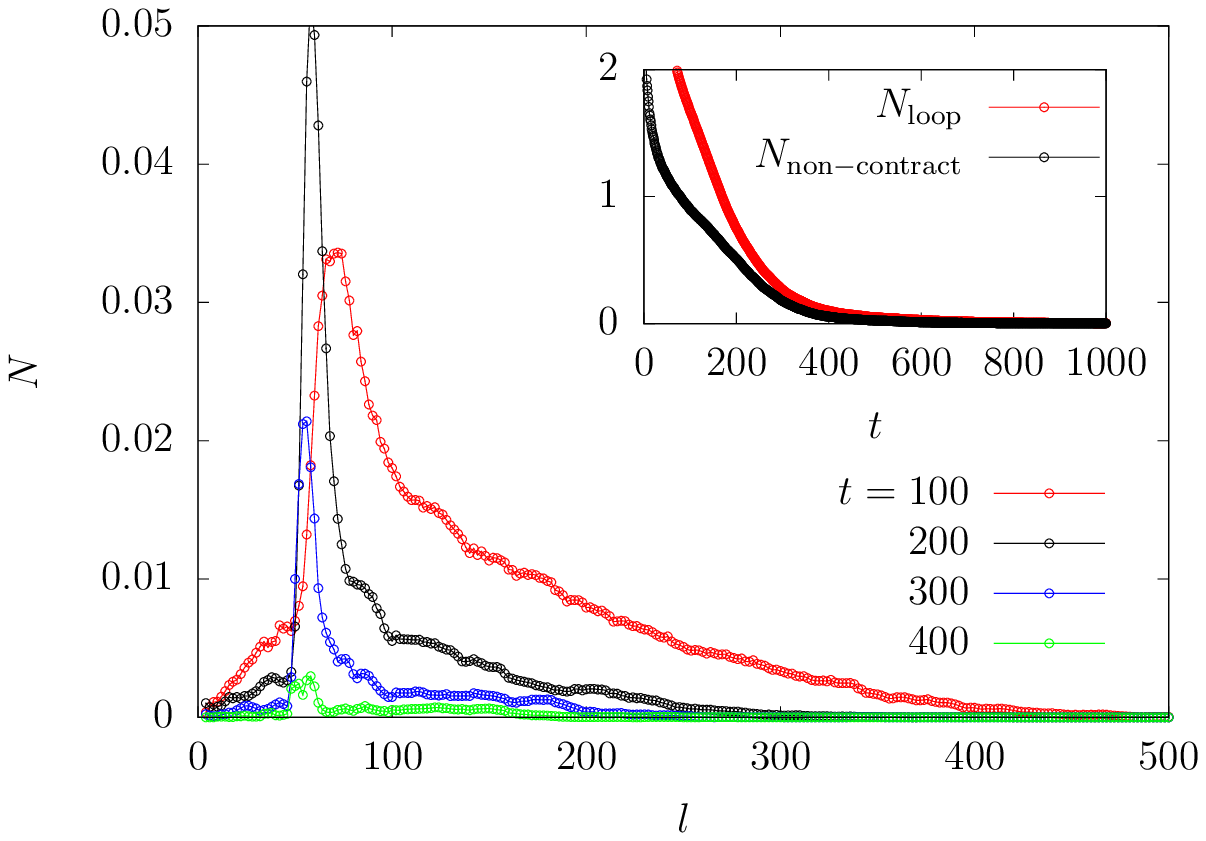}
\end{minipage}
\begin{minipage}{0.49\linewidth}
\centering
(d)\\
\includegraphics[width=0.95\linewidth]{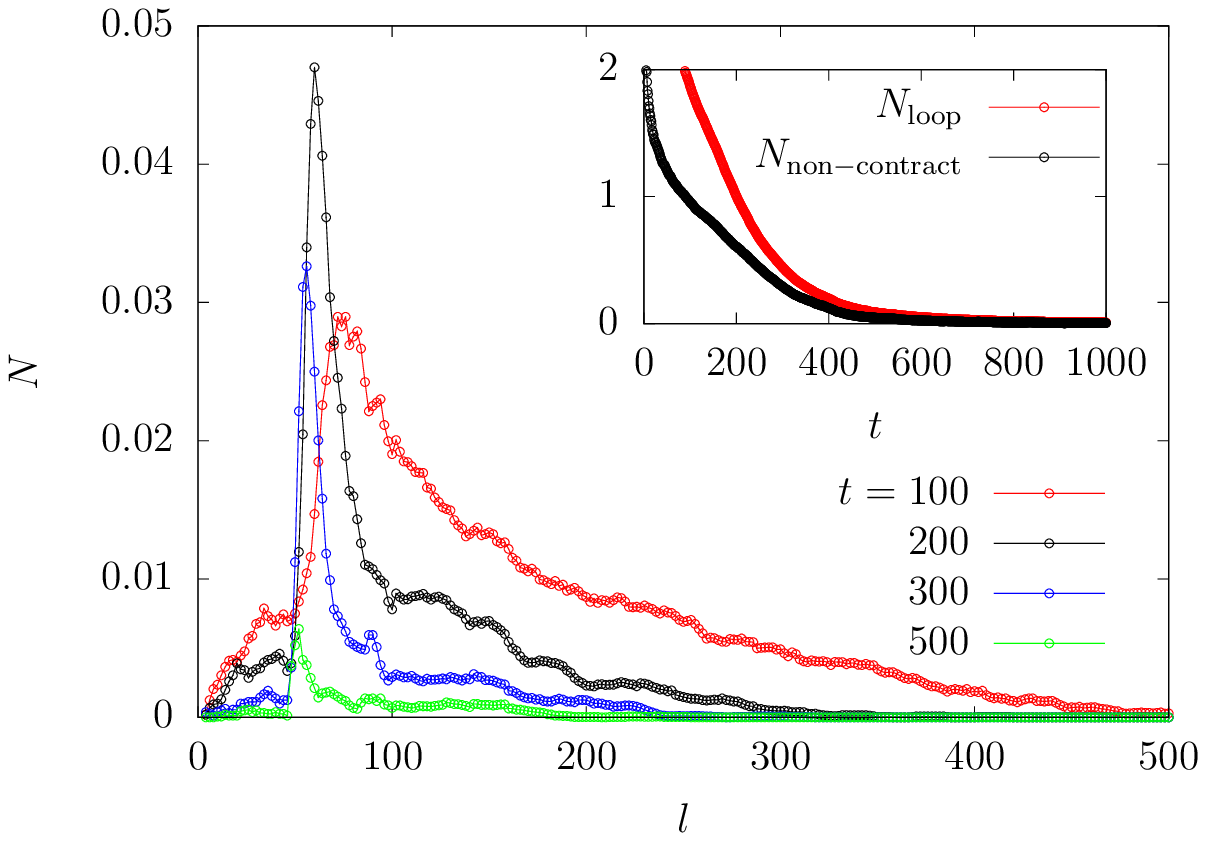}
\end{minipage}
\caption{\label{fig:fast-size-final}
(Color online.)
Late epochs - saturation.
Time dependent length number densities $N(l,t)$  in the 
late stages of the scaling regimes $t = 100$, $200$, $300$, $400$ of (a) over-damped 
Langevin dynamics \eqref{eq:over-damped-Langevin},
(b) under-damped dynamics \eqref{eq:under-damped-Langevin}, (c) ultra-relativistic limit of the under-damped Langevin equation \eqref{eq:under-damped-ultra-relativistic},
and (d) non-relativistic limit of under-damped Langevin equation \eqref{eq:under-damped-non-relativistic},
with the maximal reconnection rule to identify the vortex loops, although this choice is irrelevant at these times.
In the insets the total number of loops and the number of non-contractible loops as functions of time.
The linear system size is $L = 100$. }
\end{figure}

Figure \ref{fig:fast-size-power} shows the number density of vortex loops $N(l,t)$ during the early stages of the 
dynamic scaling regime $10 \leq t \leq 30$.
The short length scales, that we will see are bounded from above by $\xi\sub{d}(t)$, 
are weighted in such a way that $N(l,t) \simeq l$ and the pre-factor decreases with time. 
The algebraic behaviour $N(l,t) \propto l^{-2.17}$ at intermediate $l$, $\xi\sub{d}(t) \leq l \leq L^2$, 
gets narrower as time elapses, while the 
$ l^{-1}$ tail at long $l$ as seen in $N\up{(S)}(l)$ in equilibrium at high temperature, see Fig~\ref{fig:vortex-size} (c),
remains. 
At $t = 30$ and after, see Fig.~\ref{fig:fast-size-final}, 
the algebraic behaviour of $N\up{(M)}(l,t) \propto l^{-2.17}$ is almost totally wiped out while the $N(l,t) \propto l^{-1}$
tail has support over shorter lengths than initially. 


The small $l$ weight is clearly time dependent. In Sec.~\ref{subsubsec:analytic} we conjecture that it is given by 
\begin{equation}
N(l,t) \simeq \frac{l}{t^{5/2}}
. 
\label{eq:short-length-scaling}
\end{equation}
In Fig.~\ref{fig:fast-size-power} we included a straight line with the linear dependence $l$ 
that gives a very good description of  the data.
In Fig.~\ref{fig:short-length-scaling} we present data at times $10 \leq t \leq 30$ 
in the form $t^2 N(l,t)$ against $l/t^{1/2}$ and we check that the time dependence in 
Eq.~(\ref{eq:short-length-scaling}) is also very well verified.

\subsubsection{Late-time regime}

Finally we study the very late regime in which just a few vortices are left in each sample, 
see Fig.~\ref{fig:fast-vortex-one-loop} (c) and (d),  and dynamic scaling will soon break down.

Figure \ref{fig:fast-size-final} shows the number of vortex loops $N(l,t)$ at $t = 100$, $200$, $300$ and $400$.
At $100$, $\rho\sub{vortex}$ is still in the power-law regime 
(see Fig. \ref{fig:fast-rhov}) while at longer times  it has definitely left it. 
The intermediate algebraic regime with power $-2.17$ has already disappeared.
A region slightly narrower than a
decade with the power $-1$ remains at $t=50$ and long lengths.
At $t > 100$  this power has also disappeared and there 
remains a (very noisy) single peak structure,
which reflects the existence of just a few rapidly shrinking vortex loops in the samples. The insets display the total 
number of vortices and the number of non-contractible vortex loops that decay in time towards zero in both cases.

\subsection{Analytic derivation of the vortex-length number density}
\label{subsubsec:analytic}

We now focus on the dynamics after the transient 
and before finite size effects lead to saturation of the growing length ($7 \simeq t\sub{p} < t < 
t\sub{S} \simeq 100$).

We assume that after the transient $t_{\rm p}$, 
the length of each vortex is reduced at the same rate as the \textcolor{black}{dynamic correlation length $\xi_d$} grows
\begin{equation}
l(t, l_{\rm p}) \simeq \gamma_{\rm v} \sqrt{t_{\rm a} - t}
\; , 
\end{equation}
with $\gamma_{\rm v}$ a parameter, 
$t_{\rm a} =  t_{\rm p} + l_{\rm p}^2 / \gamma_{\rm v}^2$ the annihilation time at which $l(t_{\rm a}) =0$, 
and $l_{\rm p}$ the length of the vortex ring at $t_{\rm p}$, the time at the end of the transient  (say, $t\sub{p}\simeq 7$ in the 
previous Section, but note that this time could be a function of the system size, as occurs in the 
$2d$ Ising model~\cite{Blanchard14,Tartaglia16,Insalata16} or the $2d$ voter model~\cite{Tartaglia15}).

We suppose that the vortices are sufficiently long and far apart that
they evolve independently of each other. Neglecting  the fact that they break up and disappear in the 
course of evolution we use
\begin{equation}
N(l,t) \simeq \int dl_{\rm p} \, N(l_{\rm p}, t_{\rm p}) \, \delta (l - l(t, l_{\rm p}))
\end{equation}
to obtain
\begin{equation}
N(l,t) \simeq 
\frac{l \ N(\sqrt{\gamma_{\rm v}^2 (t-t_{\rm p}) + l^2}, t_{\rm p})}{\left[\gamma_{\rm v}^2 (t-t_{\rm p}) + l^2\right]^{1/2}}
 \; \label{eq:loop-distribution}
\end{equation}

The key point is to keep the full form of the $N$ evaluated at $t\sub{p}$ appearing 
in the numerator, in such a way to include the two power laws.
The double algebraic decay at $t=0$  or $t_{\rm p}$  are 
well approximated by 
\begin{equation}
N(l, t\sub{p}) \simeq l^{-\alpha_{\rm L}} \ [c_1^{n}(t) + c_2^n(t) l^{(\alpha_{\rm L}-1) n}]^{1/n}
\; .
\label{eq:guess}
\end{equation}
A sharp  cross-over between the two power laws is obtained 
for large  $n$ ($n=6$ is sufficient, see Fig.~\ref{fig:fast-size-initial-stochastic}). 
The cross-over takes place at $l^*(t) \simeq [c_1(t)/c_2(t)]^{1/(\alpha_{\rm L}-1)}$.
If $\alpha_{\rm L}=5/2$, as for the initial state, $c_1(0) \propto L^3$ and $c_2(0)$ is a finite 
constant that ensure $l^*(0) \simeq L^2$ and the 
limit forms in Eq.~(\ref{eq:limits}). At $t_{\rm p}$ the weight of the loops with 
$l > l^*(t_{\rm p})$ is not modified with respect to the initial one, and $c_2(t_{\rm p}) = c_2$.
The total number of loops diminishes in time but remains $O(L^3)$ until the very late epochs~\cite{KoCu16}.
For $N$ as in Eq.~(\ref{eq:guess}), to leading order in $L$, and 
ignoring constants, $N_{\rm loop}(t) = \int dl \, N(l,t) \simeq c_1(t) \int_{l_0}^{l^*} dl  \, l^{-\alpha_{\rm L}}
+ c_2(t) \int_{l^*}^{L^3} dl \, l^{-1} \simeq c_1(t) l_0^{1-\alpha_{\rm L}}$ 
that scales as $L^3$ if $c_1(t) =L^3 \overline c_1(t_{\rm p})$. 
According to Fig.~\ref{fig:fast-size-initial-stochastic},   $l^*(t_{\rm p}) < l^*(0)$. 

Let us first focus on long length scales. From the numerator in Eq.~\ref{eq:loop-distribution} one estimates a 
crossover at a {\it dynamic length} $l^*(t)$, 
that is advected towards smaller scales as time evolves, 
\begin{eqnarray}
{l^*}(t) \approx \sqrt{(c_1(t_{\rm p})/c_2)^{1/(\alpha_{\rm L}-1)}- \gamma_{\rm v}^2 (t-t_{\rm p})}
\; ,
\end{eqnarray}
as observed in the numerical  data.
We recover
\begin{equation}
N(l,t) \simeq c_2 \ l^{-1} \qquad \mbox{at} \qquad
l \gg l^*(t)
\end{equation}
independently of time. Instead, for  $l < l^*(t)$ we  need to 
correct Eq.~(\ref{eq:loop-distribution}) to take into account  the annihilation of vortices with 
short length that implies $N_{\rm loop}(t) \simeq t^{-\zeta}$, 
see the insets in Fig.~\ref{fig:fast-size-final}, and the consequent 
reduction of the averaged length size $\langle l \rangle \simeq t^{-\zeta+1/2}$. We enforce this scaling heuristically, 
by simply multiplying Eq.~(\ref{eq:loop-distribution}) by $(\gamma_{\rm v}^2 t)^{-\zeta}$.
Proceeding in this way, and taking $t\gg t_{\rm p}$, 
\begin{equation}
(\gamma_{\rm v}^2 t)^{\zeta+\alpha_{\rm L}/2} \ N(l,t)  
\simeq 
\frac{ c_1(t_{\rm p}) \ \ l/ (\gamma_{\rm v}^2 t)^{1/2} }{\left[ 1+ l^2/(\gamma_{\rm v}^2 t) \right]^{(1+\alpha_{\rm L})/2}}
\; . 
\label{eq:Pell-corr}
\end{equation}
Finally, this regime can also be split in two 
\begin{eqnarray}
(\gamma_{\rm v}^2 t)^{\zeta+\frac{\alpha_{\rm L}}{2}} \  \frac{N(l,t)}{L^3}  
\simeq 
\left\{
\begin{array}{ll}
[l/(\gamma_{\rm v}^2 t)^{1/2}] 
& \;  l \ll \xi_{\rm d}(t) \;\;\;\;\;
\\
{[l/(\gamma_{\rm v}^2 t)^{1/2}]}^{-\alpha_{\rm L}} 
& \; l \gg \xi_{\rm d}(t) \;\;\;\;\;
\end{array}
\right.
\end{eqnarray}
For $\alpha_{\rm L} \simeq 2.17$ and $\zeta \simeq 1.1$, the exponent in the 
left-hand-side is close to $2$, the  value used in the scaling checked in Fig.~\ref{fig:short-length-scaling}, as 
well as the linear growth of $N$ as a function of $l$ for length-scales  that are shorter than $\xi\sub{d}$, that is 
highlighted with a dashed line in Fig.~\ref{fig:fast-size-power}.

This argument is crude as it treats the vortices as being independent. Still, it gives a rather accurate description of the data
in the intermediate time regime, see Fig.~\ref{fig:short-length-scaling}, for lengths that are $l\ll L^2$.

\section{Conclusions}
\label{sec:conclusions}

We presented a detailed study of the equilibrium properties and stochastic dynamic evolution 
of relativistic bosons at finite chemical potential in three dimensions.
We modelled the system with a U(1)-invariant complex field theory and its dynamics with various non-conserved
order parameter equations of Langevin type. These models have been used to describe, in various limits, 
properties of type II superconductors, magnetic materials and aspects of cosmology and are 
thus of interest to a vast variety of physicists.

Let us start by listing what we have done in this paper and later briefly discuss our results.

(i) 
We used four Langevin-like equations to study the statics and dynamics of the U(1) 
complex field theory in three dimensions.

(ii)
We characterised the geometrical and statistical properties of the vortex tangle in equilibrium 
at all temperatures, paying special attention to the influence of the microscopic dynamics and the 
two reconnection criteria.

(iii)
We analysed the out of equilibrium relaxation after an infinitely rapid quench from equilibrium above the 
thermodynamic instability to zero temperature and we analysed our results in terms 
of dynamic scaling and the geometric structure of the evolving vortex network. 

The main conclusions drawn from the analysis above are the following.

(i)
 We demonstrated that the four microscopic dynamic equations show no difference in the 
 equilibrium states reached. Indeed, we revisited the equilibrium properties of the model to 
 establish the second order phase transition and its critical properties 
with the four dynamic algorithms and we checked that they are all very accurate in
finding the correct criticality.
 
 (ii) 
We explained the influence that the reconnection criteria  can have on the 
statistical properties of the vortex network.

(iii) 
 We showed that at high temperatures the equilibrium vortex loop configurations 
 share the statistics of fully-packed loop models, with lines shorter than 
 $L^2$ behaving as Gaussian random walks, and longer lines  
 appearing with a weight proportional to $l^{-1}$. 
 
(iv) Moreover, we confirmed~\cite{Bittner,Kajantie} that the thermodynamic transition does  not 
 coincide with the threshold for line percolation. We found that this geometric threshold and 
 independently of the reconnection rule, the algebraic decay of the 
 number density of vortex lengths is characterized by the 
 exponent $\alpha\up{(M,S)}\sub{L} \simeq 2.17$ implying the fractal dimension
$D\sub{L}\up{(M,S)} =d/(\alpha\sub{L}\up{(M,S)}-1) \simeq 2.56$, the same 
values for the maximal and stochastic rules.

Next we turned to the analysis of the evolution after sudden quenches.

(v)
We used the dynamic scaling hypothesis to 
extract the dynamic growing length from the analysis of the dynamic structure
factor. We found that all dynamic evolutions yield data in good agreement with 
the expected dynamic exponent $1/z\sub{d}=1/2$ apart from the over-damped
Langevin equation that obtains a smaller value   $1/z\sub{d}\simeq 0.43$
for the length and time-scales used. The fact that this equation overestimates
 $z\sub{d}$ had already been found in~\cite{Mondello} and we do not have a simple 
 explanation for it.

Our main results concern the out of equilibrium evolution of the vortex tangle. 

(vi) 
We showed that the network present in the initial state evolves towards a situation in
which  the strings present three length-scale regimes: 
loops that are shorter than the growing length 
$\Delta x \ll l \ll \xi\sub{d}(t) = t^{1/2}$,
loops that are longer than the growing length but still shorter than $L^2$,
$\xi\sub{d}(t) \ll l \ll L^2$, and  very long loops $L^2 \ll l$, 
behave very differently.
In the first length-scale regime the lines feel the microscopic dynamics, they are smooth curves in $3d$ 
space, and the length distribution satisfies dynamic
scaling with respect to the growing length $\xi\sub{d} \simeq t^{1/z\sub{d}}$.
In the intermediate length regime  
the length statistics is very close, actually indistinguishable from, the one at the geometric threshold (although the 
evolution is done at zero temperature) and the lines behave as self-seeking random walks, in the 
sense that their fractal dimension is smaller than 2. In the very long length regime the statistics is 
the one of the longest loops in fully-packed loop models.

(vii) Finally, we gave a (rough) analytic argument to derive the functional form of the number density of vortex lengths during the 
time-evolution of the system.

This work suggests a number of possible lines for future research. The shape of the individual 
filaments could be examined by computing, for example, the local curvature and torsion (see, e.g.,~\cite{Taylor14}
for this kind of analysis in random superposition of waves).
We did not give a quantitative 
estimate of the system-size dependence of the time needed to reach the regime in which vortices with 
intermediate lengths have algebraic statistics, numerically indistinguishable from the ones at the percolating threshold 
(as did in~\cite{Blanchard14} 
for the $2d$ Ising model or in~\cite{Tartaglia15} for the voter model where the critical percolation state is reached). 
This possibly diverging time-scale 
would give rise to a new length-scale to take into account in corrections of dynamic scaling, as 
applied to the description of correlation functions~\cite{Blanchard14}.
In a separate publication we will present the analysis of the dynamics after
finite-rate quenches~\cite{Antunes}. We will follow the analysis in~\cite{Biroli,Jelic} to 
characterise the number of topological defects and their statistical 
properties out of equilibrium.
Quenched randomness is known to modify the relaxation dynamics of single (directed) elastic lines~\cite{Kolton0,Iguain}
and ensembles of such lines in interaction~\cite{Kolton,Pleimling}.
The effect of quenched disorder on the dynamics of domain walls in $2d$ coarsening systems~\cite{Sicilia08,Insalata16}
has interesting universal properties with respect to the clean limit. An investigation of the effect of random fields and energies
in models with loops is also an interesting line of research. The effect of external potentials, as the ones used to
Bose-Einstein condensates, should also be a relevant case of study~\cite{Jackson09,Cockburn10}.
Finally, we think that these results are prompt for experimental observation~\cite{Walmsley08,Bradley06,Paoletti}.

\appendix

\section{Fokker-Planck}
\label{app:FP}

Here, we prove that the (uncommon) under-damped Langevin equation~\eqref{eq:under-damped-Langevin-general} takes the system to 
the equilibrium ensemble average in Eq.~\eqref{eq:statistical-average-under-damped}.
We consider the general real functional $f(\psi, \psi^\ast, \phi, \phi^\ast)$.
The Ito's lemma gives
\begin{align}
df = \frac{\delta f}{\delta \psi} d\psi + \frac{\delta f}{\delta \psi^\ast} d\psi^\ast + \frac{\delta f}{\delta \phi} d\phi + \frac{\delta f}{\delta \phi^\ast} d\phi^\ast + 2 \gamma\sub{L} T \frac{\delta^2 f}{\delta \phi \delta \phi^\ast} dt.
\end{align}
Introducing the probability density functional $P(\psi,\psi^\ast,\phi,\phi^\ast,t)$ we obtain
\begin{align}
\begin{split}
&\quad \frac{\partial}{\partial t} \int D\psi\: D\psi^\ast\: D\phi\: D\phi^\ast\: f P \\
&
\qquad
= \int D\psi\: D\psi^\ast\: D\phi\: D\phi^\ast\: 
\bigg\{ c^2 \frac{\delta f}{\delta \psi} (\phi + i \mu \psi) + c^2 \frac{\delta f}{\delta \psi^\ast} (\phi^\ast - i \mu \psi^\ast) 
\nonumber
\\
&\phantom{=}
\qquad\qquad
- \frac{\delta f}{\delta \phi} \bigg[ \frac{\delta H}{\delta \psi^\ast} + \gamma\sub{L} c^2 (\phi + i \mu \psi) \bigg] - 
\frac{\delta f}{\delta \phi^\ast} \bigg[ \frac{\delta H}{\delta \psi} + \gamma\sub{L} c^2 (\phi^\ast - i \mu \psi^\ast) \bigg] 
+ 2 \gamma\sub{L} T \frac{\delta^2 f}{\delta \phi \delta \phi^\ast} \bigg\} P 
\\
&
\qquad
= \int D\psi\: D\psi^\ast\: D\phi\: D\phi^\ast\: f \bigg\{ - c^2 (\phi + i \mu \psi) \frac{\delta}{\delta \psi} - c^2 (\phi^\ast - i \mu \psi^\ast) \frac{\delta}{\delta \psi^\ast} + 2 \gamma\sub{L} c^2 \\
&\phantom{=}
\qquad\qquad
+ \bigg[ \frac{\delta H}{\delta \psi^\ast} + \gamma\sub{L} c^2 (\phi + i \mu \psi) \bigg] \frac{\delta}{\delta \phi} 
+ \bigg[ \frac{\delta H}{\delta \psi} + \gamma\sub{L} c^2 (\phi^\ast - i \mu \psi^\ast) \bigg] 
\frac{\delta}{\delta \phi^\ast} + 2 \gamma\sub{L} T \frac{\delta^2}{\delta \phi \delta \phi^\ast} \bigg\} P.
\end{split}
\end{align}
Imposing that this relation holds for arbitrary $f$, we obtain the Fokker-Planck equation
\begin{align}
\begin{split}
\frac{\partial P}{\partial t} &= \bigg[ - c^2 (\phi + i \mu \psi) \frac{\delta}{\delta \psi} - c^2 (\phi^\ast - i \mu \psi^\ast) \frac{\delta}{\delta \psi^\ast} + 2 \gamma\sub{L} c^2 \\
&
+ \bigg\{ \frac{\delta H}{\delta \psi^\ast} + \gamma\sub{L} c^2 (\phi + i \mu \psi) \bigg\} \frac{\delta}{\delta \phi} + \bigg\{ \frac{\delta H}{\delta \psi} + \gamma\sub{L} c^2 (\phi^\ast - i \mu \psi^\ast) \bigg\} \frac{\delta}{\delta \phi^\ast} + 2 \gamma\sub{L} T \frac{\delta^2}{\delta \phi \delta \phi^\ast} \bigg] P,
\end{split}
\nonumber
\end{align}
with the steady solution $P \propto e^{- H / T}$. 

\section{Dependence on the discretisation mesh}
\label{app:mesh}

Under the scale transformation $\Vec{x} \to \lambda \Vec{x}$ 
the energy functional \eqref{eq:statistical-average} changes as
\begin{align*}
\int d^dx\: \bigg\{ |\nabla \psi|^2 - g \rho |\psi|^2 + \frac{g}{2} |\psi|^4 \bigg\}
& \;\; \stackrel{d \Vec{x} \to \lambda d\Vec{x} }{\longrightarrow} \;\;
\lambda^{d-2} \int d^dx\: \bigg\{ |\nabla \psi|^2 - \lambda^2 g \rho |\psi|^2 + \frac{\lambda^2 g}{2} |\psi|^4 \bigg\}
\end{align*}
($\psi(\Vec x) \to \psi(\lambda\Vec x)$).
Therefore, we obtain the same statistical properties for a model with space rescaled as $\Vec{x} \to \lambda \Vec{x}$, 
and parameters transformed as
$g \to g/\lambda^2$, and $T \to \lambda^{2-d} T $.

In the main text we called $\Delta x$ the space discretization mesh. The equilibrium correlation length at $T=0$
in the mean-field approximation is $\xi = (g\rho)^{-1/2}$. In the limit in which the ratio between these two parameters squared,
$\sigma=\Delta x^2/\xi^2 = (\Delta x)^2 g\rho$, approaches infinity, the continuum model approaches the $3d$ XY  model in which the 
modulus of the field is fixed to $\rho$~\cite{Bittner}. The effective temperature felt by the model is $\Delta x^{2-d} T$. This model 
was simulated in~\cite{Kajantie} where it was found that the vortex density $\rho\sub{vortex}$ is an increasing function of temperature $T$ 
at fixed lattice spacing $\Delta x$. Therefore, $\rho\sub{vortex}$ should also increase for finer spatial resolution at fixed temperature.
We expect the same effect for the field theory at finite $\sigma$.

\section{Averaged vortex density in the infinite temperature limit}
\label{sec:inf-temp-rho}

Here, we consider the averaged vortex density $\rho\sub{vortex}$ in the limit of infinite temperature $T \to \infty$.
The flux $v_P$ across the square  plaquette $P$, with vertices at the points $A, \ B, \ C, \ D$, 
is
\begin{align}
\begin{split}
v_P &= \ 
\frac{1}{2 \pi} \bigg[ \Im \log \bigg( \frac{\psi_{B}}{\psi_{A}} \bigg) + \Im\log  \bigg( \frac{\psi_{C}}{\psi_{B}}\bigg) + \Im \log \bigg( \frac{\psi_{D}}{\psi_{C}} \bigg) + \Im \log \bigg( \frac{\psi_{A}}{\psi_{D}} \bigg) \bigg] \\
&\equiv \ \frac{1}{2 \pi} ( \theta_{AB} + \theta_{BC} + \theta_{CD} + \theta_{DA} ), \label{eq:flux-P}
\end{split}
\end{align}
with the complex field $\psi_{X} \equiv |\psi_{X}| e^{i \theta_{X}}$ at the positions $X= A$, $B$, $C$, and $D$.
$\theta_{XY}$ is the  phase differences 
$\theta_{XY} \equiv \theta_{Y} - \theta_{X} + F_{XY}= \Im \log(\psi_Y / \psi_X)$ 
 of the complex field $\psi$ at the positions $X$ and $Y$.
The phases $\theta_X$ and $\theta_Y$ are defined in the range $(- \pi, \pi]$.
The function $F_{XY}$ has the same form as $F_{AB}$ in Eq. \eqref{eq:phase-difference} and the phase difference $\theta_{XY}$ is also defined in the range $(-\pi,\pi]$.
In the limit of infinite temperature $T \to \infty$, the phases $- \pi < \theta_{X,Y} \leq \pi$ take uniformly 
distributed random values between $- \pi$ and $\pi$, i.e., $P_{\theta_{X}}(\theta_{X}) = P_{\theta_{Y}}(\theta_{Y}) = 1 / (2 \pi)$,
independently of the positions $X$ and $Y$,
where $P_{\theta_{X}}(\theta_{X})$ ($P_{\theta_{Y}}(\theta_{Y})$) is the probability density for $\theta_{X}$ ($\theta_{Y}$).
The probability density $P_{\theta_{XY}}(\theta_{XY})$ for the phase difference $\theta_{XY}$ becomes
\begin{align}
\begin{split}
P_{\theta_{XY}}(\theta_{XY})
&= \int_{- \pi}^\pi d\theta_{X}\: P_{\theta_{X}}(\theta_{X})
\int_{\mathrm{max}[- \pi + \theta_{X}, - \pi]}^{\mathrm{min}[\pi + \theta_{X}, \pi]} d\theta_{Y}\: P_{\theta_{Y}}(\theta_{Y}) \delta(\theta_{XY} + \theta_{X} - \theta_{Y}) \\
&\quad + \int_0^\pi d\theta_{X}\: P_{\theta_{X}}(\theta_{X})
\int_{\mathrm{max}[- 2 \pi + \theta_{X}, - \pi]}^{\mathrm{min}[- \pi + \theta_{X}, \pi]} d\theta_{Y}\: P_{\theta_{Y}}(\theta_{Y}) \delta(\theta_{XY} - 2 \pi + \theta_{X} - \theta_{Y}) \\
&\quad + \int_{- \pi}^0 d\theta_{X}\: P_{\theta_{X}}(\theta_{X})
\int_{\mathrm{max}[\pi + \theta_{X}, - \pi]}^{\mathrm{min}[2 \pi + \theta_{X}, \pi]} d\theta_{Y}\: P_{\theta_{Y}}(\theta_{Y}) \delta(\theta_{XY} + 2 \pi + \theta_{X} - \theta_{Y}) \\
%
%
%
&= \frac{1}{4 \pi^2} \Bigg\{ \int_{- \pi}^0 d\theta_{X}\:
\int_{- \pi + \theta_{X}}^{\pi + \theta_{X}} d\theta_{Y}\: \delta(\theta_{XY} + \theta_{X} - \theta_{Y}) \\
&\qquad\quad + \int_0^\pi d\theta_{X}\:
\int_{- \pi + \theta_{X}}^{\pi + \theta_{X}} d\theta_{Y}\: \delta(\theta_{XY} + \theta_{X} - \theta_{Y}) \Bigg\} \\
%
& = \frac{1}{2 \pi}.
\end{split}
\end{align}
As a result, the phase differences $\theta_{XY}$ also take uniformly distributed random values between $- \pi$ and $\pi$ independently 
of the positions $X$ and $Y$.
(Note that $P_{\theta_{XY}}(\theta_{XY}) = (2 \pi - |\theta_{XY}|) / (4 \pi^2)$ when the range of $\theta_{XY}$ is not $- \pi < \theta_{XY} \leq \pi$ but $- 2 \pi < \theta_{XY} \leq 2 \pi$ with $F_{XY} = 0$ for arbitrary $\theta_{X}$ and $\theta_{Y}$.)

We now consider the flux $v_P$ in Eq. \eqref{eq:flux-P}.
Since $\theta_{DA}$ takes the form in Eq. \eqref{eq:flux-condition}, the condition $v_P = 0$, i.e., that no vortex pierces the plaquette, is 
$- \pi \leq \theta_{AB} + \theta_{BC} + \theta_{CD} < \pi$, and it occurs  with probability
\begin{align}
& P(v_P=0) =
\int_{-\pi}^\pi d\theta_{AB}\: P_{\theta_{AB}}(\theta_{AB}) \int_{-\pi}^\pi d\theta_{BC}\: 
P_{\theta_{BC}}(\theta_{BC}) \int_{\mathrm{max}[-\pi-(\theta_{AB}+\theta_{BC}),-\pi]}^{\mathrm{min}[\pi-(\theta_{AB}+\theta_{BC}),\pi]} d\theta_{CD}\: P_{\theta_{CD}}(\theta_{CD}) 
\nonumber\\
& \qquad\quad\quad\;\;
= \frac{1}{8 \pi^3} \Bigg\{ \int_{-\pi}^\pi d\theta_{AB}\: \int_{-\pi}^{-\theta_{AB}} d\theta_{BC}\: \int_{-\pi-(\theta_{AB}+\theta_{BC})}^{\pi} d\theta_{CD} 
\nonumber\\
&\qquad\qquad\qquad\qquad 
+ \int_{-\pi}^\pi d\theta_{AB}\: \int_{-\theta_{AB}}^{\pi} d\theta_{BC}\: \int_{-\pi}^{\pi-(\theta_{AB}+\theta_{BC})} d\theta_{CD} \Bigg\} 
\nonumber\\
&
\qquad\quad\quad\;\;
= \frac{2}{3} \; .
\end{align}
The averaged vortex density $\rho\sub{vortex}$ equals the probability that a vortex pierces a plaquette,  
and $\rho\sub{vortex} = 1- P(v_P=0) = 1/3$, in the limit of the infinite temperature.

\acknowledgements
We thank I. Carusotto,  J. T. Chalker, P. Comaron, F. Larcher, M. Picco, N. P. Proukakis and H. Takeuchi
for very useful discussions.
This research was  supported in part by the National Science Foundation under 
Grant No. PHY11-25915 and by KAKENHI (22740219, 22340114, and 22103005), 
Global COE Program ``the Physical Sciences Frontier", the Photon Frontier Network Program, MEXT, Japan, 
and the IRSES European Project ``SoftActive". LFC is a member of the Institut Universitaire de France.

\end{document}